\documentclass{article}

\usepackage{arxiv}

\usepackage[utf8]{inputenc} 
\usepackage[T1]{fontenc}    
\usepackage{hyperref}       
\usepackage{url}            
\usepackage{booktabs}       
\usepackage{amsfonts}       
\usepackage{nicefrac}       
\usepackage{microtype}      
\usepackage{lipsum}		
\usepackage{subfiles}

\usepackage{amsmath, mathrsfs}
\usepackage{mathtools}
\usepackage{nccmath}
\usepackage{caption}
\usepackage{subcaption}
\usepackage{tabularx}

\usepackage{tikz}
\usetikzlibrary{positioning}
\usepackage{array}
\usepackage{notoccite}
\usetikzlibrary{shapes.geometric, arrows}
\usepackage[numbers]{natbib}
\graphicspath{ {./graphic/} }

\title{Non-Intrusive Parametric Model Order Reduction with Error Correction Modeling for
	Changing Well Locations Using a Machine Learning Framework}

\author{
  Hardikkumar Zalavadia \\
  Harold Vance Department of Petroleum Engineering\\
  Texas A\&M University\\
  College Station, TX 77843-3116 \\
  \texttt{hzalavad@tamu.edu} \\
   \And
 Eduardo Gildin \\
  Harold Vance Department of Petroleum Engineering\\
  Texas A\&M University\\
  College Station, TX 77843-3116 \\
  \texttt{egildin@tamu.edu} \\
}

\date{}

\renewcommand{\headeright}{}
\renewcommand{\undertitle}{}

\begin{document}
\maketitle

\begin{abstract}
The objective of this paper is to develop a non-intrusive Parametric Model Order Reduction (PMOR) methodology for the problem of changing well locations in an oil field, that can eventually be used for well placement optimization to gain significant computational savings. In the past for reservoir development applications, majority of focus was laid on development of MOR for changing controls of the wells for well control optimization. In this work, we propose a proper orthogonal decomposition (POD) based PMOR strategy that is non-intrusive to the simulator source code, as opposed to the convention of using POD as a simulator intrusive procedure, and hence extends its applicability to any commercial simulator. The non-intrusiveness of the proposed technique stems from formulating a novel Machine Learning (ML) based framework used with POD. The features of ML model (Random Forest used here) are designed such that they take into consideration the temporal evolution of the state solutions and thereby avoiding simulator access for time dependency of the solutions. The proposed PMOR method is global in the sense that a single reduced order model can be used for all the well locations of interest in the reservoir. We address a major challenge of explicit representation of the well location change as a parameter by introducing geometry-based features and flow diagnostics inspired physics-based features. An error correction model based on reduced model solutions is formulated later to correct for discrepancies in the state solutions at well gridblocks.

The proposed methodology is applied to a homogeneous reservoir first, to analyze the and
validate the proposed idea. It was observed that the global PMOR could predict the overall
trend in the pressure and saturation solutions at the well blocks but some bias was observed
that resulted in discrepancies in prediction of quantities of interest (QoI) like oil production
rates and water cut. Thus, the error correction model proposed using Artificial Neural Networks
(ANN) that considers the physics based reduced model solutions as features, proved to reduce
the error in QoI significantly. This workflow is also applied to a heterogeneous channelized
reservoir using a section of SPE10 model that showed promising results in terms of accuracies.
Speed-ups of about 50x-100x were observed for different cases considered when running the
test scenarios. The proposed workflow for reduced order modeling is “non-intrusive” and hence can increase its applicability to any simulator used. Also, the method is formulated such that all the
simulation time steps are independent and hence can make use of parallel resources very
efficiently and also avoid stability issues that can result from error accumulation over
timesteps.

\end{abstract}


\section{Introduction} \label{sec:1.1}

Numerical simulations involve solving a large-scale system of equations derived from the Partial Differential Equations (PDEs) of complex engineering models. This is the case with reservoir simulations that are used extensively in Closed Loop Reservoir Management (CLRM) \cite{Jansen2009} and Closed Loop Field Development (CLFD) \cite{Shirangi2015} workflows. CLRM which is based on optimal continuous operation of existing wells has been an area of significant research. The workflow of CLRM involves well control optimization based on the current geological knowledge, using these settings to collect data for some time period and finally performing history matching to integrate this data to update the geological models for consistency with observed data. CLFD which is an extension to CLRM, involves optimization of well number, type, location, and controls based on the current geological knowledge and then, sequentially drilling new wells and collecting data, and a history matching step from the available data to update geological information; repeated until the optimal number of wells are drilled. These workflows, repeated during the development of reservoir, can help make better reservoir management decisions than the heuristic approaches.

The optimization and history matching steps during the development cycle require rapid and repeated fine scale reservoir simulation runs, which correspond to solving a discretized parametric PDE which is time dependent and highly nonlinear in nature. Accurate high fidelity computational models can therefore incur substantial computational costs. To avoid this bottle neck of costly evaluations, researchers have developed various surrogate modeling techniques that are approximations of high-fidelity models, which aim to provide large computational savings while preserving the accuracy. One such technique is Reduced Order Modeling (ROM), that include intrusive or non-intrusive projection-based methods. They do not approximate the physics and rely on reducing the dimensionality of the state-space of fine scale models \cite{Antoulas2005, Hinze2005}, and thus are usually computationally cheaper than coarsened models \cite{King1998, March2012} and more robust than the data-driven models \cite{Kennedy2001, Knill1999}. 

ROM approaches have been applied to reservoir simulation mainly within the context of well control optimization problems with fixed well configurations and have reached a good level of maturity to achieve significant computational saving. Initial work on ROM for the well-control optimization problem started with the application of method called Proper Orthogonal Decomposition (POD) \cite{Doren2006, Cardoso2008}. However, there is not much speed-up reported by using this POD strategy since it only targets the linear solver. Full Jacobian and residual matrices are still computed during each iteration in simulator. In order to overcome this limitation of POD, other techniques were developed for fast computation of non-linear terms in the reservoir simulator. One such technique developed was Trajectory Piecewise Linear (TPWL) applied in conjunction with POD for production optimization \cite{Cardoso2009}. This is accomplished by linearizing the states around previously saved states using Taylor series expansions. Computational speed-ups of the $\mathbb{O}(10^2-10^3)$ in the online testing stage were shown in \cite{He2011, Cardoso2009}. Later, TPWL was extended to include the quadratic terms for more accurate representation of non-linear functions and thus called Trajectory Piecewise Quadratic (TPWQ) method \cite{Trehan2016a}. This method proved to be more accurate as compared to TPWL but the overhead time increases significantly because it involves third-order matrix-tensor products. Another such method called Discrete Empirical Interpolation Method (DEIM) \cite{Chaturantabut2010} was also introduced applied in combination with POD for changing well control optimization problem \cite{Gildin2013, Ghasemi2015, Sorek2017}, which can provide higher accuracies as it aims at reducing the dimensions of discretized parametric PDEs by computing the non-linear terms at discrete locations in the spatial domain and then interpolating them to the rest of the locations using projection based interpolation. However, less speed-ups have been reported for POD-DEIM as compared to POD-TPWL as a result of linearization performed in the latter and it is also more invasive with respect to the simulator as compared to POD-TPWL. Other methods were later implemented for this application of changing well controls, such as, POD - Trajectory based DEIM (TDEIM) \cite{Tan2017} which is a combination of TPWL and DEIM techniques to utilize their benefits and avoid their shortcomings;  bilinear approximation and quadratic bilinear formulation \cite{Ghasemi2014, Ghasemi2016} and Gauss-Newton with Approximated Tensors (GNAT) method \cite{Jiang2019}. Recently, non-intrusive methods have been developed for well control optimization problem, namely Dynamic Mode Decomposition with control (DMDc) and input-output DMD (ioDMD) in \cite{Zalavadia2019}.

Well location optimization in the context of surrogate or low complexity models have been developed in \cite{Centilmen1999, Taware2012}. However, until now, for projection based reduced order models, the focus mainly has been developing them for well control optimization problem to the best of our knowledge. But these ROM workflows were not meant for integrated schemes like Closed Loop Field Development which involve changing reservoir or well configurations. This challenge is the main objective of this work. For complex non-linear problems with non-affine parameter dependence, as mentioned before, direct (simulator invasive) use of POD do not provide major computational advantage. This is because of the cost to compute projection coefficients of the non-linear functions that depend on the dimension of high fidelity model. As the dimension of a system increases, the computational advantage of using pure POD method decreases. Other alternative methods were also developed that still uses POD implementation, but the coefficients are not obtained by projection process, rather by interpolation over the parameter domain of reduced order basis \cite{Casenave2015}. Since ROBs belong to non-linear matrix manifolds, standard interpolation techniques cannot be applied as they would not preserve the characteristics of the ROBs, unless a large number of samples are used \cite{Barthelmann2000, Amsallem2011}. To overcome this issue, researchers have developed ROB method using POD and then employed Machine Learning techniques to interpolate the basis coefficients. \cite{Hesthaven2018} developed POD-NN where he used Neural Networks (NN) for high dimensional coefficient prediction. \cite{Swischuk2018} compared performance of different ML techniques to predict the basis coefficients for an aerodynamic and a structural problem. It considered the use of particular solutions in the POD expansion as a way to embed physical constraints that must be preserved during coefficient prediction. These methods however, did not consider the time-dependent problems. Such a non-intrusive ROM scheme have been applied here for changing well locations which is highly non-linear in nature and also involves time dependency. In \cite{Zalavadia2018}, a local Parametric Model Order Reduction (PMOR) technique based on error map prediction was developed to choose an appropriate basis from a dictionary of pre-existing ROBs for a new well location. However, this method is still limited to use for the cases when there is no water cut observed at the producers and for less permeability contrast reservoirs. In the proposed work here, we seek to develop a method that is not limited by these restrictions using a non-intrusive global PMOR technique.

We begin with a section introducing governing reservoir simulation equations followed by concepts leading to projection based parametric ROM. We then show motivation behind using non-intrusive global PMOR technique. Here, we discuss about parameterizing the states of the reservoir using POD-based model order reduction followed by a way of addressing the physical constraints in the POD method. In the next section, we formulate the non-intrusive PMOR problem using machine learning methods with a brief discussion on the ML methods and feature selection techniques used. This formulation is then validated using some case studies for prediction of pressure and saturation states at new well locations. In the next section, errors in the well block states are modeled using ML algorithms to correct for the errors in the output quantities of interest (QoI) like oil, water production rates and water-cut followed by some case studies explaining the validity of proposed method. 

\section{Reservoir Simulation} \label{sec:1.2}

The governing equations for a two phase isothermal immiscible incompressible oil-water system is obtained by combining the mass conservation law and the Darcy's law for two phases. This governing equation considering the gravity effects and constant porosity $\varphi$ is a PDE of the form: 

\begin{subequations}
	\label{eq:2}
	\begin{align}
	\varphi \frac{\partial S_l}{\partial{t}}= \nabla\cdot(\textbf{v}_l)-q_l\label{eq:2_1} \\
	\textbf{v}_l = \lambda_l K(\nabla p_l-\rho_l g\nabla z)\label{eq:2_2}
	\end{align}
\end{subequations}

where, $l$ indicates liquid, oil (o) or water (w), $\varphi$ is the porosity of the reservoir, $\rho_l$ indicates density of the fluid per unit volume, and $\widetilde{m_l}$ (mass flow rate) $=\rho_l q_l$ denotes the external sources and sinks, $q_l$ being the volumetric flow rate. $\textbf{v}_l$ here represents the Darcy velocity of liquid $(l)$. $\lambda_l = k_{rl}(S_l)/\mu_l$, where $\mu_l$, $k_{rl}$ are the viscosity of fluids and relative permeability of fluids, respectively and K is the absolute permeability. $p_l$  and $S_l$ are the pressure and saturation of each phase respectively. $g$ is the gravity acceleration constant and $z$ accounts for vertical coordinates. Equations (\ref{eq:2_1}) and (\ref{eq:2_2}) for oil and water, along with the constraints, saturation equation ($S_o+S_w=1$) and capillary equation $(p_{cow}=p_o-p_w)$ completes the oil-water model. In these equations, $p_o$ and $S_w$ are considered to be the primary unknowns, and once they are solved, the rest of the unknowns $p_w$ and $S_o$ are computed easily from the constraint equations. 

In order to solve the PDE, we perform spatial discretization of this equation that result in the following parametric system of ODE in time to solve for the state variables $p_o$ and $S_w$ represented by $\textbf{x}$:
\begin{equation}
\begin{aligned}
-\textbf{D}(\textbf{x},\zeta) \dot{\textbf{x}}(t,\zeta) + \textbf{T}(\textbf{x},\zeta)\textbf{x}(t,\zeta) + \textbf{G}(\textbf{x},\zeta) + \textbf{Q}(\textbf{x},t,\zeta) = \textbf{R}(\textbf{x},\zeta) \label{eq:3}
\end{aligned}
\end{equation}
$\textbf{x}(t,\zeta) = [P_o,S_w] \in R^{N_d}$, is a state vector with $N_d$ degrees of freedom and $\zeta \in \mathbb{P} \subset R^{N_\zeta}$ is a vector of parameters. $\dot{\textbf{x}}(t,\zeta)$ is the derivative of the states with respect to time, $\textbf{D}$ is the accumulation matrix, $\textbf{T}$ is the transmissibility matrix, $\textbf{G}$ is the gravity vector, $\textbf{Q}$ contains the (volumetric) sources and sinks terms $(q_l)$ and $\textbf{R}$ is the residual vector. We satisfy $\textbf{R}$ = 0 upon convergence to the solution for each time step. We consider the fully-implicit procedure to solve equation (\ref{eq:3}). The source/sink volumetric term correspond to production/injection from a well and is written as the Peaceman equation:
\begin{equation}
\begin{aligned}
q_l=WI_l \thinspace \lambda_l(S_l) \thinspace(p_o-p_{wf})\label{eq:4}
\end{aligned}
\end{equation}
Here, $WI_l$ is the well index associated with each phase, $p_o$  is the well-block pressure of the oil phase and $p_{wf}$ is the bottom hole flowing pressure. Thus, $\textbf{Q}(\textbf{x},\zeta)$ is a vector with $q_l$ value at the indices corresponding to the well grid blocks and 0 elsewhere.

The parameter $\zeta$ of the system depends on the problem under consideration. For well control optimization, $\zeta=p_{wf}\in R^{N_t}$ for each well, where $N_t$ is the number of simulation time steps. For well placement optimization, $\zeta$ represents the indices in sparse vector $\textbf{Q}$, corresponding to spatial location of the well to be optimized in the source/sink vector. Since in this paper we are interested in the changing well location problem, we consider the latter. 
\begin{equation}
\label{eq:5}
\zeta=\begin{dcases}
p_{wf} \enspace or \enspace q_{inj},& \text{for well control optimization} \\
indices \enspace in \enspace vector \enspace \textbf{Q},              & \text{for well location optimization}
\end{dcases}
\end{equation}

Using the fully implicit method, for a given system parameter, at each time step, this non-linear system of equations (\ref{eq:3}), is solved using Newton’s method:
\begin{ceqn}
	\begin{align}
	\textbf{J}(\zeta)^{n+1}\thinspace \boldsymbol{\delta}(\zeta)^{n+1}=-\textbf{R}(\zeta)^{n+1}\label{eq:6}
	\\
	\textbf{x}(\zeta)^{n+1}=\textbf{x}(\zeta)^n+\boldsymbol{\delta}(\zeta)^{n+1}\label{eq:7}
	\end{align}
\end{ceqn}
Here, $n$ and $n+1$ represent the previous and current time levels respectively and $\textbf{J}^{n+1}=\frac{\partial \textbf{R}^{n+1}}{\partial \textbf{x}^{n+1}}$ is the Jacobian matrix. We satisfy $\textbf{R}^{n+1} = 0$ upon convergence to the solution for each time step, thus requires multiple newton iterations (see (\cite{Ertekin2001}) for detailed derivation). The size of the Jacobian matrix is $N_d$ x $N_d$ and that of the residual is $N_d$ x $1$. The degrees of freedom in a reservoir simulation usually ranges from hundreds to millions. Thus, the size of the system matrices (Jacobian and Residual) significantly increases the time to solve the linear system in the Newton method. For a detailed explanation on oil-water flow simulation, one can refer to some classical literature (\cite{Aziz1979, Ertekin2001, Chen2006, Lie2018}).

\section{Projection based Model Order Reduction} \label{sec:1.3}

In this section, we introduce projection based MOR technique that projects a high dimensional reservoir simulation equation onto a subspace of much lower dimension, thus reducing both the number of equations and variables involved. In this work, projection based reduced order modeling is the method of choice for the problem of changing well location.

\subsection{Galerkin projection}
The main idea of projection based MOR is to generate a dynamical system of much lower dimension $r$ as compared to the full order dimension $N_d$, such as in equation (\ref{eq:6}), while still retaining its dominant dynamical properties. One way to achieve this is Galerkin projection that is employed here. We begin with a state space representation of the reservoir simulation equation for two phase flow to explain Galerkin projection. Referring to equation (\ref{eq:3}), at convergence, neglecting gravity, it can be re-written as:
\begin{equation}
\begin{aligned}
\dot{\textbf{x}}(t,\zeta) = \textbf{D}(\textbf{x},\zeta)^{-1}\textbf{T}(\textbf{x},\zeta)\textbf{x}(t,\zeta) - \textbf{D}(\textbf{x},\zeta)^{-1}\textbf{Q}(\textbf{x},t,\zeta)\label{eq:13}
\end{aligned}
\end{equation}
which can be written as a state space form:
\begin{equation}
\begin{aligned}
\dot{\textbf{x}}(t,\zeta) = \textbf{A}(\textbf{x},\zeta)\textbf{x}(t,\zeta) +\textbf{B}(\textbf{x},\zeta)\textbf{u}(\textbf{x},t,\zeta)\label{eq:14}
\end{aligned}
\end{equation}
with initial condition $\textbf{x}(0,\zeta) = \textbf{x}^0(\zeta)$. The states here represent pressure and saturation for two phase flow i.e., $\textbf{x}(t,\zeta) = [P_o,S_w]$. Here $\textbf{A} = \textbf{D}(\textbf{x},\zeta)^{-1}\textbf{T}(\textbf{x},\zeta)$  and $\textbf{B} = -\textbf{D}(\textbf{x},\zeta)^{-1}$.

The first step in Galerkin projection is to define a trial basis $\Phi(\zeta) \in R^{N_d \times r}$, referred to as ROB (Reduced Order Basis), which is full rank and describes the subspace $\mathcal{S}_\Phi(\zeta)$. The state vector $\textbf{x}$ can then be decomposed as sum of two orthogonal components, one in $\Phi(\zeta)$ and the other in $\Phi^\perp(\zeta)$ which can be written as:
\begin{equation}
\begin{aligned}
{\textbf{x}}(t,\zeta) = \Phi(\zeta) \thinspace \textbf{x}_r(t, \zeta) +  \Phi^\perp(\zeta) \thinspace \tilde{\textbf{x}}_r(t, \zeta) \label{eq:15}
\end{aligned}
\end{equation}
where, $\textbf{x}_r(t, \zeta) \in R^r$ and $\tilde{\textbf{x}}_r(t, \zeta) \in R^{N_d-r}$. For a reduced representation of the state space, we neglect components of the state in $\Phi^\perp(\zeta)$. Thus, 
\begin{equation}
\begin{aligned}
{\textbf{x}}(t,\zeta) \approx \Phi(\zeta) \thinspace \textbf{x}_r(t, \zeta) \label{eq:16}
\end{aligned}
\end{equation}
Thus, $\textbf{x}_r \in R^r$ represent the components of the state vector $\textbf{x} \in R^{N_d}$ in the subspace $\Phi(\zeta)$. We can now write the equation (\ref{eq:14}) as:
\begin{equation}
\begin{aligned}
\Phi(\zeta)\dot{\textbf{x}}_r(t,\zeta) = \textbf{A}(\textbf{x},\zeta)\Phi(\zeta)\textbf{x}_r(t,\zeta) +\textbf{B}(\textbf{x},\zeta)\textbf{u}(\textbf{x},t,\zeta)\label{eq:17}
\end{aligned}
\end{equation}

The next step is to define the test basis, which for Galerkin projection is the same as the trial basis i.e., $\Phi(\zeta)$. This test basis is left multiplied in equation (\ref{eq:17}) to obtain lower order system equation.
\begin{equation}
\begin{aligned}
\Phi(\zeta)^T\Phi(\zeta)\dot{\textbf{x}}_r(t,\zeta) = \Phi(\zeta)^T\textbf{A}(\textbf{x},\zeta)\Phi(\zeta)\textbf{x}_r(t,\zeta) +\Phi(\zeta)^T\textbf{B}(\textbf{x},\zeta)\textbf{u}(\textbf{x},t,\zeta)\label{eq:18}
\end{aligned}
\end{equation}
The columns of $\Phi(\zeta)$ are orthonormal i.e., $\Phi(\zeta)^T\Phi(\zeta)=I_r$ and the projector is $\Pi_{\Phi(\zeta), \Phi(\zeta)} = \Phi(\zeta)\Phi(\zeta)^T$. Let $\hat{\textbf{x}}$ be the approximated full state after projection. The model reduction error is then,
\begin{equation}
\begin{aligned}
\epsilon_{ROM}(t,\zeta) &= \textbf{x}(t,\zeta) - \hat{\textbf{x}}(t,\zeta)\\ &= \textbf{x}(t,\zeta) - \Pi_{\Phi(\zeta), \Phi(\zeta)}\textbf{x}(t,\zeta) +\Pi_{\Phi(\zeta), \Phi(\zeta)}\textbf{x}(t,\zeta) - \hat{\textbf{x}}(t,\zeta)\\ &= (I_n - \Pi_{\Phi(\zeta), \Phi(\zeta)})\textbf{x}(t,\zeta) + \Phi(\zeta)(\Phi(\zeta)^T \textbf{x}(t,\zeta) - \textbf{x}_r(t,\zeta))\\&= \epsilon_{\Phi(\zeta)^\perp}(t,\zeta) + \epsilon_{\Phi(\zeta)}(t,\zeta)  \label{eq:19}
\end{aligned}
\end{equation}

The first term $\epsilon_{\Phi(\zeta)^\perp}(t,\zeta)$ corresponds to the projection error that result from neglecting the state projection on the orthogonal subspace and the second term $\epsilon_{\Phi(\zeta)}(t,\zeta)$ is the result of error from solving a dynamical system that is different than the original one. Since the error components are orthogonal, we can write the following equality:

\begin{equation}
\begin{aligned}
\parallel\epsilon_{ROM}(t,\zeta)\parallel_2^2 \thinspace=\thinspace \parallel \epsilon_{\Phi(\zeta)^\perp}(t,\zeta) \parallel_2^2 + \parallel \epsilon_{\Phi(\zeta)}(t,\zeta) \parallel_2^2 \label{eq:20}
\end{aligned}
\end{equation}
This is used for an a priori indication of the ROM error that is indicative of the quality of the basis $\Phi(\zeta)$. The orthogonal component of the projection error can be computed by just knowing the fine scale solution. Thus, if $\Phi(\zeta)$ is not a suitable basis for parameter $\zeta$, $\epsilon_{\Phi(\zeta)^\perp}$ will be large and the full ROM error $\epsilon_{ROM}(t,\zeta)$ will be even larger. 

Also, it can be shown that Galerkin projection can lead to unstable solutions in the case of non-symmetric system matrix $A$, leading to unphysical system. Thus, it is very important for an appropriate choice of basis $\Phi(\zeta)$ for stability of the system. One way to address this issue was proposed in \cite{Carlberg2017, He2015} using least-squares Petrov-Galerkin (LSPG) method, which computes a different test basis than the trial basis. However, the computational cost associated with LSPG is higher and hence for the current scope of work, we rely on Galerkin projection method. 

\subsection{Proper Orthogonal Decomposition}

Proper Orthogonal Decomposition (POD) constructs a basis of dimension $r$ by orthogonal transformation of the data observations such that it represents the data in certain least square optimal sense. The data can thus be represented by a linear combination of the basis vectors which are called the basis functions. POD can be implemented for infinite dimensional or finite dimensional data and does not assume the data source which can be from a linear or non-linear system.

POD is typically employed on the state solutions of the system. As a first step to generate the reduced basis, the full order system is solved for a given parameter $\zeta$, which is called the training step, to generate an ensemble of snapshots which are basically the state solutions gathered at all simulation time steps by solving equations (\ref{eq:6} and \ref{eq:7}). More details on this can be found in \cite{Gildin2013}. Snapshot matrix $\textbf{X}(\zeta)$ is defined as:
\begin{equation}
\begin{aligned}
\textbf{X}(\zeta)=[\textbf{x}^1 && \textbf{x}^2  && \textbf{x}^3 &&  ...  &&\textbf{x}^{N_t}] , && \textbf{X}(\zeta)\in R^{N_d \times N_t}\label{eq:21}
\end{aligned}
\end{equation}

where, $N_d$  is the full order dimension of states $P_o$ and $S_w$, and $N_t$ is the total number of snapshots collected over a time period. From now, we denote $P_o$ and $S_w$ as $\textbf{p}$ and $\textbf{S}$ respectively. In order to project a fine scale system to a low dimensional space, the projection basis \{$\phi_i$\}$_{i=1}^r$ is obtained by solving the minimization problem:
\begin{equation}
\begin{aligned}
\underset{\phi_i}{\text{min}} \displaystyle\sum_{j=1}^{N_t}\parallel \textbf{x}^j - \displaystyle\sum_{i=1}^{r} ({\textbf{x}^j}^T \phi_i)\phi_i\parallel\label{eq:22}
\end{aligned}
\end{equation}
The solution to this minimization problem is given by the SVD of snapshot matrix $\textbf{X}(\zeta)$ \cite{Hinze2005}, and selecting first $r$ columns of the left projection matrix. 
\begin{ceqn}
	\begin{align}
	\textbf{X}(\zeta) = \textbf{U}(\zeta)\mathbf{\Lambda}(\zeta)\textbf{V}(\zeta)^T\label{eq:23}
	\end{align}
\end{ceqn}
where, $\textbf{U}$ and $\textbf{V}$ are the left and right singular matrices respectively and $\mathbf{\Lambda}$ is a diagonal matrix with eigenvalues in decreasing order. The $r$ columns of $\textbf{U}$ are usually selected by the fraction of energy to be captured (generally more than 90\%): 
\begin{ceqn}
	\begin{align}
	E = \frac{\displaystyle\sum_{i=1}^{r} \sigma_i}{\displaystyle\sum_{i=1}^{N_d} \sigma_i}\label{eq:24}
	\end{align}
\end{ceqn}
where, $\sigma_i$  is the $i^{th}$ diagonal element of $\mathbf{\Lambda}$. Thus, for a fixed parameter, the states span the space $\Phi(\zeta)=[\textbf{U}_1(\zeta) \enspace \textbf{U}_2(\zeta)\enspace ... \enspace \textbf{U}_r(\zeta)] \in R^{N_d\times r}$ and for each state:
\begin{ceqn}
	\begin{align}
	\textbf{p}(\zeta) \approx \Phi_p(\zeta)\textbf{p}_r(\zeta), && \textbf{S}(\zeta) \approx \Phi_s(\zeta)\textbf{S}_r(\zeta)  \label{eq:25} 
	\end{align}
\end{ceqn}
Thus, for a given parameter, we may write:
\begin{ceqn}
	\begin{align}
	\begin{bmatrix}
	\textbf{p} \\
	\textbf{S} \\
	\end{bmatrix} =
	\begin{bmatrix}
	\Phi_p &\enspace 0 \\
	0 &\enspace \Phi_s\\
	\end{bmatrix} 
	\begin{bmatrix}
	\textbf{p}_r \\
	\textbf{S}_r \\
	\end{bmatrix} \label{eq:26}
	\end{align}
\end{ceqn}
which takes a short notation:
\begin{ceqn}
	\begin{align}
	\textbf{X} = \Phi \textbf{X}_r\thinspace, \enspace\enspace r \ll N_d \label{eq:27}
	\end{align}
\end{ceqn}
Here, $\Phi$ is a diagonal matrix with pressure basis and saturation basis as its diagonal elements. Using Galerkin projection, the linear system of equations in equation (\ref{eq:6}) can be written as follows:
\begin{ceqn}
	\begin{align}
	\Phi(\zeta)^{T} \textbf{J}(\zeta)^{n+1} \Phi(\zeta)\thinspace \boldsymbol{\delta}_{r}(\zeta)^{n+1} = -\Phi(\zeta)^{T} \textbf{R}(\zeta)^{n+1}  \label{eq:28} 
	\end{align}
\end{ceqn}
which leads to a system of reduced order equations:
\begin{ceqn}
	\begin{align}
	\textbf{J}_{r}(\zeta)^{n+1} \boldsymbol{\delta}_{r}(\zeta)^{n+1} = -\textbf{R}_{r}(\zeta)^{n+1}  \label{eq:29} 
	\end{align}
\end{ceqn}
and after solving the equation (\ref{eq:29}), the reduced state space variables are updated as:
\begin{ceqn}
	\begin{align}
	\textbf{x}_{r}(\zeta)^{n}+\boldsymbol{\delta}_{r}(\zeta)^{n+1} = \textbf{x}_{r}(\zeta)^{n+1}  \label{eq:30} 
	\end{align}
\end{ceqn}

Thus, $r$ being of much smaller magnitude compared to the fine scale degrees of freedom $N_d$, POD has shown to achieve significant speed-ups at the linear solver level in reservoir simulation applications by greatly reducing the dimensions of the huge Jacobian and residual matrices. However, since this method still requires us running fine scale simulations to evaluate full Jacobian and residual matrices, it poses a challenge in terms of overall computational benefit. This problem will be addressed in the next sections as we build on the  MOR strategy for changing well locations.

\subsection{Parametric Model Order Reduction concept}
Thus, as we saw above for POD, the main aim is to replace the computationally expensive solver in equation \ref{eq:6} to a reduced form of the Jacobian and Residual as shown in equation (\ref{eq:29}), where, $r \ll N_d$. To this end, we aim to construct reduced order models or reduced order basis that are an accurate approximation of the fine-scale solution in the $entire$ parameter domain i.e., for any $\zeta$ in equation (\ref{eq:3}). For clarity again, ROB is defined as the basis of lower dimension obtained by POD and the ROBs are used to construct ROMs which are lower dimensional system matrices the accuracy of which is reflected as the predicted output of the system. Thus, ROM error refers to the error in output predicted by the reduced system.

Thus, for any parameter $\zeta \in \mathbb{P}$ representing well location we require the condition that the ROM error is less than a specified threshold, which is defined mathematically as:
\begin{ceqn}
	\begin{align}
	\parallel \textbf{X}(t, \zeta) - \Phi(\zeta)\textbf{X}_r(t,\zeta)\parallel < \eta  \label{eq:31} 
	\end{align}
\end{ceqn}
Here, $\eta$ is some maximum allowable error tolerance between fine scale and ROM solution with $\Phi(\zeta)$ obtained by POD in this work. We note again that, the parameter $\zeta$ depends on the optimization problem considered. 

In the following sections, we develop MOR techniques that are developed in the context of POD concept for fast computation of simulations for changing well locations during well location optimization. 

\section{Motivation for non-intrusive global PMOR using machine learning} \label{sec:1.4}
In this section, we discuss about the motivation behind developing the non-intrusive global PMOR method using machine learning techniques. It was shown in the work \cite{Zalavadia2018} that the ROB corresponding to one well location is only a valid basis for fewer locations when compared to the ROB obtained by concatenating solutions from different well locations. This is the reason behind considering the global ROB strategy for well location problem. The global ROB can be obtained by random sampling in the parameter space or using greedy algorithms. For the former case, fine scale simulations for randomly distributed parameters are computed and using POD, reduced basis is obtained. It is highly likely that the distribution contains many unnecessary samples or neglects some important samples. Whereas, using greedy algorithms, the samples are carefully chosen using an optimality criteria and usually lead to much better samples. However, the computational expense involved with a greedy sampling procedure for certain kind of problems, especially, highly non-linear dynamical system like reservoir simulation, make it infeasible to use. Thus, for this work, we just focus on random sampling of the candidate well locations. 

Most of the PMOR methods for well control optimization that have been relies on the simulator source code. Methods like POD-TPWL \cite{Cardoso2009} are intrusive to some extent as it requires access to Jacobian and residual matrices, that is not easily available for commercial simulators and POD-DEIM \cite{Sorek2017} is highly intrusive to the source code. One of the non-intrusive methods that has been used for changing well controls is DMD \cite{Bao2017} and DMDc \cite{Zalavadia2019}. We propose a new non-intrusive PMOR technique that has been applied in \cite{Swischuk2018, Hesthaven2018} for problems that are time independent or steady state systems. We extend this technique to time dependent problems which is the case for reservoir simulation where states evolve over time. One of the other main advantages of proposing this strategy is, it is extremely fast compared to if global ROB is used within a simulator to project non-linear functions. As discussed before, the reason is, as more parameters are used for concatenating solutions for a global ROB, we expect the basis dimension to increase monotonically and hence projection of non-linear functions become slower. The proposed PMOR technique does not involve computing the non-linear functions and thus projecting them, and can make efficient use of parallel computational facilities as will be discussed in the later section (\ref{sec:1.5}).

The use of machine learning has revolutionized the field of decision making in numerous areas of applications, where it learns complex models entirely from huge data sets. However, the availability of extensive amount of data for engineering applications can be a bottleneck, as it usually involves running a large number of fine scale simulations. Another challenge is to ensure the ML models understand the physics of the system like conservation laws. Thus, blending the positives of both MOR and ML in that, MOR provides a low dimensional representation of the dynamics of fluid flow and hence represent the physics of the system and ML provides a complex mapping of the input parameters (well locations in our case) to the MOR parameters as will be discussed in the next section, is the main motivation of using this strategy. 

\subsection{Parameterizing states of reservoir using POD}
We begin by introducing the parameterization of the reservoir states using proper orthogonal decomposition. Let us consider the states $\textbf{x}(t,\zeta) \in R^{N_d}$ at time $t \in \mathbb{T}$ and $\zeta \in \mathbb{P} \subset R^{N_\zeta}$, a vector of input parameters. Each column in $\textbf{x}$ collected over time is also referred to as snapshots. $\textbf{x}$ represents the pressure and saturation states for our case, i.e. $x =[p,s]$ and $\zeta$ corresponds to the well location. In order to construct the global ROB, we collect the states snapshots from different well locations and hence can be written as:
\begin{ceqn}
	\begin{align}
	S = \{\textbf{x}(t_i, \zeta_j)|i=1,...,N_t,j = 1,...,N_\zeta\}\label{eq:5_1}
	\end{align}
\end{ceqn}
Thus, we have a total of $N_t N_\zeta$ snapshots, with each of the $N_\zeta$ parameter has $N_t$ snapshots all concatenated in a single snapshot matrix $S$. The $N_\zeta$ parameters can be selected using experimental design sampling methods like Lattice Hypercube Sampling (LHS), full factorial design etc. or randomly. Thus, POD entails obtaining a reduced order basis (ROB) by taking a singular value decomposition of the snapshot matrix $S$ and then selecting first few vector $r$ (reduced dimension) of the left singular matrix by certain energy criteria on the singular values. This ROB is represented by $\mathbf{\Phi}$ = $\{\phi_1(x), \phi_2(x), ... , \phi_r(x)\}$, where $\phi_i(x), i=1,...,r,$ are called the reduced basis functions and we seek solutions at any parameter that are a linear combination of these basis functions. We assume that these basis functions span a space $\Phi_{rb}$ called the reduced basis space:
\begin{ceqn}
	\begin{align}
	\Phi_{rb} = span\{\phi_1(x), \phi_2(x), ... , \phi_r(x)\}\label{eq:5_2}
	\end{align}
\end{ceqn}
where, $r\ll N_d$. Thus, for any $\zeta \in \mathbb{P}$, we seek a solution $x(\zeta)$:
\begin{ceqn}
	\begin{align}
	\tilde{\textbf{x}}(t,\zeta) = \sum_{i=1}^{r}c_{\textbf{x}i}(t,\zeta)\phi_{\textbf{x}i}(x)\label{eq:5_3}
	\end{align}
\end{ceqn}
Here, $\textbf{x}$ correspond to $p$ (pressure) or $s$ (saturation). Note that $\textbf{x}$ here denote the states and $x$ denote the spatial variable. Thus, we have
\begin{subequations}
	\begin{align}
	\tilde{\textbf{p}}(t,\zeta) = \sum_{i=1}^{r_p}c_{pi}(t,\zeta)\phi_{pi}(x)\label{eq:5_4_1}\\
	\tilde{\textbf{s}}(t,\zeta) = \sum_{i=1}^{r_s}c_{si}(t,\zeta)\phi_{si}(x)\label{eq:5_4_2}
	\end{align}
	\label{eq:5_4}
\end{subequations}

$c_{pi}(t,\zeta)$ and $c_{si}(t,\zeta)$ represent the basis coefficients or POD expansion coefficients for pressure and saturation respectively, and $\tilde{\textbf{p}}(t,\zeta)$ and $\tilde{\textbf{s}}(t,\zeta)$ represent the approximated pressure and saturation solutions at a parameter $\zeta$ and time $t$. Note that the magnitudes of pressure and saturation scale very differently, with pressures usually in the range of thousands of psi and saturation as a fraction between 0 and 1, we compute the global ROBs of pressure and saturation separately as denoted by $\Phi_{pi}(\textbf{x})$ and $\Phi_{si}(\textbf{x})$. Thus, the dimension of both these basis can be different as depicted by $r_p$ and $r_s$ for pressure and saturation respectively. From now on, we just use $r$ for both the states for ease of notations but it should be remembered that it is different for different states. 

These POD parameterizes the states in terms of the basis coefficients. These basis coefficients can be calculated as:
\begin{ceqn}
	\begin{align}
	\textbf{c}_{\textbf{x}}(\zeta) = \Phi_\textbf{x}^T\textbf{x}(\zeta) \in R^{r\times t}, \label{eq:5_5}
	\end{align}
\end{ceqn}
for both states. The matrix $\textbf{c}_{\textbf{x}}(\zeta)$ contains columns of basis coefficients at each timestep. Thus, $\textbf{c}_{\textbf{x}}(t, \zeta) = [c_{\textbf{x}1}(t,\zeta)\enspace, c_{\textbf{x}2}(t,\zeta)\enspace,...,c_{\textbf{x}r}(t,\zeta)]^T$.

\subsection{Addressing physical constraints in POD}

In this method, we also enforce physical constraints in the form of POD representation \cite{Swischuk2018}. This is done by writing the equation (\ref{eq:5_3}) in a different way as:
\begin{ceqn}
	\begin{align}
	\tilde{\textbf{x}}(t,\zeta) = \bar{\textbf{x}} + \sum_{i=1}^{r} c_{\textbf{x}i}(t,\zeta)\bar{\phi}_{\textbf{x}i}(x)\label{eq:5_6}
	\end{align}
\end{ceqn}

Here, $\bar{\textbf{x}}$ is called the particular solution which enforces certain characteristics of the spatial behavior of states during POD-based prediction. For our case, this particular solution is defined as the mean of all the snapshots corresponding to all the parameters in snapshot matrix. By doing so, as expected, the accuracy of state prediction increased using the formulation proposed in the next section (\ref{sec:1.5}). The procedure to calculate the global basis functions remain the same as above, except now, the POD basis $\bar{\Phi} =  \{\bar{\phi}_1(x), \bar{\phi}_2(x), ... , \bar{\phi}_r(x)\}$ is computed on the mean subtracted snapshot matrix:
\begin{ceqn}
	\begin{align}
	\tilde{S} = \{\textbf{x}(t_i, \zeta_j)-\bar{\textbf{x}}|i=1,...,N_t,j = 1,...,N_\zeta\}\label{eq:5_7}
	\end{align}
\end{ceqn}
After, the basis coefficients are obtained, the states at new parameters and a given time are computed as equations (\ref{eq:5_8_1})and (\ref{eq:5_8_2}).

\begin{subequations}
	\begin{align}
	\tilde{\textbf{p}}(t,\zeta) = \bar{\textbf{p}} + \sum_{i=1}^{r_p}c_{pi}(t,\zeta)\phi_{pi}(x)\label{eq:5_8_1}\\
	\tilde{\textbf{s}}(t,\zeta) = \bar{\textbf{s}} + \sum_{i=1}^{r_s}c_{si}(t,\zeta)\phi_{si}(x)\label{eq:5_8_2}
	\end{align}
\end{subequations}

\section{Global PMOR problem formulation} \label{sec:1.5}

\subsection{Formulation }
Once the global ROB is obtained using the method shown above, a traditional way of solving such system is solving the non-linear reservoir simulation equation online and getting the coefficients of the basis for a new parameter by projecting the non-linear Jacobian and residual functions. However, it has been shown several times in the literature that this technique shows modest improvements in the computational speedups as it requires computing the fine scale non-linear functions during each Newton iteration. Also, for well location changes, implementing the simulator intrusive POD-based online procedure caused lot of stability issues and it is very challenging to decide suitable dimensions of global basis. Changing well location as the parameter for PMOR is observed as much more challenging as compared to well control changes as even a global PMOR technique is found difficult to represent the controllability properties of a new well location. 
One way to address these issues is using non-intrusive reduced basis methods, where the basis coefficients are obtained by interpolating the ROBs or ROMs over the parameter domain. But, the reduced bases belong to non-linear, matrix manifolds and hence the standard interpolation can fail in preserving the constraints characterizing those manifolds and  requires having a large dataset \cite{Amsallem2009, Choi2015}. 
We present an alternative way to overcome this issue by multidimensional mapping of input parameters to the basis coefficients given a global ROB. We use ML techniques that have capabilities of mapping complex non-linear relationships and specially suitable for interpolation of basis coefficients where the parameters have a non-affine dependence. Similar approach was applied to steady state cases in \cite{Swischuk2018, Hesthaven2018}. We extend this technique to time dependent problems. Thus, the ML model $\mathscr{F}$ is trained to learn the relation:

\begin{equation}
\label{eq:5_8}
\begin{aligned}
\mathscr{F} : [\zeta_j,t_k] \to [c_{\textbf{x}1}(\zeta_j,t_k), c_{\textbf{x}2}(\zeta_j,t_k), ... , c_{\textbf{x}r}(\zeta_j,t_k)],  \enspace  \enspace  \enspace j = 1,..., N_\zeta \enspace and \enspace k = 1,..., N_t 
\end{aligned}
\end{equation}

\begin{equation}
\begin{aligned}
[c_{\textbf{x}i}(\zeta_j,t_k)]_{i=1}^{r} = \Phi_{\textbf{x}}^T\textbf{x}(\zeta_j, t_k), \enspace \textbf{x} = [p,s]^T \label{eq:5_9}
\end{aligned}
\end{equation}
Note that, the basis $\Phi_\textbf{x}$ is obtained from the mean subtracted snapshot matrix and the basis dimensions of pressure and saturation are different shown by $r$ here. $\zeta_j$ are the parameters representing the well location used for training which are the features of the ML model. In order to consider the temporal evolution of the coefficients, we also add time as one of the features and the coefficients as a function of time as the outputs of the ML model. 

Once this ML model is trained, we predict the POD basis coefficients for each state, pressure and saturation for a new well location, which correspond to the states in low dimensional subspace that are projected later back to the full dimensional space. For a new well location represented by parameter $\zeta^* \in \mathbb{P}$, the online procedure can be written as:

\begin{equation}
\begin{aligned}
\mathscr{F}(\zeta^*, t_k) = [c_{\textbf{x}i}(\zeta^*,t_k)]_{i=1}^{r} = \textbf{c}_\textbf{x}(\zeta^*,t_k)\label{eq:5_10}
\end{aligned}
\end{equation}

\begin{equation}
\begin{aligned}
\tilde{\textbf{x}}(\zeta^*,t_k)= \sum_{i=1}^{r} c_{\textbf{x}i}(t_k,\zeta^*)\phi_{\textbf{x}i} + \bar{\textbf{x}} \emph{} = \Phi_{\textbf{x}}c_{\textbf{x}}(\zeta^*, t_k) + \bar{\textbf{x}} \label{eq:5_11}
\end{aligned}
\end{equation}

Note that here, $\bar{\textbf{x}}$, which is the training snapshot mean, act as a physical constraint enforced on the predicted solution.

\subsection{Machine Learning and feature selection}

We are interested in understanding the relationship between the input parameters representing well locations and the basis coefficients of pressure and saturation. The inputs representing the well locations are defined in high dimensional spaces as will be shown later, and the outputs, which are the POD coefficients, also are expected to be very high dimensional based on the number of basis functions chosen that increase with increasing number of training well locations and size of the reservoir. Such high dimensional input-output relation and a highly non-linear relation that is very difficult to define explicitly calls for machine learning strategies. However, for engineering applications like the ones considered, it becomes very challenging to generate a large data set for the machine learning algorithms to capture the underlying dynamics. Thus, there is trade-off that needs to be considered between getting a good training data set and the least computational expense at building a good predictor. Thus sometimes, specially for such applications as developing MOR for changing well locations in the reservoir, it is just beneficial to use simpler ML algorithms that are faster to train and can still capture the underlying non-linearity. Thus, we consider using Random Forest (RF) regressors as our algorithm of choice, since they are much faster to train and capable of mapping complex input-output relationships while using concepts like bootstrap aggregation to avoid overfitting. Other techniques like Artificial Neural Networks (ANN) have been used for non-intrusive ROM in \cite{Hesthaven2018}. According to many applications, NN is found to be a feasible method where we have access to huge data sets and spending a lot of time in training is a viable option. We however, analyzed its performance as compared to that obtained by a fast algorithm like Random Forests (RF) and it was found very slow to train as well as did not lead to better accuracies as compared to RF. Other faster and simpler methods like kNN and multi polynomial regression were also considered but their performance was very poor for such complex problem and hence not shown here. We thus only show the results obtained with the RF models. A brief description about the Random Forest model is shown below. 

\subsection{Random Forests (RF)}
Random Forests (RF) is a supervised machine-learning technique that constructs an ensemble of decorrelated decision trees. It is based on the concept of bootstrap aggregating (bagging) to reduce the high variance from single decision trees and further using decorrelated trees to induce more randomness and variance reduction that eventually improves prediction performance for test cases.  Like ANN, RF can be used for both regression and classification. 
\begin{figure}[h]
	\centering
	\includegraphics[width=0.48\textwidth]{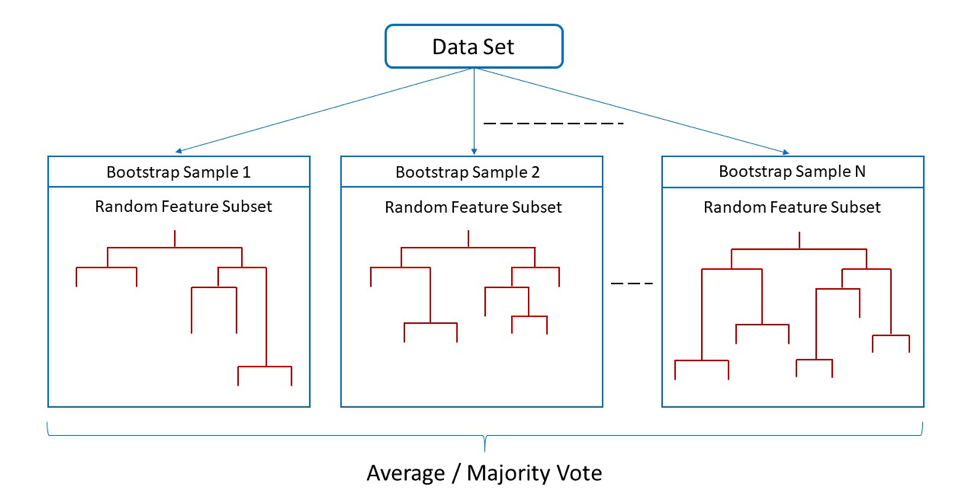}
	\caption{{Description of Random Forests model where, each of the branches from sample dataset correspond to a bootstrap sample and a random feature subset is chosen for each branch to construct a decision tree}}
	\label{fig:3}       
\end{figure}
We briefly describe the working of a RF here. The way a single decision tree is constructed is using a recursive binary splitting strategy. At each node of a tree, best split of a predictor variable is made that results in two branches and then successively splits the predictor space at each step. For the regression problem, the best split is the one that leads to the maximum possible reduction in error Residual Sum of Squares (RSS), and for the classification problem, Gini index or the cross-entropy are generally used to evaluate the best split. Bagging generates new training sets by sampling the data with replacement and then we generate an ensemble of decision trees for each of these training sets. A random forest further decorrelates the trees by selecting only a random subset of features for building each tree. During prediction, for regression, the outcome is simply averaged over all the trees in the forest and for classification, the outcome is based on majority vote. For more details on Random Forests refer to \cite{Breiman2001}. Figure \ref{fig:3} shows the working of random forests pictorially. 
For this work, the hyperparameters for the random forest regressor model are chosen to be the minimum number of splits at the leaf node ($N_l$) and the number of features to search for split ($N_{fmax}$). For all the cases shown, the best model was found through cross validation.

\subsection{Well Location Feature Determination}
Before we employ the ML techniques to predict the basis coefficients, we need to quantify the features representing the parameter $\zeta$. As we state before, the system parameter $\zeta$ for well location optimization corresponds to the index in the source/sink term vector. There is no explicit quantification of this parameter that represents changing well location in the PDE (Eq. \ref{eq:3}). Thus, defining the input (well location) - output (basis coefficient) relationship is difficult. In this section, we first define geometric features for changing well locations. Then, we introduce some physics based features using the concepts from flow diagnostics.

\subsubsection{Geometric and Physics based features}
For the input features of prediction problems (equation (\ref{eq:5_8})), one of the obvious choices for changing well locations is the coordinates of the new well locations. We also consider the distance and angle between the wells (injectors and producers) in the reservoir. These features are chosen based on the understanding that the state solution behavior to some extent is influenced by distance and orientation between the wells.

As we discussed the geometric features that correspond to changing well locations and deem important for relation to the output state solutions (as represented by basis coefficients), we now define the physics based features that are of major importance as they describe the physics of flow and their relation to the well locations.

When we consider a heterogeneous reservoir case, one obvious choice of physics based feature that need to be included is permeability of the gridblock at the well location considered. What is also of major importance for us to understand the change in physics of flow with changing well location is some representation of well connectivity and flow heterogeneity between well pairs. In order to understand such properties, we now introduce the concept of flow diagnostics and various properties that are calculated during flow diagnostics that serve as important features for our problem. 

\subsection*{Flow Diagnostics}

Flow diagnostics are computational tools based on simple and controlled numerical flow experiments to help us quickly get quantitative and qualitative information regarding the flow patterns in a reservoir model. Traditional reservoir simulations can perform this task but computationally very demanding, and in contrast, flow diagnostic measures can be obtained within seconds. Thus, they are an inexpensive and reliable alternative to rank and/or compare realizations or strategies, and ideal for interactive visualization output due to computational advantage.

The reason behind this computational advantage is because the flow diagnostic tools provide quantitative information based on the steady-state flow. Flow diagnostic tools have been developed using streamline simulation techniques \cite{Datta-Gupta2007, Datta-Gupta1995} that has shown to be very effective for various applications, but have some limitations in terms of computational complexity and extensibility. Most of the existing commercial simulators are based on finite volume methods capable of simulating fluid flow with different spatial grid geometry because of their mass conservative nature. Keeping this in mind, the flow diagnostic properties were obtained using standard finite-volume discretization in \cite{Shahvali2012, Moyner2015}. We use the flow diagnostics module of MRST (Matlab Reservoir Simulation Toolbox) \cite{Moyner2015} to evaluate these properties with the aim to aid in accurate construction of reduced order models for changing well locations. We now briefly introduce the governing equations that lead to quantitative measures of flow dynamics as explained in the works \cite{Shahvali2012, Moyner2015}. 

\subsection*{Governing Equations}

A single phase incompressible flow is considered to compute the representative flow field,
\begin{subequations}
	\label{eq:4_9}
	\begin{align}
	\nabla\cdot\textbf{v} = 0\label{eq:4_9_1} \\
	\textbf{v}= - \frac{K}{\mu}\nabla p\label{eq:4_9_2}
	\end{align}
\end{subequations}
where, $K$ is the permeability tensor, $\mu$ is the viscosity of fluid, $\textbf{v}$ is the Darcy velocity and $p$ is the phase pressure. Here, for the purpose of flow diagnostics, no flow boundary conditions are assumed and $\mu$ is set to 1. 

\subsection*{Stationary Tracer Distribution and Time of Flight}
To observe the transport properties of the flow field, we consider the equation (\ref{eq:4_10}) describing the neutral tracer transport which is injected into the injection well with a concentration $c$. For simplicity we neglect source/sink terms and assume zero concentration of tracer initially in the entire domain. The transport equation is written as:
\begin{equation}
\label{eq:4_10}
\begin{aligned}
\phi \frac{\partial}{\partial t}c + \textbf{v}\cdot\nabla c = 0
\end{aligned}
\end{equation}
where, $\phi$ is the porosity of reservoir. At late times, the equation (\ref{eq:4_10}) takes the steady state form:
\begin{equation}
\label{eq:4_11}
\begin{aligned}
\textbf{v}\cdot\nabla c = 0
\end{aligned}
\end{equation}

The stationary tracer equation gives an idea about the regions in the reservoir that are influenced by injectors and producers and ultimately help us understand swept volumes, well allocation factors etc. However, it does not give enough information about the impact of flow on the heterogeneity of the reservoir as dictated by permeability and porosity distribution. Thus, instead of considering the tracer distribution, we consider time of flight ($TOF$) coordinate $\tau(x)$ which indicates the time it takes for a tracer to travel from the nearest injector to a given point $x$ in the reservoir. $TOF$ is defined as:

\begin{equation}
\label{eq:4_12}
\begin{aligned}
\tau(x) = \int_{\psi(s)}\frac{\phi}{|\textbf{v}(x)|}ds
\end{aligned}
\end{equation}

where $\psi$ and $s$ denote the streamline and the arc length measured along the streamline, respectively. Operator identity shown in equation (\ref{eq:4_13}) is used to represent the 1D flow along streamlines for the equation (\ref{eq:4_10}) \cite{Natvig2006}.
\begin{equation}
\label{eq:4_13}
\begin{aligned}
\textbf{v}\cdot\nabla = |\textbf{v}|\cdot\frac{\partial}{\partial s}
\end{aligned}
\end{equation}

Similarly, we can derive the $TOF$ equation in Eulerian coordinates using this identity: 
\begin{equation}
\label{eq:4_14}
\begin{aligned}
\textbf{v}\cdot\nabla \tau= \phi
\end{aligned}
\end{equation}

These equations (\ref{eq:4_11} and \ref{eq:4_14}) have a hyperbolic form. They can be written in a conservative form as:

\begin{equation}
\label{eq:4_15}
\begin{aligned}
\nabla \cdot (\textbf{v}u) = b
\end{aligned}
\end{equation}
where, $u = TOF$ or tracer concentration and $b$ represents the source or sink. This form is used since it has a natural finite volume discretization that can be used to solve for $TOF$ and $c$ and is a generalization to the case where $\nabla \cdot \textbf{v} \ne 0$.

For the cases with more than one producer and one injector one can also consider the tracer distribution to identify quantities like well allocation factors, well-pair connections etc. as important features. Two quantities are calculated called the forward and backward time of flight using equation (\ref{eq:4_14}) as:
\begin{subequations}
	\label{eq:4_16}
	\begin{align}
	\textbf{v}\cdot\nabla \tau_f= \phi\\
	-\textbf{v}\cdot\nabla \tau_b= \phi
	\end{align}
\end{subequations}
The forward TOF $\tau_f$ is the time required for a tracer to reach at a given point in the reservoir after injected at an inflow boundary (injector in our case). The backward TOF $\tau_b$ is obtained by reversing the flux field which indicates the time required for a tracer released at a given location within the reservoir to reach a producer/outflow boundary. The sum of both these quantities gives the total time of flight also defined as the residence time of the particle.
\begin{equation}
\label{eq:4_17}
\begin{aligned}
\tau_f + \tau_b = \tau
\end{aligned}
\end{equation}

\subsection*{Lorenz Coefficient}

Time of flight and tracer concentration can also be used to assess displacement heterogeneity of the reservoir using a quantity called Lorenz coefficient. The heterogeneity is represented in terms of flow capacity - storage capacity ($F-\Phi$) diagrams. (Note that for the sake of consistency in notations with the literature, for this section we refer to $\Phi$ as the storage capacity and not to be confused with POD basis). This is equivalent to plotting fractional flow versus saturation for a 1D flow displacement. As defined in \cite{Shook2009}, the storage capacity $\Phi$ is defined as the cumulative pore volume as fraction of total travel time $\tau$, i.e., $\Phi(\tau) = \int_{0}^{\tau}\phi(x(\tau))d\tau$. The flow capacity is defined as the cumulative flux for increasing travel time. For an incompressible flow, $ F(\tau) = \int_{0}^{\tau}\frac{\phi(x(\tau))}{\tau}d\tau$, since the pore volume equals the product of the flux and the total travel volume. Lorenz coefficient which is the measure of displacement heterogeneity is defined by measuring how much the flow capacity deviates from the ideal piston like displacement and hence can be written as:
\begin{equation}
\label{eq:4_18}
\begin{aligned}
L_c = 2 \int_{0}^{1} (F(\Phi) - \Phi) d\Phi
\end{aligned}
\end{equation}

which is twice the area under the $F-\Phi$ curve and above the line $F=\Phi$. Thus, it is 0 which corresponds to homogeneous displacement and unity for infinitely heterogeneous displacement. For a detailed explanation on calculating $F$ and $\Phi$, the reader is referred to \cite{Shook2009} for streamline based calculations and \cite{Shahvali2012} for finite volume based diagnostics.

\subsection*{Flow diagnostic features - Example}
\begin{figure}[htb!]
	\begin{subfigure}{0.32\textwidth}
		\centering
		\includegraphics[width=\textwidth]{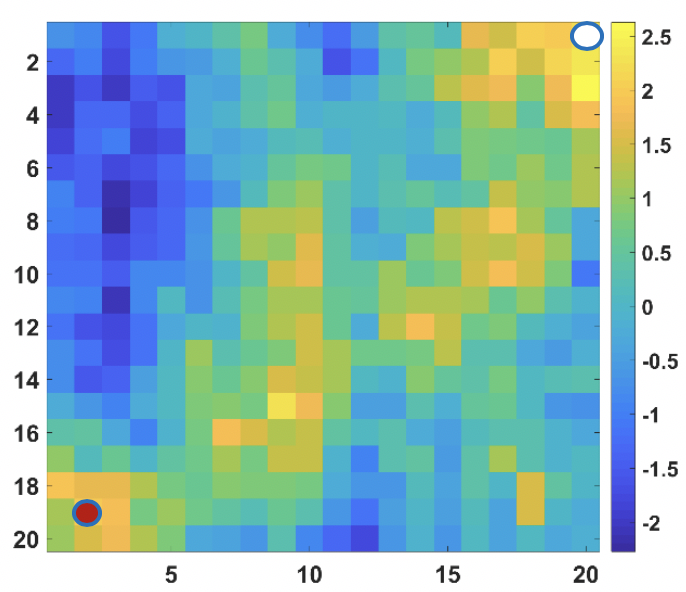}
		\caption{Normalized Permeability Field}
		\label{fig:4_9a}
	\end{subfigure}%
	~
	\centering
	\begin{subfigure}{0.38\textwidth}
		\centering
		\includegraphics[width=\textwidth]{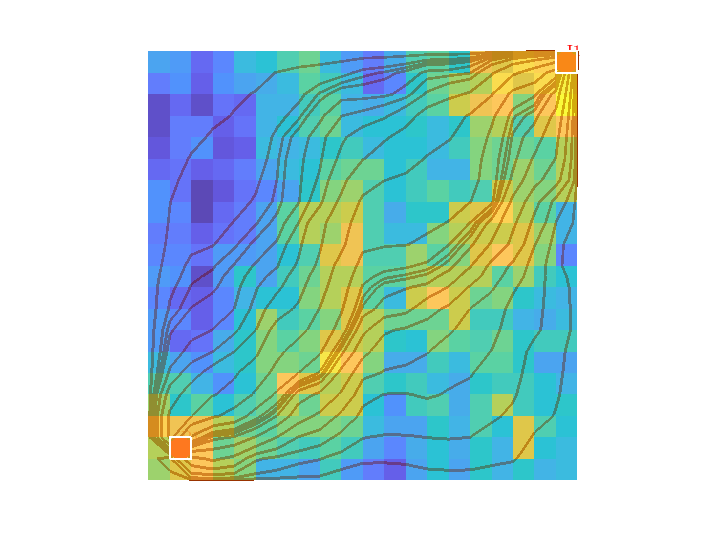}
		\caption{Streamlines}
		\label{fig:4_9b}
	\end{subfigure}%
	\caption{(a) Heterogeneous channelized permeability field with one producer in red and one injector in white (b) Streamlines imposed on the permeability field}
	\label{fig:4_9}
\end{figure}

We now show an example of the flow diagnostic features for one of the well configurations. Consider the heterogeneous permeability field as shown in figure (\ref{fig:4_9a}) with wells at location $(2, 19)$. We use the MRST flow diagnostics module that computes TOF and Lorenz coefficient based on solving equations as shown above (\ref{eq:4_16} - \ref{eq:4_18}). 

\begin{figure}
	
	\centering
	\begin{subfigure}{0.32\textwidth}
		\centering
		\includegraphics[width=\textwidth]{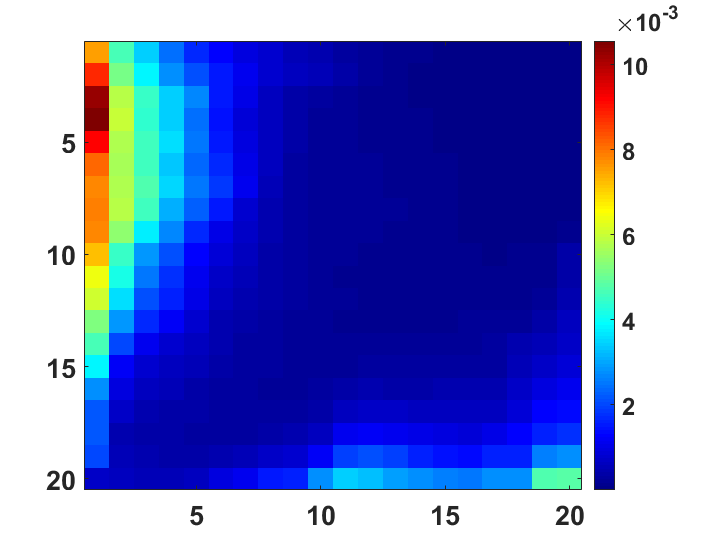}
		\caption{Total travel time}
		\label{fig:4_10a}
	\end{subfigure}%
	~
	\centering
	\begin{subfigure}{0.32\textwidth}
		\centering
		\includegraphics[width=\textwidth]{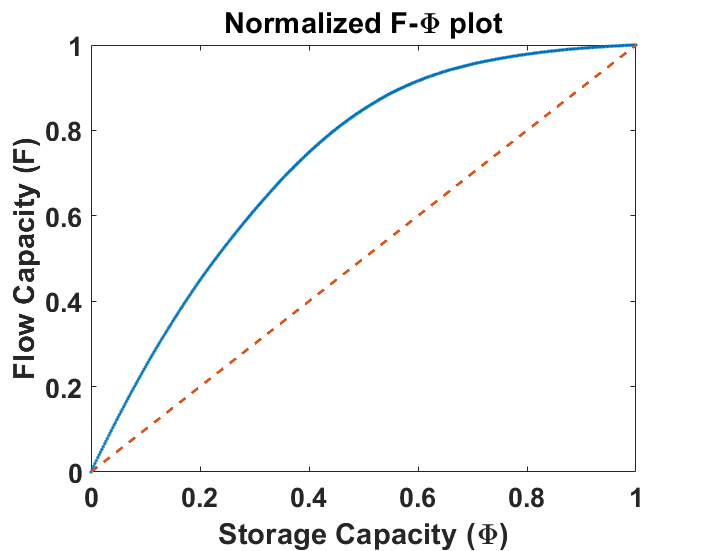}
		\caption{$F-\Phi$ plot to compute Lorenz coefficient }
		\label{fig:4_10b}
	\end{subfigure}%
	\caption{(a) The total travel time of the particle injected at injector to any point in the reservoir (b) $F - \Phi$ plot to compute Lorenz coefficient as the area under the solid $F-\Phi$ curve and the dashed line $F = \Phi$}
	\label{fig:4_10}
\end{figure}

It can be seen that the low permeability area in the reservoir that is not well connected to the injector has a higher travel time and the producer location as shown above has a very low travel time as it is located on a high permeability gridblock and connected to the injector via a channel. Thus, while considering the feature to be added for ML procedure, we use the TOF at the producer well location. Note that, since we are interested in finding time of flight at producer well location, it is sufficient to compute forward time of flight since backward time of flight will be negligible. 

The set of features used are listed in Table \ref{tab:5_1}. Note that the features correspond to information that we think best describes the well location. We also include time information to account for understanding the temporal behavior of the POD basis coefficients.

\begin{table}[htb!]
	\centering
	
	\begin{tabular}{l}
		\hline\noalign{\smallskip}
		\textbf{Feature Set for a well configuration} \\
		\noalign{\smallskip}\hline\noalign{\smallskip}
		$x$ - X coordinate of the well 
		\\
		$y$ – Y coordinate of the well
		\\
		$r$ – Distance between well and injector
		\\
		$\theta$ – Angle between well and injector
		\\
		$K$ – Permeability at the well location
		\\
		$TOF$ - Total time of flight at the producer well location
		\\
		$LC$ - Lorenz coefficient for the well configuration
		\\
		$index$ - Well gridblock number in the reservoir
		\\
		$t$ - Time at which the POD coefficients are computed
		\\
		\noalign{\smallskip}\hline
	\end{tabular}
	\caption{Geometric and physics based features for ML model construction corresponding to the well configuration}
	\label{tab:5_1}       
	\vspace*{-0.2cm}
\end{table}

\subsection{Feature Selection}

\begin{figure}[h]
	\centering
	\includegraphics[width=0.4\textwidth]{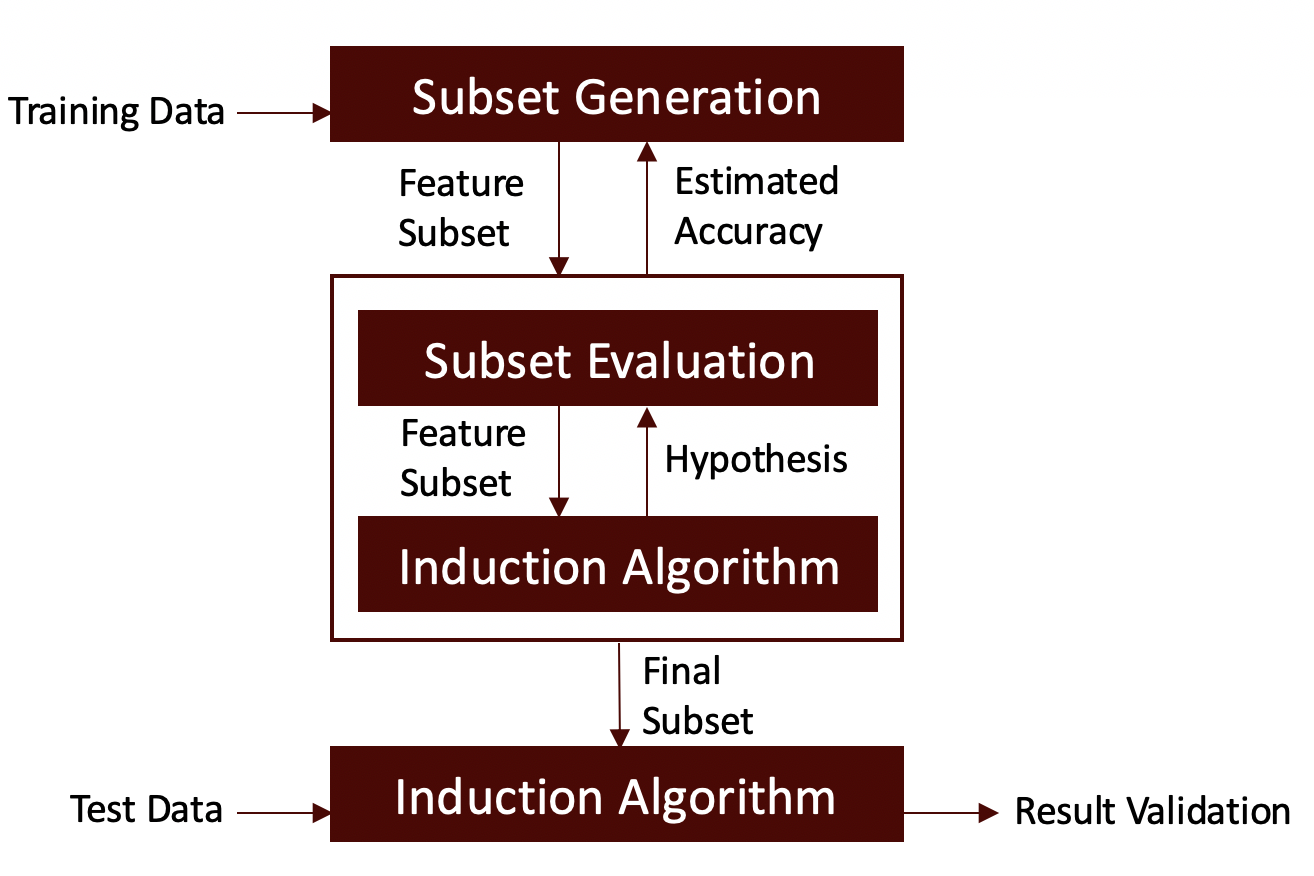}
	\caption{Wrapper approach}
	\label{fig:4_14}       
\end{figure}

The important step before training a ML model is to remove any redundant and correlated features. For this procedure we use the Wrapper approach, the procedure of which is briefly outlined below. The wrapper approach sequentially selects subset of feature space and evaluate its performance with respect to an induction algorithm (black box model) using cross-validation and subsequently adds or deletes features based on the search criteria. The search engine we use is Best First Search (BFS) using Forward selection which is more robust than the greedy hill-climbing search. Additional details about the wrapper approach can be found in \cite{Kohavi1997}. The procedure is explained in the Figure \ref{fig:4_14}. This step becomes computationally expensive as the number of features grows. But, what we observe with the feature set that we propose for the well locations is that, the BFS methodology sometimes selects all the features, depending on the data set and the black-box model used. This suggests than no features in this case are correlated. So while training the models, we use all the features and then for feature importance, we try to tune better the hyperparameters and the regularization parameters instead of spending computational effort using the wrapper approach. 

\subsection{Remarks}

As we have formulated the method, we now demonstrate its performance with some case studies. However, before that, we point out some specifics of these cases. The injector for all the cases shown here is at a fixed location and the producer location is considered as the parameter i.e. only producer can change location in the reservoir. All the new producer well locations have the same BHPs to consider only the well location as a system parameter. The workflow can be extended in the future to also include varying well BHPs as the feature but not in the current scope of work. The simulations are run long enough to observe significant water cut at most of the well locations in order to introduce complexity to the problem which was a limitation in the local ROB based workflow \cite{Zalavadia2018}. For the cases shown below, we use random sampling in the parameter space to train the ML models. Experimental design sampling techniques or greedy sampling procedures as listed before can be used for sampling the parameter space for better representative parameters, but, for simplicity and fast sampling we choose the parameters randomly for training purpose. However, the number of random samples chosen is an open challenge that needs to be addressed in the future. With random sampling we usually expect a lot more samples than something like greedy sampling procedures, in efforts to capture representative samples for global ROB. The main aim here is to show the validity of proposed methodology given the high complexity of the problem compared to that of changing well control problem and hence we use relatively large training sample sizes here.

\section{Case Study} \label{sec:1.6}
\begin{figure}
	\centering
	\begin{subfigure}{0.33\textwidth}
		\centering
		\includegraphics[width=\textwidth]{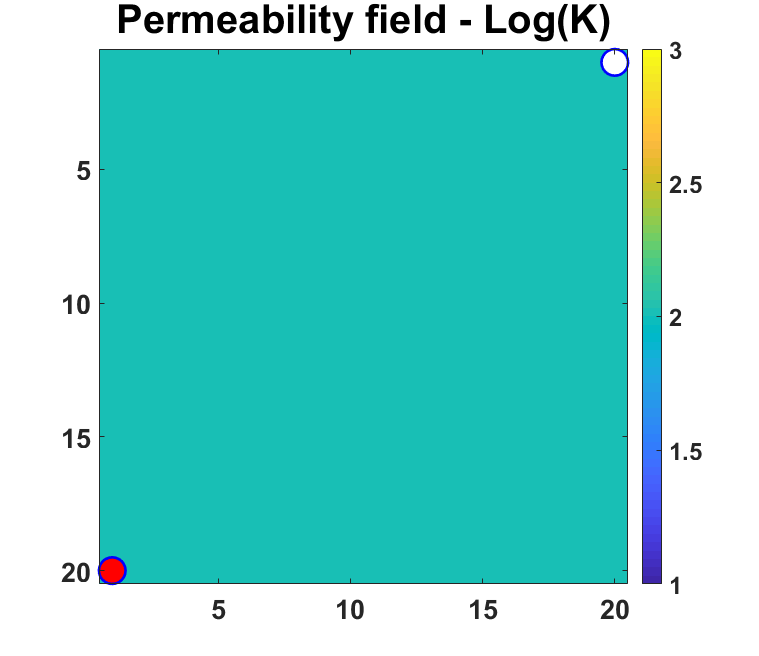}
		\caption{}
		\label{fig:5_1_1}
	\end{subfigure}%
	~
	\centering
	\begin{subfigure}{0.33\textwidth}
		\centering
		\includegraphics[width=\textwidth]{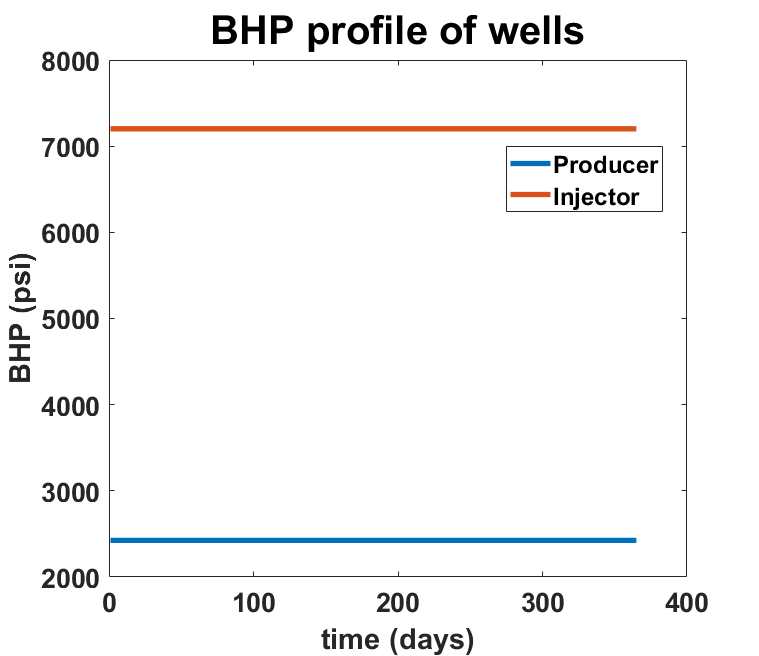}
		\caption{}
		\label{fig:5_1_2}
	\end{subfigure}%
	\caption{(a) Homogeneous permeability reservoir with 1 injector (white) and 1 producer (red). Producer is the parameter and can change location anywhere in the reservoir (b) BHP profiles of injector and producer that is set constant for all well configurations considered}
	\label{fig:5_1}
\end{figure}

In this section, we present the numerical results for the proposed non-intrusive global PMOR technique. We start with a simple example of a homogeneous reservoir model. The model has $20\times20$ gridblocks and has one injector and one producer well as shown in Figure \ref{fig:5_1_1}. The porosity is set constant to $0.2$ and the relative permeability model is defined by Corey function of degree $2$. The permeability of the field is considered to be $100 mD$. The 2-phase flow is considered incompressible neglecting the capillary and gravitational effects. All the producers are produced with a constant BHP of 2425 psi and injector injects water a constant BHP of 7200 psi. We look to change the producer locations only and the injector location is fixed at $(20,1)$ gridblock. The simulations are ran for a period of 1 year with which we observe watercut at all the producer locations. 

The problem is setup as follows. First, in order to construct a global ROB, we randomly sample the well locations.  Fine scale simulations are run for each well location and the snapshots are collected in a single snapshot matrix as shown in equation (\ref{eq:5_1}). We then follow the procedure as described before to train the ML model. Since, we have 400 grid blocks, there are 399 locations we consider where the producer can move (1 injector block). So we randomly sample 100 cases to train the ML models and the rest 299 cases are used as test cases.

\subsection{Energy of eigenvalues - Global PMOR}

Now, we discuss about an observation on energy of eigenvalues that is important in understanding the complexity of developing PMOR strategy for changing well location using this case of homogeneous reservoir. For problem of changing well controls and fixed well configuration, we usually observe higher number of saturation basis as compared to pressure basis due to the fast moving pressure front and slow saturation front. Here, for changing well location problem, as we construct the global PMOR by concatenating solutions from different well configurations, as expected,  we will see an increase in the number of basis for both pressure and saturation. However, we look how these eigenvalues decay or rather how the energy of the basis change with increasing number of parameters in the snapshot matrix. 

\begin{figure}
	\centering
	\begin{subfigure}{0.30\textwidth}
		\centering
		\includegraphics[width=\textwidth]{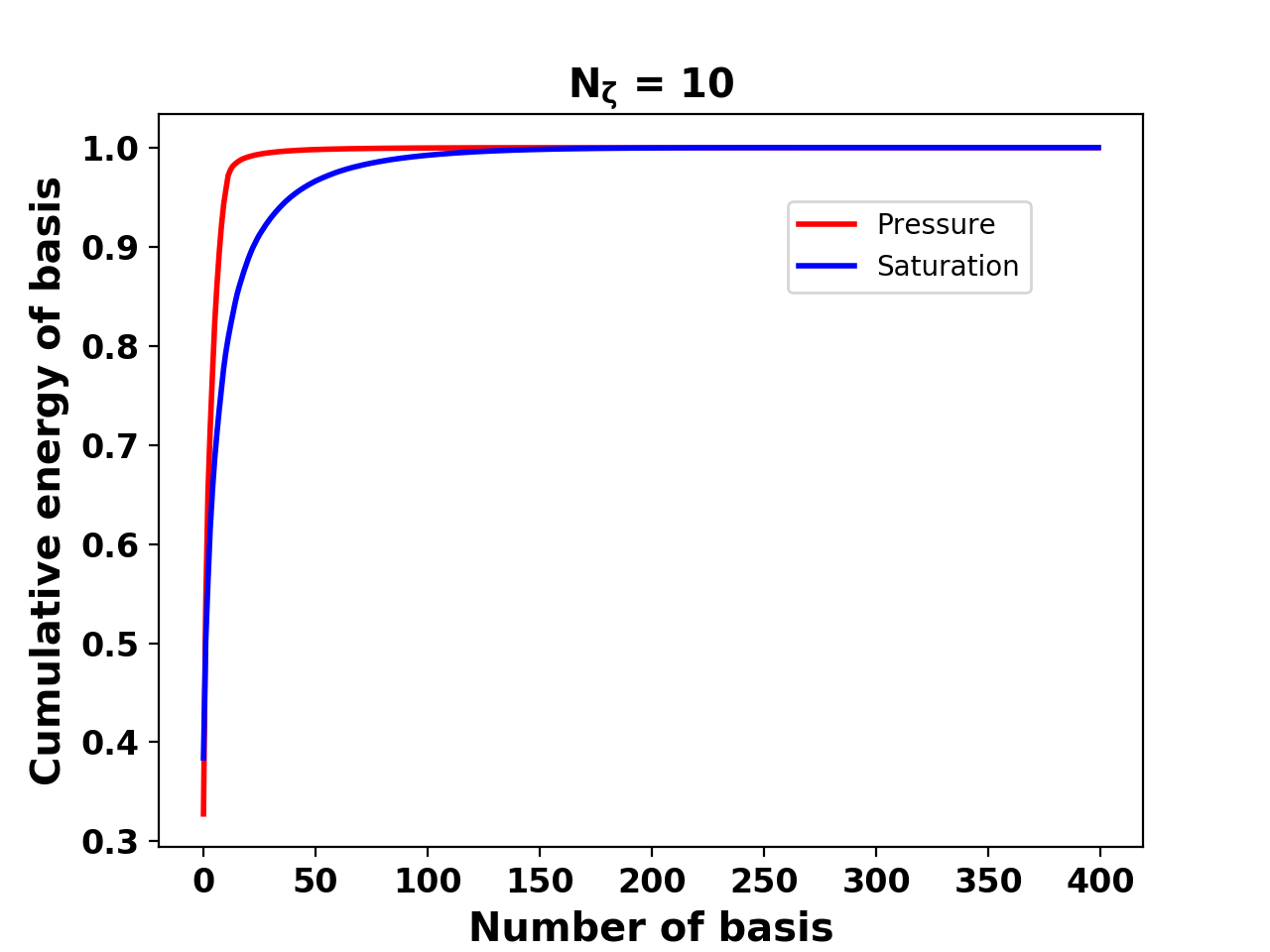}
		\caption{}
		\label{fig:5_2_1}
	\end{subfigure}%
	~
	\centering
	\begin{subfigure}{0.30\textwidth}
		\centering
		\includegraphics[width=\textwidth]{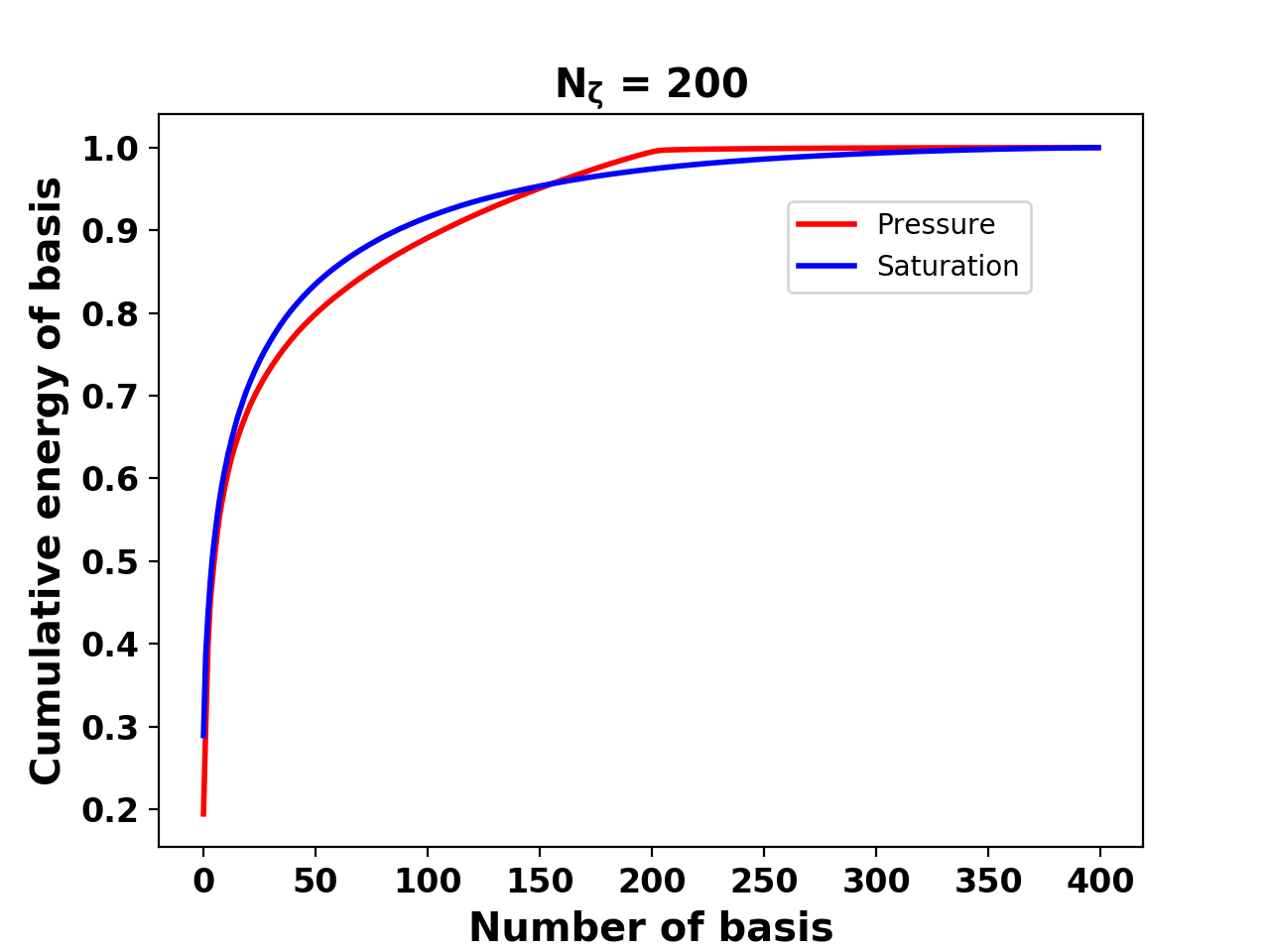}
		\caption{}
		\label{fig:5_2_4}
	\end{subfigure}%
	~
	\centering
	\begin{subfigure}{0.30\textwidth}
		\centering
		\includegraphics[width=\textwidth]{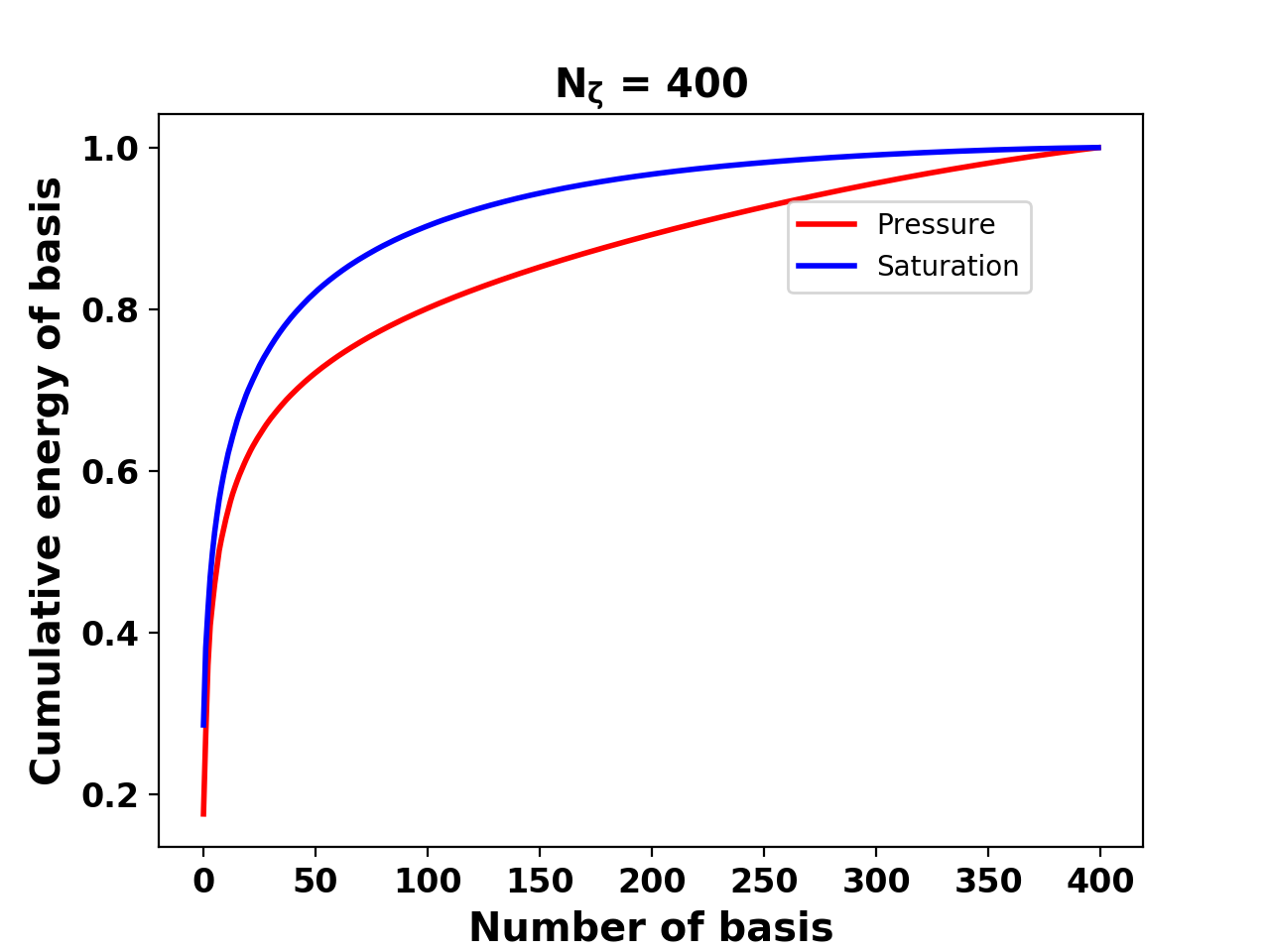}
		\caption{}
		\label{fig:5_2_6}
	\end{subfigure}%
	\captionsetup{justification=centering}
	
	\caption{Change in cumulative energy of the basis as calculated by the magnitude of eigenvalues for increasing number of parameters (well configurations) in the state snapshot matrix}
	\label{fig:5_2}
\end{figure}

PMOR for a well control usually use a lot more saturation basis as compared to pressure basis for a fixed well configuration as the pressure behavior has a fast moving front and is not expected to change significantly for for a fixed well location as compared to the saturation behavior that has a very slow moving front with major updates occurring at the saturation front. For the problem of changing well locations, at each well location we have the lowest or highest pressure point in the reservoir corresponding to the well BHP and hence there is also significant difference in the pressure behavior observed moving from one well location to the other. This  has a significant impact on the number of well configurations used in the training set. A typical energy criteria used for choosing the basis dimensions is above $90\%$ of the energy of eigenvalues from the equation (\ref{eq:24}). As can be seen in Figure \ref{fig:5_2}, we begin with $N_\zeta = 10$ where, as usual, we have a higher number of saturation basis than pressure basis satisfying more than $90\%$ energy criteria. However, after adding more parameters, these trend changes, especially after $N_\zeta = 200$. It requires significantly more number of pressure basis to capture to understand the dynamics. Note, that the pressure basis energy plateaus at a point equaling number of training parameter $N_\zeta$. As we add more training parameters, we see that the saturation basis energy profile does not change significantly. Thus, for many cases at hand, depending on the training parameters selected, we may observe higher number of pressure basis as compared to saturation basis. This observation also gives a notion on the complexity of the problem as there is a trade-off between choosing the number of training samples and the basis dimension required for a good quality basis over the domain of parameters. For example, we may choose 10 training samples and capture 99.99\% energy basis but still may not produce accurate results as each parameter has very different dynamical behavior. So it may sometimes be good to have more training samples and a lower energy criteria at the expense of running more fine scale simulations in the training phase and getting significant speedups in the testing phase.

\subsection{Results}

\begin{figure}[htb!]
	\centering
	\begin{subfigure}{0.33\textwidth}
		\centering
		\includegraphics[width=\textwidth]{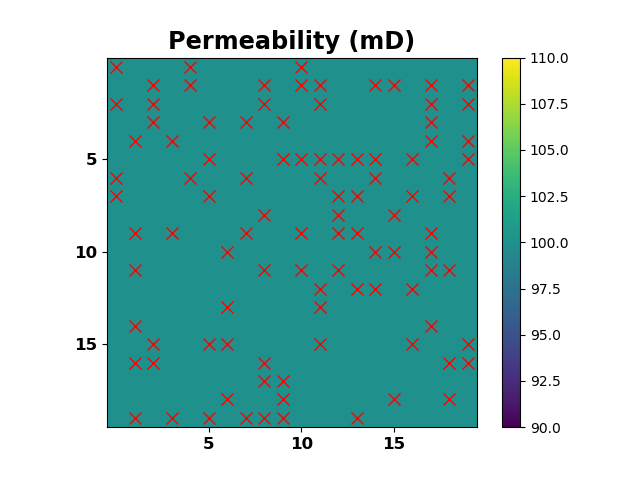}
		\caption{}
		\label{fig:5_3_1}
	\end{subfigure}%
	~
	\centering
	\begin{subfigure}{0.29\textwidth}
		\centering
		\includegraphics[width=\textwidth]{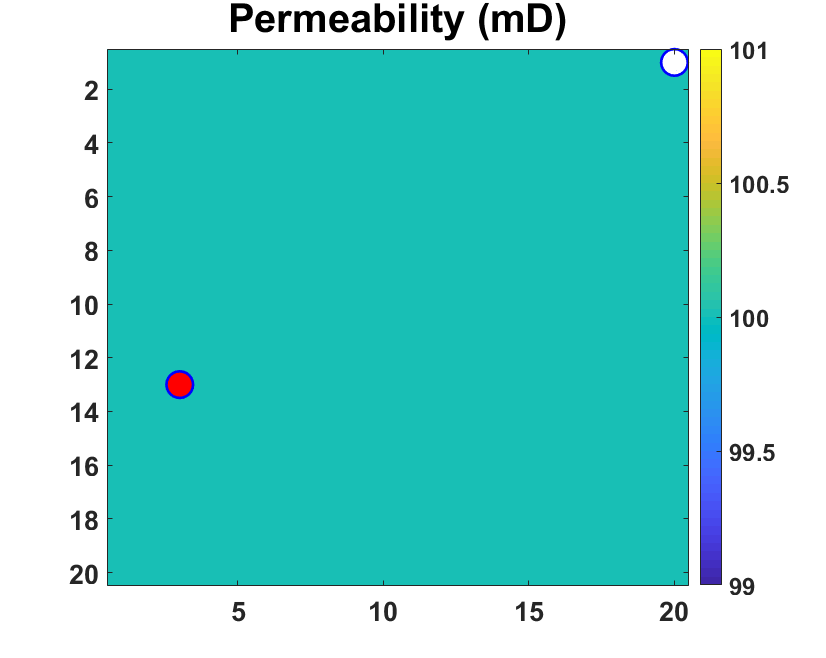}
		\caption{}
		\label{fig:5_3_2}
	\end{subfigure}%
	\caption{(a) Training samples of producer well locations shown by red crosses (b) Test case 1 well location at (3,13) in homogeneous reservoir}
	\label{fig:5_3}
\end{figure}

We now show the prediction results for new well locations not included in the training set. Figure \ref{fig:5_3} shows the training samples of producer well locations each simulated one at time and the new well location in the first test set for which we predict the POD coefficients. 

First, we use Random Forest regressor to compute the POD basis coefficients. For pressure, the basis dimension is 99 and that for saturation is 113, which corresponds to the output dimension for ML models of pressure and saturation respectively. The outputs are predicted using the relation in equation (\ref{eq:5_10}).  and for this example the optimal tuning parameters for RF regressor found are shown in Table \ref{tab:5_2}.

\begin{table}[htb!]
	\begin{center}
		
		\begin{tabular}{lclclc|c|}
			\hline\noalign{\smallskip}
			\textbf{} & \textbf{RF Regression} & \textbf{Train Accuracy} & \textbf{Test Accuracy} \\
			\noalign{\smallskip}\hline\noalign{\smallskip}
			\textbf{Pressure}& $N_{fmax}$=3, $N_l$=2 &\enspace\enspace\enspace\enspace\enspace\enspace 99.83 & 98.18\\
			\textbf{Saturation} & $N_{fmax}$=3, $N_l$=3 & \enspace\enspace\enspace\enspace\enspace\enspace98.89 & 93.21\\
			\noalign{\smallskip}\hline
		\end{tabular}
	\end{center}
	\caption{Hyperparameters chosen by 5-fold Cross Validation for Random Forest Regressor using 100 training samples}
	\label{tab:5_2}       
	\vspace*{-1em}
\end{table}

\begin{figure}[htb!]
	\centering
	\begin{subfigure}{0.5\textwidth}
		\centering
		\includegraphics[width=\textwidth]{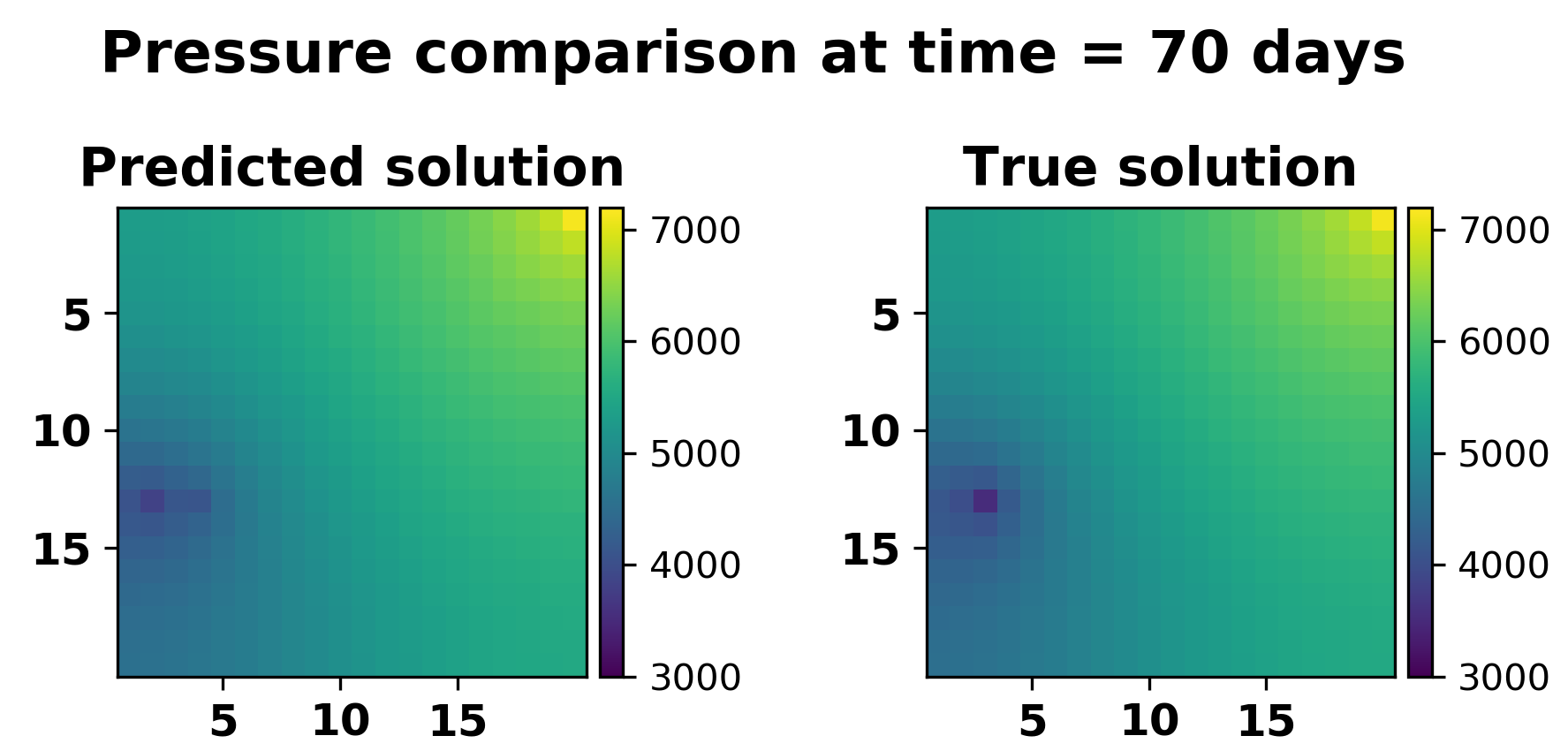}
		\caption{}
		\label{fig:5_4_1}
	\end{subfigure}%
	~
	\centering
	\begin{subfigure}{0.5\textwidth}
		\centering
		\includegraphics[width=\textwidth]{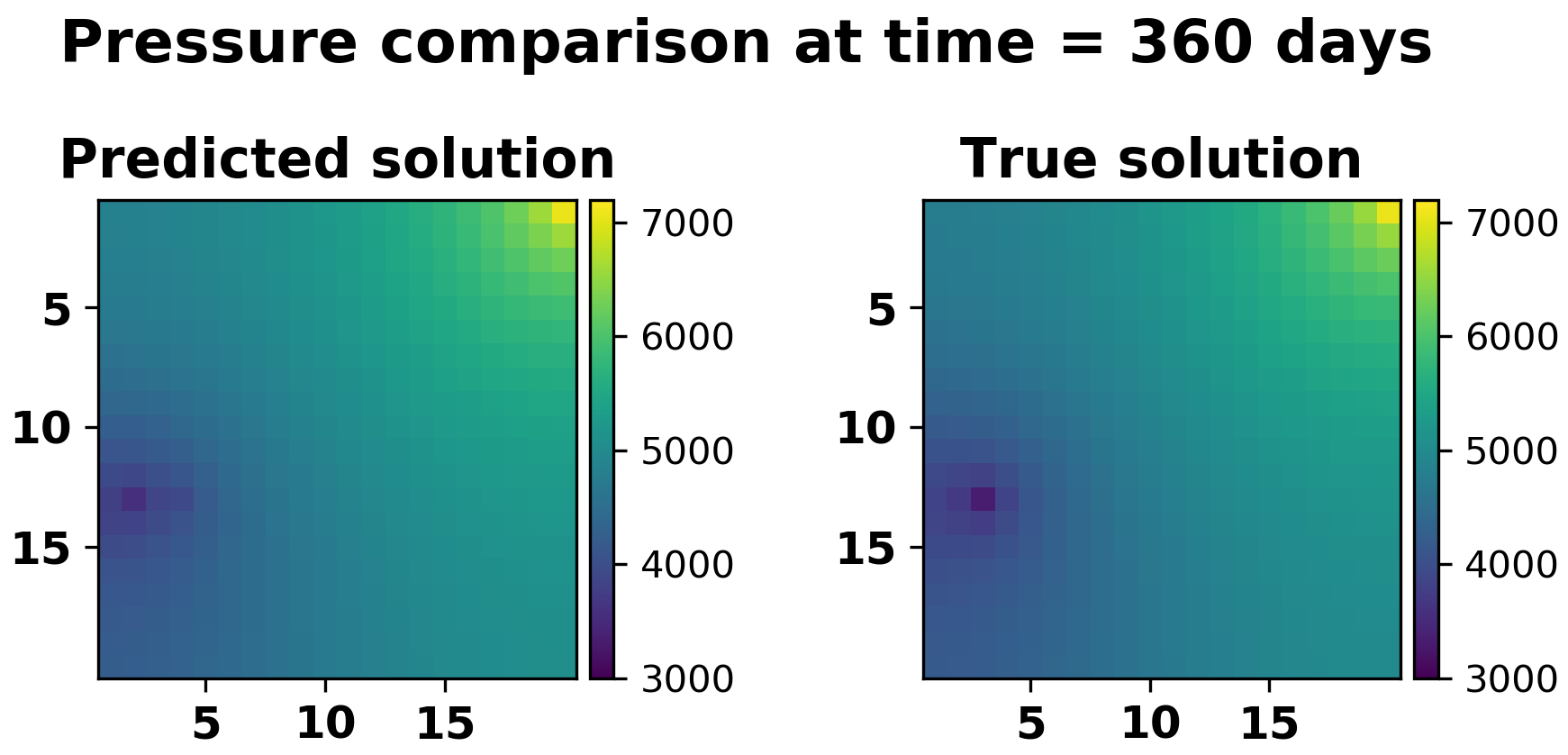}
		\caption{}
		\label{fig:5_4_2}
	\end{subfigure}%
	\caption{Pressure solution comparison at (a) Time = 70 days and (b) Time = 360 days}
	\label{fig:5_4}
\end{figure}

\begin{figure}[htb!]
	\centering
	\begin{subfigure}{0.5\textwidth}
		\centering
		\includegraphics[width=\textwidth]{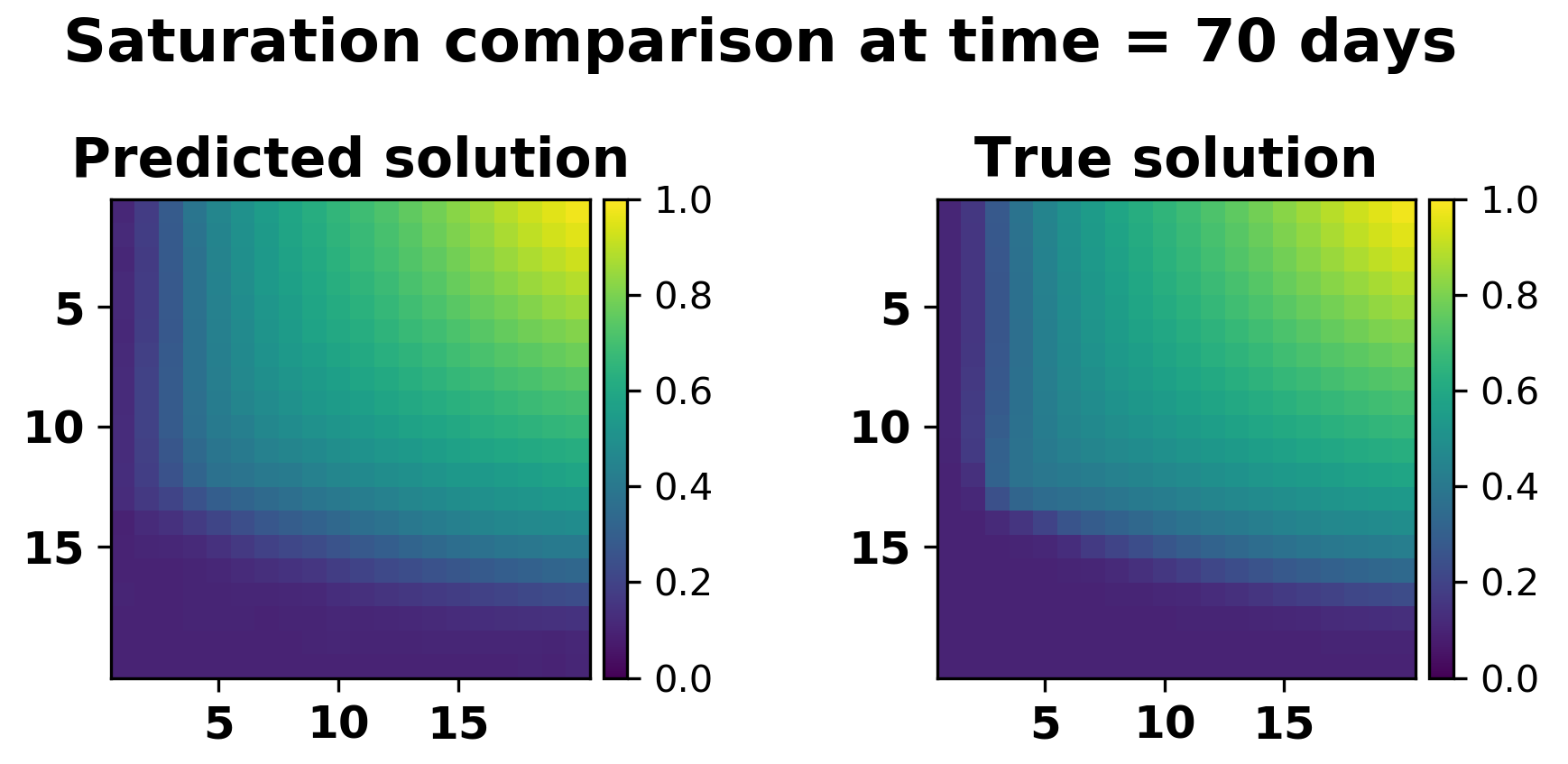}
		\caption{}
		\label{fig:5_5_1}
	\end{subfigure}%
	~
	\centering
	\begin{subfigure}{0.5\textwidth}
		\centering
		\includegraphics[width=\textwidth]{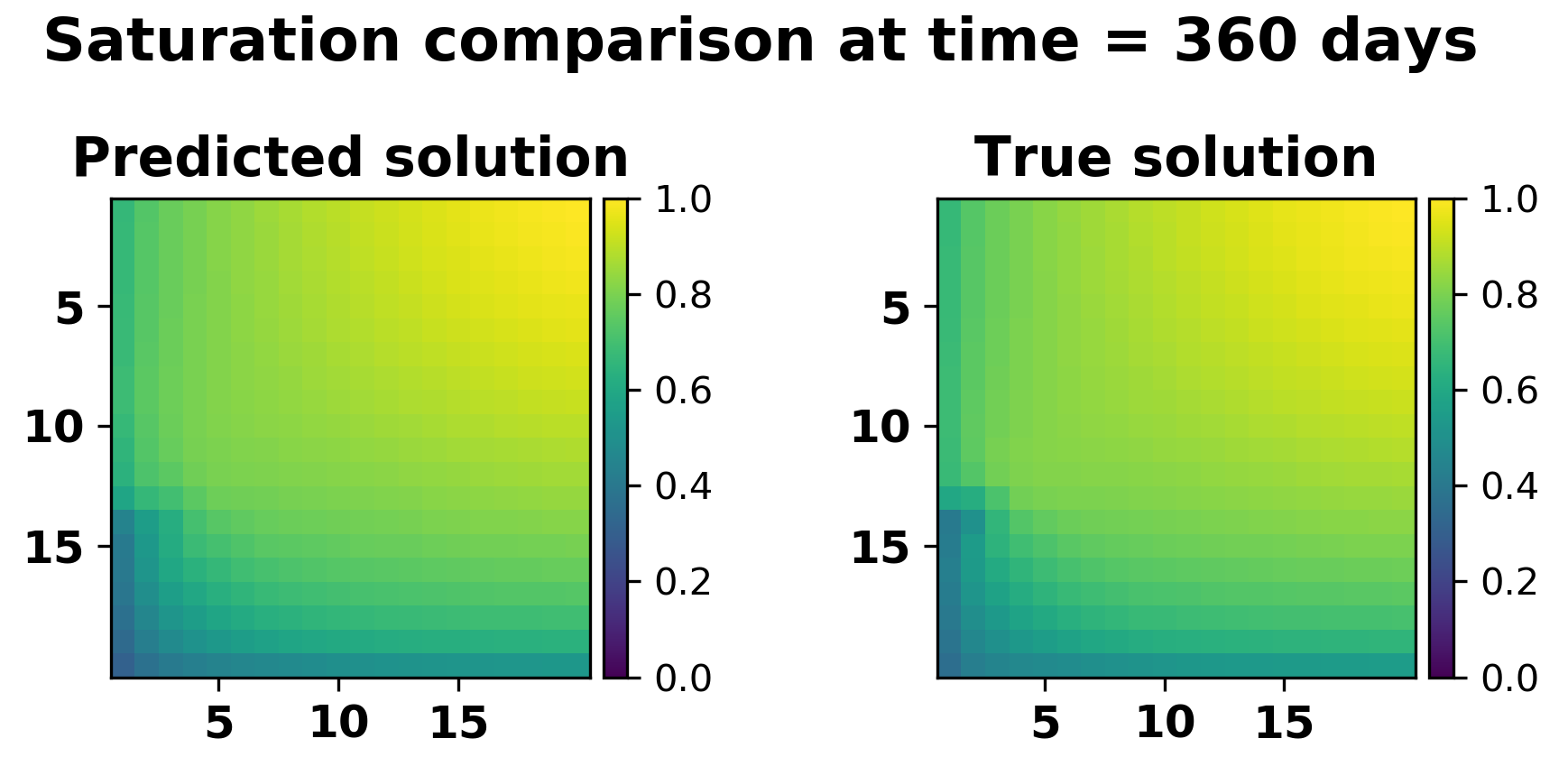}
		\caption{}
		\label{fig:5_5_2}
	\end{subfigure}%
	\caption{Saturation solution comparison at (a) Time = 70 days and (b) Time = 360 days}
	\label{fig:5_5}
\end{figure}

\begin{figure}[htb!]
	\centering
	\begin{subfigure}{0.25\textwidth}
		\centering
		\includegraphics[width=\textwidth]{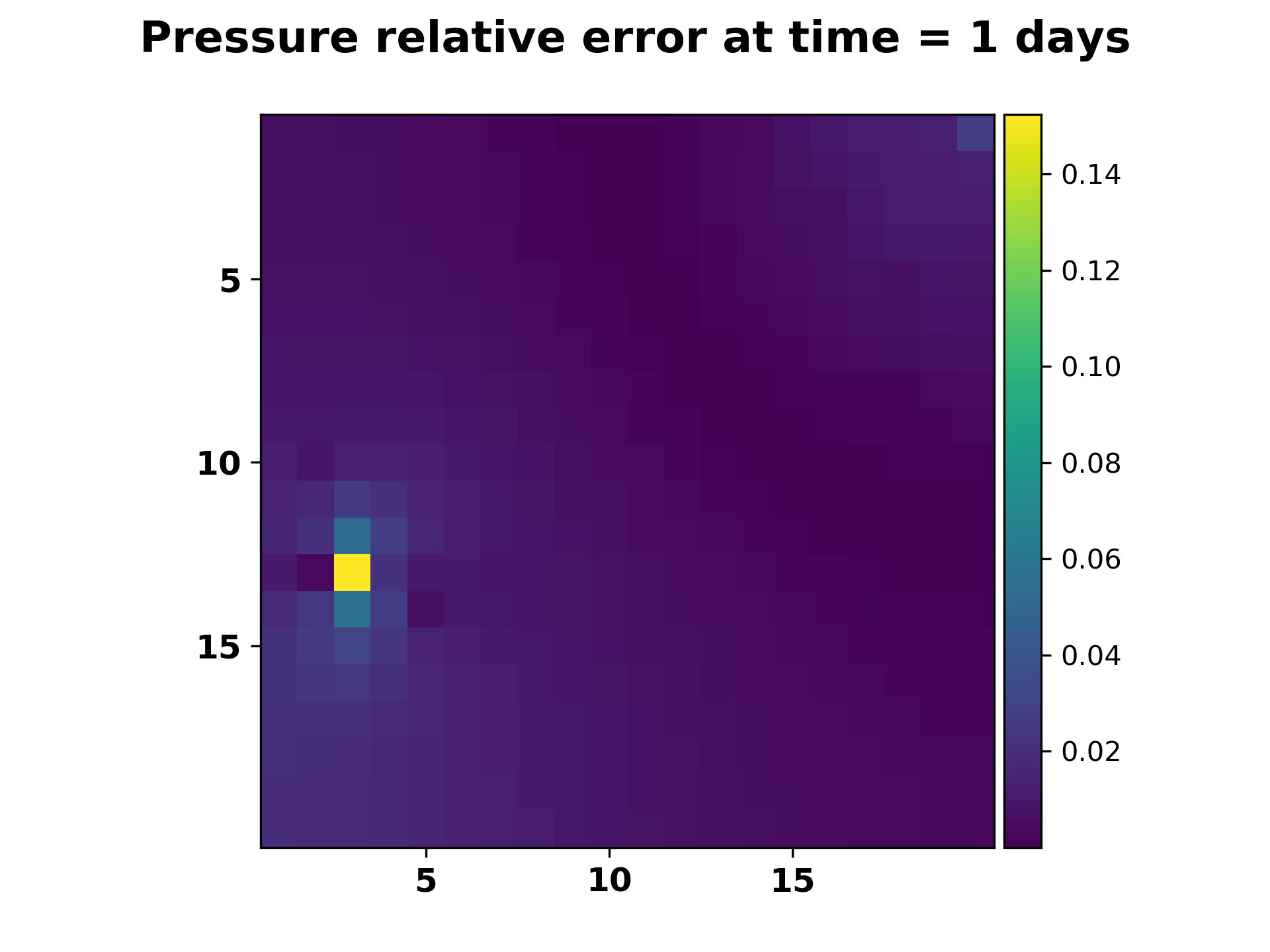}
		\caption{}
		\label{fig:5_6_1}
	\end{subfigure}%
	~
	\centering
	\begin{subfigure}{0.25\textwidth}
		\centering
		\includegraphics[width=\textwidth]{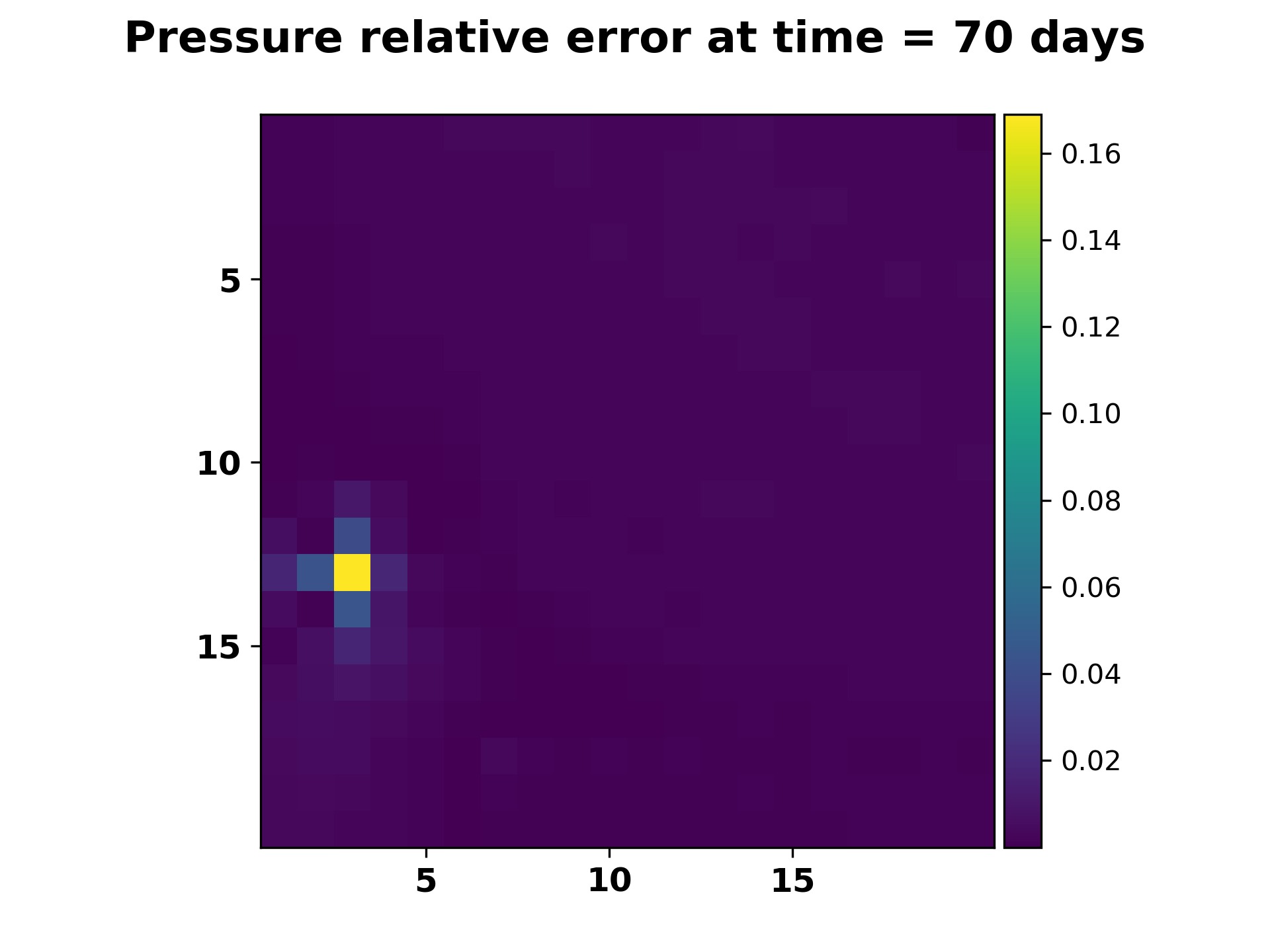}
		\caption{}
		\label{fig:5_6_2}
	\end{subfigure}
	~
	\centering
	\begin{subfigure}{0.25\textwidth}
		\centering
		\includegraphics[width=\textwidth]{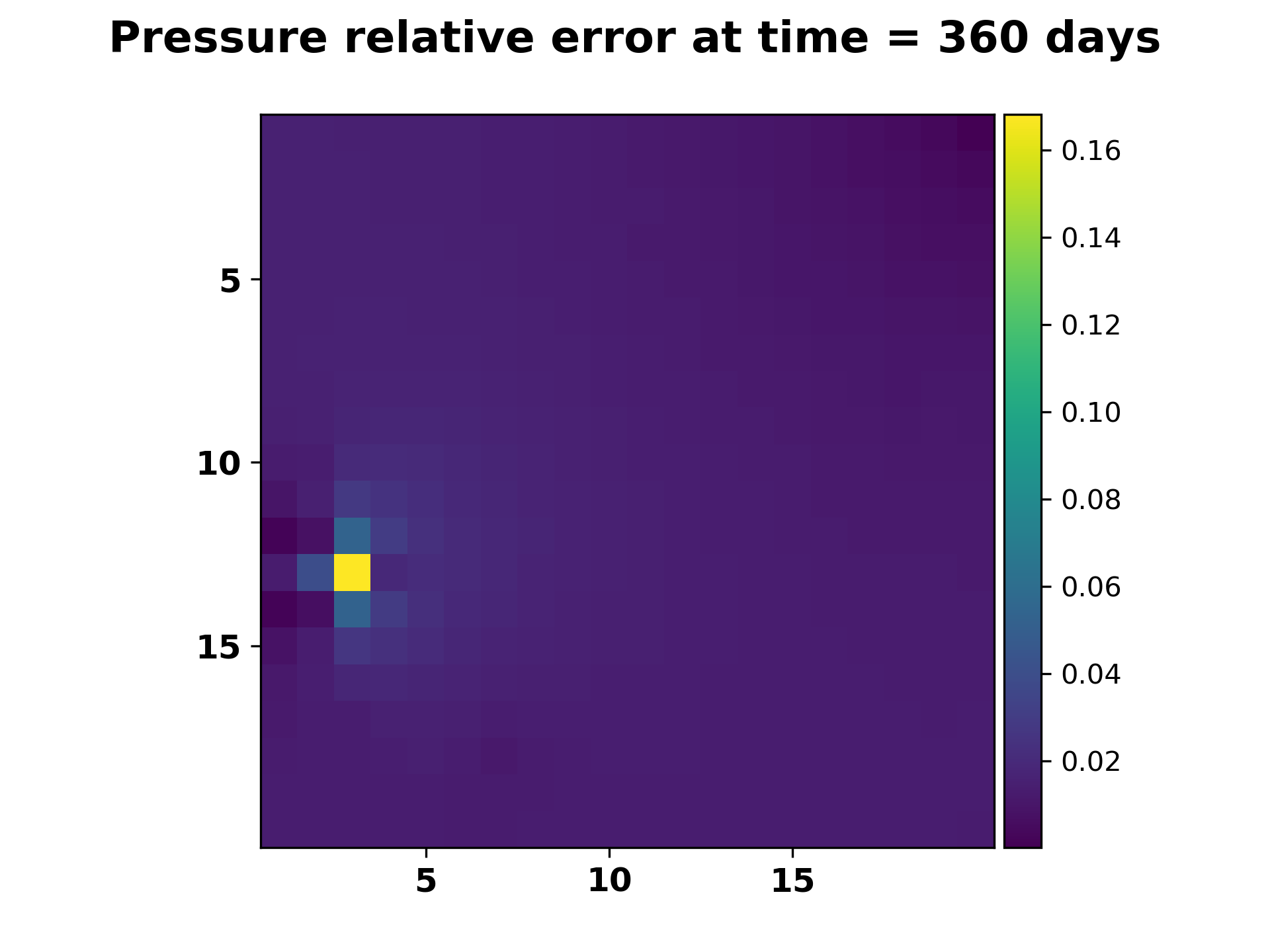}
		\caption{}
		\label{fig:5_6_3}
	\end{subfigure}%
	\caption{Relative error in pressure at time = (a) 1 day , (b) 70 days and (c) 360 days}
	\label{fig:5_6}
\end{figure}

\begin{figure}[htb!]
	\centering
	\begin{subfigure}{0.25\textwidth}
		\centering
		\includegraphics[width=\textwidth]{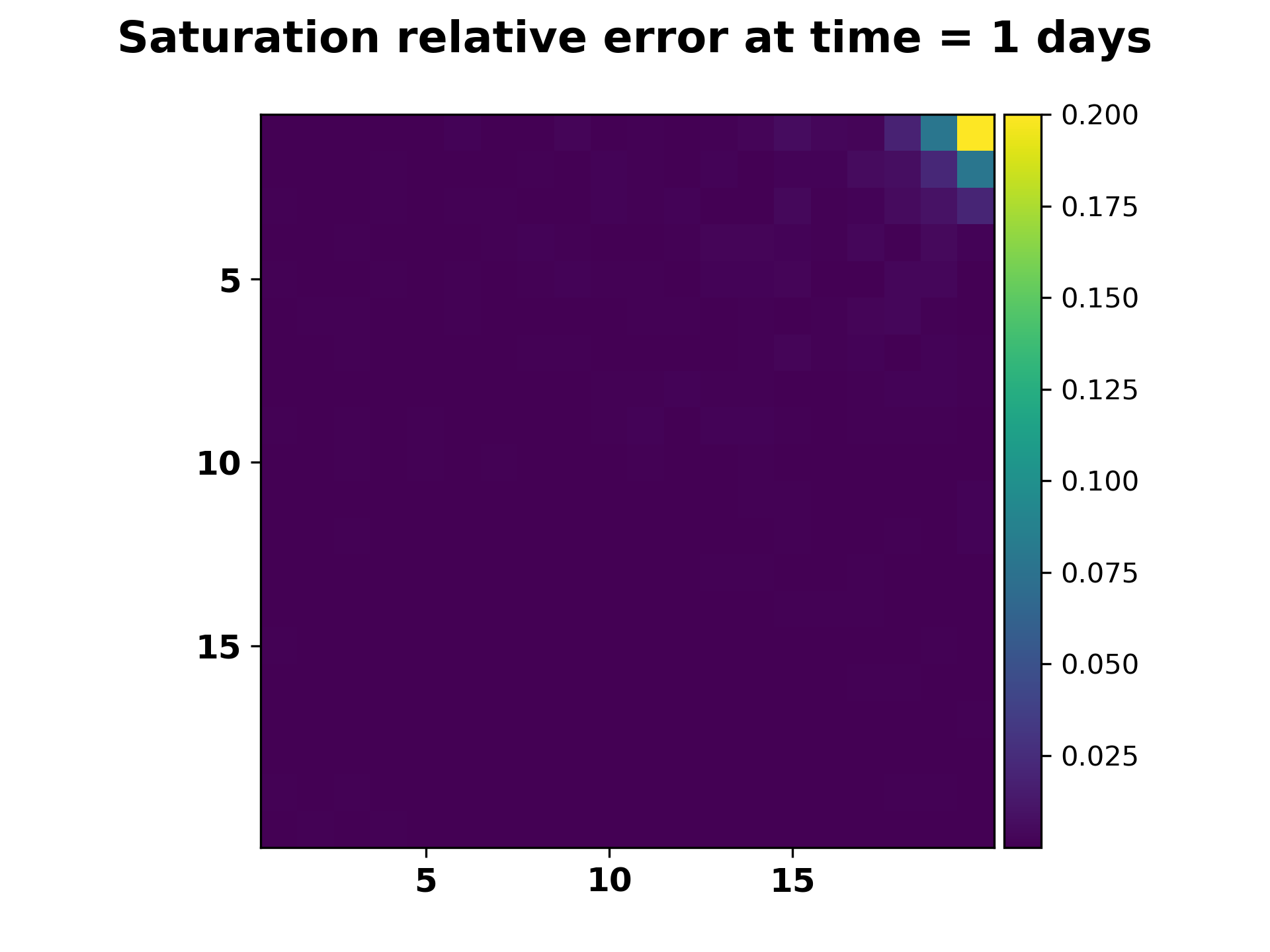}
		\caption{}
		\label{fig:5_7_1}
	\end{subfigure}%
	~
	\centering
	\begin{subfigure}{0.25\textwidth}
		\centering
		\includegraphics[width=\textwidth]{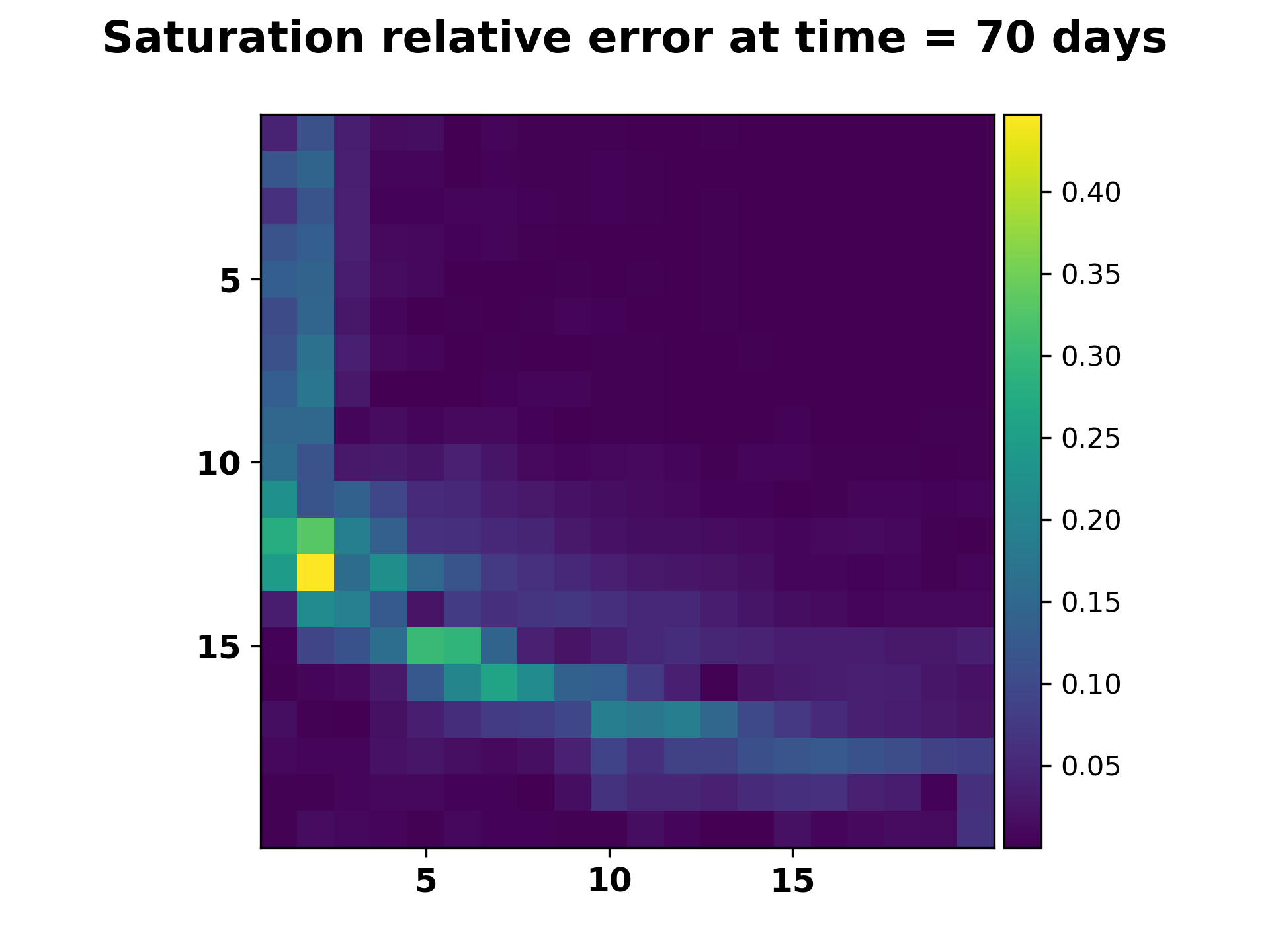}
		\caption{}
		\label{fig:5_7_2}
	\end{subfigure}
	~
	\centering
	\begin{subfigure}{0.25\textwidth}
		\centering
		\includegraphics[width=\textwidth]{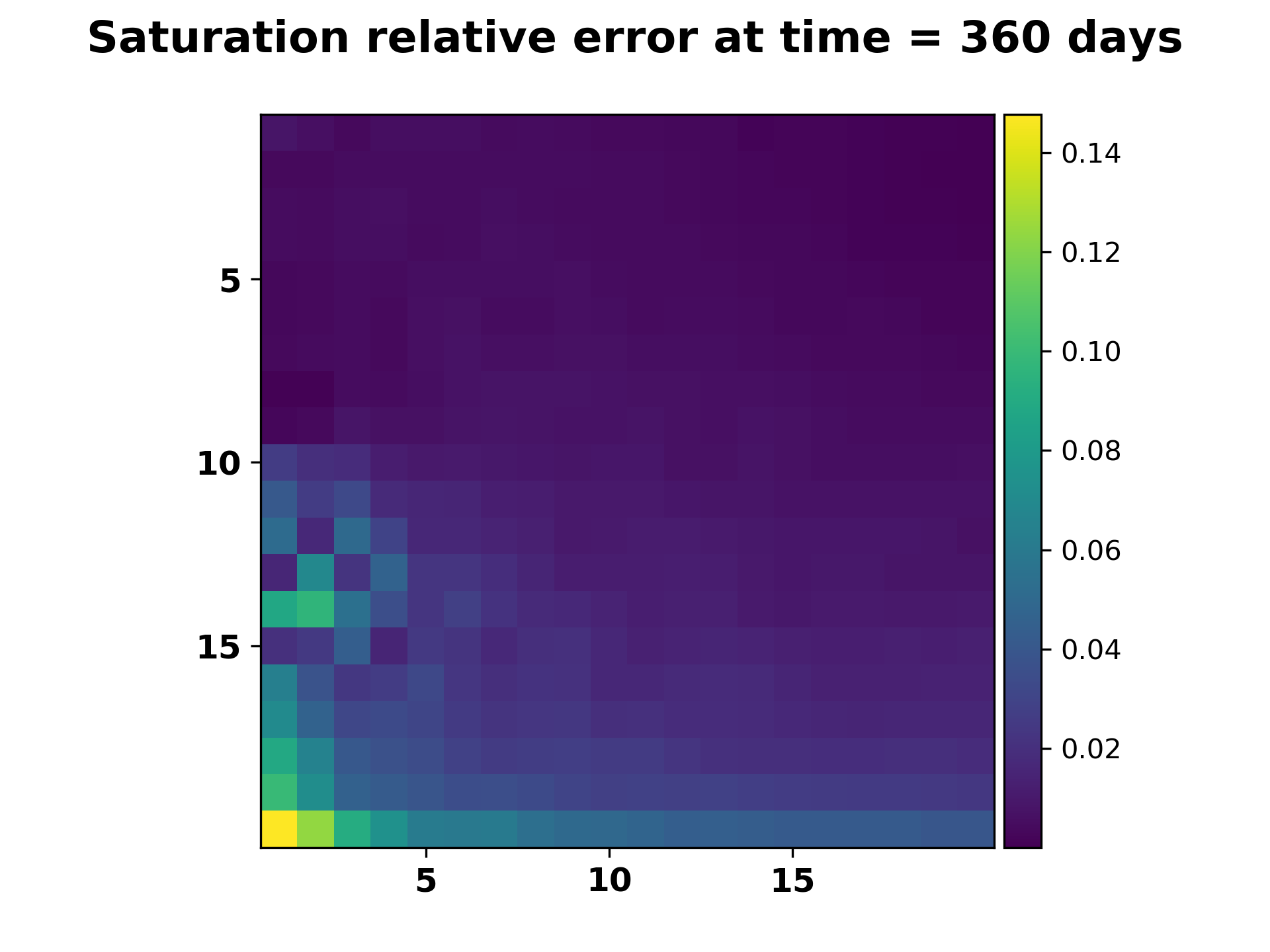}
		\caption{}
		\label{fig:5_7_3}
	\end{subfigure}%
	\caption{Relative error in saturation at time = (a) 1 day , (b) 70 days and (c) 360 days}
	\label{fig:5_7}
\end{figure}

Figures \ref{fig:5_4} and \ref{fig:5_5} show the comparison between predicted and true states (pressure and saturation) for the new well configuration at two different times. We also show the error in states along time. As can be seen in Figures \ref{fig:5_6} and \ref{fig:5_7}, prediction of the overall state behavior is fairly accurate but there is discrepancy in the solution close to the producer well location for pressure, and at the fluid front for saturation.

\begin{figure}[htb!]
	\centering
	\begin{subfigure}{0.30\textwidth}
		\centering
		\includegraphics[width=\textwidth]{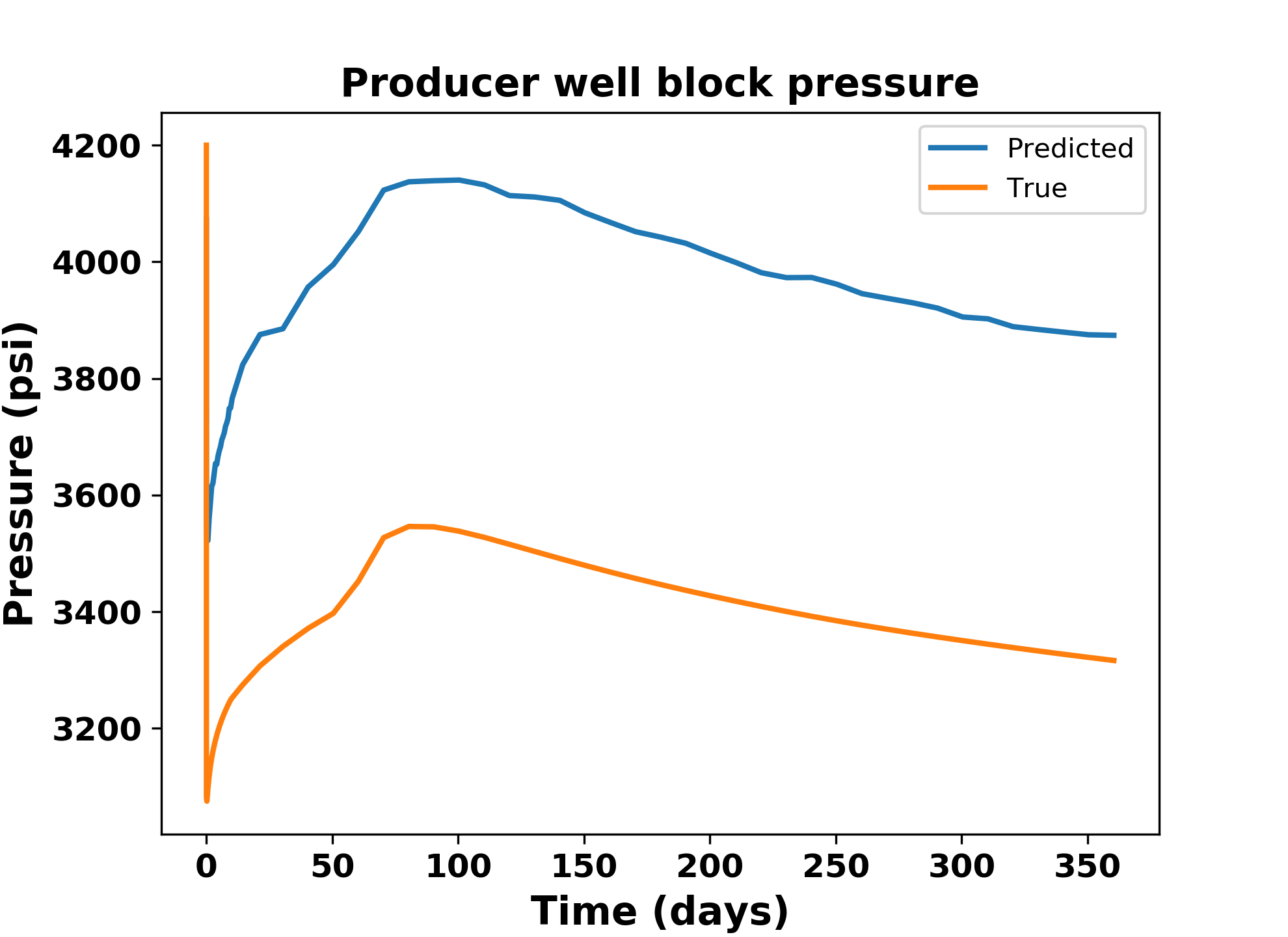}
		\caption{}
		\label{fig:5_8_1}
	\end{subfigure}%
	~
	\centering
	\begin{subfigure}{0.30\textwidth}
		\centering
		\includegraphics[width=\textwidth]{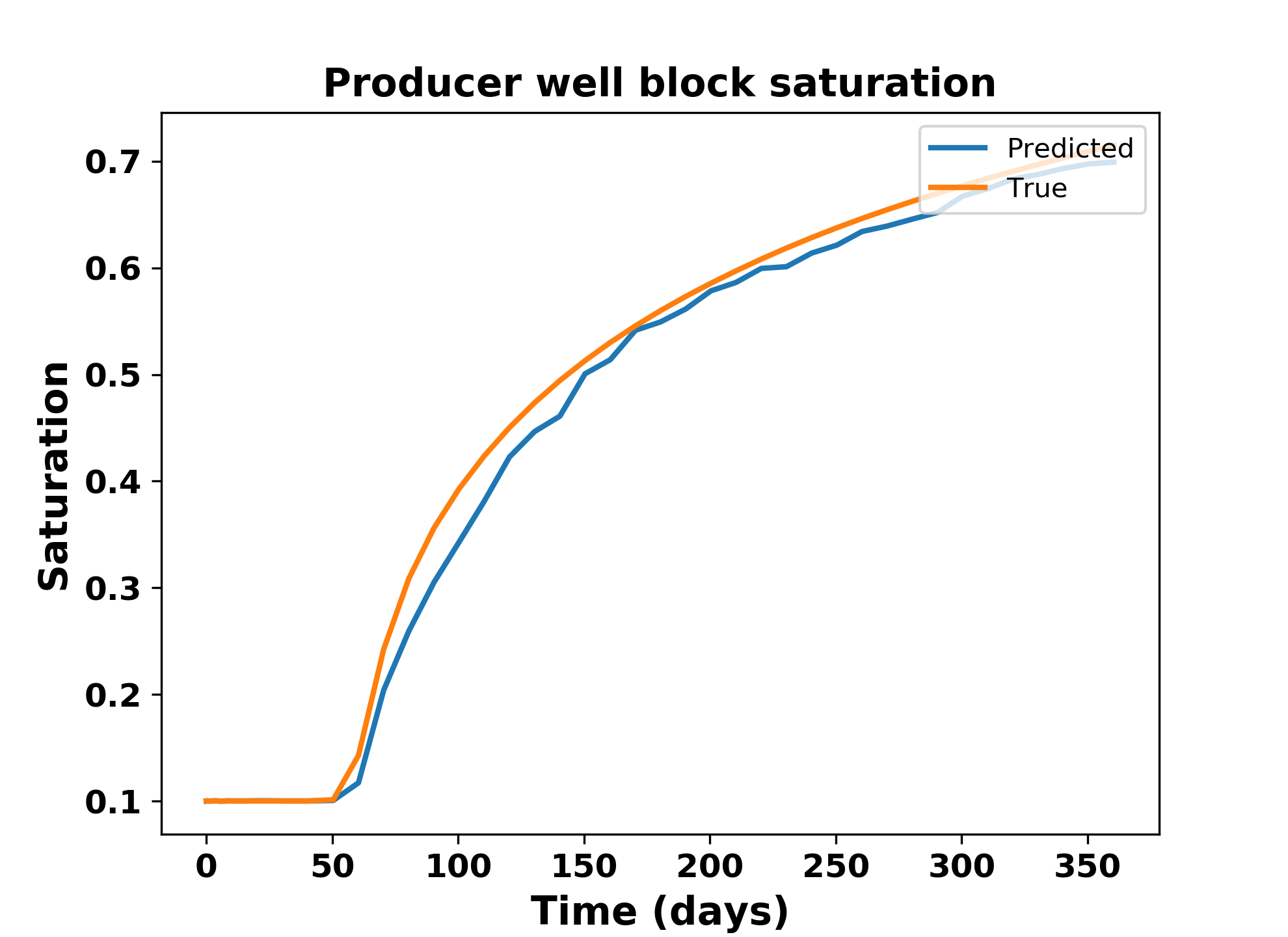}
		\caption{}
		\label{fig:5_8_2}
	\end{subfigure}%
	\caption{Well block state solution comparison (a) Pressure and (b) Saturation}
	\label{fig:5_8}
\end{figure}

\begin{figure}[htb!]
	\centering
	\begin{subfigure}{0.30\textwidth}
		\centering
		\includegraphics[width=\textwidth]{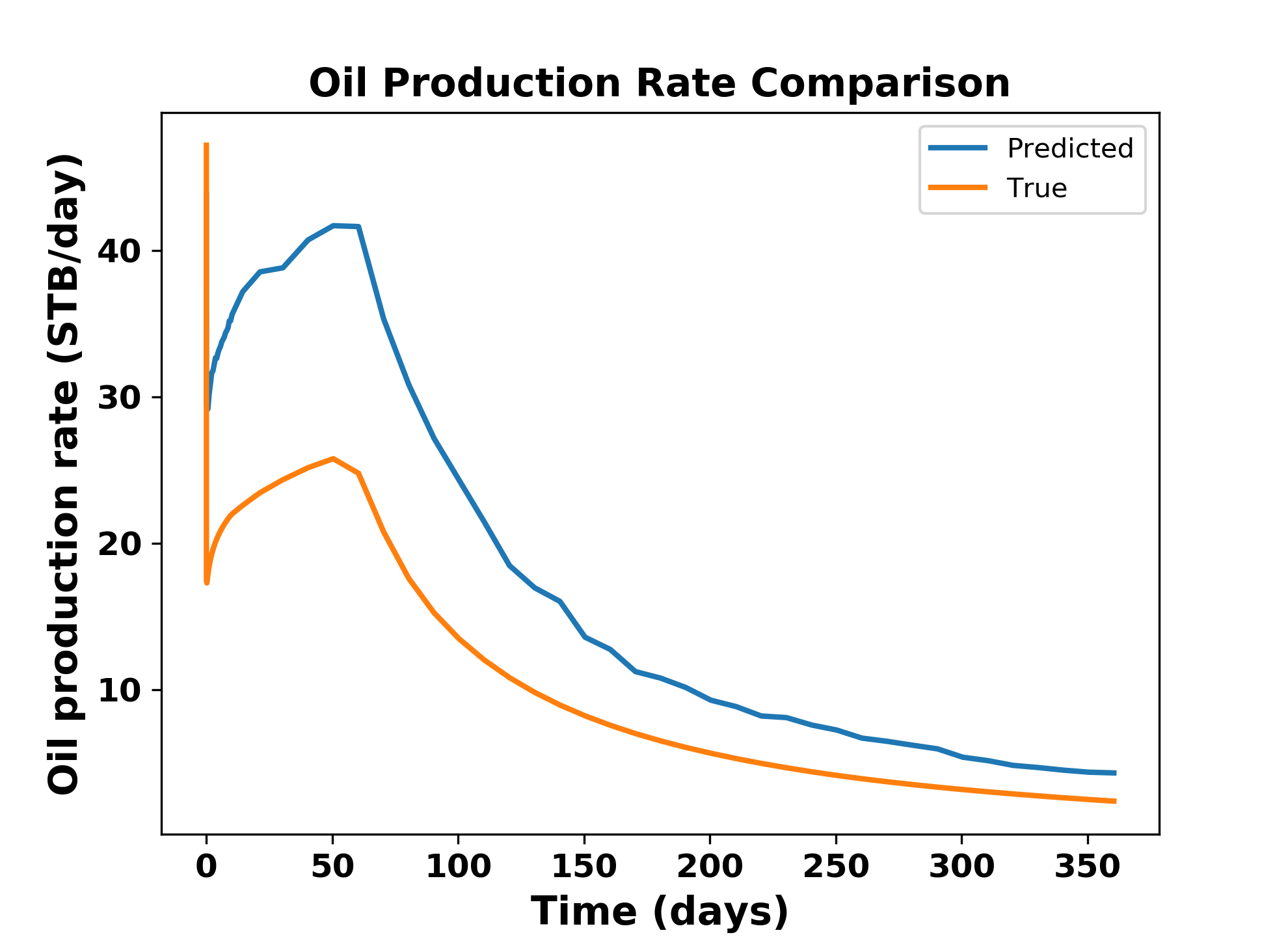}
		\caption{}
		\label{fig:5_9_1}
	\end{subfigure}%
	~
	\centering
	\begin{subfigure}{0.30\textwidth}
		\centering
		\includegraphics[width=\textwidth]{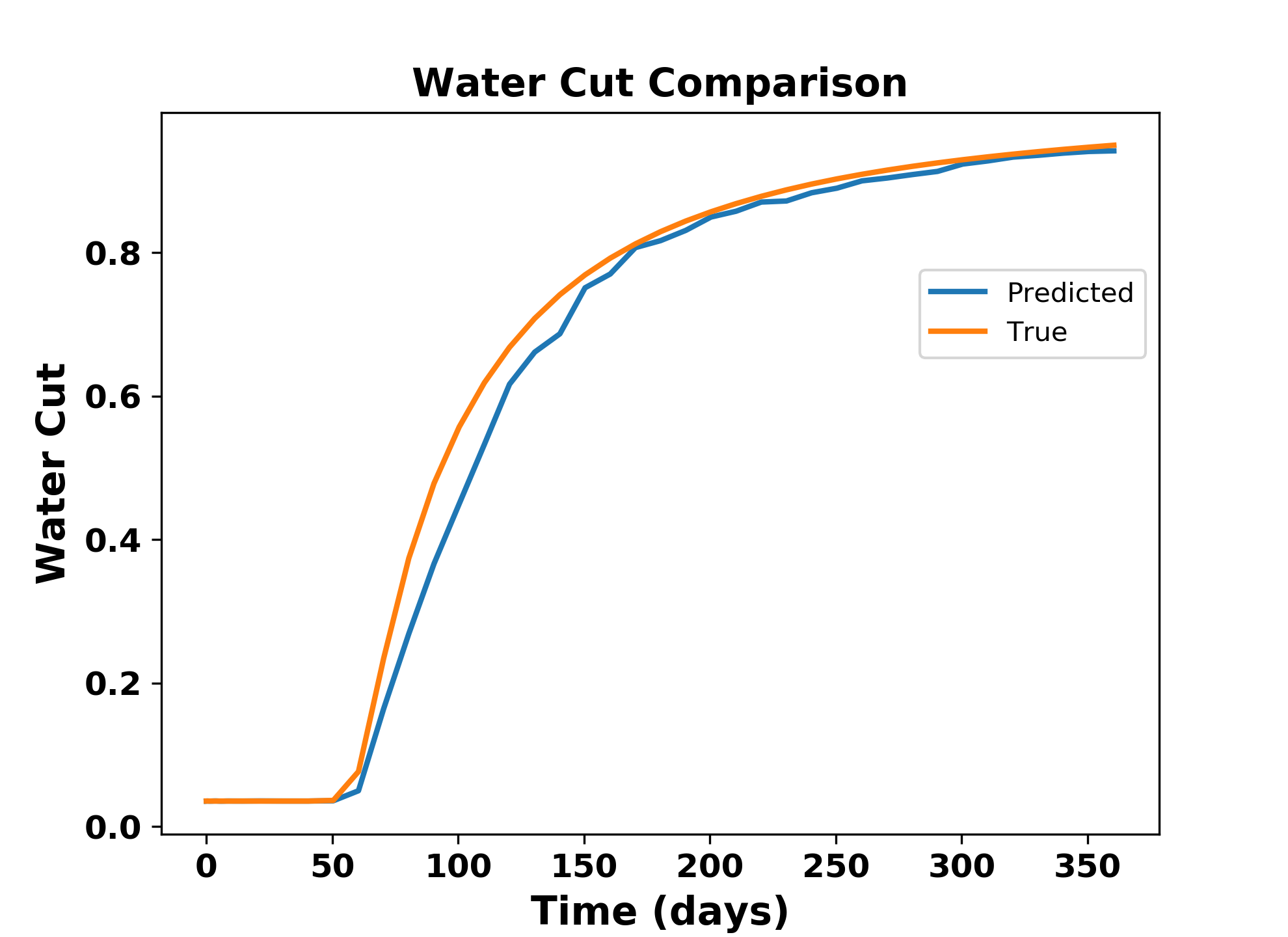}
		\caption{}
		\label{fig:5_9_2}
	\end{subfigure}%
	\caption{Quantities of Interest comparison (a) Oil production rate and (b) Water cut}
	\label{fig:5_9}
\end{figure}

The quantities of interest that are the outputs of the simulation like fluid production rates, water cut etc. are solely dependent on the state quantities at the well blocks through the Peaceman equation (\ref{eq:4}). Thus, for this prediction results we expect some error in these quantities due to higher prediction error close to the wells. To quantify these discrepancies, we show the pressure and saturation solution comparison at the producer well block in Figure \ref{fig:5_8}. As we can see, the predicted solution captures the overall trend of dynamical behavior of the states but has kind of a bias in the solution. Figures \ref{fig:5_9_1} and \ref{fig:5_9_2} show the comparison of oil production rate and water cut between the true and predicted solution. 

\begin{figure}[!ht]
	\centering
	\includegraphics[scale=0.26]{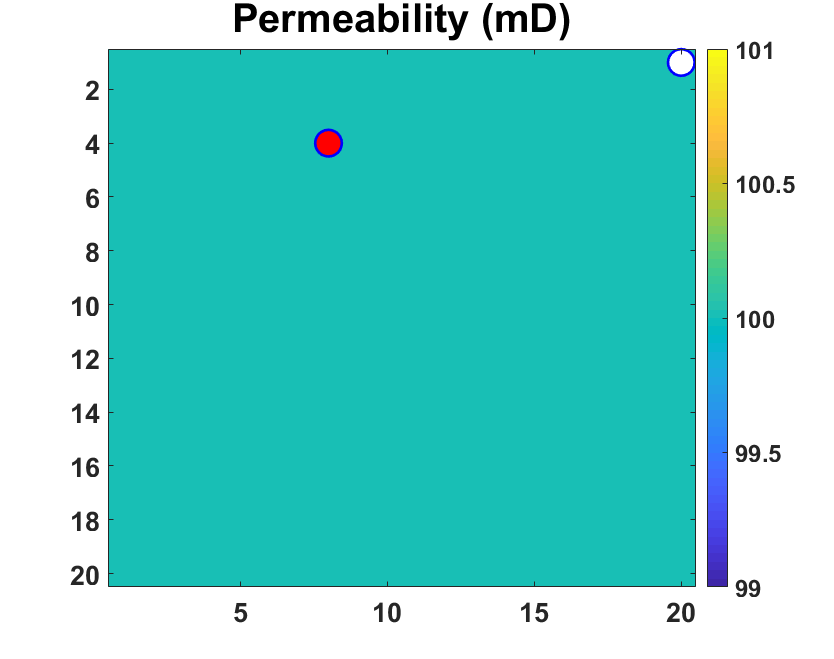}
	\caption{Test case 2 with producer at (8,4) grid block and injector at (20,1) gridblock in homogeneous permeability reservoir}
	\label{fig:5_13_1}
\end{figure}

\begin{figure}[htb!]
	\centering
	\begin{subfigure}{0.45\textwidth}
		\centering
		\includegraphics[width=\textwidth]{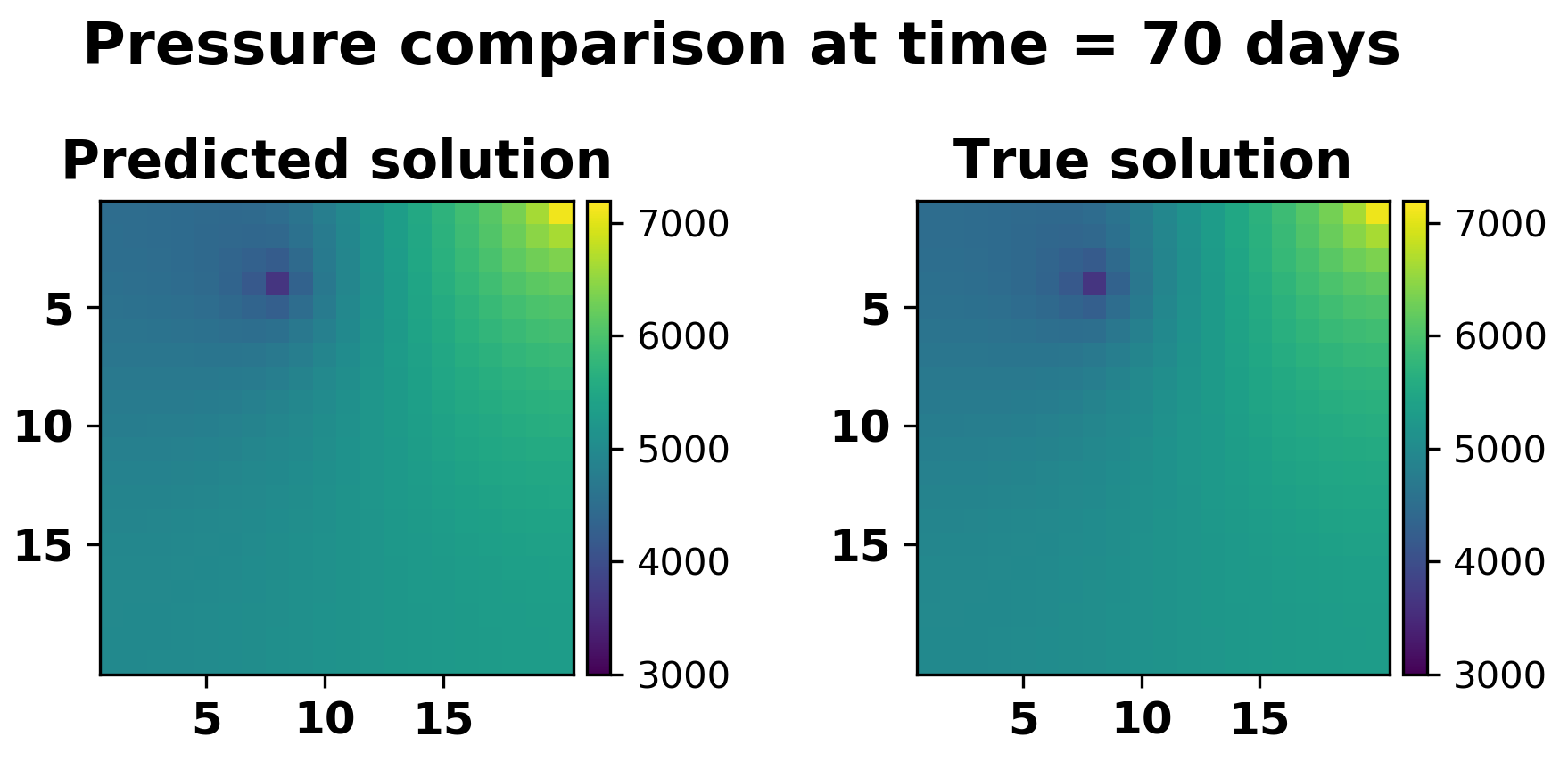}
		\caption{}
		\label{fig:5_13_2}
	\end{subfigure}%
	~
	\centering
	\begin{subfigure}{0.45\textwidth}
		\centering
		\includegraphics[width=\textwidth]{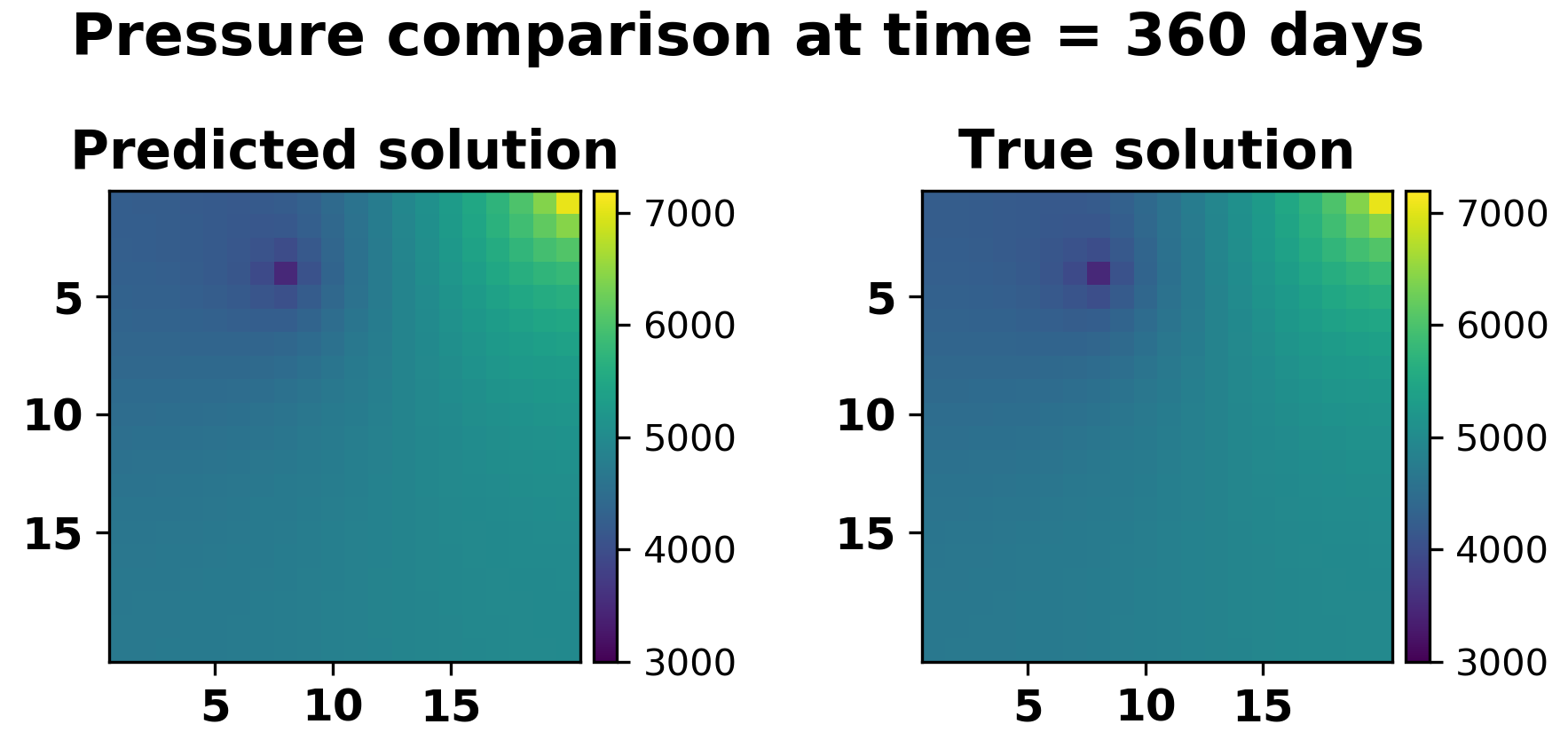}
		\caption{}
		\label{fig:5_13_3}
	\end{subfigure}%
	\caption{Pressure solution comparison at (a) Time = 70 days and (b) Time = 360 days}
	\label{fig:5_13}
\end{figure}

\begin{figure}[htb!]
	\centering
	\begin{subfigure}{0.45\textwidth}
		\centering
		\includegraphics[width=\textwidth]{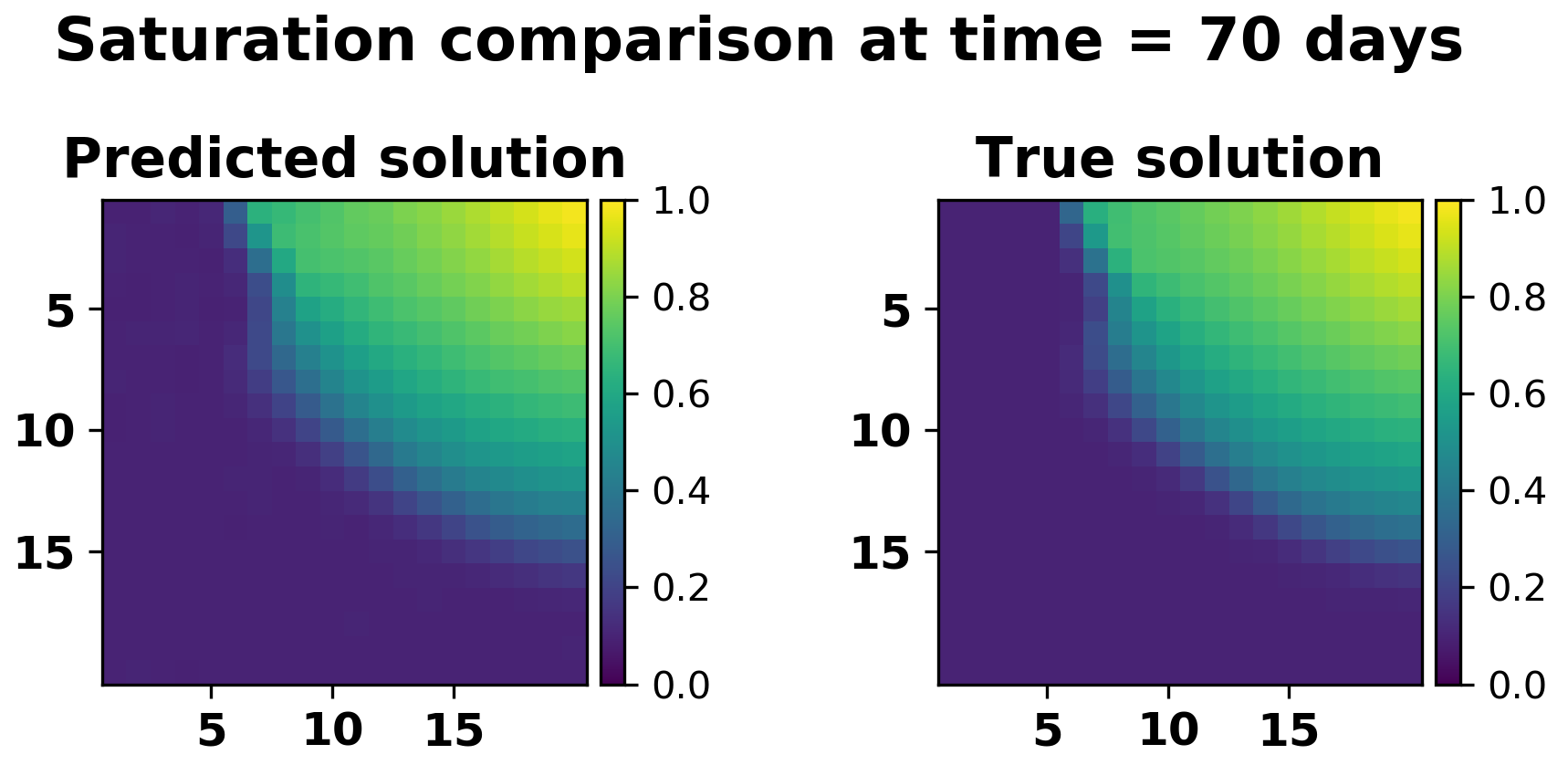}
		\caption{}
		\label{fig:5_13_4}
	\end{subfigure}%
	~
	\centering
	\begin{subfigure}{0.45\textwidth}
		\centering
		\includegraphics[width=\textwidth]{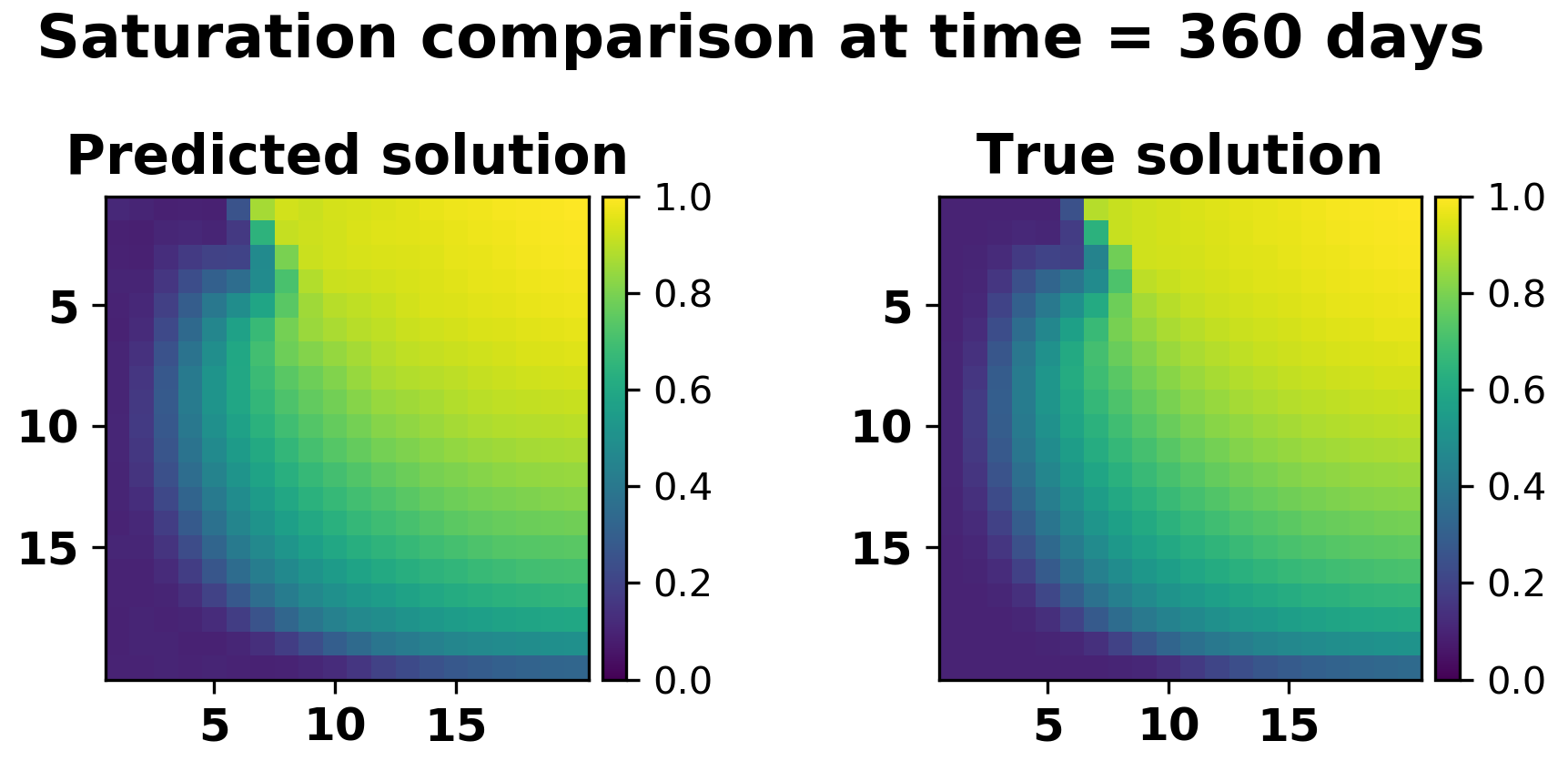}
		\caption{}
		\label{fig:5_13_5}
	\end{subfigure}%
	\caption{Saturation solution comparison at (a) Time = 70 days and (b) Time = 360 days}
	\label{fig:5_13_6}
\end{figure}

\begin{figure}[htb!]
	\centering
	\begin{subfigure}{0.3\textwidth}
		\centering
		\includegraphics[width=\textwidth]{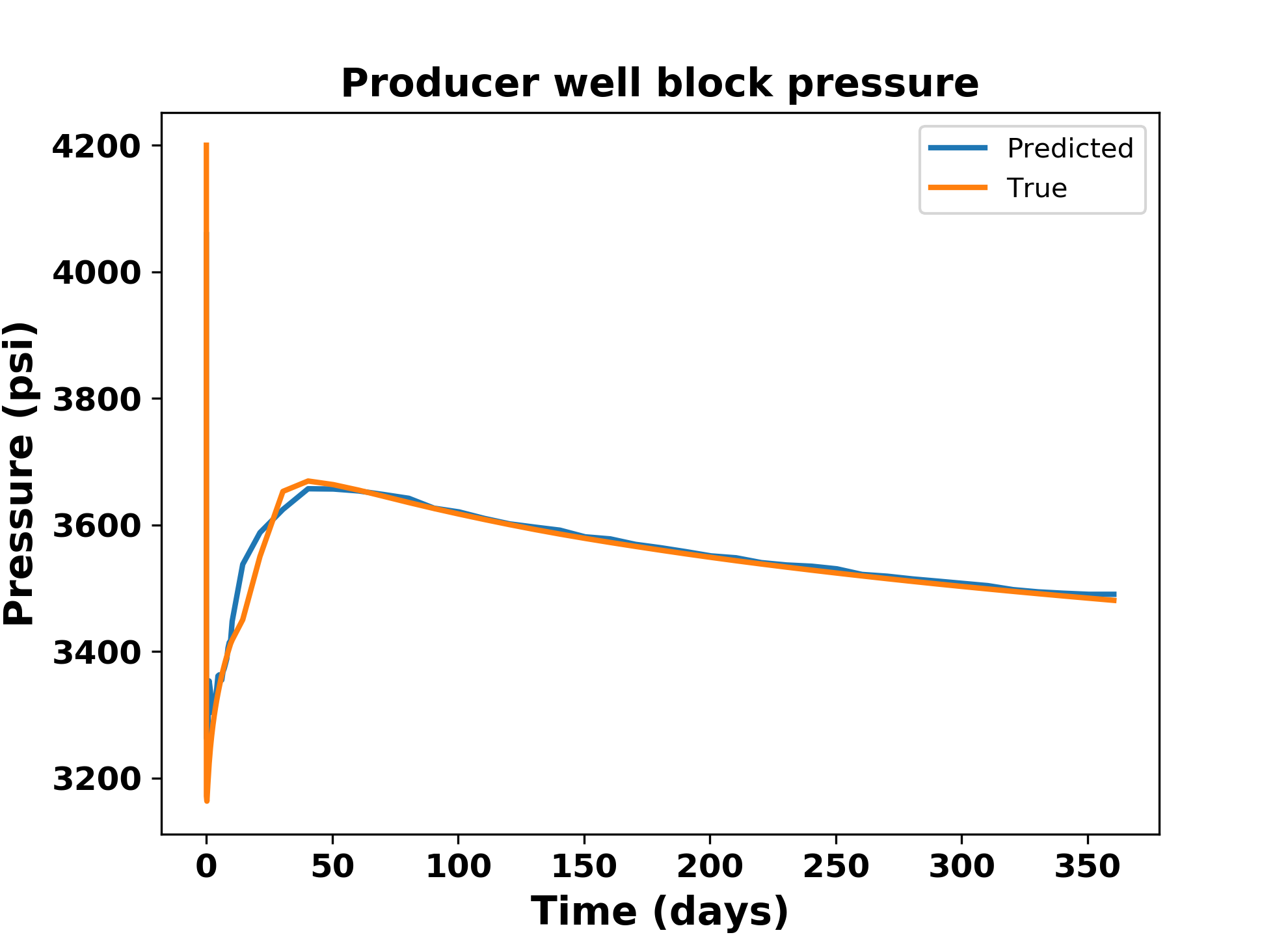}
		\caption{}
		\label{fig:5_14_1}
	\end{subfigure}%
	~
	\centering
	\begin{subfigure}{0.3\textwidth}
		\centering
		\includegraphics[width=\textwidth]{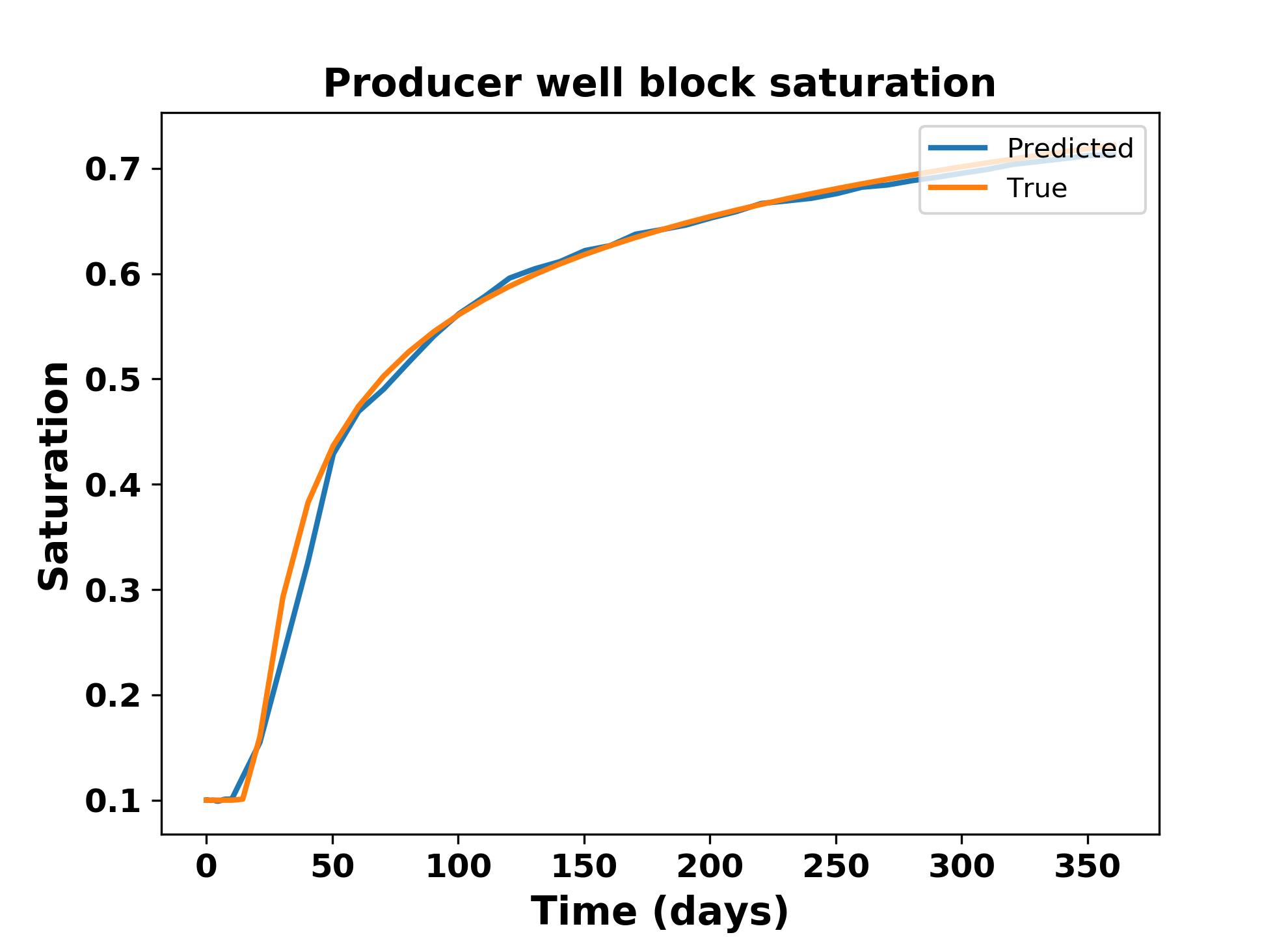}
		\caption{}
		\label{fig:5_14_2}
	\end{subfigure}%
	\caption{Well block state solution comparison (a) Pressure and (b) Saturation}
	\label{fig:5_14}
\end{figure}

\begin{figure}[htb!]
	\centering
	\begin{subfigure}{0.3\textwidth}
		\centering
		\includegraphics[width=\textwidth]{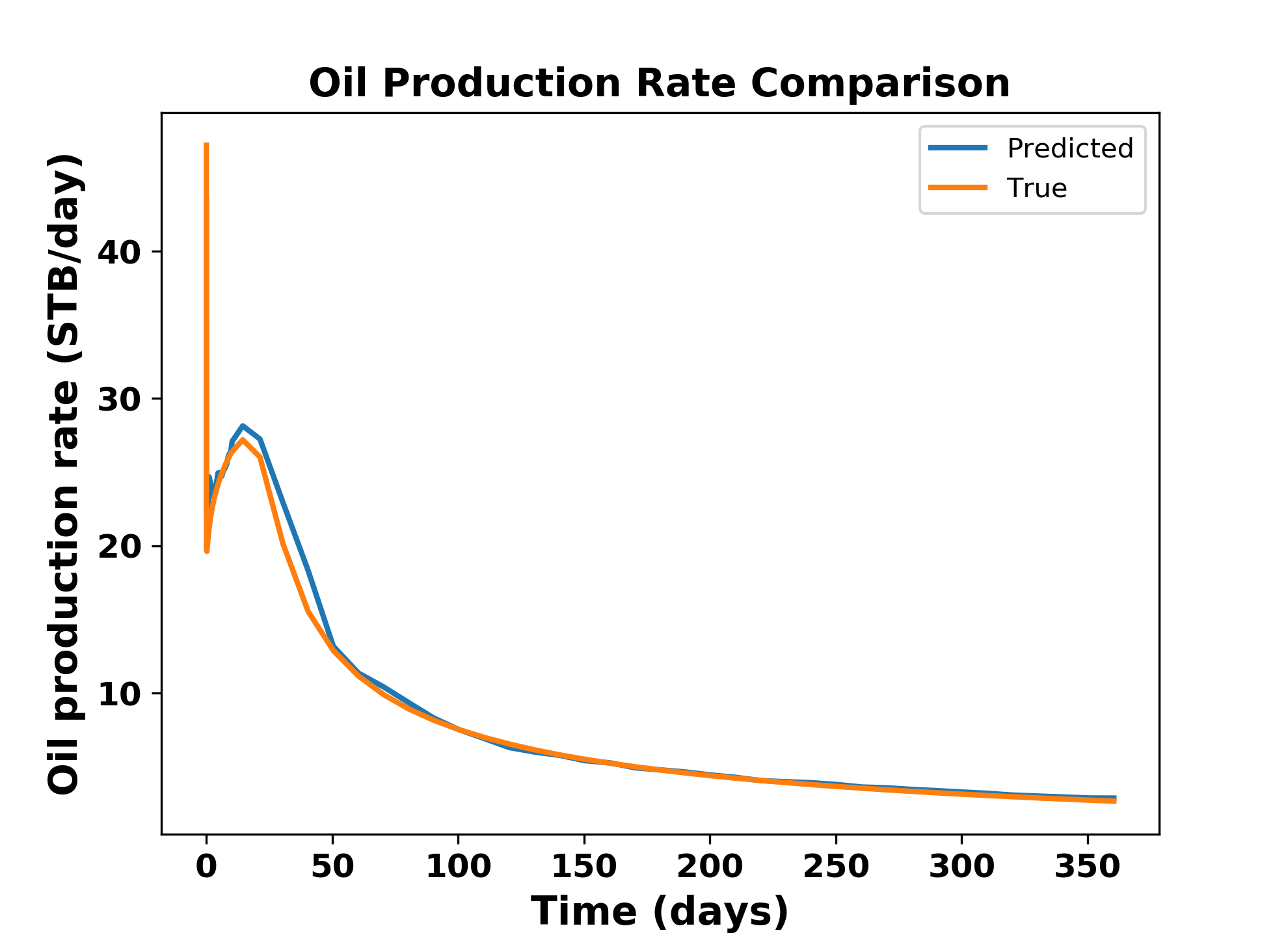}
		\caption{}
		\label{fig:5_15_1}
	\end{subfigure}%
	~
	\centering
	\begin{subfigure}{0.3\textwidth}
		\centering
		\includegraphics[width=\textwidth]{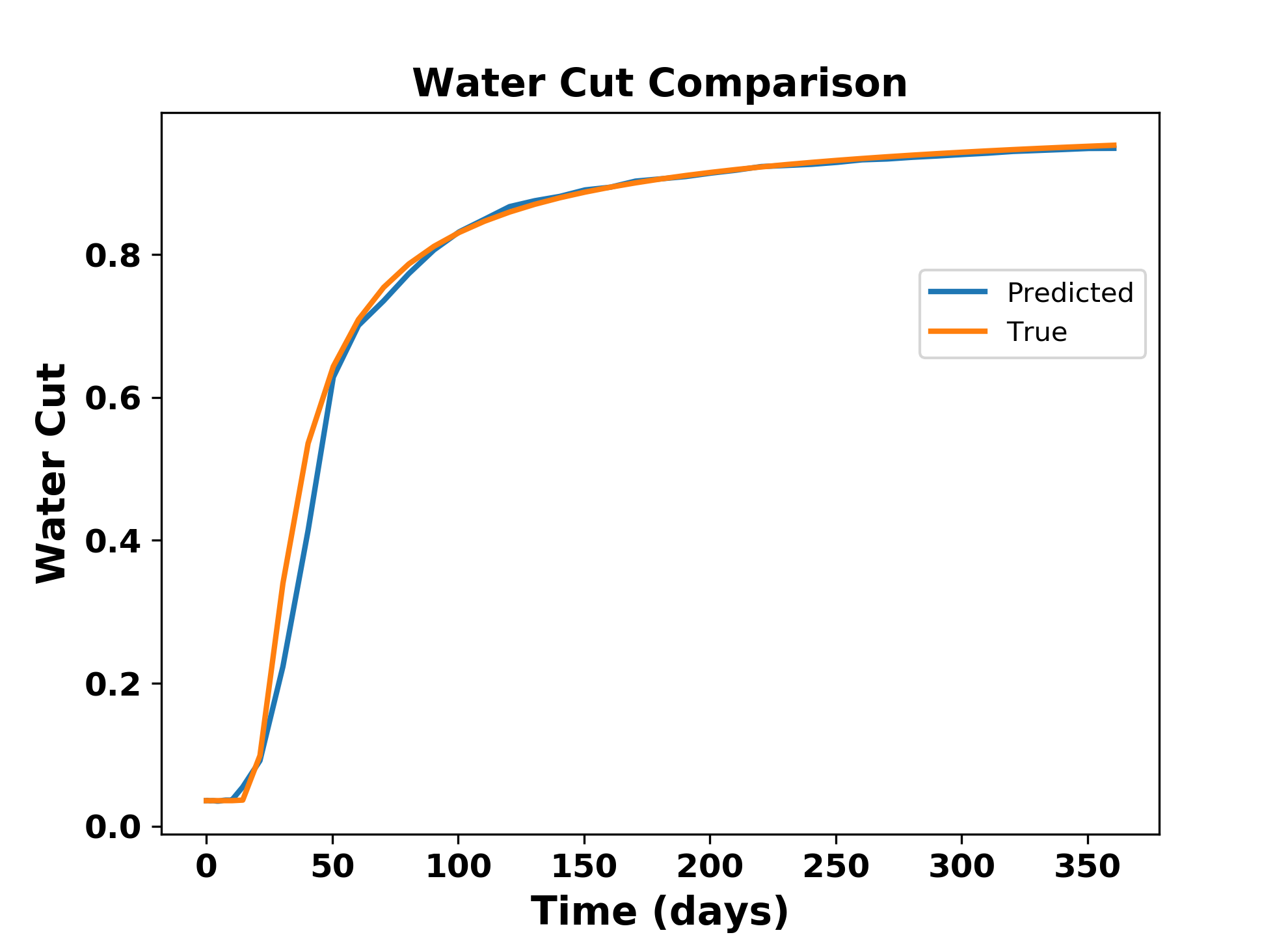}
		\caption{}
		\label{fig:5_15_2}
	\end{subfigure}%
	\caption{Quantities of Interest comparison (a) Oil production rate and (b) Water cut}
	\label{fig:5_15}
\end{figure}

We also consider a second test case with new producer well location as shown in Figure \ref{fig:5_13_1} to validate the method. Figures \ref{fig:5_13} and \ref{fig:5_13_6} shows the pressure and saturation comparison between ML predicted and true solutions. Both the solutions show a very good agreement visually. For a more detailed analysis, we also plot the well block pressure and saturation in Figure \ref{fig:5_14_1} and (\ref{fig:5_14_2}). This case shows that ML predicts the solution with a very good accuracy which eventually reflects in the Figure \ref{fig:5_15}, where we plot the oil production rate and water cut. 

\begin{figure}[htb!]
	\centering
	\begin{subfigure}{0.3\textwidth}
		\centering
		\includegraphics[width=\textwidth]{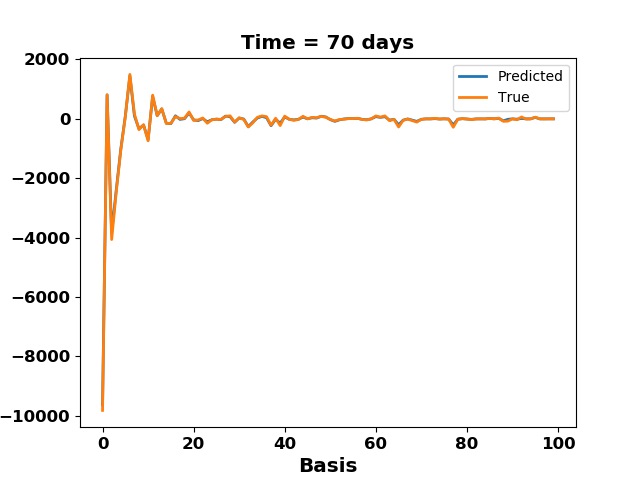}
		\caption{}
		\label{fig:5_10_1}
	\end{subfigure}%
	~
	\centering
	\begin{subfigure}{0.3\textwidth}
		\centering
		\includegraphics[width=\textwidth]{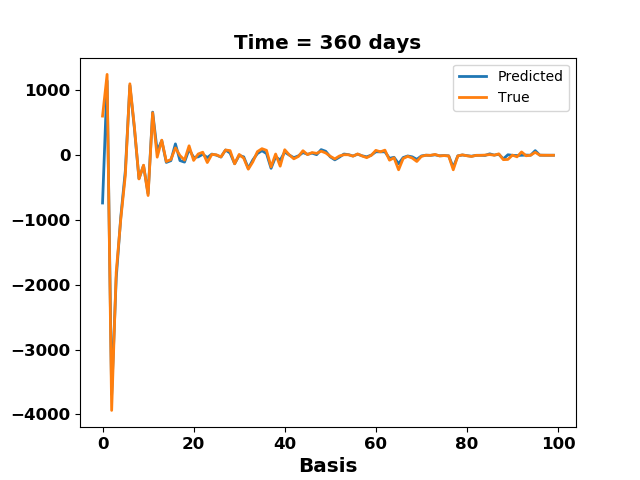}
		\caption{}
		\label{fig:5_10_2}
	\end{subfigure}%
	\caption{True and ML predicted pressure basis coefficient comparison at time =  (a) 70 days and (b) 360 days}
	\label{fig:5_10}
\end{figure}

\begin{figure}[htb!]
	\centering
	\begin{subfigure}{0.3\textwidth}
		\centering
		\includegraphics[width=\textwidth]{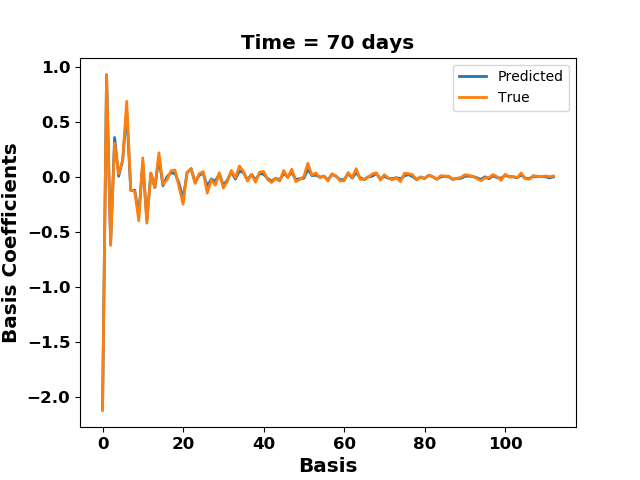}
		\caption{}
		\label{fig:5_11_1}
	\end{subfigure}%
	~
	\centering
	\begin{subfigure}{0.3\textwidth}
		\centering
		\includegraphics[width=\textwidth]{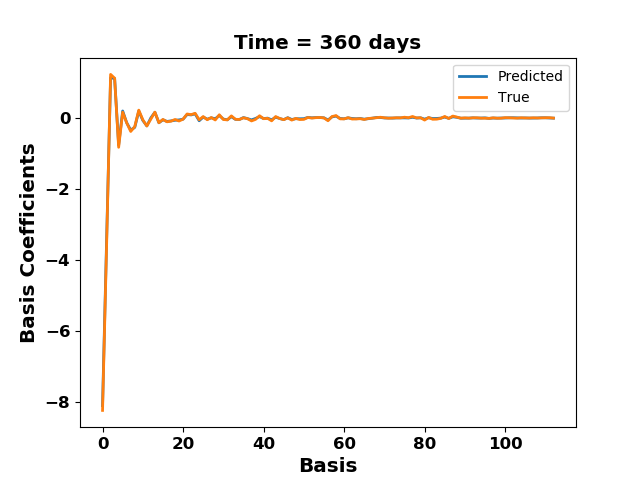}
		\caption{}
		\label{fig:5_11_2}
	\end{subfigure}%
	\caption{True and ML predicted saturation basis coefficient comparison at time =  (a) 70 days and (b) 360 days}
	\label{fig:5_11}
\end{figure}

After plotting the results from other test cases, we found that for some cases the prediction is reasonably accurate, similar to that in test case 2. But there are many well configurations that show discrepancies as the test case 1 which overall captures the solution trend but there is some bias associated. So, we need to analyze the reason behind the solution discrepancy which can either be due to the error in machine learning model or due to the quality of global basis. In order to get an intuition about these two factors, we first analyze the machine learning model performance. In Figures \ref{fig:5_10} and \ref{fig:5_11}, we compare the predicted and true basis coefficients for pressure and saturation respectively for different timesteps for test case 1. These show that the ML model predicts the coefficients with a good accuracy except for some instances like the first pressure basis at time 360 days. The true coefficients for a new parameter $\zeta^*$ are obtained by projecting the fine scale simulation on the subspace spanned by global basis $\Phi$ by computing $\Phi_\textbf{x}^T \textbf{x}(\zeta^*)$. 

Thus, the performance of ML model suggests that the discrepancies near the well locations are less likely due to the errors in ML model, specially the fact in understanding that the new producer well location should be the lowest pressure point in the reservoir. Thus, now we analyze the quality of the global basis $\Phi$ used on test case 1. Here, we look back at Section \ref{sec:1.3}, where we introduced the concept of MOR error that has a orthogonal component which basically results from neglecting the state projection on the orthogonal subspace. At a given time instant for a new parameter $\zeta^*$, it can be written as:

\begin{equation}
\centering
\begin{aligned}
\epsilon_{\Phi^\perp}(t,\zeta^*) &= (I_n - \Pi_{\Phi, \Phi})\textbf{x}(t,\zeta^*)\\
\Pi_{\Phi, \Phi} &= \Phi\Phi^T\label{eq:5_12}
\end{aligned}
\end{equation}

\begin{figure}[htb!]
	\centering
	\begin{subfigure}{0.6\textwidth}
		\centering
		\includegraphics[width=\textwidth]{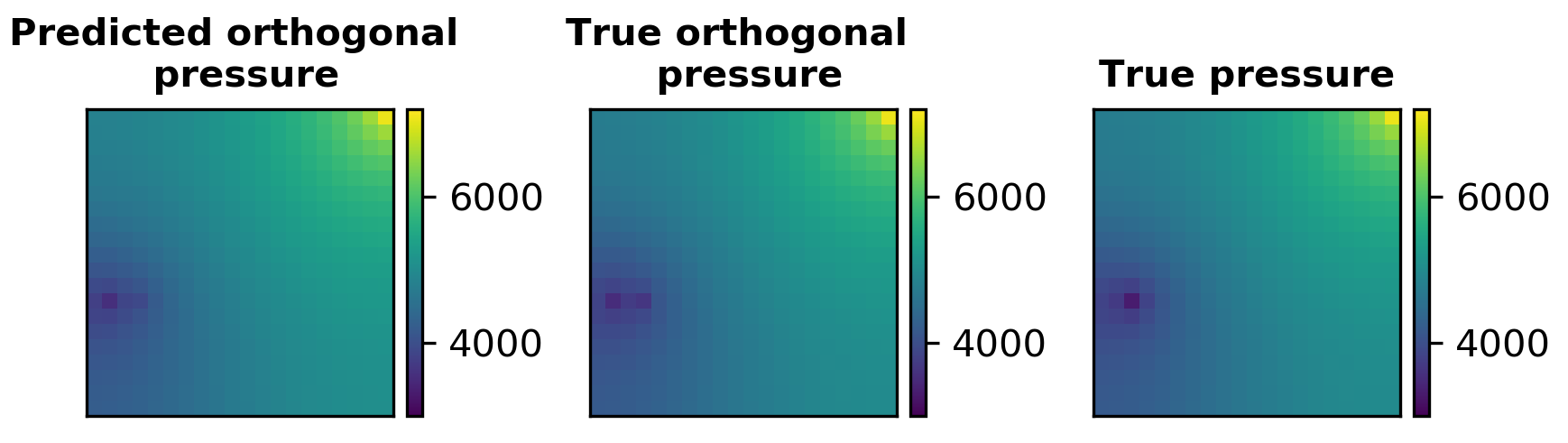}
		\caption{}
		\label{fig:5_12_1}
	\end{subfigure}%
	~\\
	\centering
	\begin{subfigure}{0.6\textwidth}
		\centering
		\includegraphics[width=\textwidth]{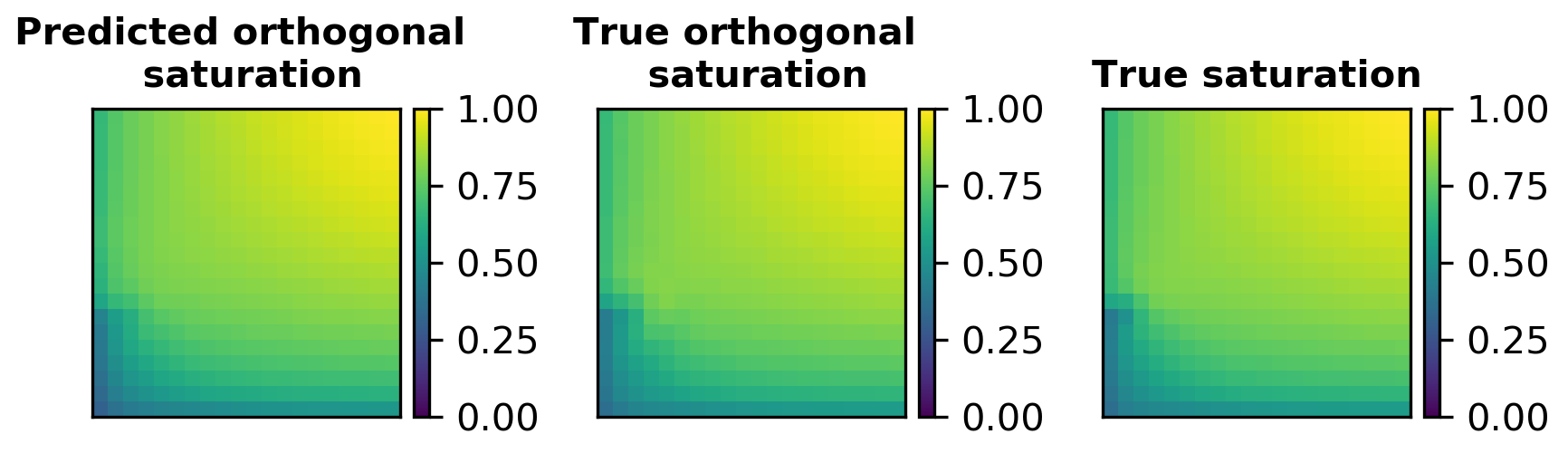}
		\caption{}
		\label{fig:5_12_2}
	\end{subfigure}%
	\caption{Comparison of predicted orthogonal and true orthogonal solutions with true solution at time = 360 days for (a) Pressure and (b) Saturation}
	\label{fig:5_12}
\end{figure}

Thus, the deviation of the product $\Phi \Phi^T$ from the identity matrix gives an a priori estimation about the quality of basis for the new parameter. To visualize this for our case, we project the true coefficients obtained by $\Phi_\textbf{x}^T \textbf{x}(\zeta^*)$ back to the fine scale domain. This is basically performing the operation $ \Pi_{\Phi, \Phi}\textbf{x}(t,\zeta^*)$ which we refer to here as the true orthogonal solution. The Figures \ref{fig:5_12_1} and \ref{fig:5_12_2} show the predicted orthogonal solution, true orthogonal solution and true fine scale solution for pressure and saturation respectively at the end of simulation for test case 1. As can be seen, the true orthogonal pressure and saturation solutions differ from the true solutions around the producer well. This is clearly visible for the case of pressure solution. However, the predicted orthogonal pressure and saturation solutions are in a very good agreement to their true orthogonal solutions. This is an indication of the quality of basis that is the main reason behind the higher errors around producer well locations. This indicates that even with a global basis of parameters, it is very difficult to preserve the controllability properties for all the new well locations. The global basis here proves to be a good quality basis for some cases like the test case 2. Thus, the best possible way to alleviate this problem is to try getting the most representative set of parameters in the training sample set which can be a future direction of research. 

\section{Modeling error correction} \label{sec:1.7}
As we saw that finding a good global basis for the entire domain of parameters is very difficult, we get solution discrepancies although it captures the overall trend of the solution. So with the given information about the reduced order model, we propose an error correction method to adjust the solution discrepancies. The error correction model takes into account the information about the reduced order model solution at a new parameter and then predicts the correction required in the quantity of interest. This model is constructed using machine learning techniques to account for high dimensional feature space and mapping the complex non-linear relationship between inputs and outputs. The quantity of interest can be the oil/water production rate, water cut or the well block states. For the current work, we just consider the well block states (pressure and saturation) as the QoI to be corrected as these states directly transfer to the correction in production rates and water cut. At a given instant in time, the error in the well block states is:
\begin{equation}
\begin{aligned}
&\Delta{\textbf{x}_{wb}}(\zeta, t_k) = (\textbf{x}_{wb}(\zeta, t_k))_{fine}  - (\textbf{x}_{wb}(\zeta, t_k))_{PMOR} \\
&where, \enspace \enspace k = 0, 1, ..., N_t \enspace and \enspace \zeta \in \mathbb{P}\label{eq:5_13}
\end{aligned}
\end{equation}

Thus, we are interested in determining $\Delta{\textbf{x}_{wb}}$ for a new well location over time given the reduced model solution obtained in equation (\ref{eq:5_11}). We also note that, the saturation map comparison at different times during prediction (Figure \ref{fig:5_5}) and comparison of the predicted injection rates with true injection rates showed good agreement. This information helps us consider the pore volumes injected predicted by the reduced model as a good reduced model information. The formulation of the error correction model is thus given by:
\begin{equation}
\begin{aligned}
\mathscr{E}(\zeta, t_k, \tilde{\textbf{x}}, PVI_{r}) \to \Delta{\textbf{x}} \label{eq:5_14}
\end{aligned}
\end{equation}
This model is constructed using ML techniques, where, the $\zeta$ correspond to the well location parameter and hence represented by the same features as used in global PMOR formulation. $\tilde{\textbf{x}}$ correspond to the PMOR predicted solution at all the gridblocks in the reservoir and $PVI_{r}$ is the pore volumes injected, predicted by the reduced order model. However, since we are interested in correction of states at the well gridblock only, it is worthwhile considering the relation between the reduced solution at gridblock location $\tilde{\textbf{x}}_{wb}$ to the correction $\Delta{\textbf{x}_{wb}}$. By this assumption, the formulation changes to:
\begin{equation}
\begin{aligned}
\mathscr{E}(\zeta, t_k, \tilde{\textbf{x}}_{wb}, PVI_{r}) \to \Delta{\textbf{x}_{wb}} \label{eq:5_15}
\end{aligned}
\end{equation}
This formulation is thus a global error model that is used for the entire parameter space. Similar idea was implemented to construct local error models for POD-TPWL method \cite{Trehan2016_ML} that showed promising results for well control changes but can be computationally expensive to construct. However, the global error model is used here to avoid further computational complexity to the already expensive PMOR training procedure. Including the reduced order model solution to the input features achieves good accuracy as we take into account the physics of ROMs rather than constructing completely data driven error models as in \cite{Knill1999, Kennedy2001}. 
Similar to the cases of constructing error maps using ML, we have the single output here corresponding to state error at well gridblock. For such cases, NN proved to be a better model at mapping the complex input-output relation and is faster to train for a single output system rather than using it for the prediction of POD coefficients in global PMOR formulation with high dimensional outputs.A brief description of Neural Networks is provided below. 

\subsection{Neural Networks}
Artificial neural networks are nonlinear statistical models that detect pattern in the data by discovering the input-output relationships, used for both regression and classification. The choice of ANN model is motivated by its capability to capture highly nonlinear complex relationship between the input features and the output. A feed-forward neural network used here consists of $L$ layers with each layer consisting of predefined nodes and an input layer consisting of the features or independent variables used for prediction. Each of these layers have an associated transfer function and the nodes are connected to the nodes from previous layers by weights. A simple mathematical description is as follows:
\begin{ceqn}
	\begin{align}
	Z^{l+1} = \Theta^{l+1} X^{l} \label{eq:NN_19}
	\end{align}
\end{ceqn}
Here, $\Theta^{l+1}$ is the weight matrix connecting the nodes between layer $l$ and $l+1$. $X^l$ is the output of the layer $l$ and $Z^{l+1}$ a simple linear regression of these outputs. Then, the output of the layer $l+1$ is given by:
\begin{ceqn}
	\begin{align}
	X^{l+1} = G(Z^{l+1}+B^{l+1}) \label{eq:NN_20}
	\end{align}
\end{ceqn}
$B^{l+1}$ is the bias term added to each node. The nonlinearity of the model arises from the activation function $G$. Most widely used activation functions include sigmoid, hyperbolic-tangent and rectified-linear unit (ReLU). For the regression problem here, we use hyperbolic-tangent activation in the hidden layers (as they work better in most cases than a sigmoid function) and simple linear regression in the output layer.

\begin{figure}
	\centering
	\includegraphics[width=0.48\textwidth]{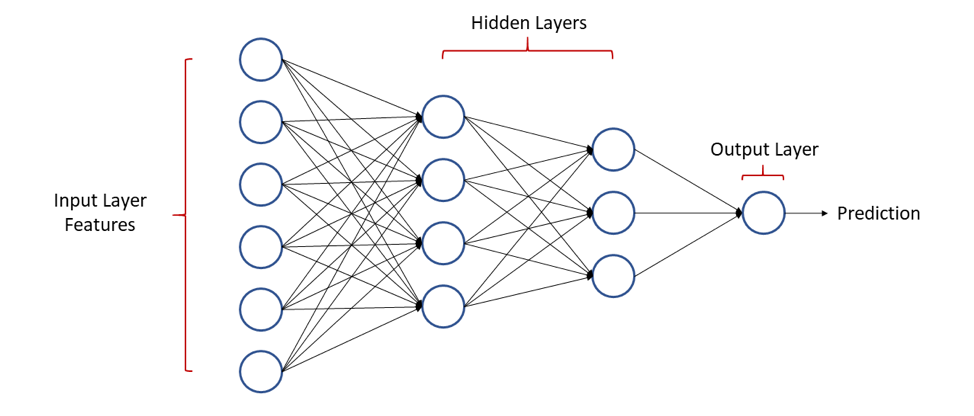}
	\caption{\textbf{Schematic of a feedforward Artificial Neural Network}}
	\label{fig:NN_2}       
\end{figure}

The model is trained using backpropagation and optimization algorithms to adjust the weights $\Theta$ of the network that minimize the cost function. The detailed description on ANN and backpropagation algorithm can be found in (\citealp{Goodfellow2016}). Fig. \ref{fig:NN_2} shows a schematic of ANN.  The neural network usually tend to overfit the data and thus, we use a regularized cost function for regression given by:
\begin{ceqn}
	\begin{align}
	J{(\theta}) = \frac{1}{2m}\Bigg[\displaystyle\sum_{i=1}^{m}\bigg(h_\theta(x^i)-y^i\bigg)^2+\lambda\displaystyle\sum_{j=1}^{n}\theta_{j}^2\Bigg] \label{eq:NN_21}
	\end{align}
\end{ceqn}

Here, $m$ represents total number of data examples, $h_\theta(x)$ represent the predicted output of the network and y is the expected output. $\theta$ represents each of the elements of the weight matrix $\Theta$ in each layer and $\lambda$ is the regularization parameter to control feature importance and prevent overfitting. We use $\lambda$ and number of units in each layer as the tuning parameters while training the ANN. The values of these hyperparameters are chosen here using K-fold cross validation technique.  

\section{Case study} \label{sec:1.8}
\subsection{Homogeneous reservoir model}
We use the error correction model now in efforts to minimize the solution discrepancies at the well locations. We use the same case study as before for the homogeneous reservoir model and changing producer well locations. In order to construct the error model, we sample randomly other 100 well locations to generate the data set that account for running 100 fine scale simulations. In order for the ML models to capture the underlying behavior of the system, it requires many data points especially to understand such complex relationships. We realize the computational expense associated with fine scale simulation runs, but better sampling strategies should be the research focus in the future, that can reduce the number of sampling points. Also, the current case is a small model and hence the number of sampling points relative to the size of the problem is high here, however, as we move to the bigger reservoir models, we expect much lower sample points relative to the reservoir size. 

\begin{figure}[htb!]
	\centering
	\begin{subfigure}{0.27\textwidth}
		\centering
		\includegraphics[width=\textwidth]{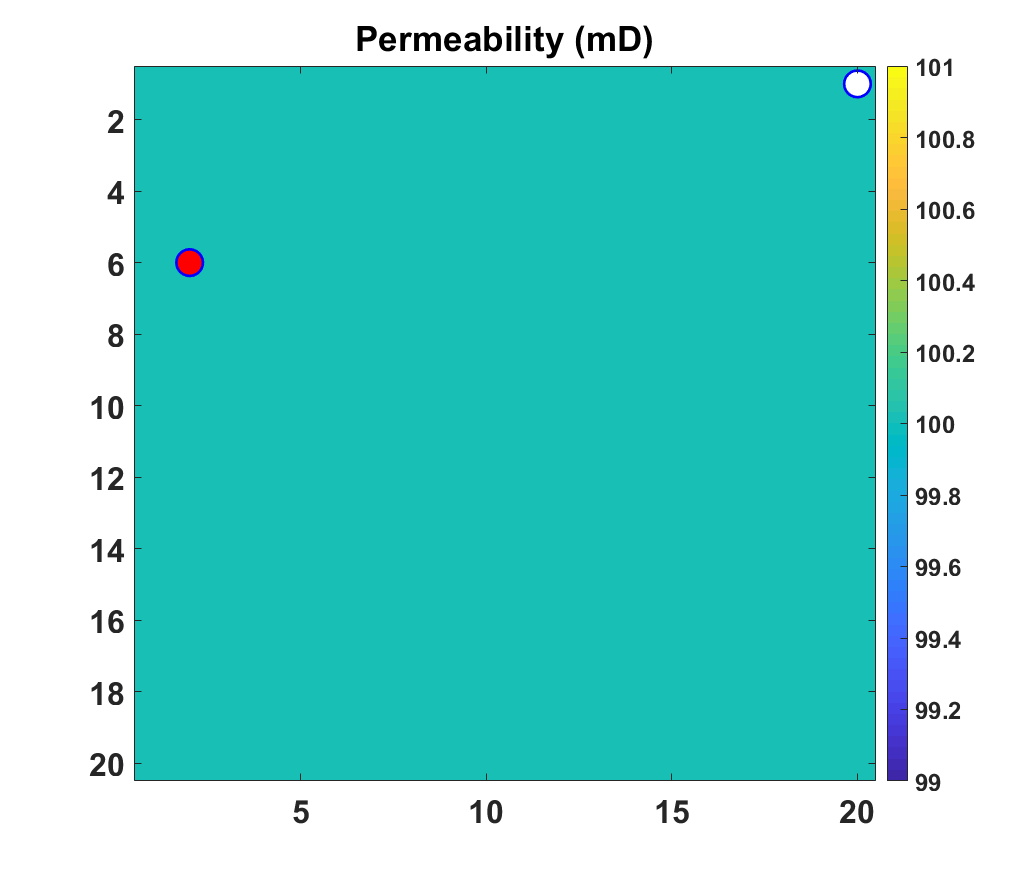}
		\caption{}
		\label{fig:5_16_1}
	\end{subfigure}%
	~
	\centering
	\begin{subfigure}{0.33\textwidth}
		\centering
		\includegraphics[width=\textwidth]{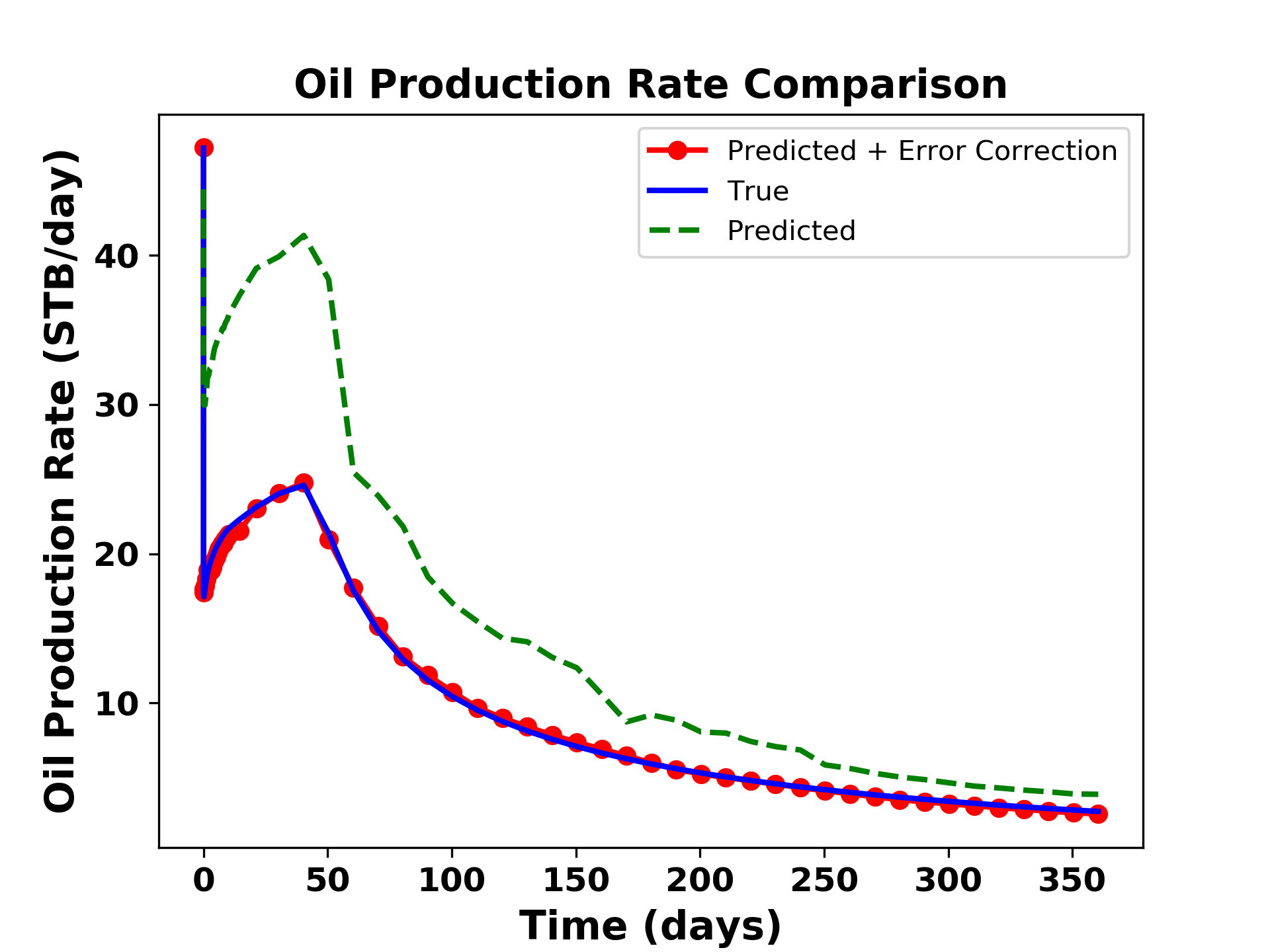}
		\caption{}
		\label{fig:5_16_2}
	\end{subfigure}
	~
	\centering
	\begin{subfigure}{0.33\textwidth}
		\centering
		\includegraphics[width=\textwidth]{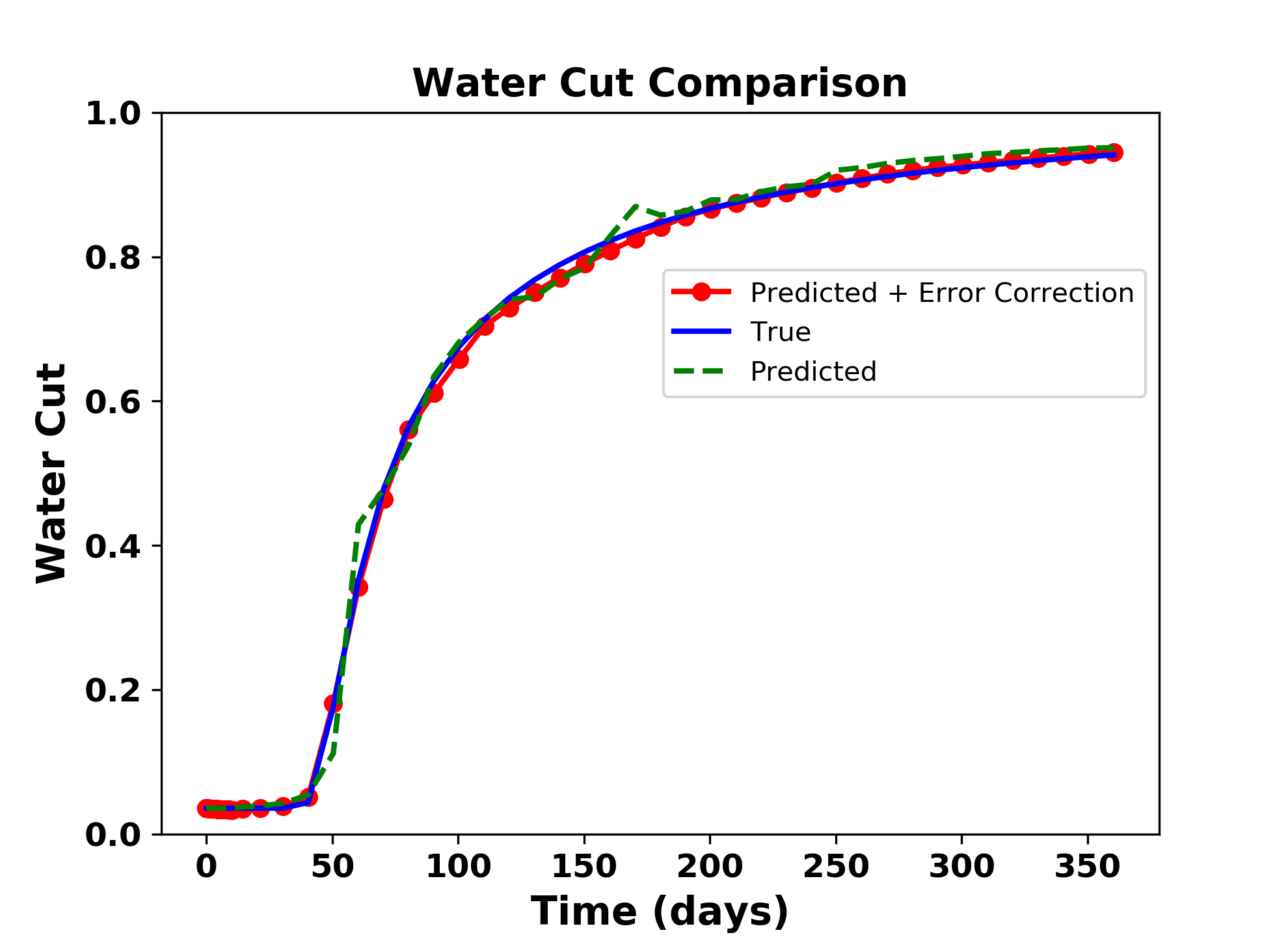}
		\caption{}
		\label{fig:5_16_3}
	\end{subfigure}%
	\caption{(a) Test case with producer well location at (2,6) (b) Comparison of oil production rate and (c) Comparison of water cut, predicted using global PMOR method alone using 100 samples (dotted green line) and after implementation of error correction model (red circled line) with the true solution (blue line)}
	\label{fig:5_16}
\end{figure}

\begin{figure}[htb!]
	\centering
	\begin{subfigure}{0.27\textwidth}
		\centering
		\includegraphics[width=\textwidth]{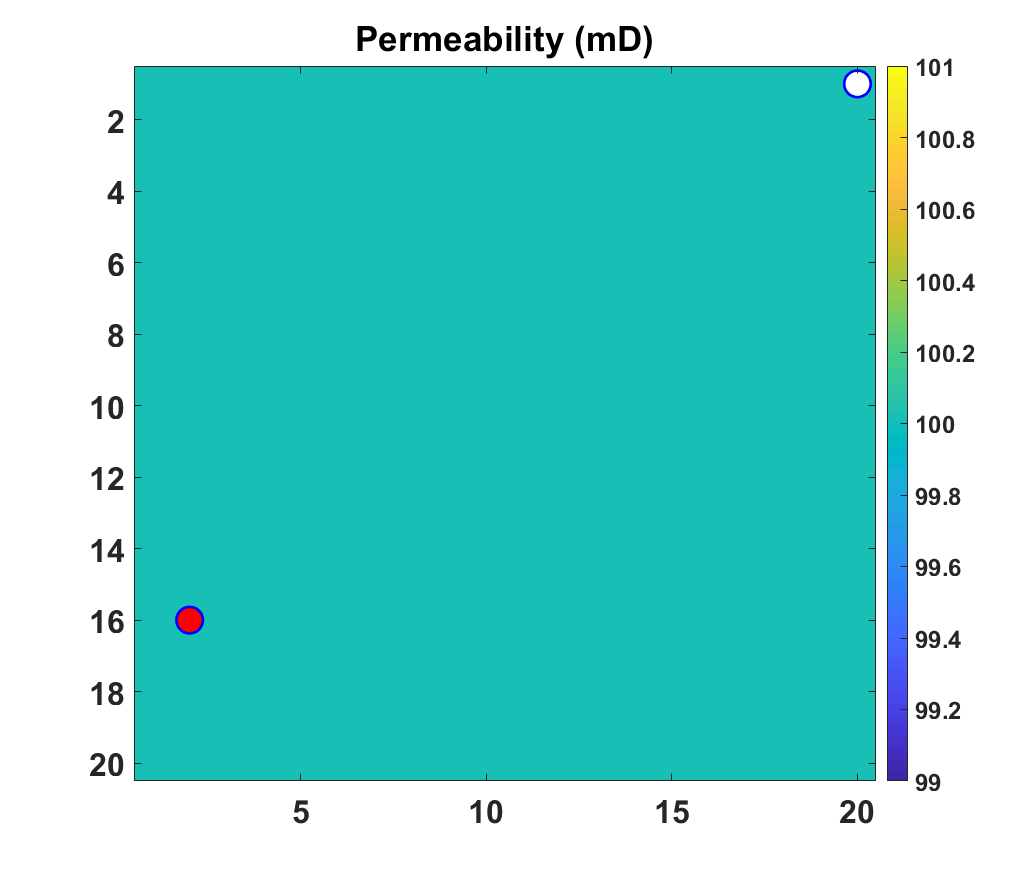}
		\caption{}
		\label{fig:5_17_1}
	\end{subfigure}%
	~
	\centering
	\begin{subfigure}{0.33\textwidth}
		\centering
		\includegraphics[width=\textwidth]{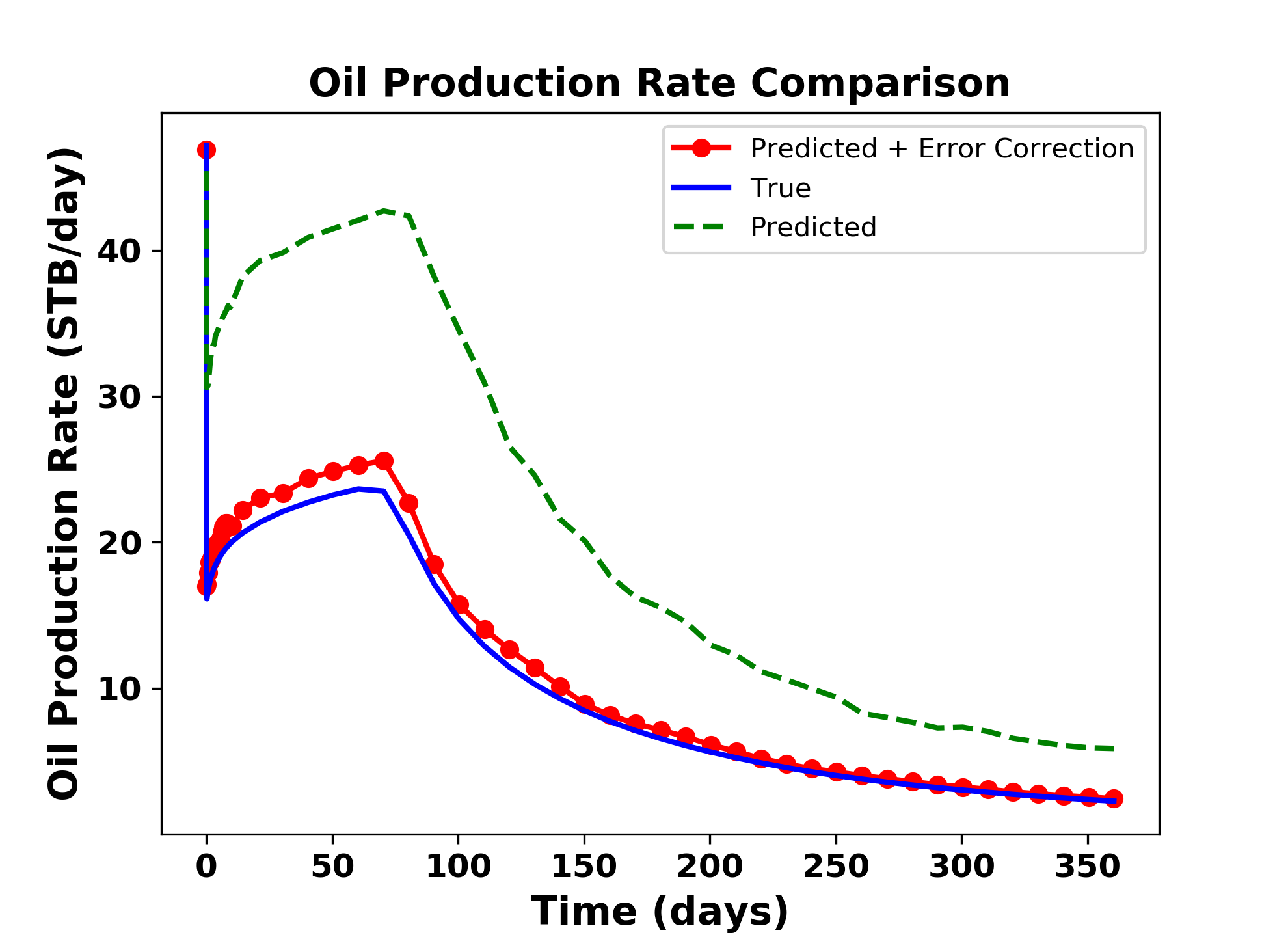}
		\caption{}
		\label{fig:5_17_2}
	\end{subfigure}
	~
	\centering
	\begin{subfigure}{0.33\textwidth}
		\centering
		\includegraphics[width=\textwidth]{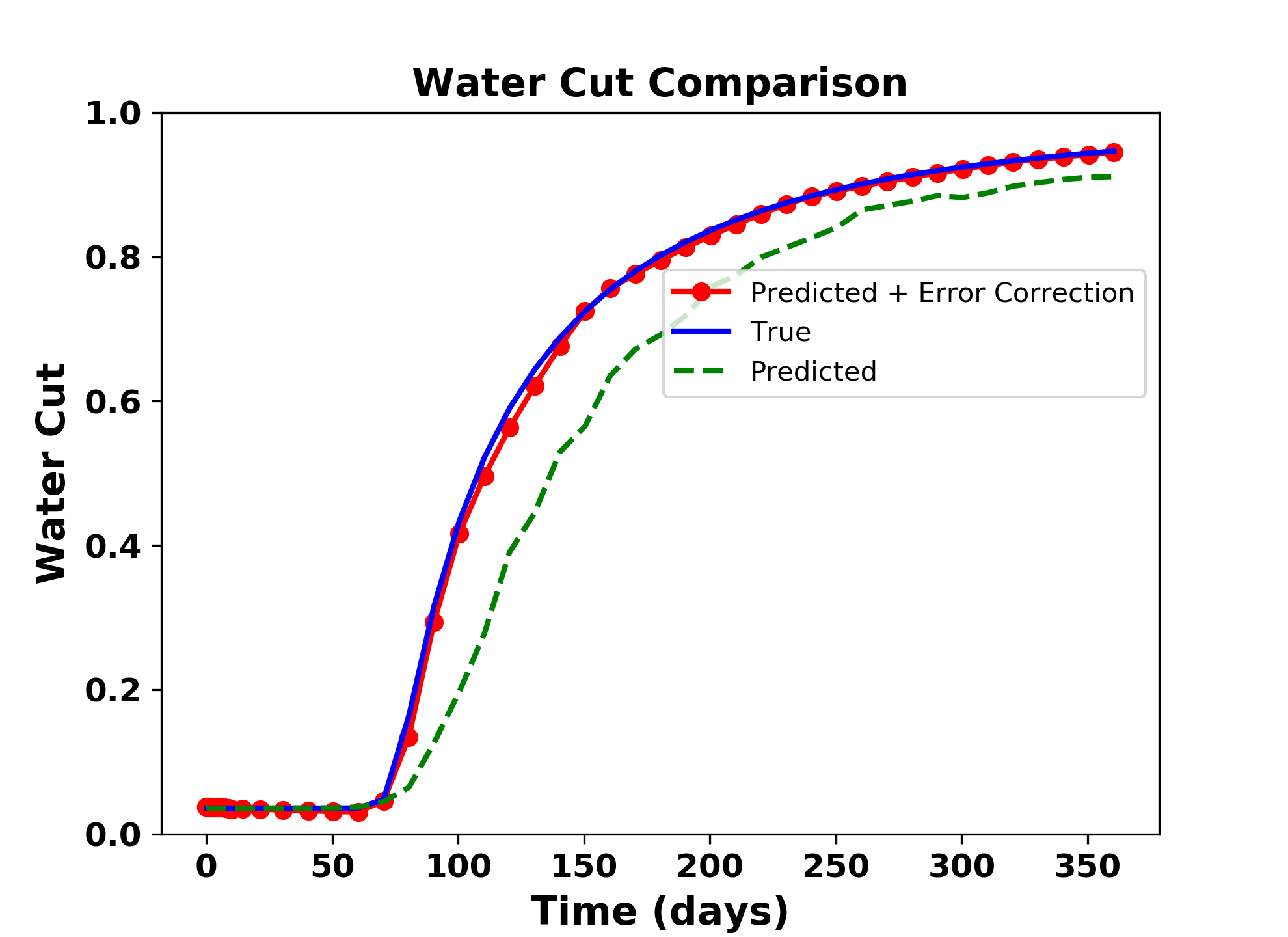}
		\caption{}
		\label{fig:5_17_3}
	\end{subfigure}%
	\caption{(a) Test case with producer well location at (3,16) (b) Comparison of oil production rate and (c) Comparison of water cut, predicted using global PMOR method alone (dotted green line) and after implementation of error correction model (red circled line) with the true solution (blue line)}
	\label{fig:5_17}
\end{figure}

\begin{figure}[htb!]
	\centering
	\begin{subfigure}{0.27\textwidth}
		\centering
		\includegraphics[width=\textwidth]{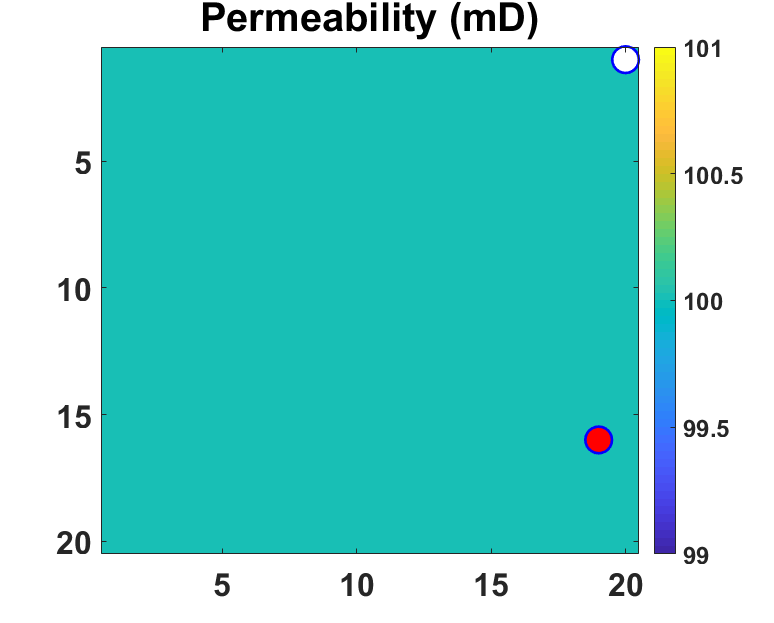}
		\caption{}
		\label{fig:5_18_1}
	\end{subfigure}%
	~
	\centering
	\begin{subfigure}{0.33\textwidth}
		\centering
		\includegraphics[width=\textwidth]{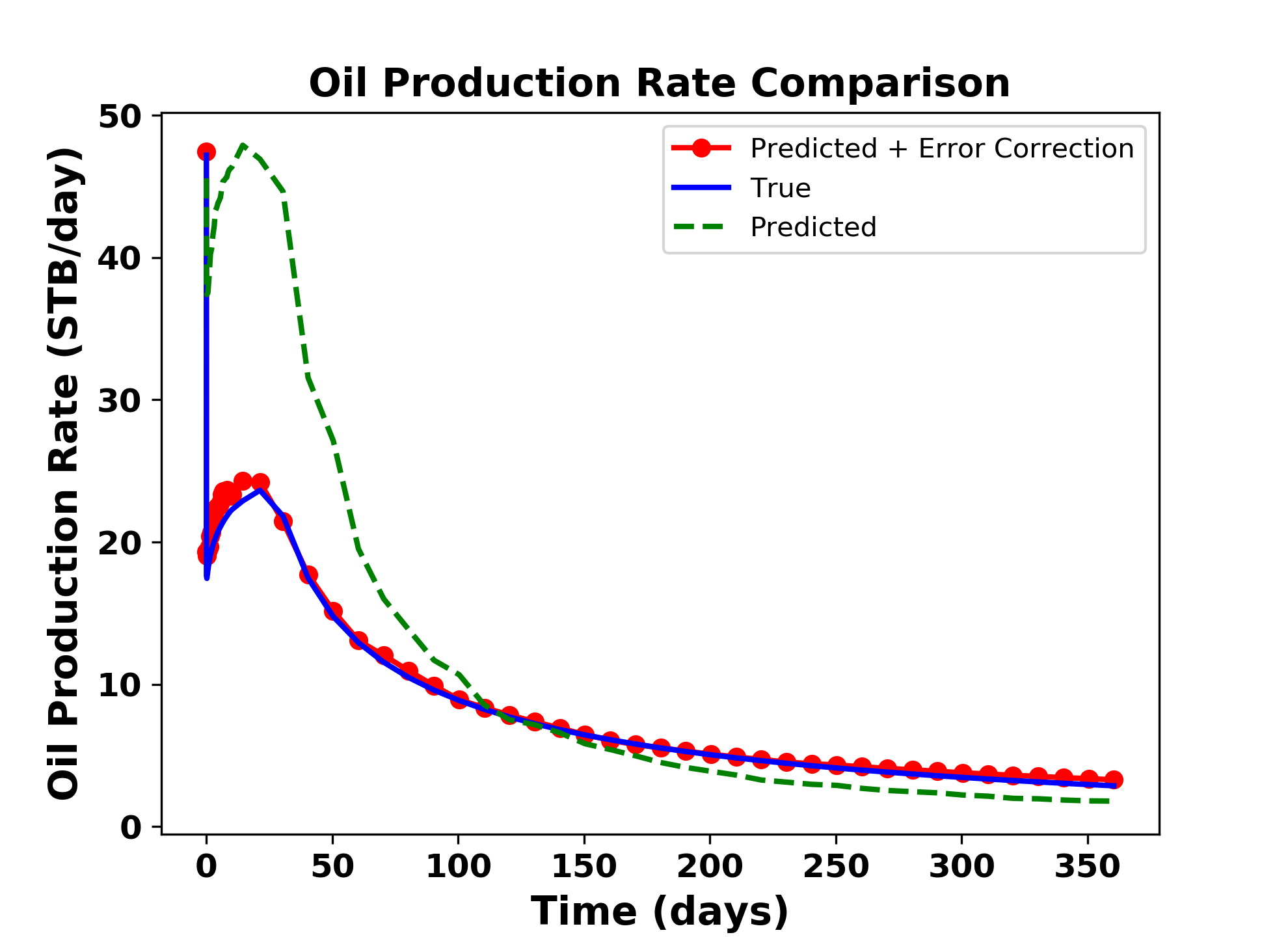}
		\caption{}
		\label{fig:5_18_2}
	\end{subfigure}
	~
	\centering
	\begin{subfigure}{0.33\textwidth}
		\centering
		\includegraphics[width=\textwidth]{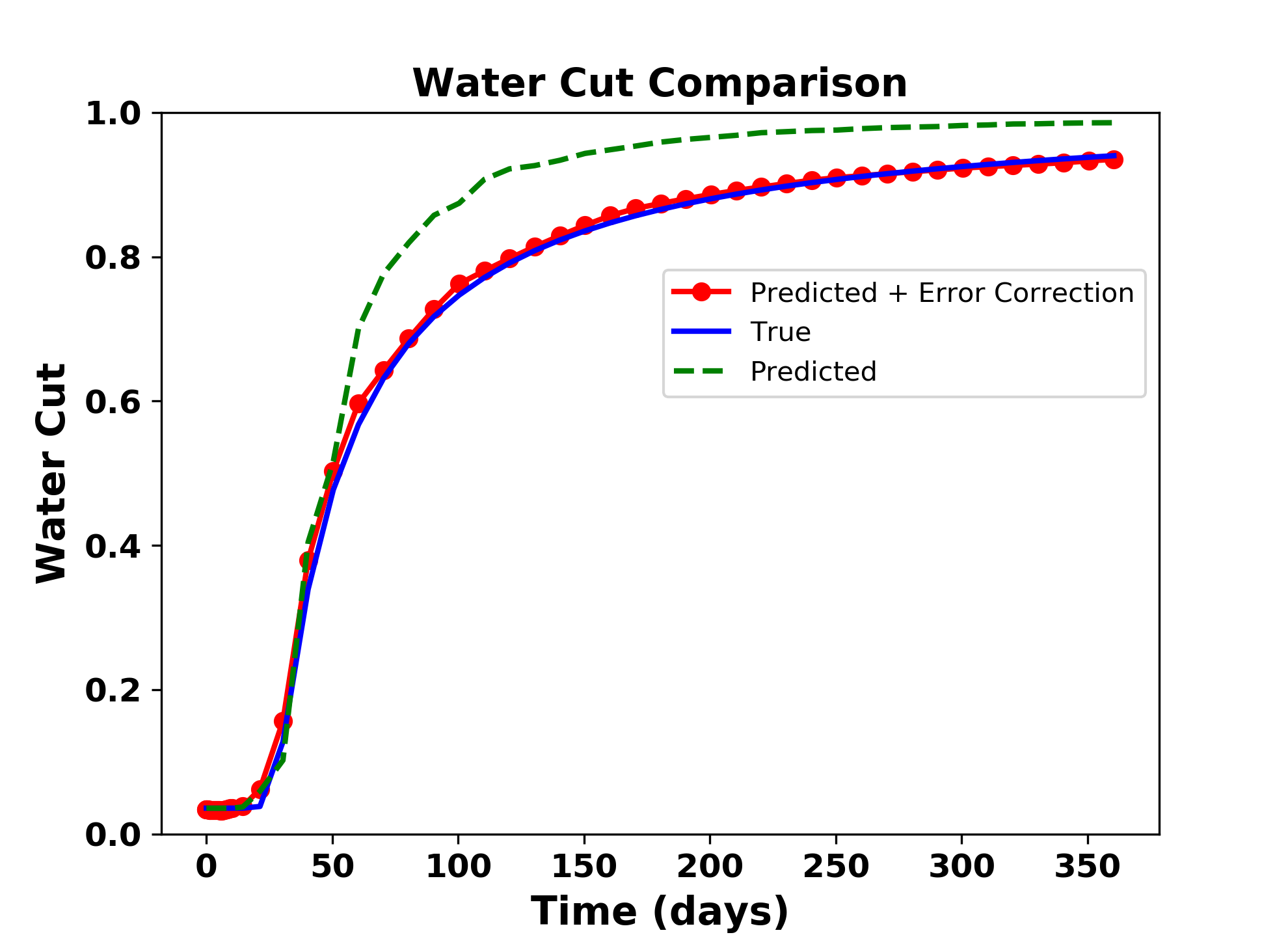}
		\caption{}
		\label{fig:5_18_3}
	\end{subfigure}%
	\caption{(a) Test case with producer well location at (16,19) (b) Comparison of oil production rate and (c) Comparison of water cut, predicted using global PMOR method alone using 100 samples (dotted green line) and after implementation of error correction model (red circled line) with the true solution (blue line)}
	\label{fig:5_18}
\end{figure}

So for the reservoir description as shown in Figure \ref{fig:5_1}, we have 100 well configurations used to construct the global PMOR model and 100 other well configurations for training the error correction ML model. Thus, we have 199 test cases in total to test the performance of this methodology. Figures \ref{fig:5_16}, \ref{fig:5_17}, \ref{fig:5_18} show the comparison of the results obtained by just implementing the non-intrusive global PMOR method, its implementation with the error correction models and the true solution obtained by fine scale simulation. 

It can be observed from these test cases that, the proposed method shows a very good accuracy with the solution trend captured by global PMOR and then adding error correction to the solution to get a much improved accuracy. This was examined for the other test cases as well. The predicted results accurately captures the water breakthrough time which is different for different test cases, as can be seen in the plots of water cut comparison.

\begin{figure}
	\centering
	\begin{subfigure}{0.45\textwidth}
		\centering
		\includegraphics[width=\textwidth]{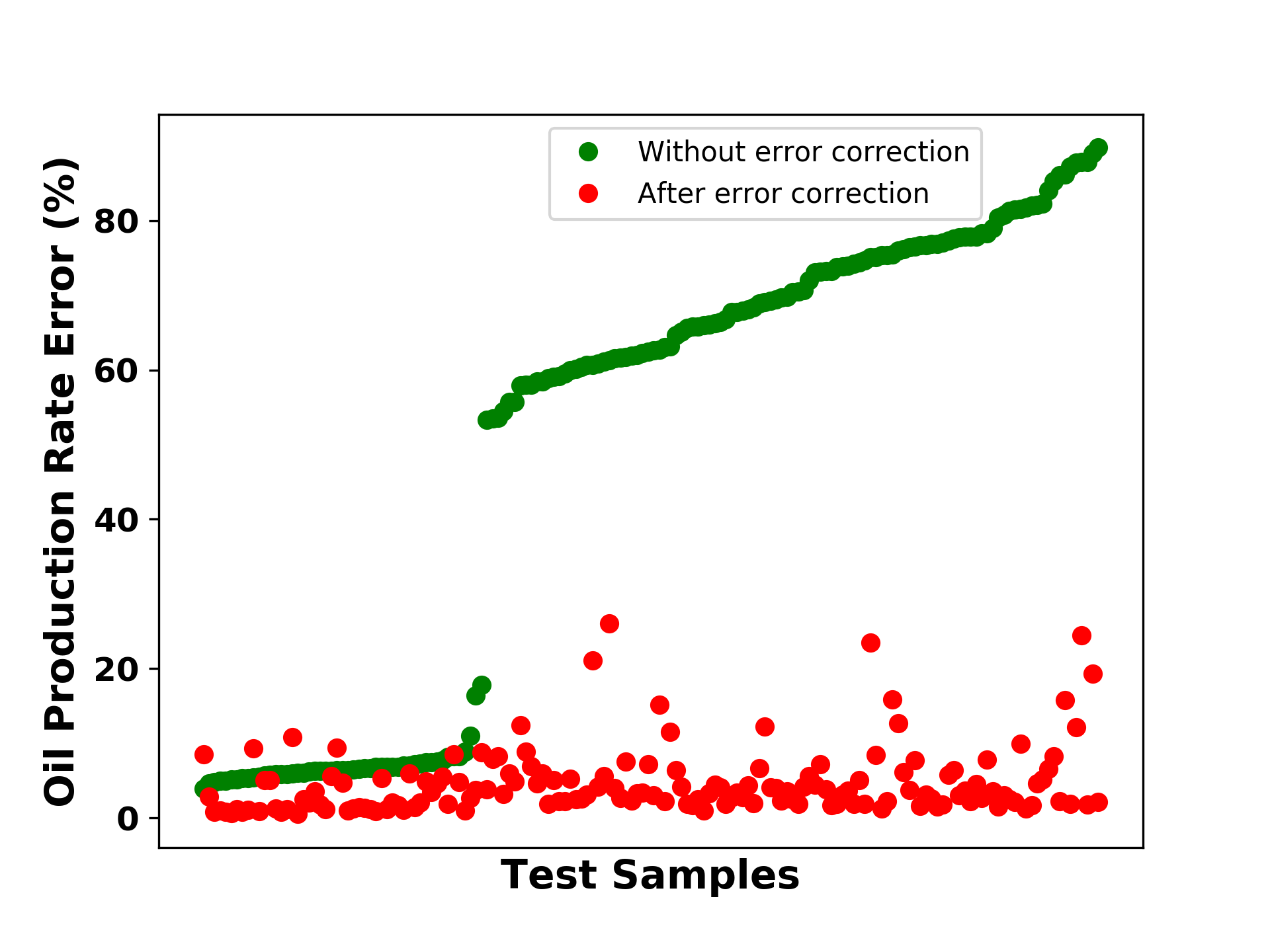}
		\caption{}
		\label{fig:5_16a}
	\end{subfigure}%
	~
	\centering
	\begin{subfigure}{0.45\textwidth}
		\centering
		\includegraphics[width=\textwidth]{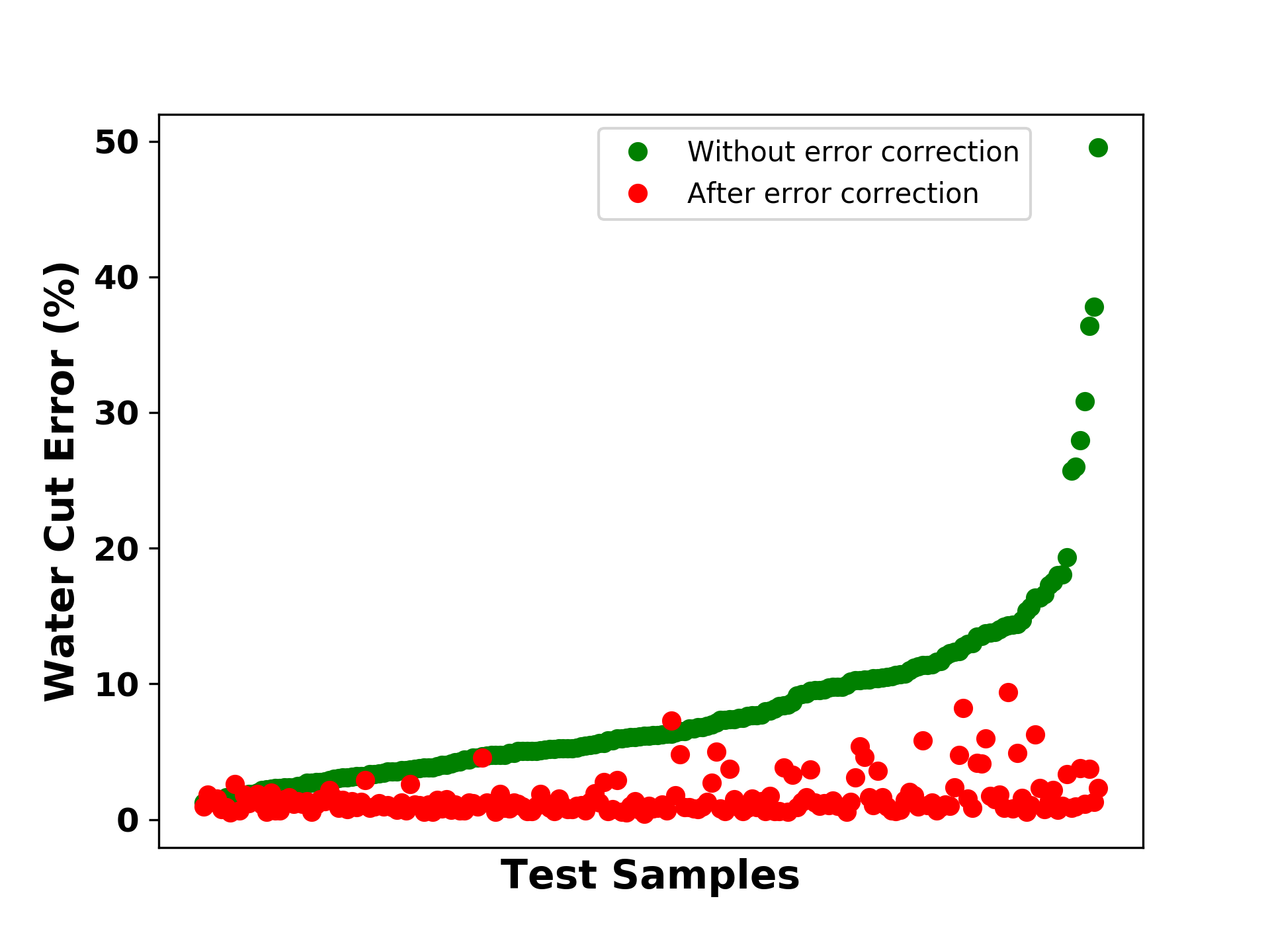}
		\caption{}
		\label{fig:5_16b}
	\end{subfigure}%
	\caption{Error in prediction of (a) Oil Production Rate (b) Water Cut, for all the test cases before and after the error correction of the solutions }
	\label{fig:5_16_err}
\end{figure}

We predicted the solutions for all the test cases and show the errors in prediction for all these cases in Figure \ref{fig:5_16_err} for oil production rate and water cut. In this figure, the errors shown in red are those obtained by POD coefficient prediction without error correction. These are arranged in increasing order for all the test cases. The blue dots show error after correcting the solutions in the second step for each corresponding red dot. This gives an intuition of the behavior of proposed method for all kinds of test cases. This plot shows that most of the test cases show a very good accuracy after error correction even when the predicted solutions just based on ML estimated POD coefficients have large errors. There are a very few test cases that show errors increasing after correction, that most probably can be attributed to random sampling, and should be solved after efficient sampling techniques. The average accuracy for all the test cases can be found in Table \ref{tab:homo_acc1}. 

\begin{table}
	\begin{center}
		
		\begin{tabular}{lclcl}
			\hline\noalign{\smallskip}
			\textbf{} & \textbf{Average Accuracy (\%)} \\
			\noalign{\smallskip}\hline\noalign{\smallskip}
			\textbf{Oil production rate }& 95.28\\
			\textbf{Water cut} & 98.4\\
			\noalign{\smallskip}\hline
		\end{tabular}
	\end{center}
	\caption{Homogeneous reservoir case 1: Average accuracy of oil production rate and water cut for all test samples}
	\label{tab:homo_acc1}       
	\vspace*{-1em}
\end{table}

Now, we show the same example, but using lesser number of training sample well locations for predicting basis coefficients and construction of global ROB (this should be a lower quality basis than before) and using same number of training points for error correction model. Since it is challenging to pick the number of sample points for training and ultimately choosing appropriate basis dimensions, such an analysis is useful to determine if the error correction model can also provide reasonable results even if the global ROB is not a very good representation of the parameter domain. The ML based PMOR model is thus constructed with 50 sample well locations for this example.

Table \ref{tab:5_3} shows the optimum hyperparameters and the corresponding train and test accuracies obtained from the trained Random Forest model to predict the basis coefficients for both pressure and saturation. As can be expected, the train accuracy decreases as the ML model does not have a large data set to learn the physics which eventually also shows a decrease in the test accuracy as compared to the previous test case.  
\begin{table}[htb!]
	\begin{center}
		
		\begin{tabular}{lclclc|c|}
			\hline\noalign{\smallskip}
			\textbf{} & \textbf{RF Regression} & \textbf{Train Accuracy} & \textbf{Test Accuracy} \\
			\noalign{\smallskip}\hline\noalign{\smallskip}
			\textbf{Pressure}& $N_{fmax}$=2, $N_l$=2 &\enspace\enspace\enspace\enspace\enspace\enspace 99.55 & 97.55\\
			\textbf{Saturation} & $N_{fmax}$=2, $N_l$=2 & \enspace\enspace\enspace\enspace\enspace\enspace98.71 & 90.79\\
			\noalign{\smallskip}\hline
		\end{tabular}
	\end{center}
	\caption{Hyperparameters chosen by 5-fold Cross Validation for Random Forest Regressor using 50 training samples}
	\label{tab:5_3}       
	\vspace*{-1em}
\end{table}

\begin{figure}[htb!]
	\centering
	\begin{subfigure}{0.27\textwidth}
		\centering
		\includegraphics[width=\textwidth]{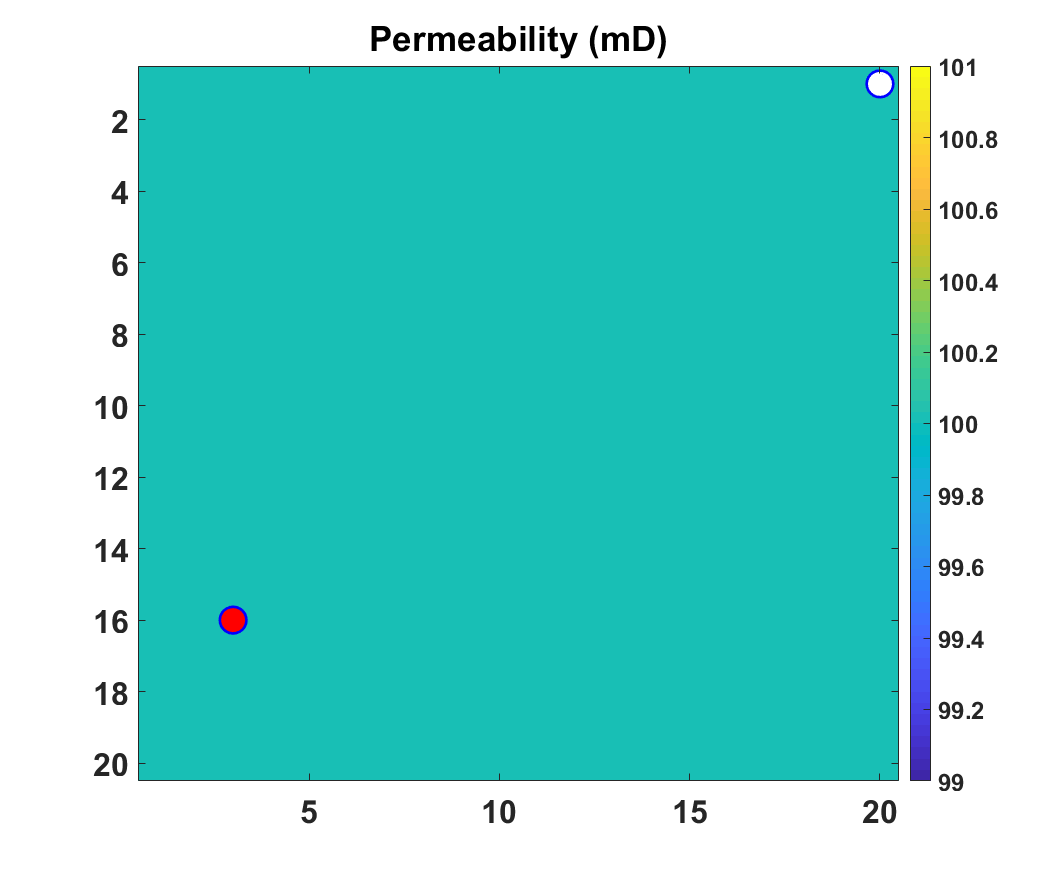}
		\caption{}
		\label{fig:5_homo2_1}
	\end{subfigure}%
	~
	\centering
	\begin{subfigure}{0.33\textwidth}
		\centering
		\includegraphics[width=\textwidth]{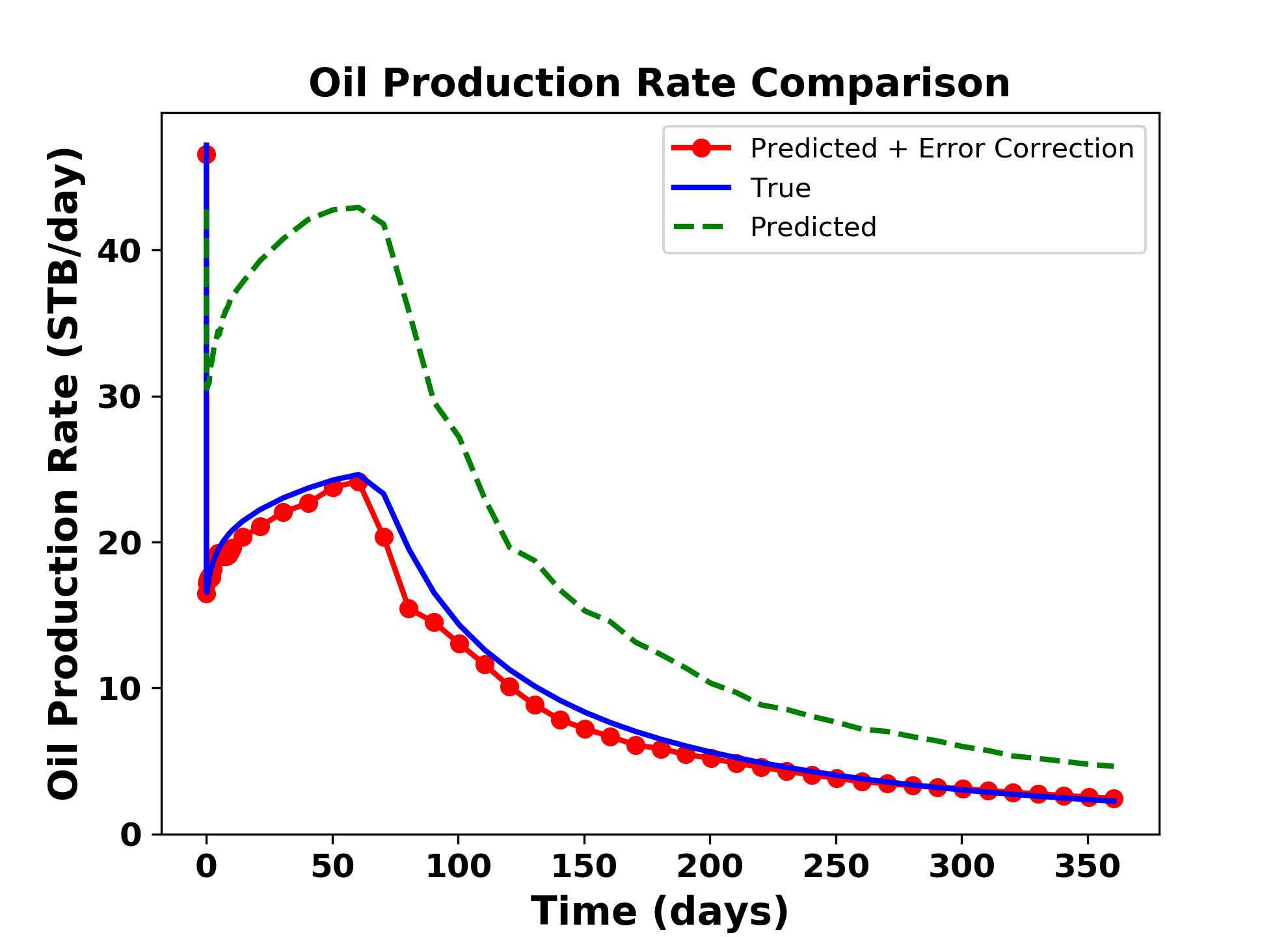}
		\caption{}
		\label{fig:5_homo2_2}
	\end{subfigure}
	~
	\centering
	\begin{subfigure}{0.33\textwidth}
		\centering
		\includegraphics[width=\textwidth]{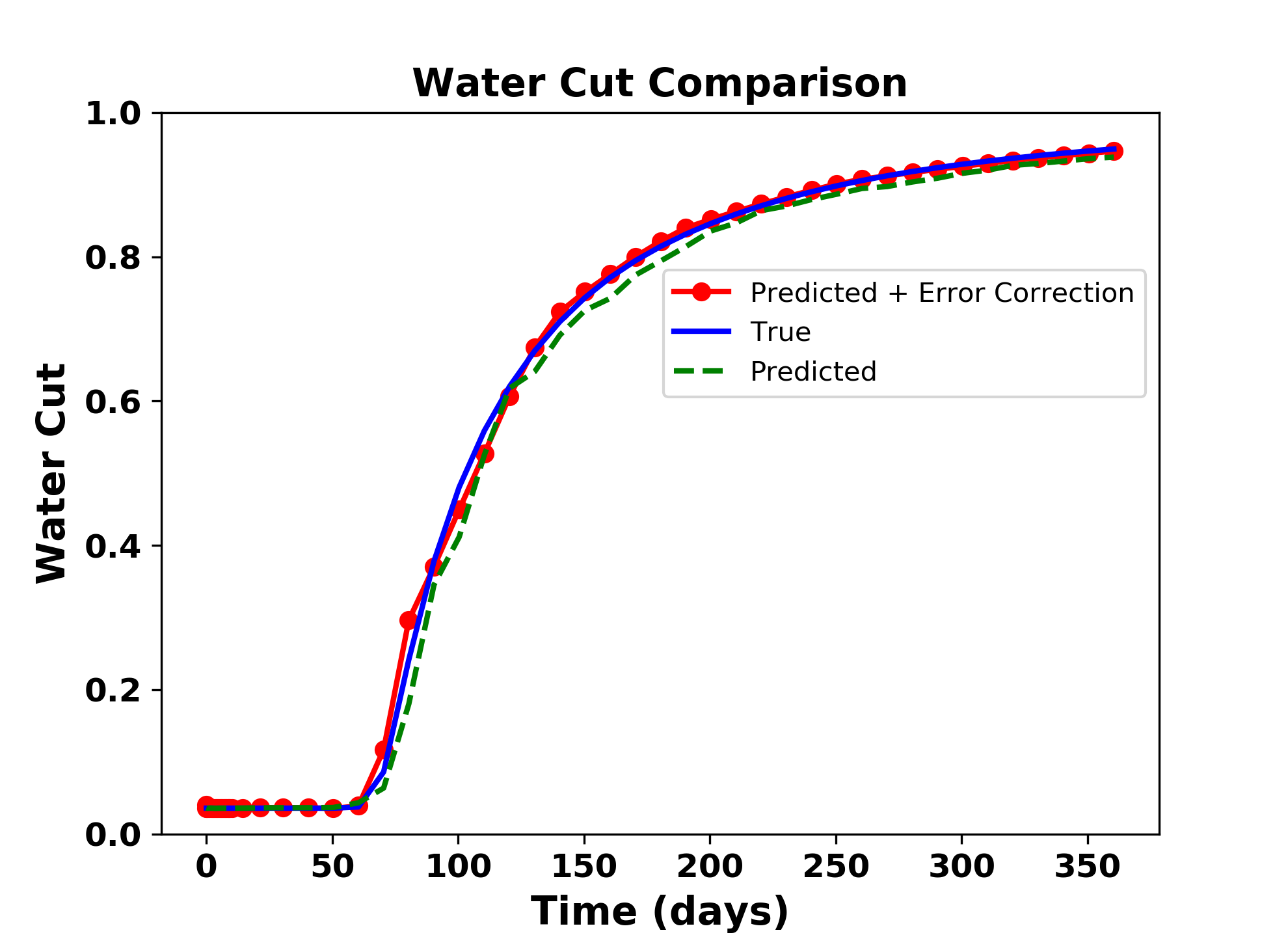}
		\caption{}
		\label{fig:5_homo2_3}
	\end{subfigure}%
	\caption{(a) Test case with producer well location at (3,16) (b) Comparison of oil production rate and (c) Comparison of water cut, predicted using global PMOR method alone using 50 samples (dotted green line) and after implementation of error correction model (red circled line) with the true solution (blue line)}
	\label{fig:5_19}
\end{figure}

\begin{figure}[htb!]
	\centering
	\begin{subfigure}{0.27\textwidth}
		\centering
		\includegraphics[width=\textwidth]{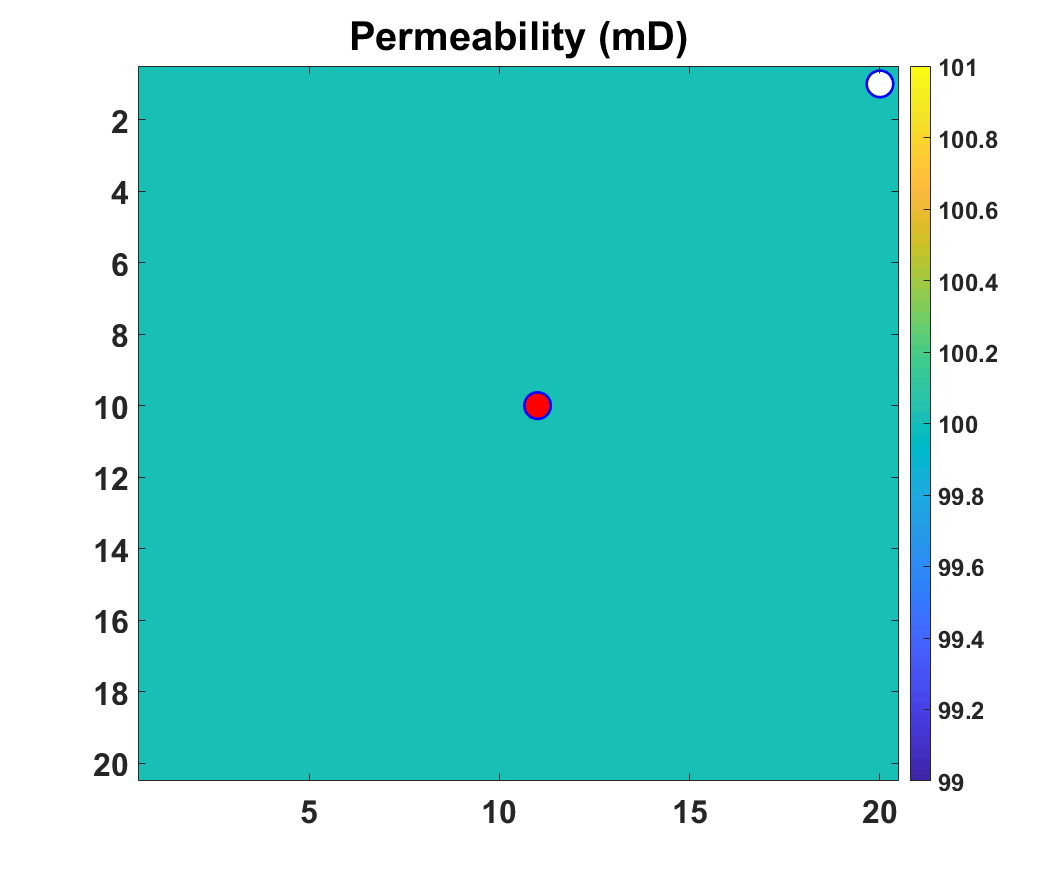}
		\caption{}
		\label{fig:5_homo2_4}
	\end{subfigure}%
	~
	\centering
	\begin{subfigure}{0.33\textwidth}
		\centering
		\includegraphics[width=\textwidth]{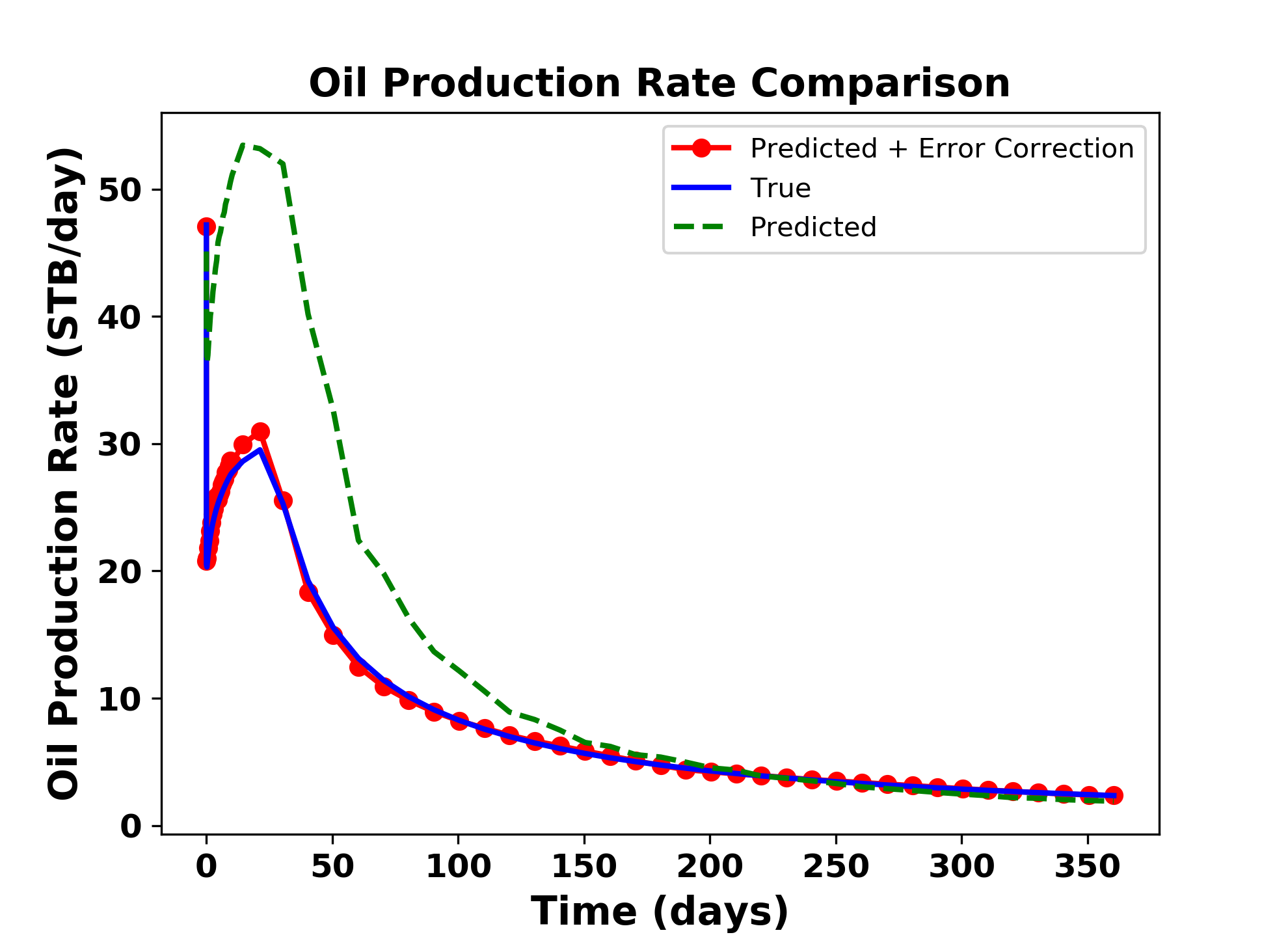}
		\caption{}
		\label{fig:5_homo2_5}
	\end{subfigure}
	~
	\centering
	\begin{subfigure}{0.33\textwidth}
		\centering
		\includegraphics[width=\textwidth]{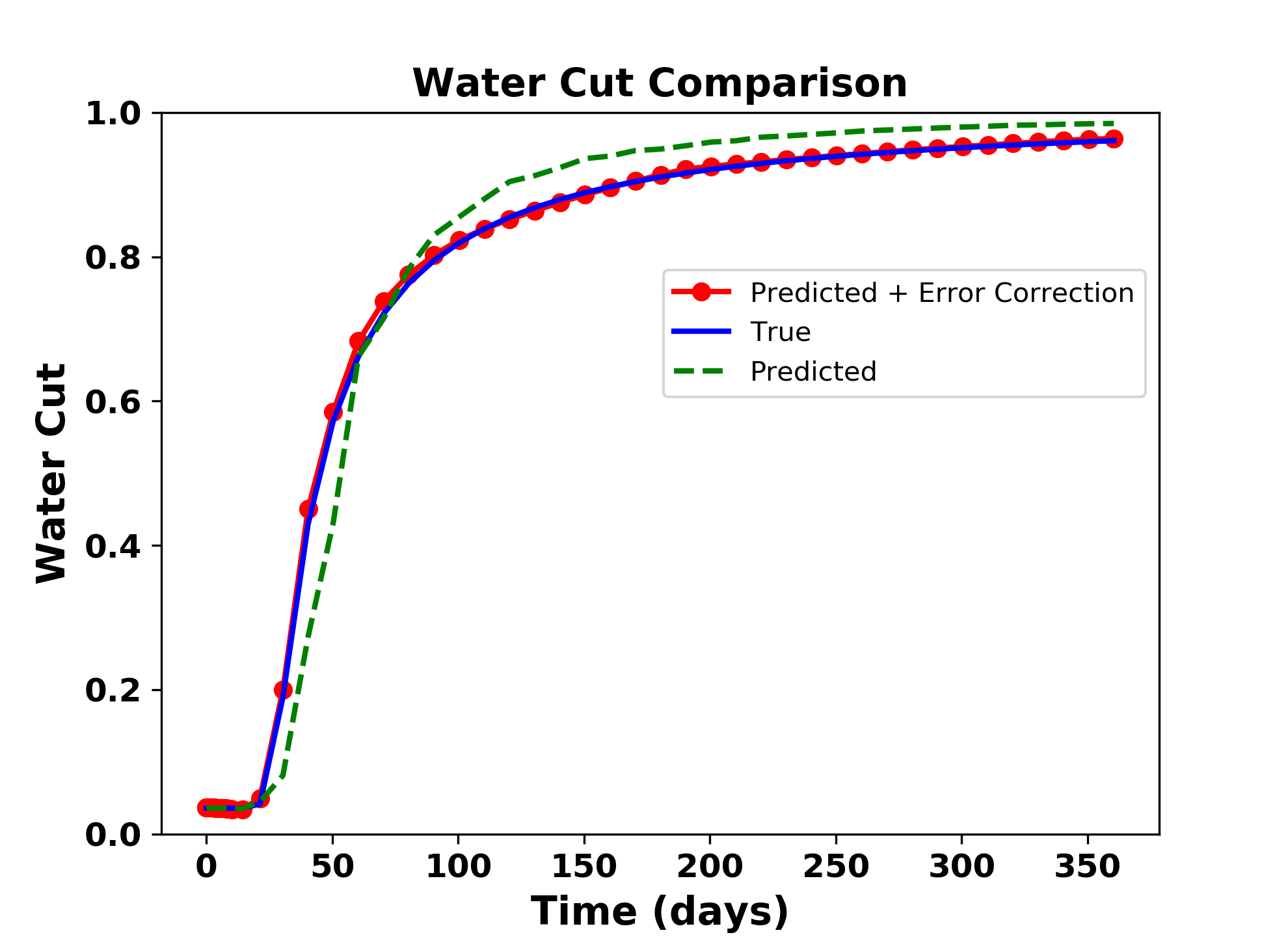}
		\caption{}
		\label{fig:5_homo2_6}
	\end{subfigure}%
	\caption{(a) Test case with producer well location at (11,10) (b) Comparison of oil production rate and (c) Comparison of water cut, predicted using global PMOR method alone using 50 samples (dotted green line) and after implementation of error correction model (red circled line) with the true solution (blue line)}
	\label{fig:5_20}
\end{figure}

\begin{figure}
	\centering
	\begin{subfigure}{0.45\textwidth}
		\centering
		\includegraphics[width=\textwidth]{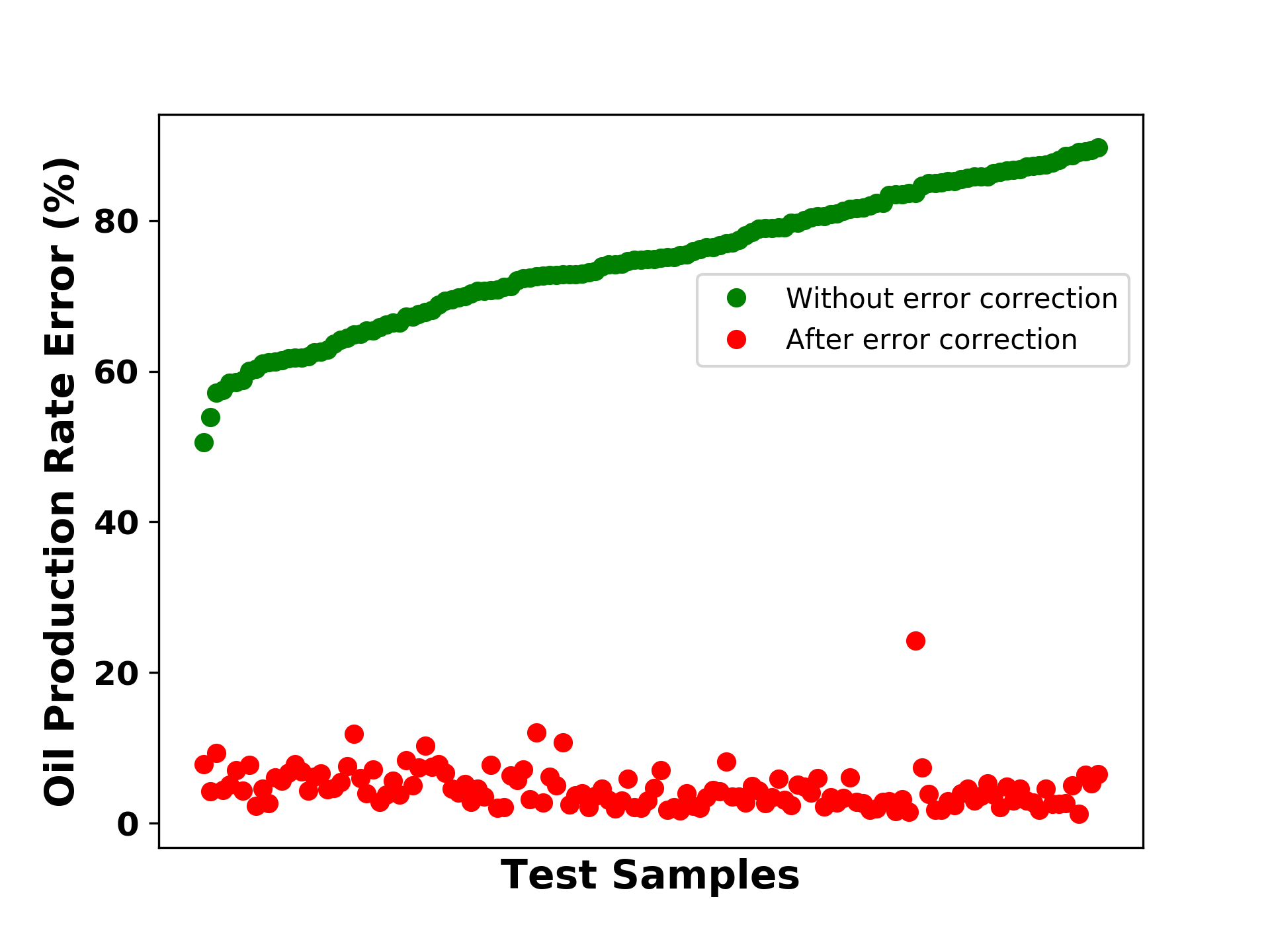}
		\caption{}
		\label{fig:5_17a}
	\end{subfigure}%
	~
	\centering
	\begin{subfigure}{0.45\textwidth}
		\centering
		\includegraphics[width=\textwidth]{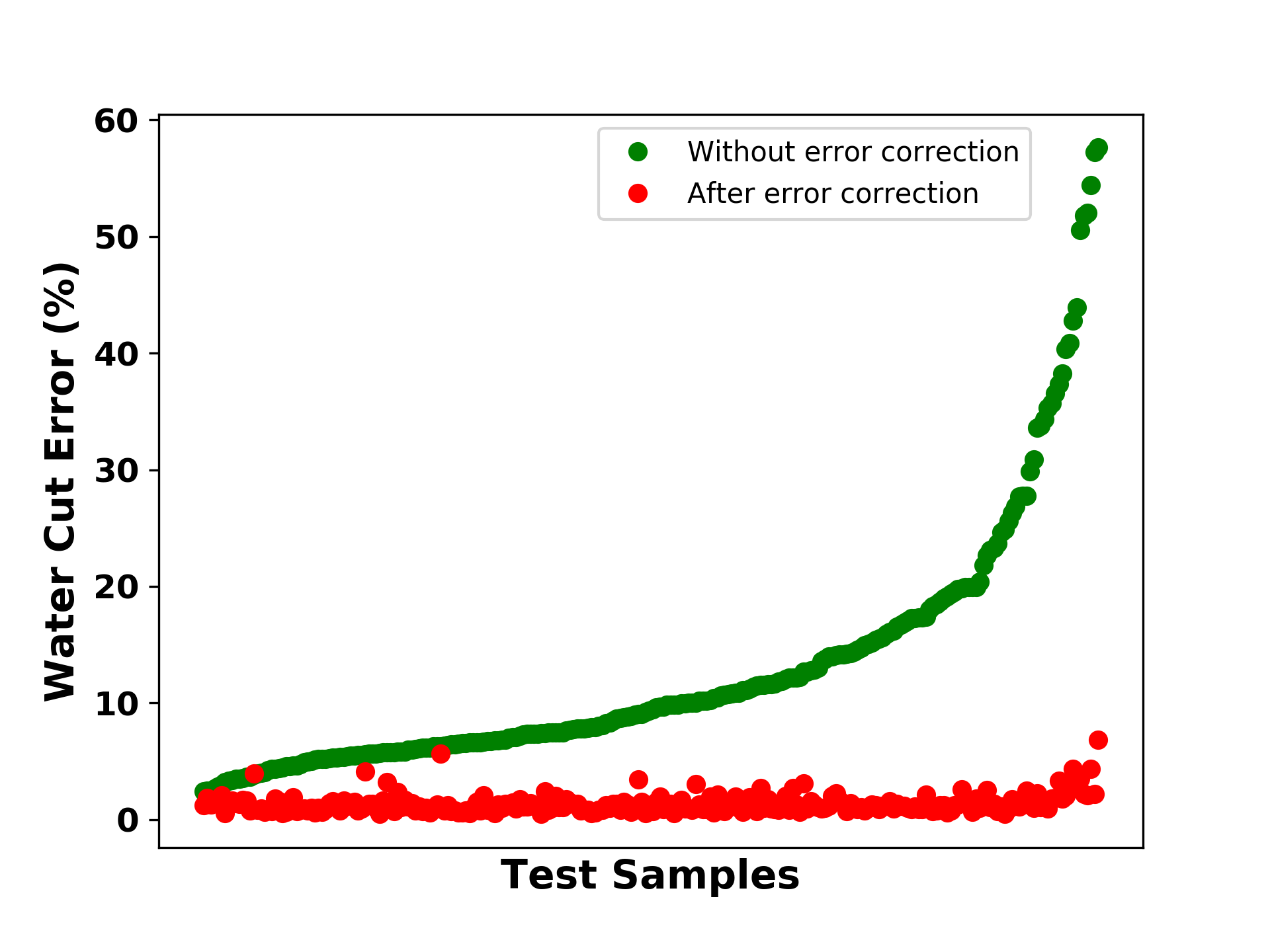}
		\caption{}
		\label{fig:5_17b}
	\end{subfigure}%
	\caption{Error in prediction of (a) Oil Production Rate (b) Water Cut, for all the test cases before and after the error correction of the solutions }
	\label{fig:5_17_err}
\end{figure}

Figures \ref{fig:5_19} and \ref{fig:5_20} show two of the test cases for this example by comparing the oil production rates and water cut. Fig. \ref{fig:5_17_err} shows the errors in prediction of oil production rates and water cut for all these cases. These results show a good agreement in the solutions thus showing the validity of error correction model for the cases when the global basis is not very accurate for the entire parameter domain but still can capture the overall trend of the solutions. However, these observations may not hold true when the quality of global basis is compromised enough to explain the appropriate physics of flow.

\begin{table}
	\begin{center}
		
		\begin{tabular}{lclcl}
			\hline\noalign{\smallskip}
			\textbf{} & \textbf{Average Accuracy (\%)} \\
			\noalign{\smallskip}\hline\noalign{\smallskip}
			\textbf{Oil production rate }& 95.2\\
			\textbf{Water cut} & 98.3\\
			\noalign{\smallskip}\hline
		\end{tabular}
	\end{center}
	\caption{Homogeneous reservoir case 2: Average accuracy of oil production rate and water cut for all test samples}
	\label{tab:homo_acc2}       
	\vspace*{-1em}
\end{table}

\subsection{Heterogeneous reservoir model - single well case}
\begin{figure}[htb!]
	\centering
	\includegraphics[width=0.8\textwidth]{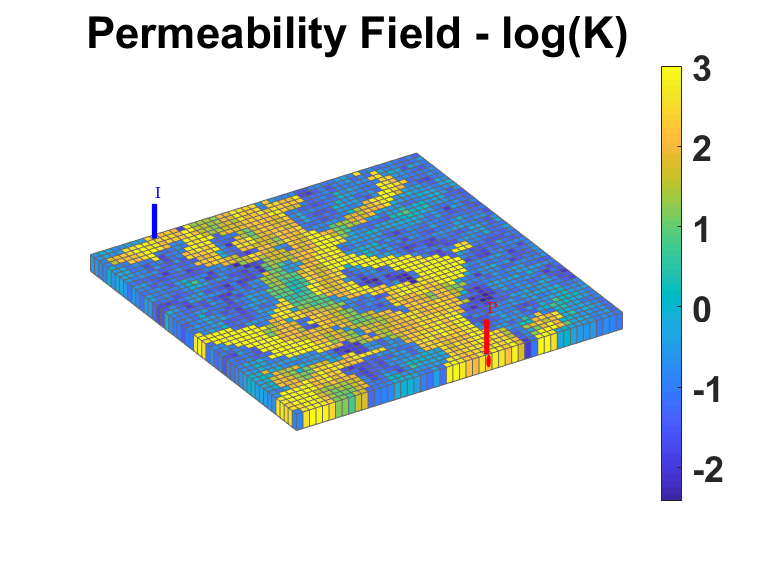}
	\caption{Heterogeneous log permeability field (section of SPE10 model layer 50) with one producer and one injector}
	\label{fig:5_21}       
\end{figure}

The methodology shows promising results for a small homogeneous reservoir model. Now we apply the same methodology to a heterogeneous reservoir model which is a section of layer 50 of the SPE10 benchmark model. Again, this is a two-phase flow (oil-water) reservoir model with one injector well and one producer well as shown in Figure \ref{fig:5_21}. The reservoir model is discretized with a Cartesian grid of size 20 ft $\times $ 20 ft $\times$ 50 ft, and it contains 2500 (50$\times$50) active cells. We neglect the capillary and gravity effects. The initial reservoir pressure is 4200 psi and the initial water saturation is considered 0. The injector and producer are BHP controlled at a constant pressures of 7000 psi and 2500 psi respectively. For the current scope of work, we believe, this model with about 5 orders of range in permeability and an injector and a producer, should be a good case with reasonable complexity to demonstrate the validity of proposed global PMOR strategy. Again we only consider that the producer is the parameter of interest which means the injector location is fixed and the producer changes location in the reservoir. The simulation is run for a duration of 3 years. 

\begin{figure}[htb!]
	\centering
	\includegraphics[width=0.7\textwidth]{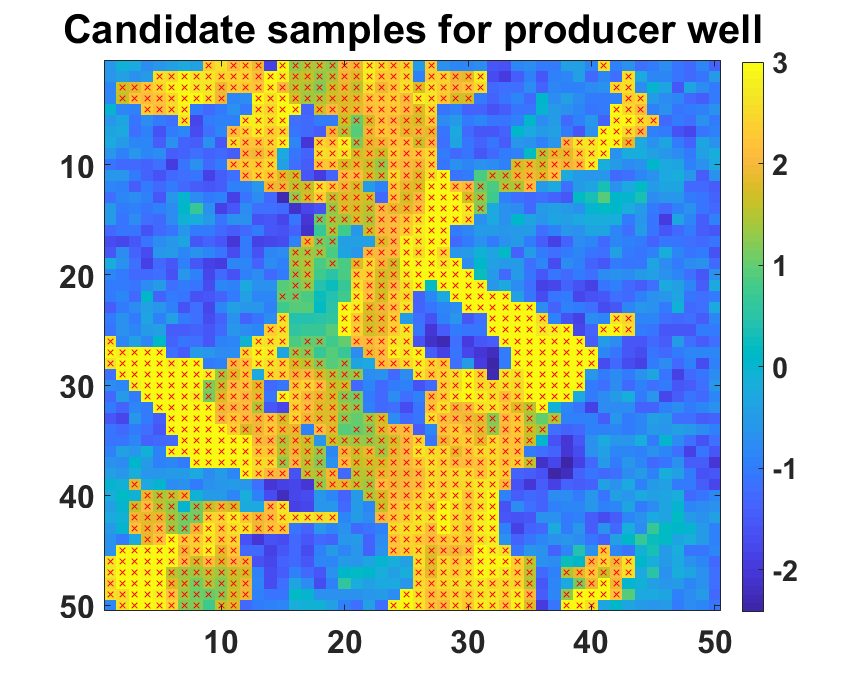}
	\caption{Samples considered as potential well locations shown by red crosses. Injector location is fixed at location (10,1) (not shown here) and hence not sampled.}
	\label{fig:5_22}       
\end{figure}

For the training of global PMOR model, we consider the producer well locations only on the high permeability gridblocks majority of which lie on the channel connecting the injector and producer. We use all the gridblocks with greater than 10 mD as candidate well locations as shown in Figure \ref{fig:5_22}. This is employed to observe the water cut for most of the well locations. Thus, here this global PMOR strategy is employed on a subset of parameter domain. These are around 1000 sample locations out of which we use 200 randomly sampled locations for training the ML based global PMOR model and 100 samples chosen for training the error correction model. The rest of the samples are test cases for which the accuracy of the PMOR model is evaluated. 

\begin{figure}
	\centering
	\begin{subfigure}{0.45\textwidth}
		\centering
		\includegraphics[width=\textwidth]{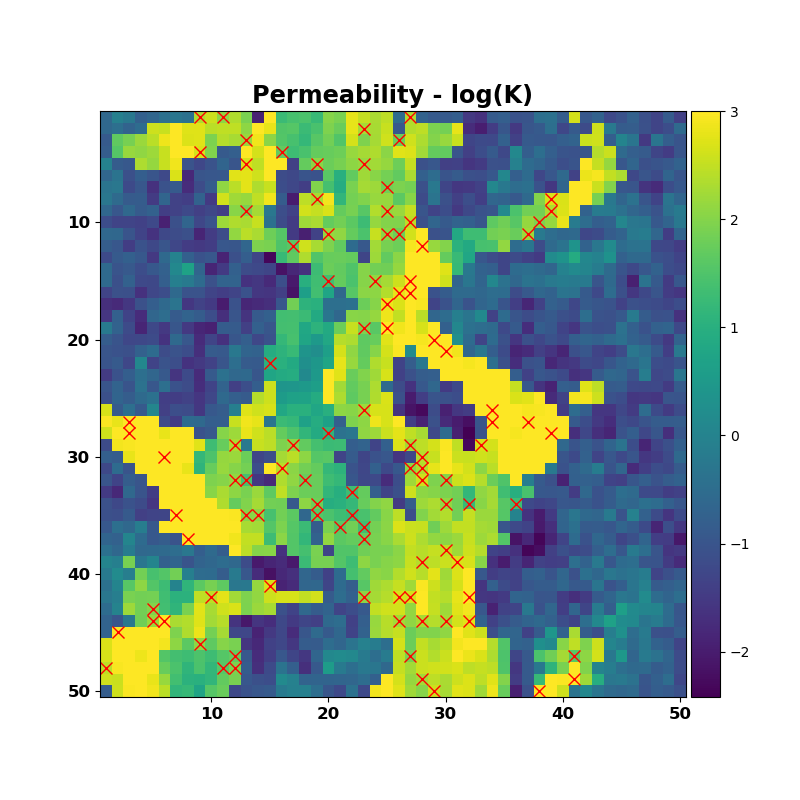}
		\caption{}
		\label{fig:5_23_1}
	\end{subfigure}%
	~
	\centering
	\begin{subfigure}{0.55\textwidth}
		\centering
		\includegraphics[width=\textwidth]{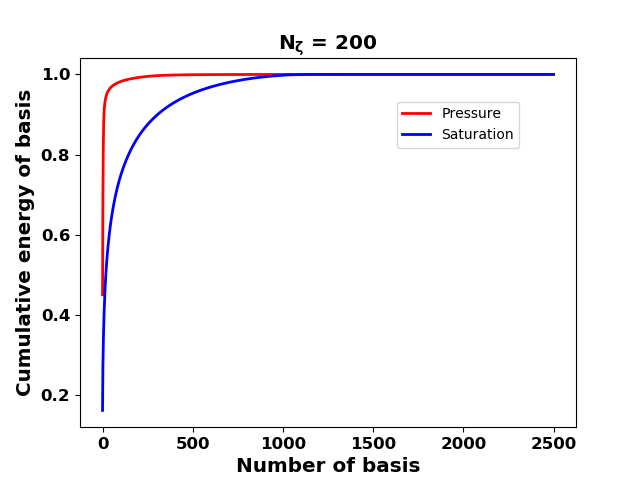}
		\caption{}
		\label{fig:5_23_2}
	\end{subfigure}%
	\caption{(a) Training samples randomly chosen for basis coefficient prediction (b) Cumulative energy of eigenvalues for pressure (red) and saturation basis (blue)}
	\label{fig:5_23}
\end{figure}

We show the randomly sampled data set for training the global PMOR ML model in Figure \ref{fig:5_23_1}. Figure \ref{fig:5_23_2} shows the cumulative energy of singular values for the given sample set. For the first example, we use 99\% energy criteria for pressure basis dimension and 90\% energy criteria for saturation basis dimension. Later in the section, we analyze the accuracy of the method with changing energy criteria for these states. We observe the steep cumulative energy curve for pressure and a more concave curve for saturation. This is expected since all the well locations are just considered on the high permeability area of the reservoir. The observations about energy curves follows the same explanation as that of the homogeneous case if we consider well locations throughout the spatial domain of reservoir.  

\begin{figure}[htb!]
	\centering
	\includegraphics[width=0.5\textwidth]{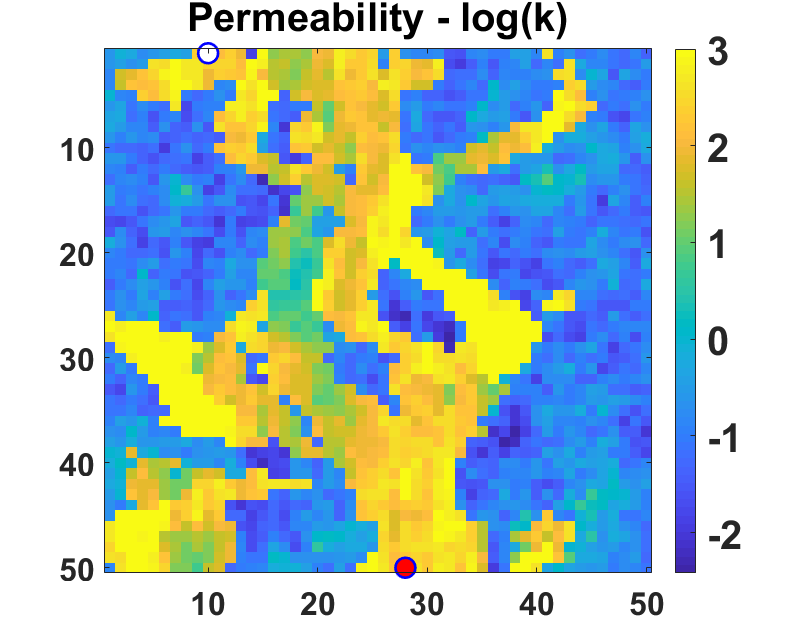}
	\caption{Test case with producer well at gridblock (28,50)}
	\label{fig:5_24_case1}       
\end{figure}

\begin{figure}[htb!]
	\centering
	\begin{subfigure}{0.6\textwidth}
		\centering
		\includegraphics[width=\textwidth]{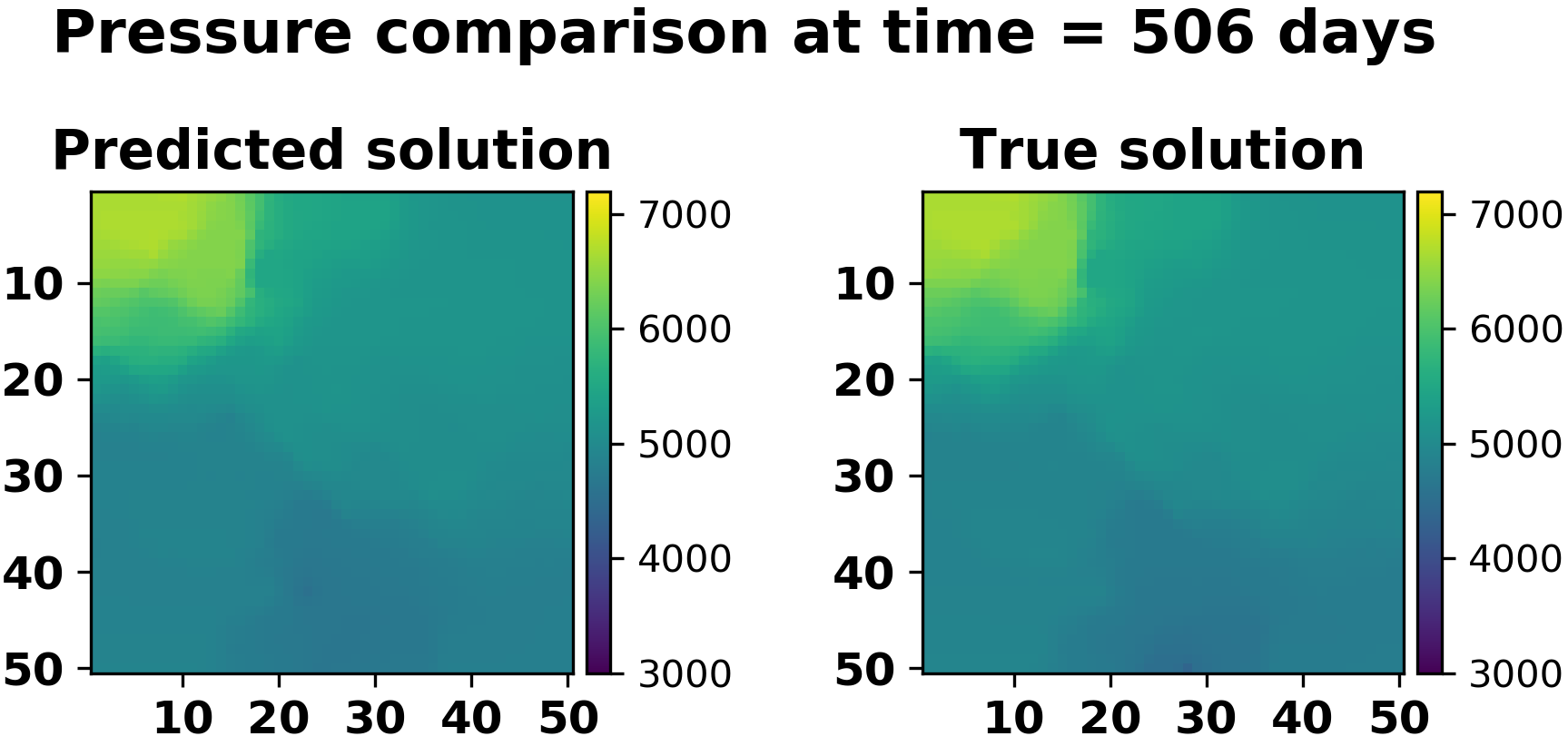}
		\caption{}
		\label{fig:5_24_1}
	\end{subfigure}%
	~\\
	\centering
	\begin{subfigure}{0.6\textwidth}
		\centering
		\includegraphics[width=\textwidth]{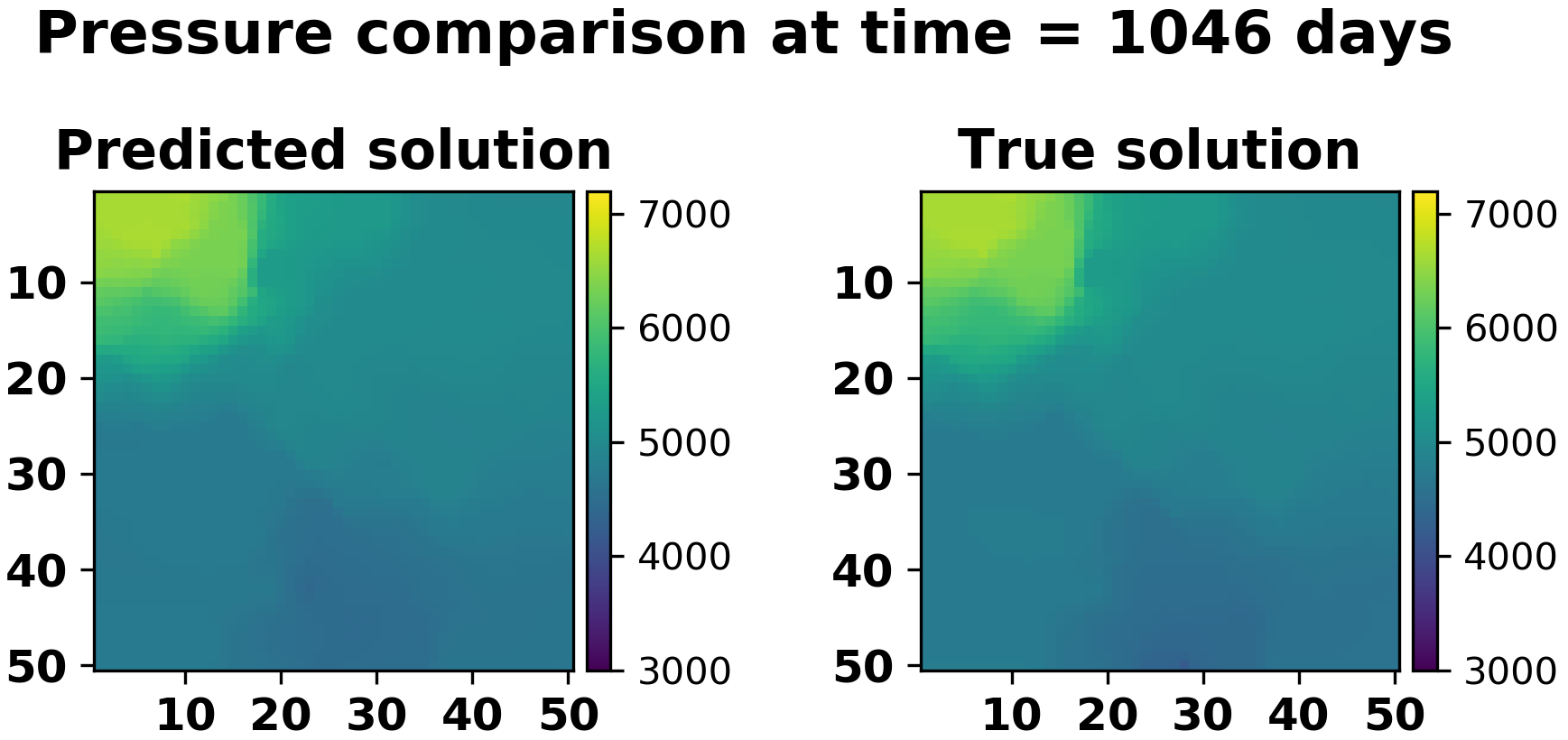}
		\caption{}
		\label{fig:5_24_2}
	\end{subfigure}%
	\caption{Pressure solution comparison at (a) Time = 506 days and (b) Time = 1026 days}
	\label{fig:5_24_a}
\end{figure}

\begin{figure}[htb!]
	\centering
	\begin{subfigure}{0.6\textwidth}
		\centering
		\includegraphics[width=\textwidth]{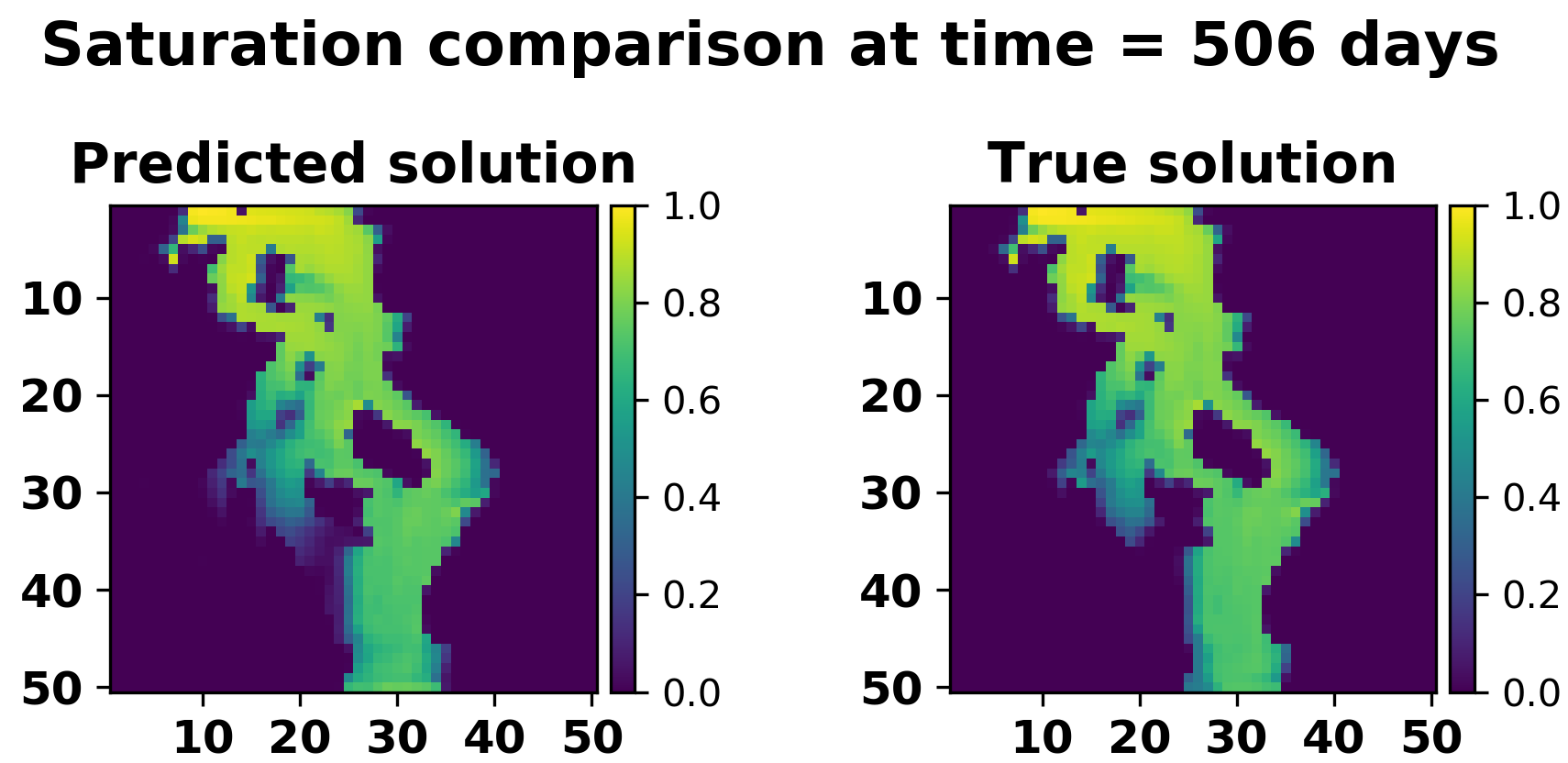}
		\caption{}
		\label{fig:5_24_3}
	\end{subfigure}%
	~\\
	\centering
	\begin{subfigure}{0.6\textwidth}
		\centering
		\includegraphics[width=\textwidth]{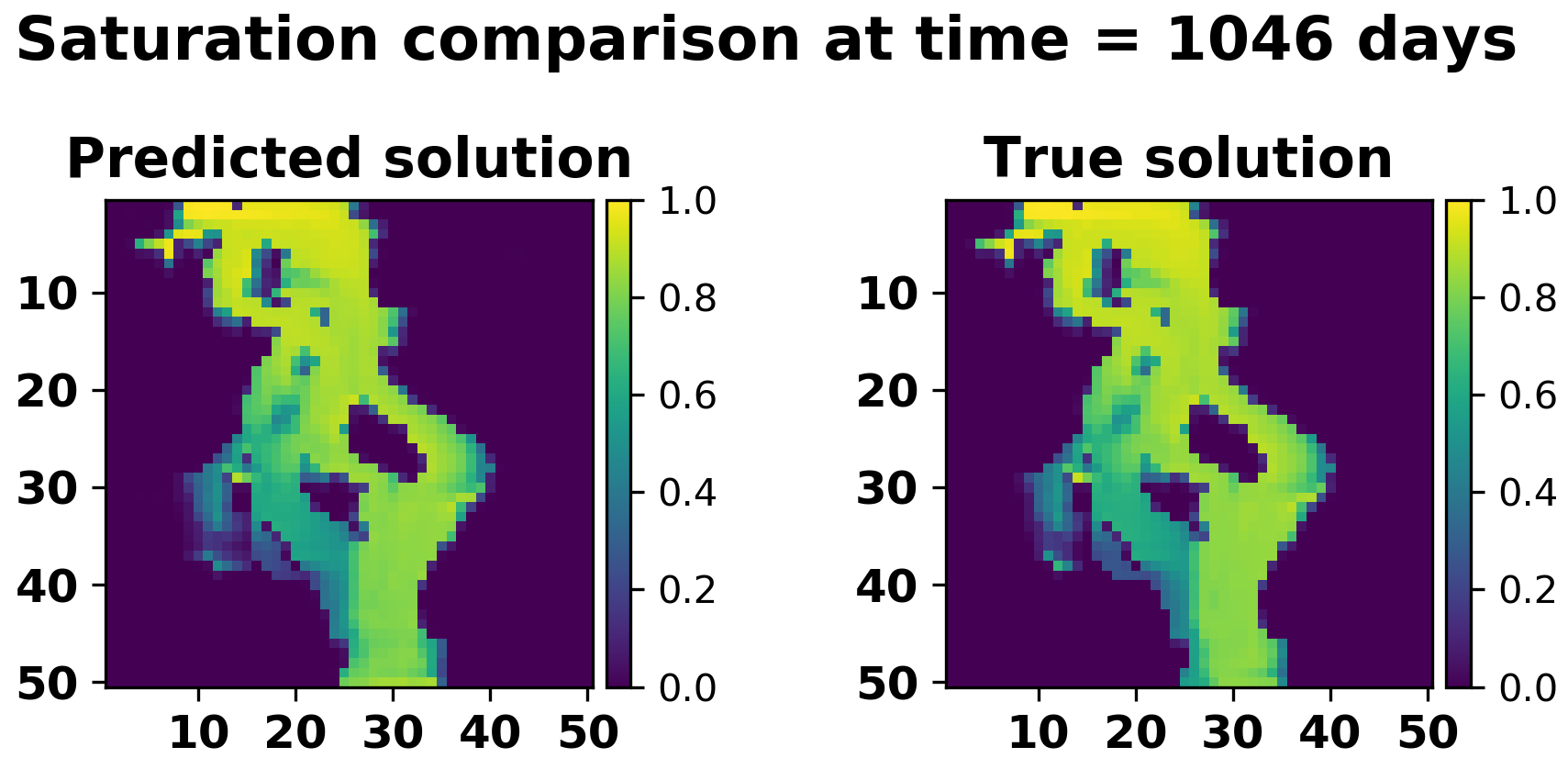}
		\caption{}
		\label{fig:5_24_4}
	\end{subfigure}%
	\caption{Saturation solution comparison at (a) Time = 506 days and (b) Time = 1026 days}
	\label{fig:5_24_b}
\end{figure}

\begin{figure}
	\centering
	\begin{subfigure}{0.50\textwidth}
		\centering
		\includegraphics[width=\textwidth]{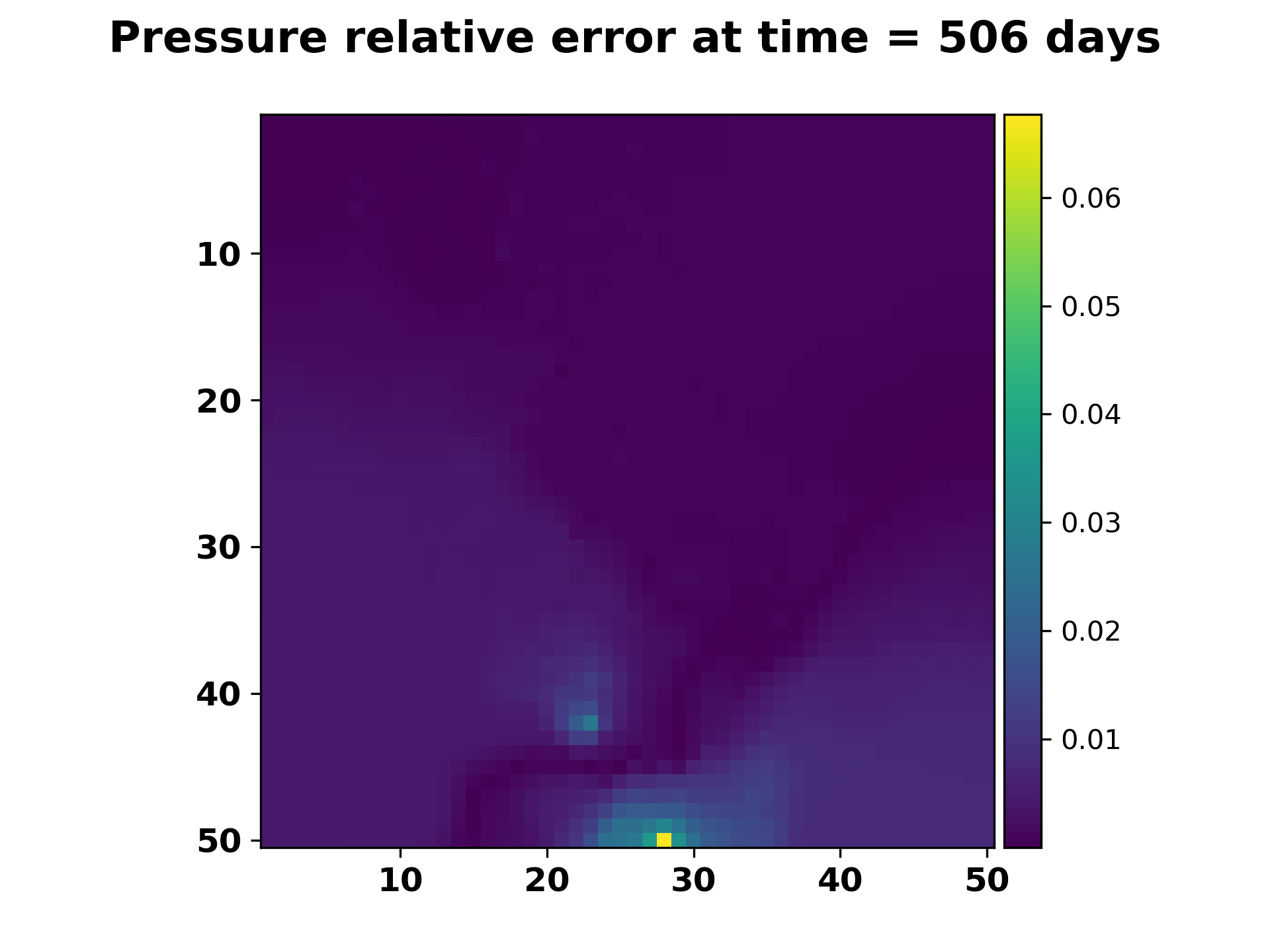}
		\caption{}
		\label{fig:5_25_1}
	\end{subfigure}%
	~
	\centering
	\begin{subfigure}{0.50\textwidth}
		\centering
		\includegraphics[width=\textwidth]{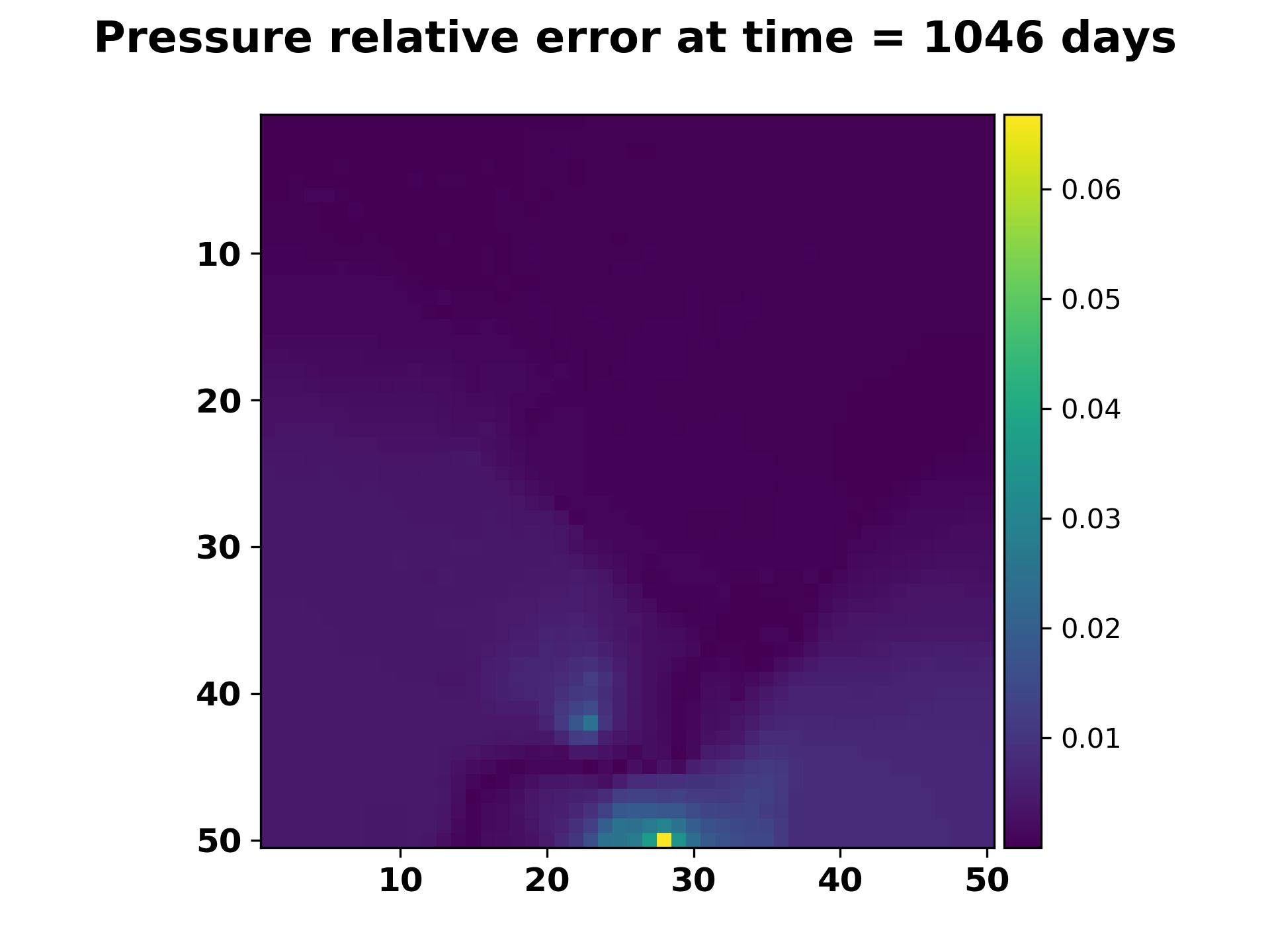}
		\caption{}
		\label{fig:5_25_2}
	\end{subfigure}%
	\caption{Relative error in pressure at time = (a) 506 days (b) 1026 days}
	\label{fig:5_25}
\end{figure}

Since, we consider the heterogeneous case here, an additional feature corresponding to the well block permeability is added which was not included for the homogeneous model before. We first consider a test case as shown in Figure \ref{fig:5_24_case1} with producer well location at (28,50). In Figures \ref{fig:5_24_a} and \ref{fig:5_24_b}, we show pressure and saturation solutions obtained by predicted POD basis coefficients with the true solutions for two of the test cases at different times respectively. We also show the error in pressure solution in Figure \ref{fig:5_25} to verify the same observations as in the homogeneous case where the maximum error is obtained at producer well locations. By observing the saturation solutions, we can see that there are small saturation errors near the waterfront. The time at which all the solutions are reported are chosen to be somewhere in the middle and at the end of simulation run time. As can be observed, the error is very small throughout the spatial domain and maximum at the well location which is the same observation as before. 

\begin{figure}[htb!]
	\centering
	\begin{subfigure}{0.45\textwidth}
		\centering
		\includegraphics[width=\textwidth]{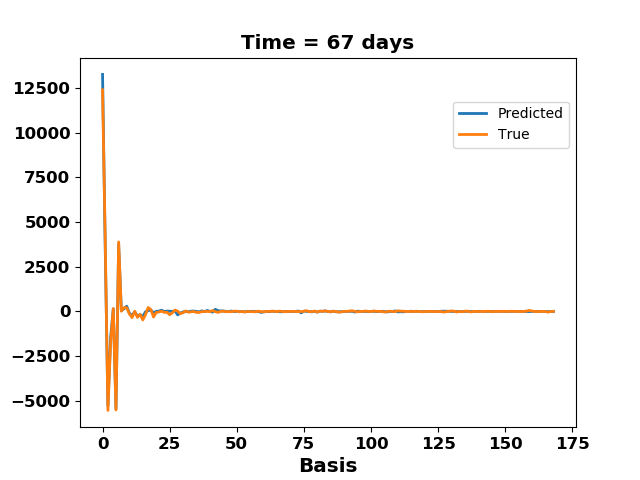}
		\caption{}
		\label{fig:5_26_1}
	\end{subfigure}%
	~
	\centering
	\begin{subfigure}{0.45\textwidth}
		\centering
		\includegraphics[width=\textwidth]{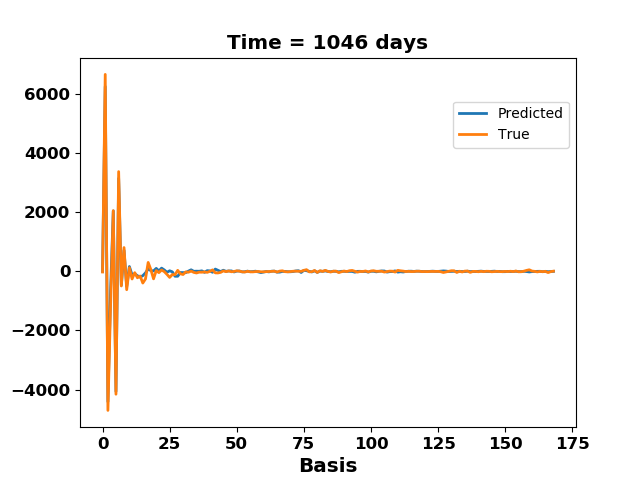}
		\caption{}
		\label{fig:5_26_2}
	\end{subfigure}%
	\caption{True and ML predicted pressure basis coefficient comparison at time =  (a) 506 days and (b) 1026 days}
	\label{fig:5_26}
\end{figure}

\begin{figure}[htb!]
	\centering
	\begin{subfigure}{0.45\textwidth}
		\centering
		\includegraphics[width=\textwidth]{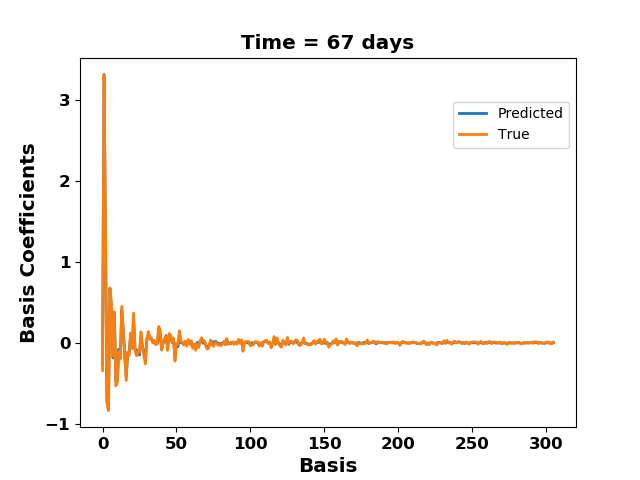}
		\caption{}
		\label{fig:5_27_1}
	\end{subfigure}%
	~
	\centering
	\begin{subfigure}{0.45\textwidth}
		\centering
		\includegraphics[width=\textwidth]{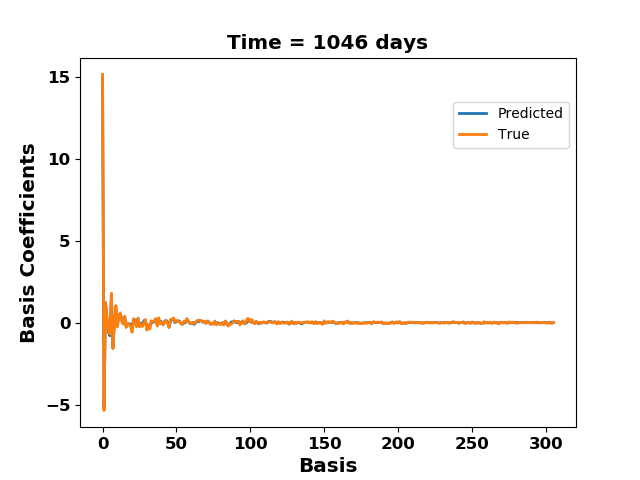}
		\caption{}
		\label{fig:5_27_2}
	\end{subfigure}%
	\caption{True and ML predicted pressure basis coefficient comparison at time =  (a) 506 days and (b) 1026 days}
	\label{fig:5_27}
\end{figure}

\begin{figure}[htb!]
	\centering
	\includegraphics[width=\textwidth]{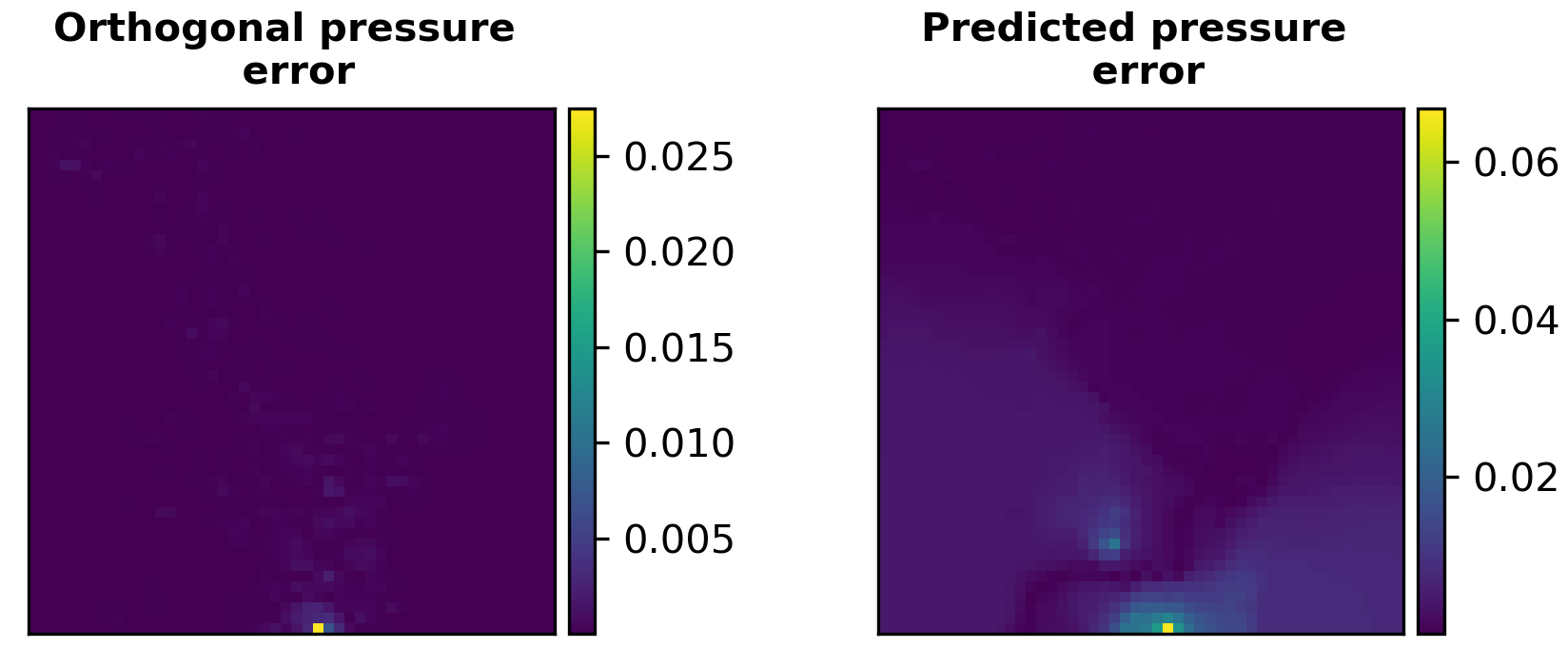}
	\caption{Comparison of relative error in pressure due to orthogonal projection (left) and ML predicted solution (right)}
	\label{fig:5_28}
\end{figure}

\begin{figure}[htb!]
	\centering
	\begin{subfigure}{0.45\textwidth}
		\centering
		\includegraphics[width=\textwidth]{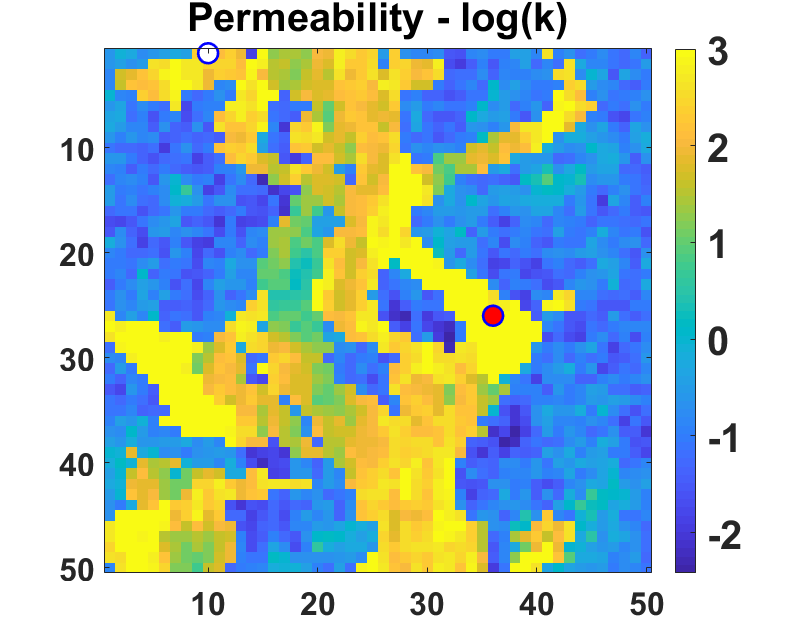}
		\caption{}
		\label{fig:5_29_1}
	\end{subfigure}%
	~\\
	\centering
	\begin{subfigure}{0.45\textwidth}
		\centering
		\includegraphics[width=\textwidth]{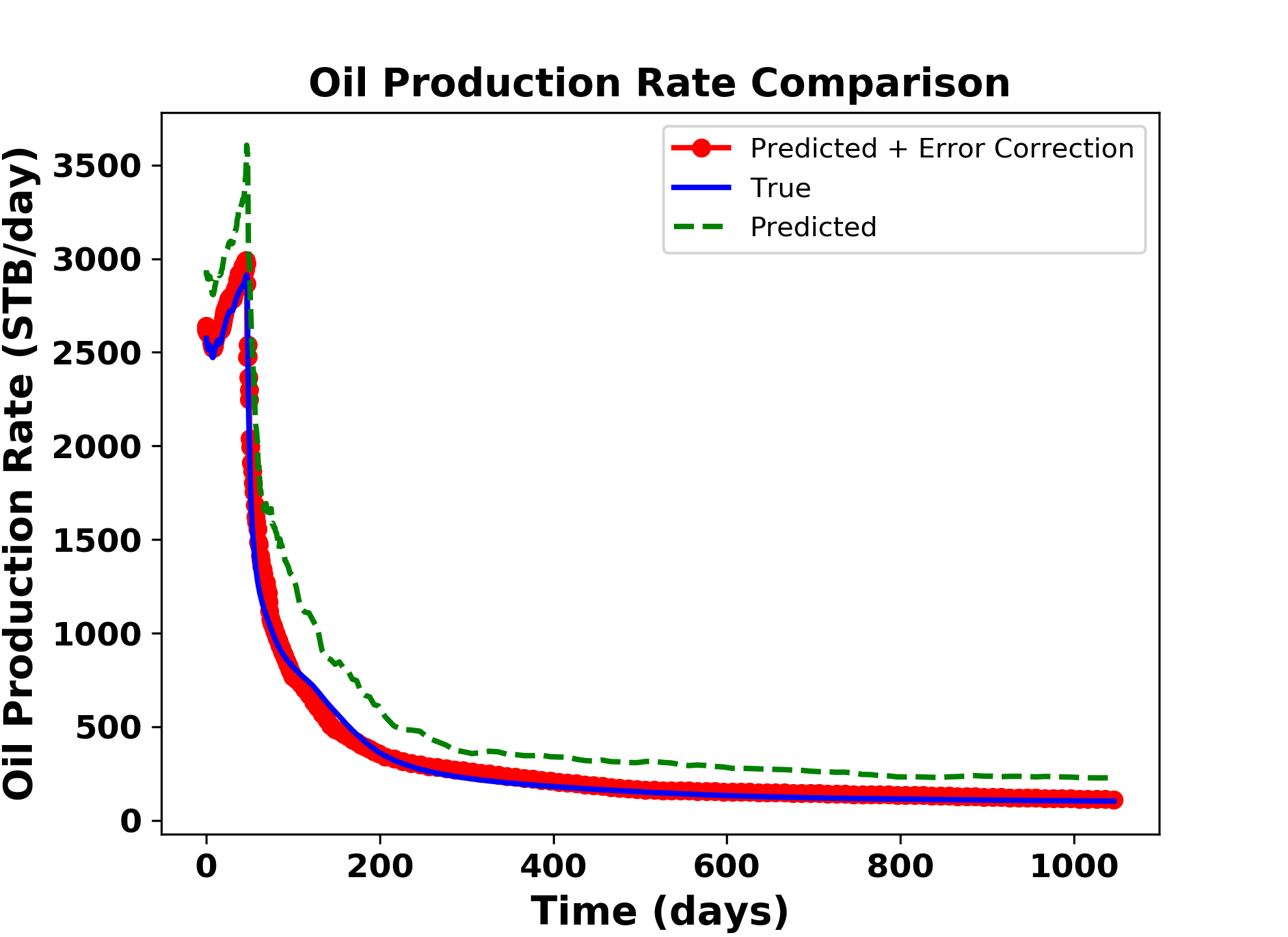}
		\caption{}
		\label{fig:5_29_2}
	\end{subfigure}
	~
	\centering
	\begin{subfigure}{0.45\textwidth}
		\centering
		\includegraphics[width=\textwidth]{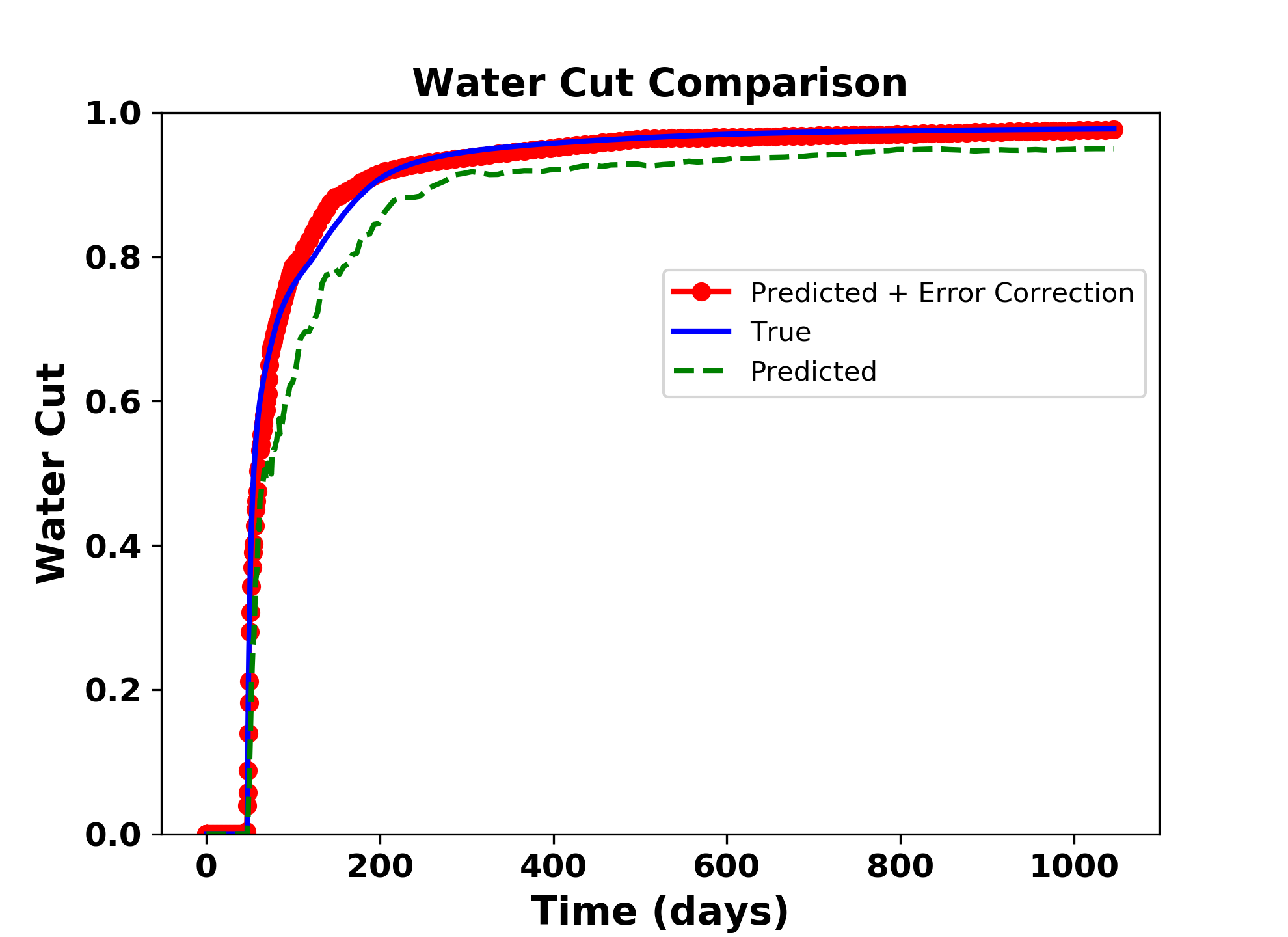}
		\caption{}
		\label{fig:5_29_3}
	\end{subfigure}%
	\caption{(a) Test case 1 with producer well location at (36,26) (b) Comparison of oil production rate and (c) Comparison of water cut, predicted using global PMOR method alone (dotted green line) and after implementation of error correction model (red circled line) with the true solution (blue line)}
	\label{fig:5_29}
\end{figure}

\begin{figure}[htb!]
	\centering
	\begin{subfigure}{0.45\textwidth}
		\centering
		\includegraphics[width=\textwidth]{Chapter5_24_case1.png}
		\caption{}
		\label{fig:5_30_1}
	\end{subfigure}%
	~\\
	\centering
	\begin{subfigure}{0.45\textwidth}
		\centering
		\includegraphics[width=\textwidth]{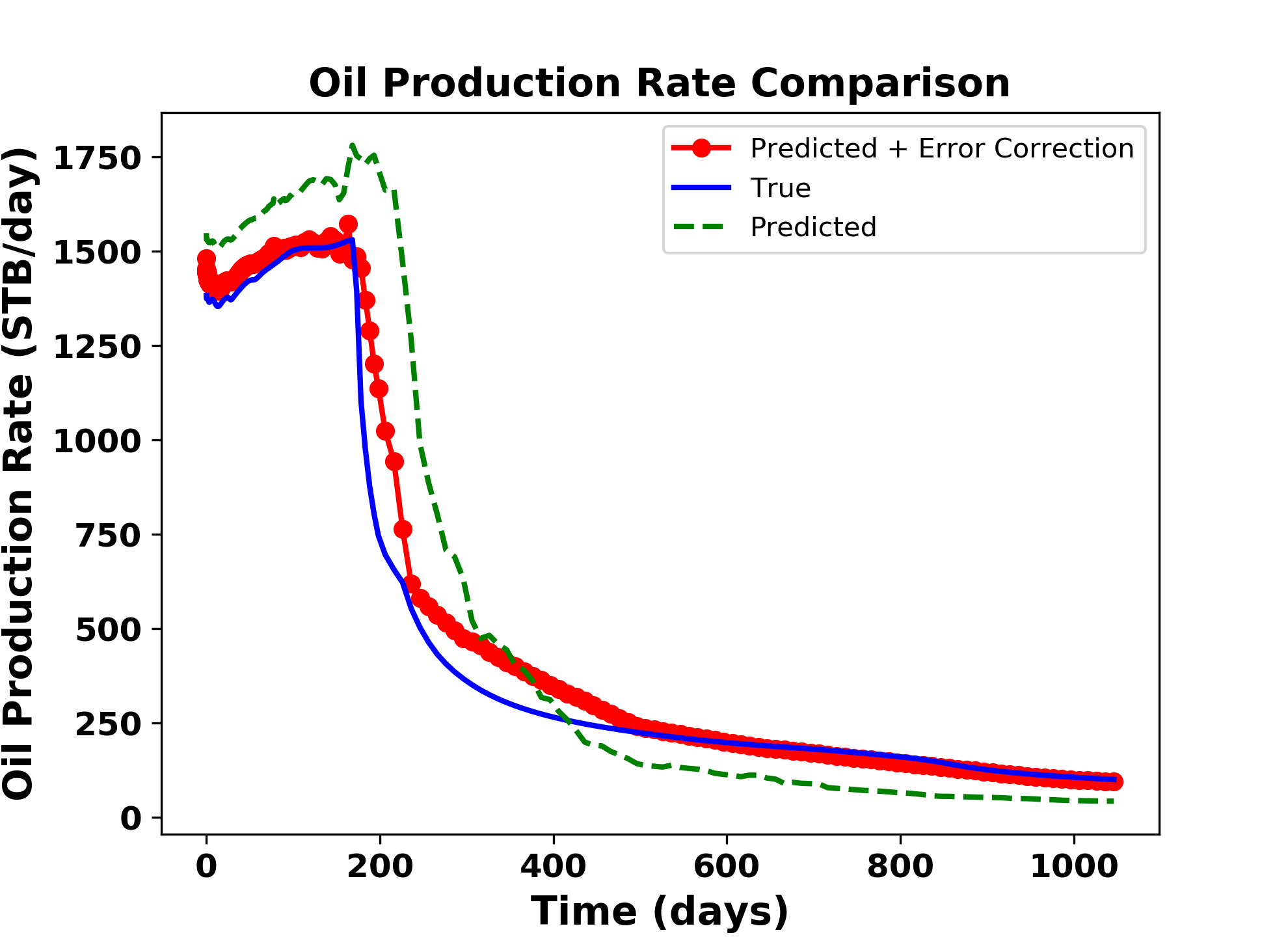}
		\caption{}
		\label{fig:5_30_2}
	\end{subfigure}
	~
	\centering
	\begin{subfigure}{0.45\textwidth}
		\centering
		\includegraphics[width=\textwidth]{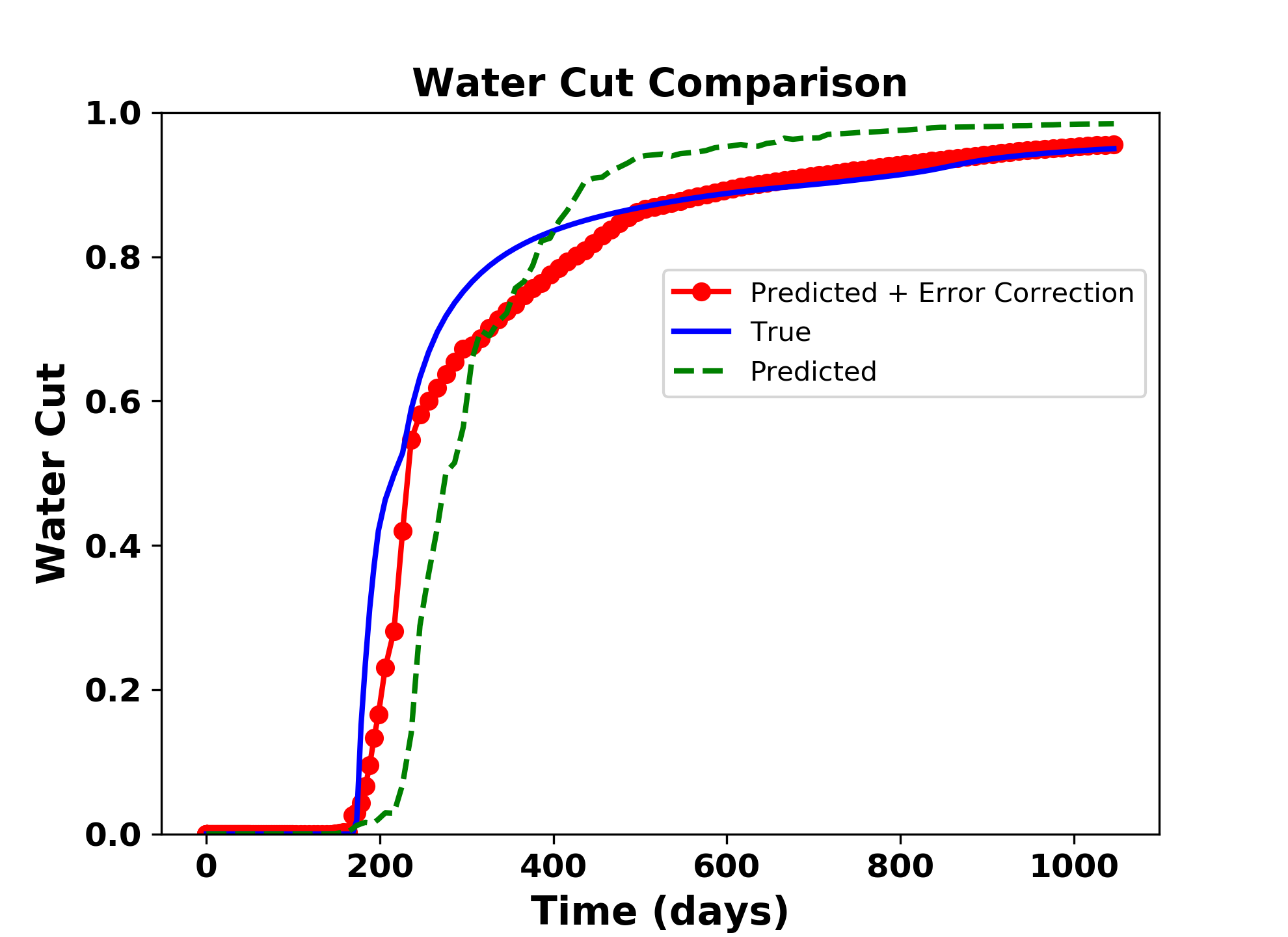}
		\caption{}
		\label{fig:5_30_3}
	\end{subfigure}%
	\caption{(a) Test case 2 with producer well location at (28,50) (b) Comparison of oil production rate and (c) Comparison of water cut, predicted using global PMOR method alone (dotted green line) and after implementation of error correction model (red circled line) with the true solution (blue line)}
	\label{fig:5_30}
\end{figure}

\begin{figure}[htb!]
	\centering
	\begin{subfigure}{0.45\textwidth}
		\centering
		\includegraphics[width=\textwidth]{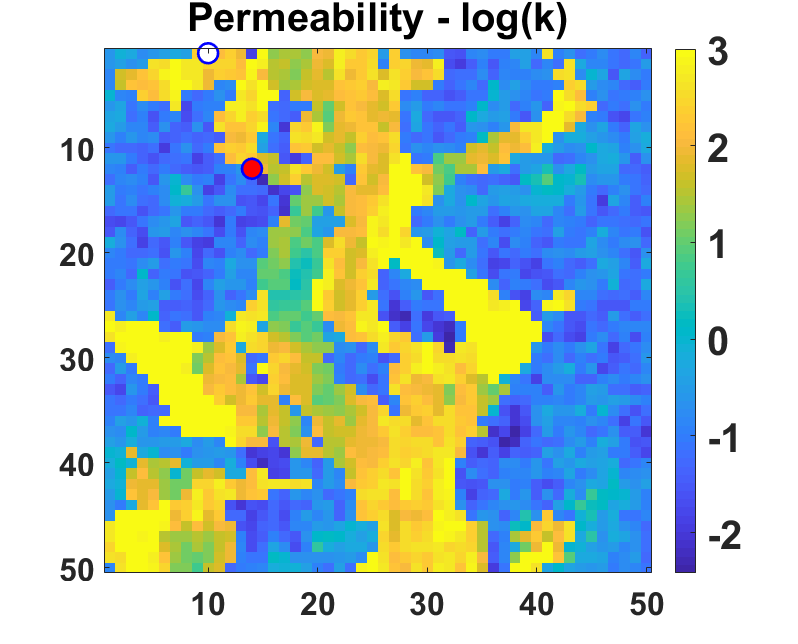}
		\caption{}
		\label{fig:5_31_1}
	\end{subfigure}%
	~\\
	\centering
	\begin{subfigure}{0.45\textwidth}
		\centering
		\includegraphics[width=\textwidth]{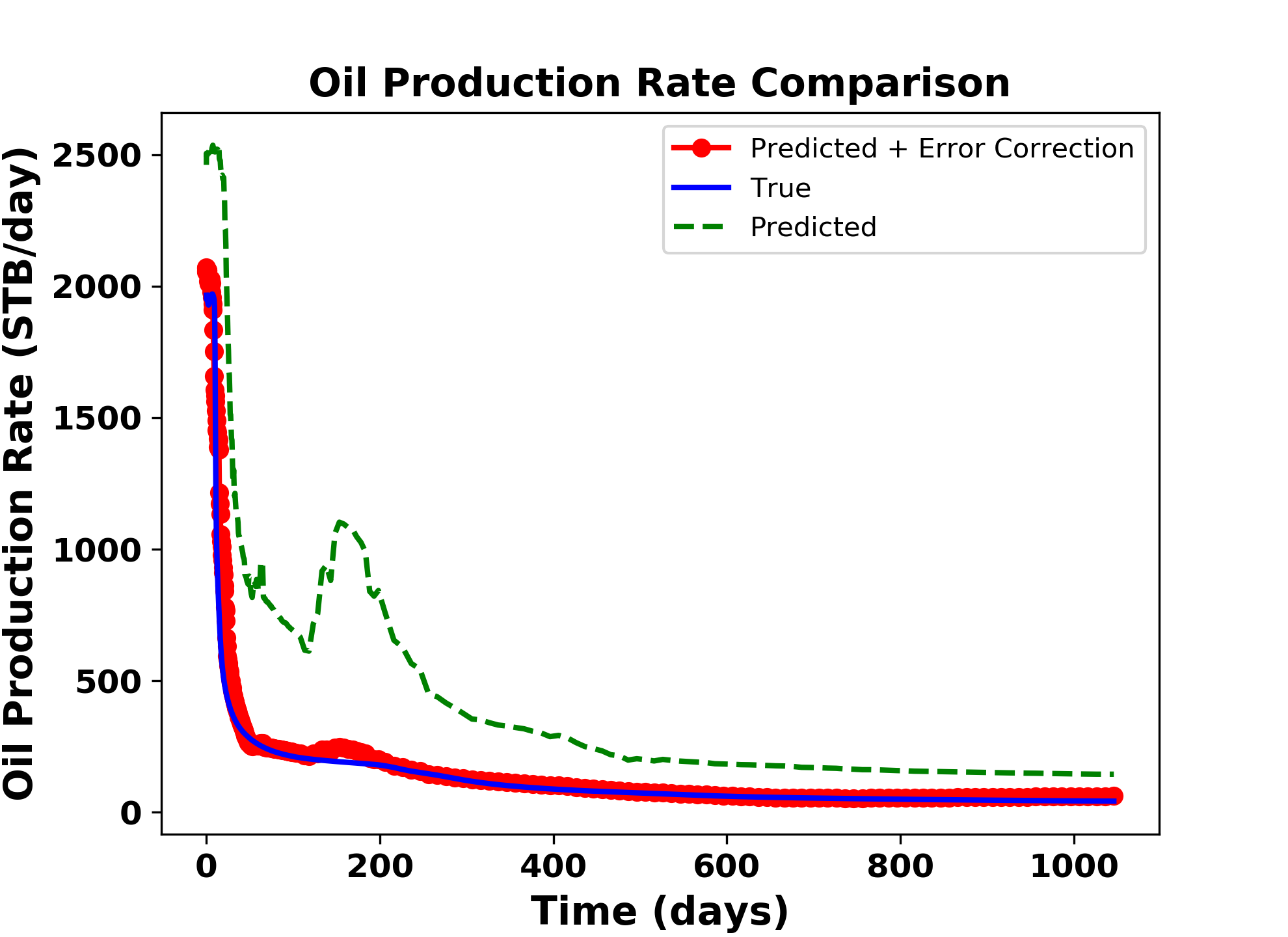}
		\caption{}
		\label{fig:5_31_2}
	\end{subfigure}
	~
	\centering
	\begin{subfigure}{0.45\textwidth}
		\centering
		\includegraphics[width=\textwidth]{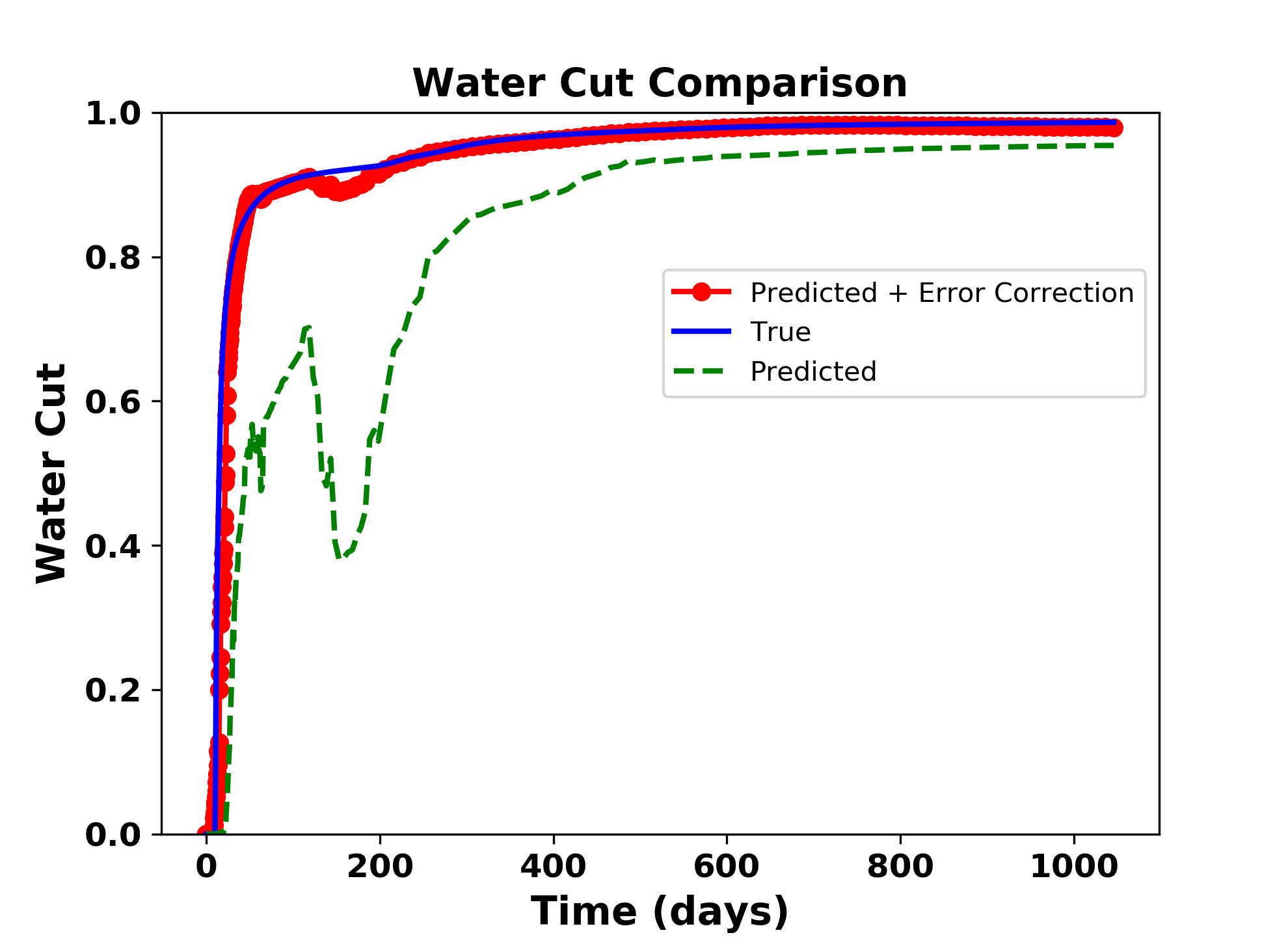}
		\caption{}
		\label{fig:5_31_3}
	\end{subfigure}%
	\caption{(a) Test case 3 with producer well location at (14,12) (b) Comparison of oil production rate and (c) Comparison of water cut, predicted using global PMOR method alone (dotted green line) and after implementation of error correction model (red circled line) with the true solution (blue line)}
	\label{fig:5_31}
\end{figure}

Here, we also show the comparison of basis coefficients for pressure and saturation as predicted by the ML algorithm at the same timesteps to determine the validity of ML models (Figures (\ref{fig:5_26}) and (\ref{fig:5_27})r). We can see that the ML model produces a very small error in prediction of the coefficients for some of the pressure and saturation basis. Hence to check if the discrepancy in the predicted solution is due to the ML model or the quality of basis, we also compare the plots of relative error in ML predicted solution and the orthogonal projection error computed using equation (\ref{eq:5_12}). This comparison for pressure in Figure \ref{fig:5_28} shows that the orthogonal projection error is maximum at the well location. Hence majority of the error in the output quantity of interest that is dependent on well block states is expected due to the quality of basis. Thus, it is in line with eour discussion for homogeneous case. As we see that it challenging to find a single global basis that is representative of the controllability properties at all the well locations, we use the error correction formulation to adjust for the error in state solutions at the gridblock. Figures \ref{fig:5_29}, \ref{fig:5_30} and \ref{fig:5_31} show the test cases for different well locations comparing the oil production rates and watercut. We can see that the corrected solutions for some cases like test case 2 may not be very accurate but still shows much improvement after error correction. Similarly, for test case 3, where we see a lot of irregularities in the solutions obtained just by coefficient prediction, show significant improvement after error correction.  

\begin{figure}[htb!]
	\centering
	\begin{subfigure}{0.45\textwidth}
		\centering
		\includegraphics[width=\textwidth]{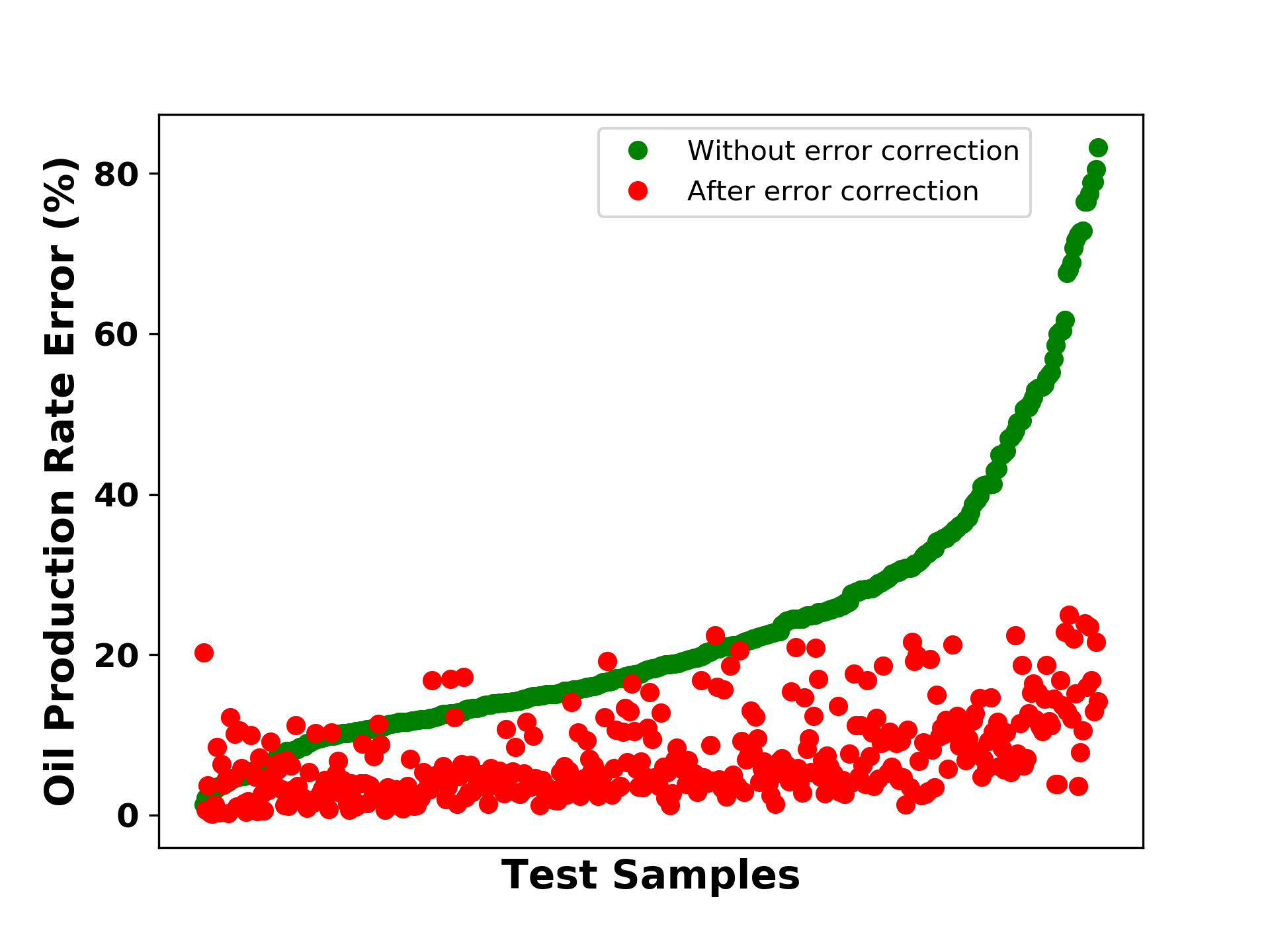}
		\caption{}
		\label{fig:5_31_1box}
	\end{subfigure}%
	~
	\centering
	\begin{subfigure}{0.45\textwidth}
		\centering
		\includegraphics[width=\textwidth]{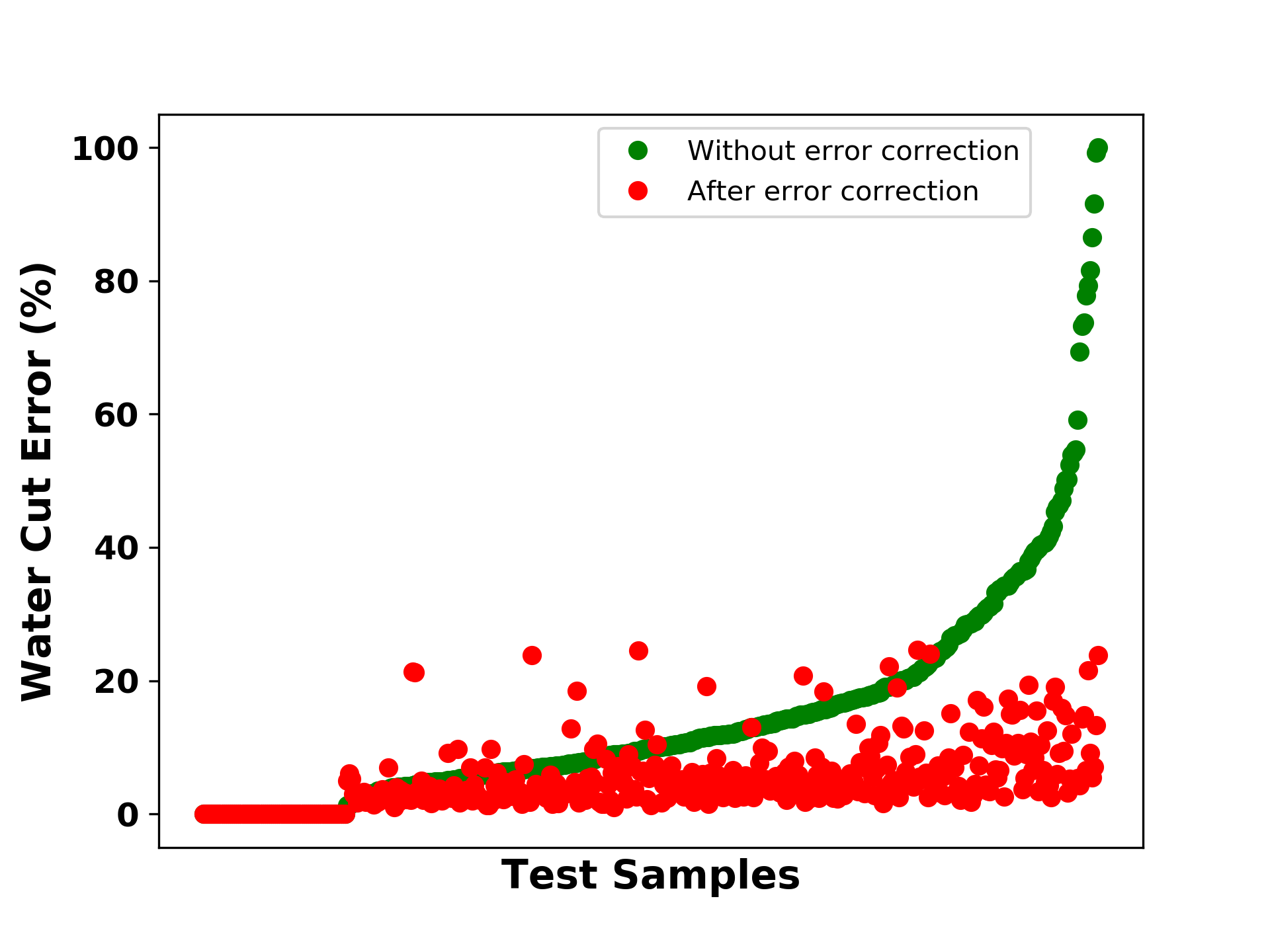}
		\caption{}
		\label{fig:5_31_2box}
	\end{subfigure}%
	\caption{Error in prediction of (a) Oil Production Rate (b) Water Cut, for all the test cases before and after the error correction of the solutions }
	\label{fig:5_31_box}
\end{figure}

For a better intuition about the results for this case study, we show a plot of accuracy in prediction for oil production rate and water cut for all the test cases (Figure \ref{fig:5_31_box}). The samples are arranged in increasing order of errors obtained just from predicted basis coefficients and for each sample we show the error after employing the error correction model. It can be observed that after error correction, the accuracy of predictions in oil production rate and water cut increase significantly. There are some test cases that produces higher error after correction which were found to be the cases where no water cut was observed and very low oil production rates. A better sampling strategy can solve this issue. The average accuracy for all test cases after error correction is shown in Table \ref{tab:hetero_acc1}. 

\begin{table}[htb!]
	\begin{center}
		
		\begin{tabular}{lclcl}
			\hline\noalign{\smallskip}
			\textbf{} & \textbf{Average Accuracy (\%)} \\
			\noalign{\smallskip}\hline\noalign{\smallskip}
			\textbf{Oil production rate }& 92.6\\
			\textbf{Water cut} & 95.04\\
			\noalign{\smallskip}\hline
		\end{tabular}
	\end{center}
	\caption{Heterogeneous reservoir case 1: Average accuracy of oil production rate and water cut for all test samples}
	\label{tab:hetero_acc1}       
	\vspace*{-1em}
\end{table}

Table \ref{tab:time_1well_highdim} shows the time comparison in seconds for these cases between fine scale simulations run in matlab and the proposed PMOR model with error correction, both run on a local 8 core machine. We see the speedups of about 50$\times$ for these test cases. This does not include time required to train the PMOR models. 

\begin{table}[htb!]
	\begin{center}
		
		\begin{tabular}{lclclc|}
			\hline\noalign{\smallskip}
			\textbf{} & \textbf{Fine scale simulation} & \textbf{PMOR + Error correction}  \\
			\noalign{\smallskip}\hline\noalign{\smallskip}
			\textbf{Test Case 1}& 65 seconds & \enspace \enspace\enspace \enspace\enspace \enspace1.3 seconds\\
			\textbf{Test Case 2} & 52 seconds & \enspace \enspace\enspace \enspace\enspace \enspace1 seconds\\
			\noalign{\smallskip}\hline
		\end{tabular}
	\end{center}
	\caption{Time (seconds) comparison for the two test cases between fine scale simulation and reduced order model with error correction}
	\label{tab:time_1well_highdim}       
	\vspace*{-1em}
\end{table}

\begin{figure}[htb!]
	\centering
	\begin{subfigure}{0.45\textwidth}
		\centering
		\includegraphics[width=\textwidth]{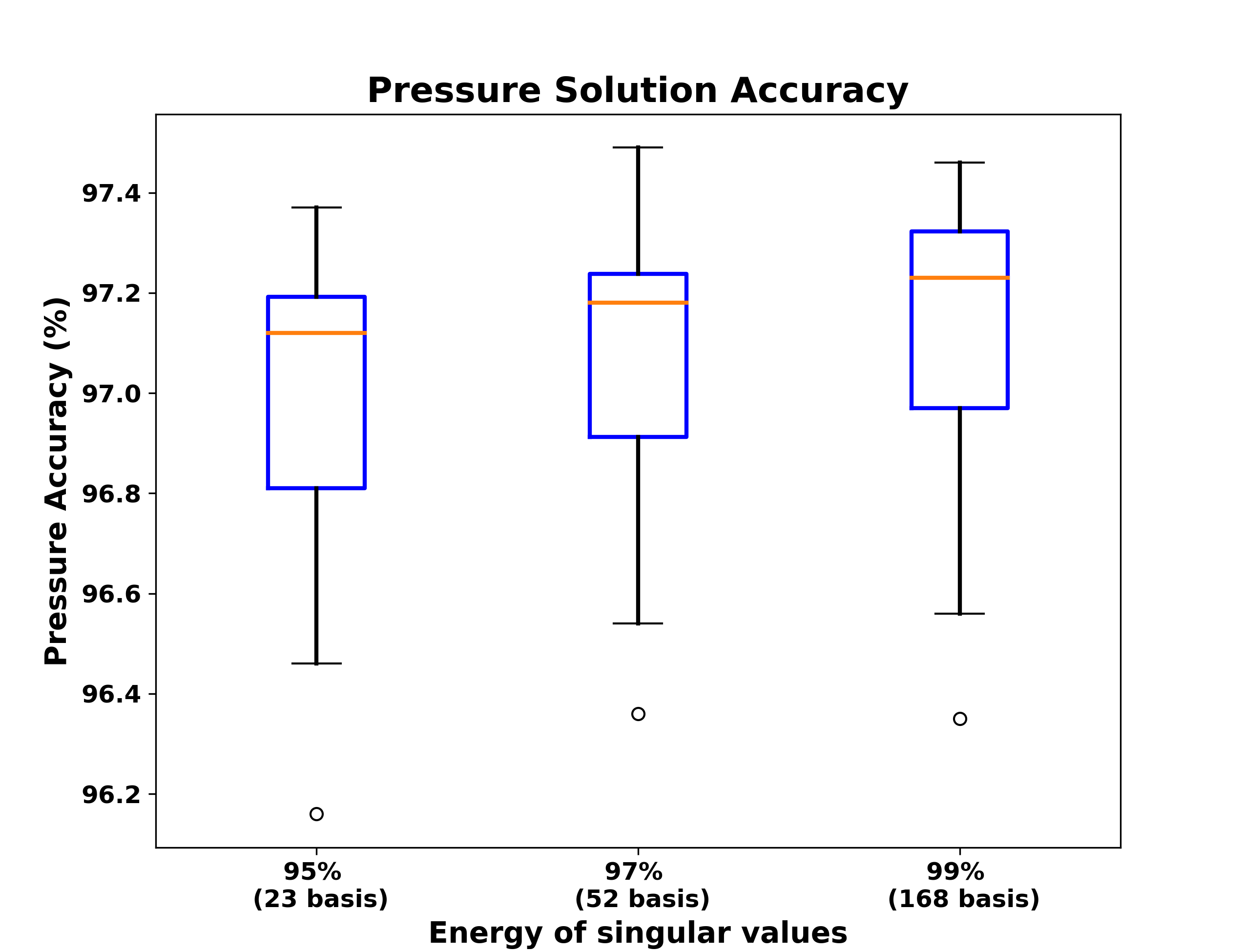}
		\caption{}
		\label{fig:5_32}
	\end{subfigure}%
	~
	\centering
	\begin{subfigure}{0.45\textwidth}
		\centering
		\includegraphics[width=\textwidth]{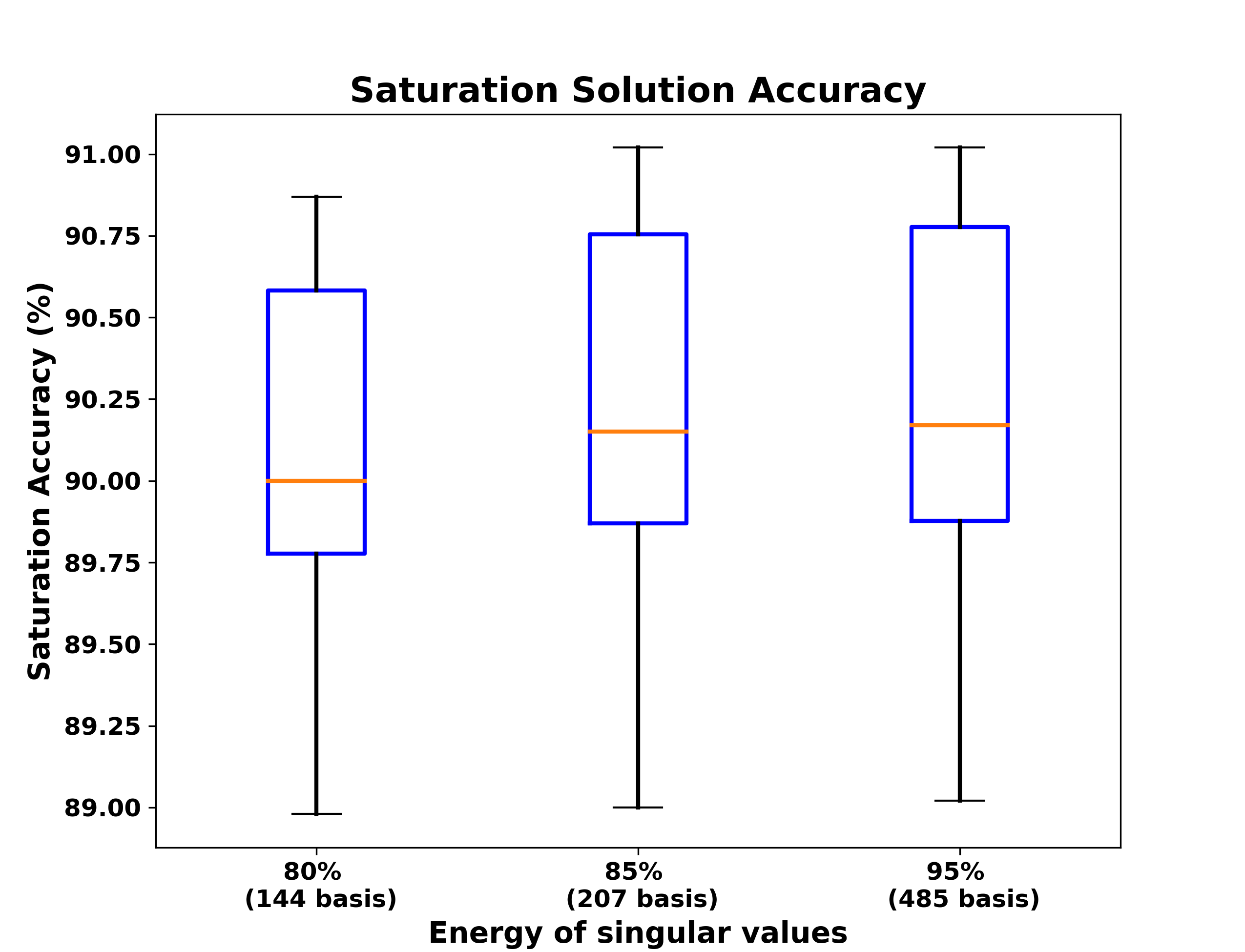}
		\caption{}
		\label{fig:5_33}
	\end{subfigure}%
	\caption{Box plot of accuracies in pressure and saturation using predicted POD coefficients for increasing basis dimensions}
	\label{fig:5_32_box}
\end{figure}

It is also important to decide on the energy criteria used for choosing the basis dimensions, since it is not trivial to make this decision a priori. As discussed before, there is a trade-off in accuracy and computational time between the decision to select the number of training samples and the basis dimensions. To get another perspective, we also show a box plot of the pressure and saturation solution accuracies as predicted from the basis coefficients for a range of energy criteria (Figure \ref{fig:5_32_box}). Each box is constructed with 10 different training sets. As can be expected, with increasing number of basis, we see an increase in the accuracies of solutions. However, we also see a drastic reduction in the number of basis with decreasing percentage of singular values selected, especially for pressure. The accuracy shown here is the that when compared at all the gridblocks, which is not very different in magnitude between different basis dimensions. But it was observed that the accuracy of solutions at the well gridblock increases significantly with increasing basis dimension. 

\begin{figure}[htb!]
	\centering
	\begin{subfigure}{0.45\textwidth}
		\centering
		\includegraphics[width=\textwidth]{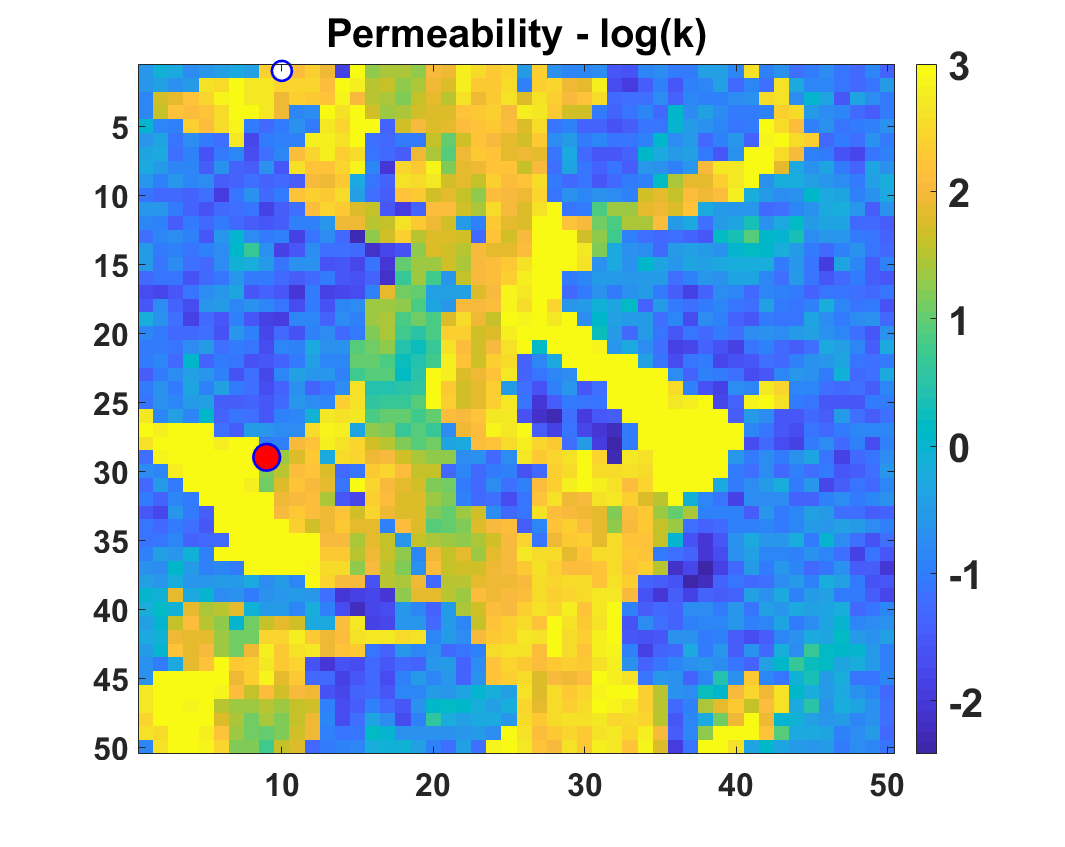}
		\caption{}
		\label{fig:5_34_1}
	\end{subfigure}%
	~\\
	\centering
	\begin{subfigure}{0.40\textwidth}
		\centering
		\includegraphics[width=\textwidth]{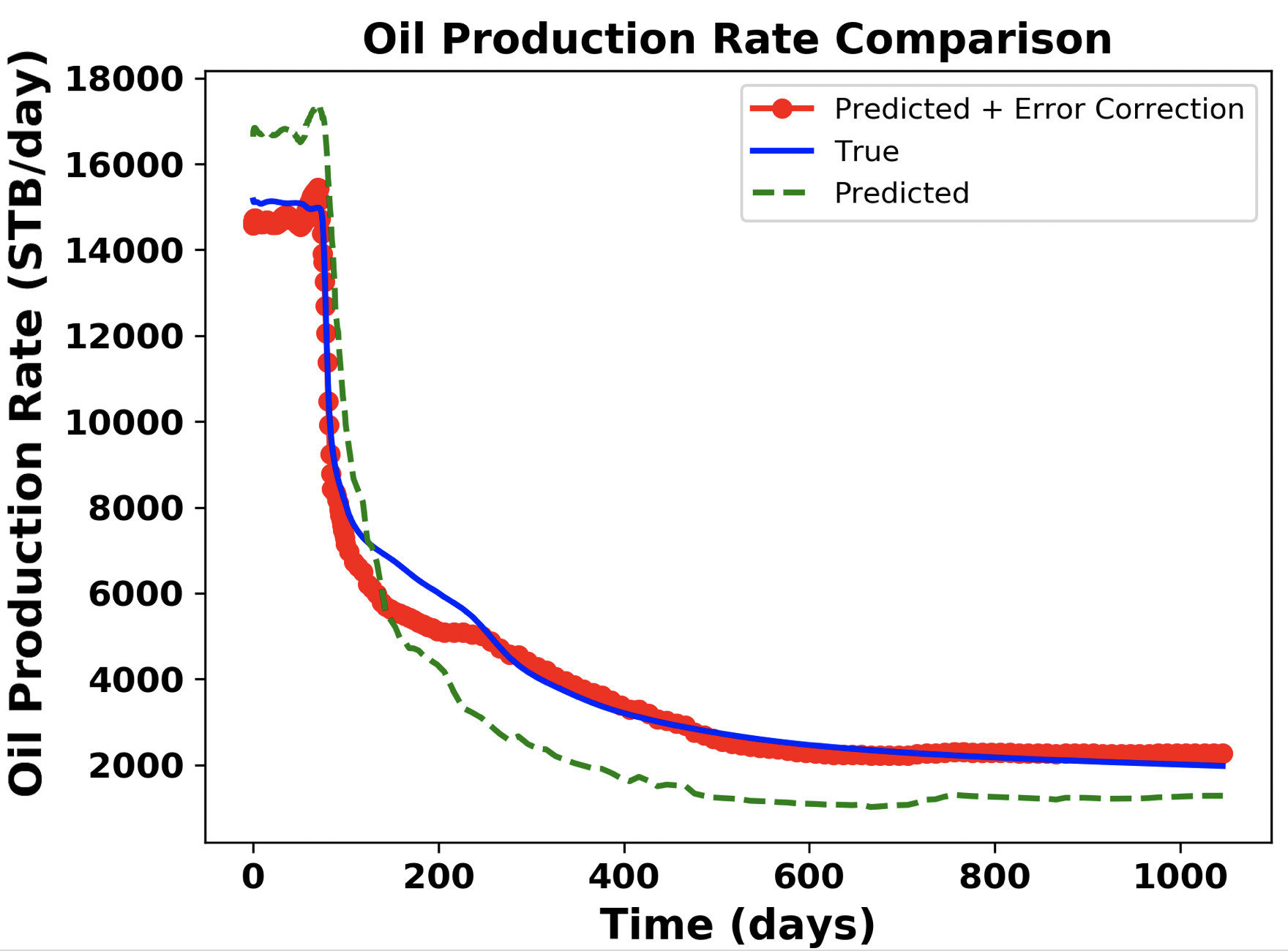}
		\caption{}
		\label{fig:5_34_2}
	\end{subfigure}
	~
	\centering
	\begin{subfigure}{0.45\textwidth}
		\centering
		\includegraphics[width=\textwidth]{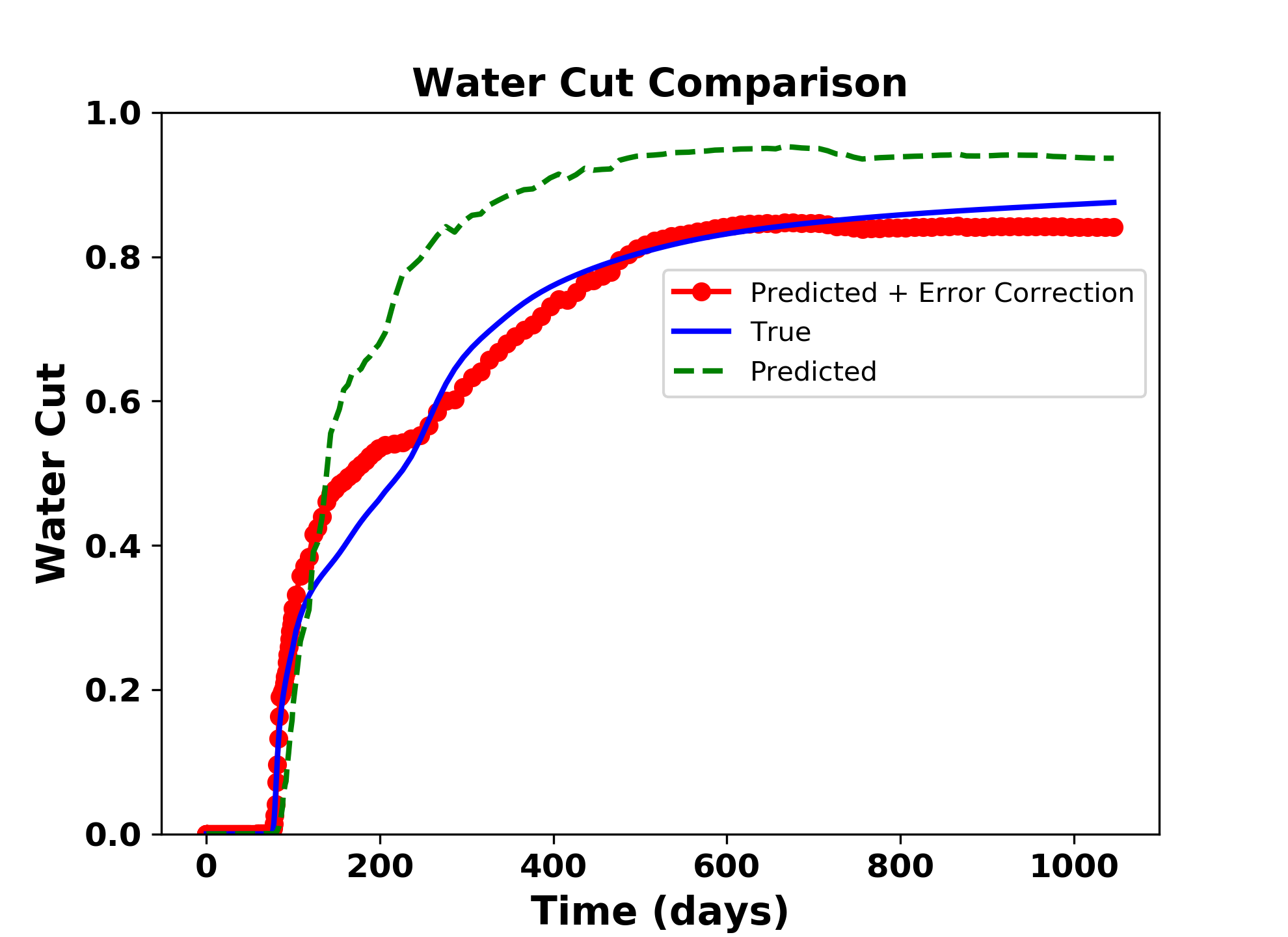}
		\caption{}
		\label{fig:5_34_3}
	\end{subfigure}%
	\caption{(a) Test case 1 with producer well location at (9,29) (b) Comparison of oil production rate and (c) Comparison of water cut, predicted using global PMOR method alone (dotted green line) and after implementation of error correction model (red circled line) with the true solution (blue line)}
	\label{fig:5_34}
\end{figure}

\begin{figure}[htb!]
	\centering
	\begin{subfigure}{0.45\textwidth}
		\centering
		\includegraphics[width=\textwidth]{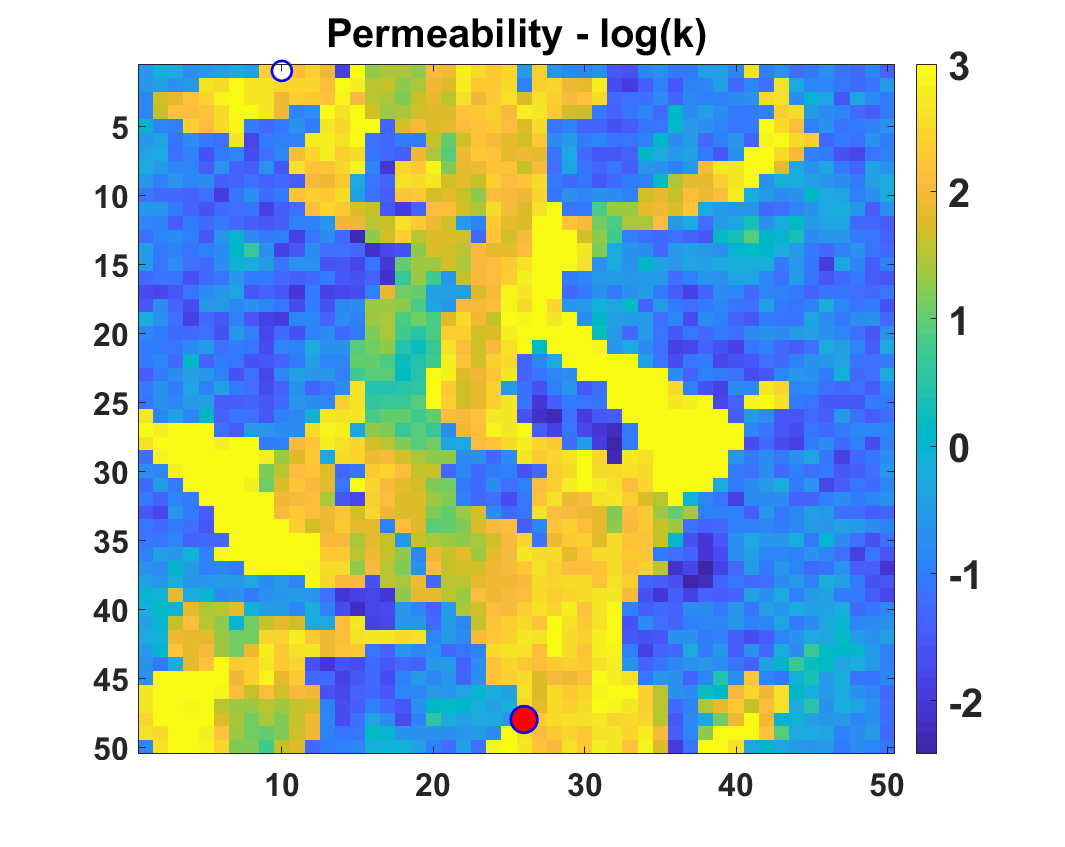}
		\caption{}
		\label{fig:5_35_1}
	\end{subfigure}%
	~\\
	\centering
	\begin{subfigure}{0.45\textwidth}
		\centering
		\includegraphics[width=\textwidth]{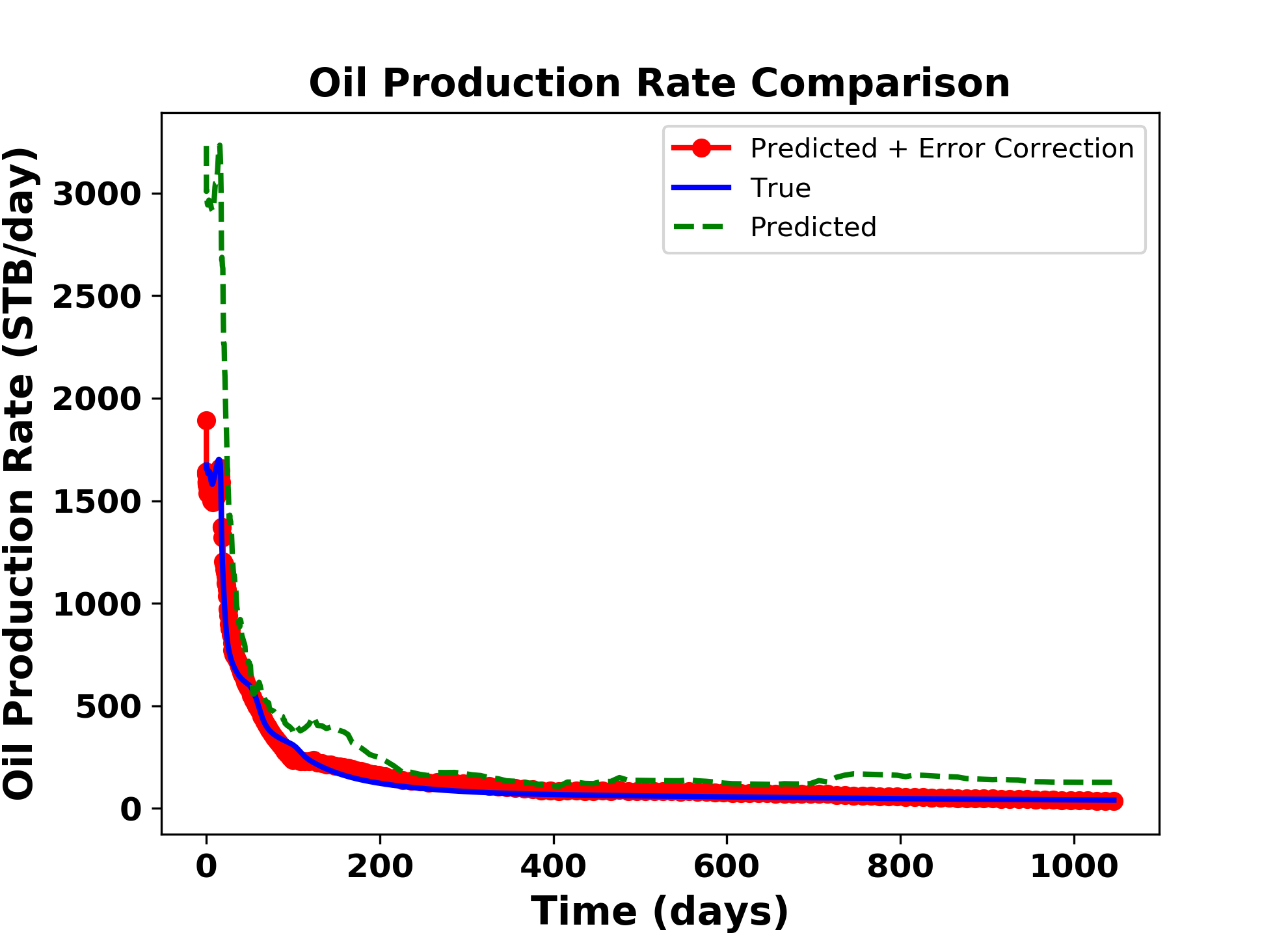}
		\caption{}
		\label{fig:5_35_2}
	\end{subfigure}
	~
	\centering
	\begin{subfigure}{0.45\textwidth}
		\centering
		\includegraphics[width=\textwidth]{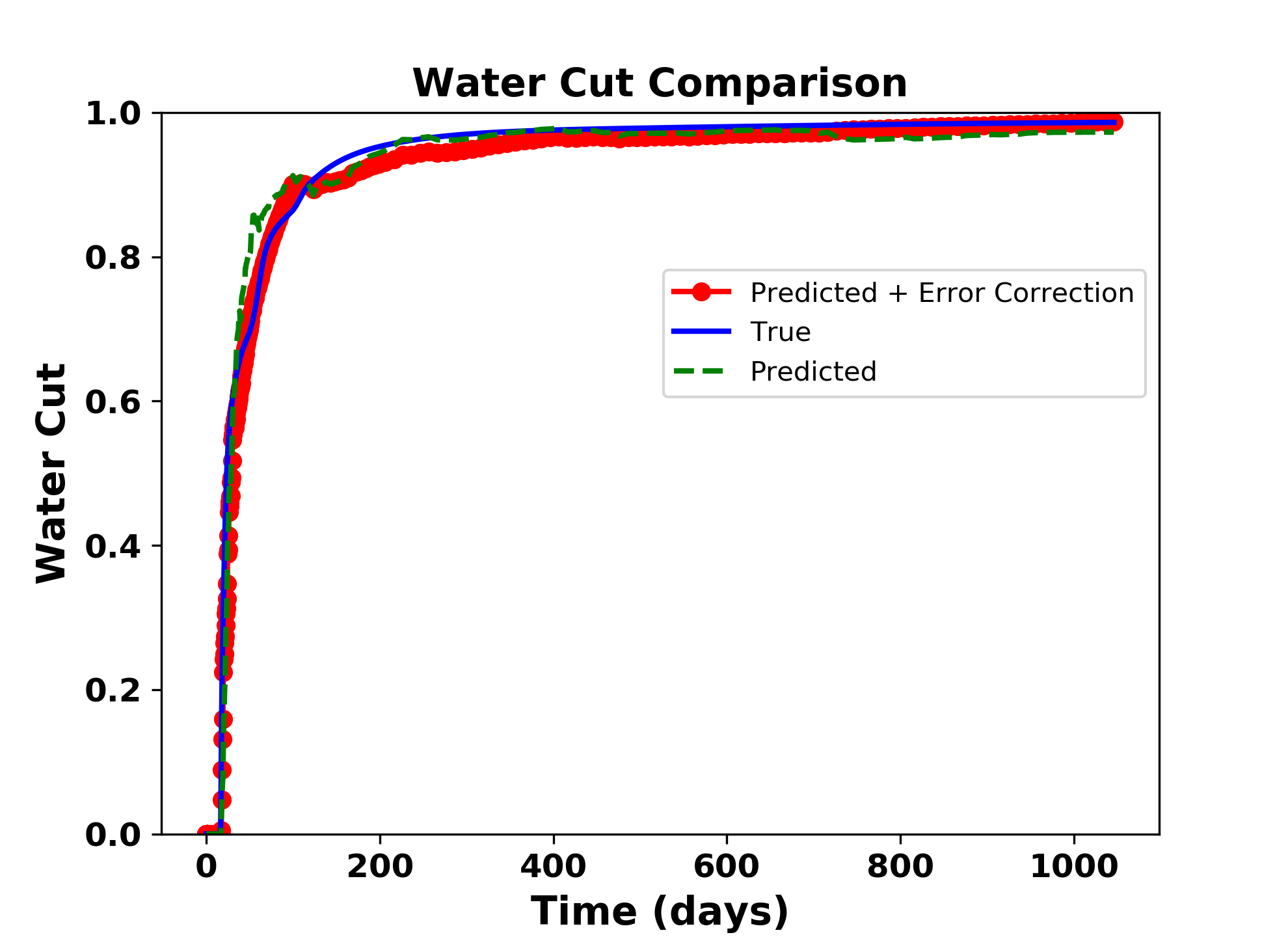}
		\caption{}
		\label{fig:5_35_3}
	\end{subfigure}%
	\caption{(a) Test case 2 with producer well location at (26,48) (b) Comparison of oil production rate and (c) Comparison of water cut, predicted using global PMOR method alone (dotted green line) and after implementation of error correction model (red circled line) with the true solution (blue line)}
	\label{fig:5_35}
\end{figure}

\begin{figure}[htb!]
	\centering
	\begin{subfigure}{0.45\textwidth}
		\centering
		\includegraphics[width=\textwidth]{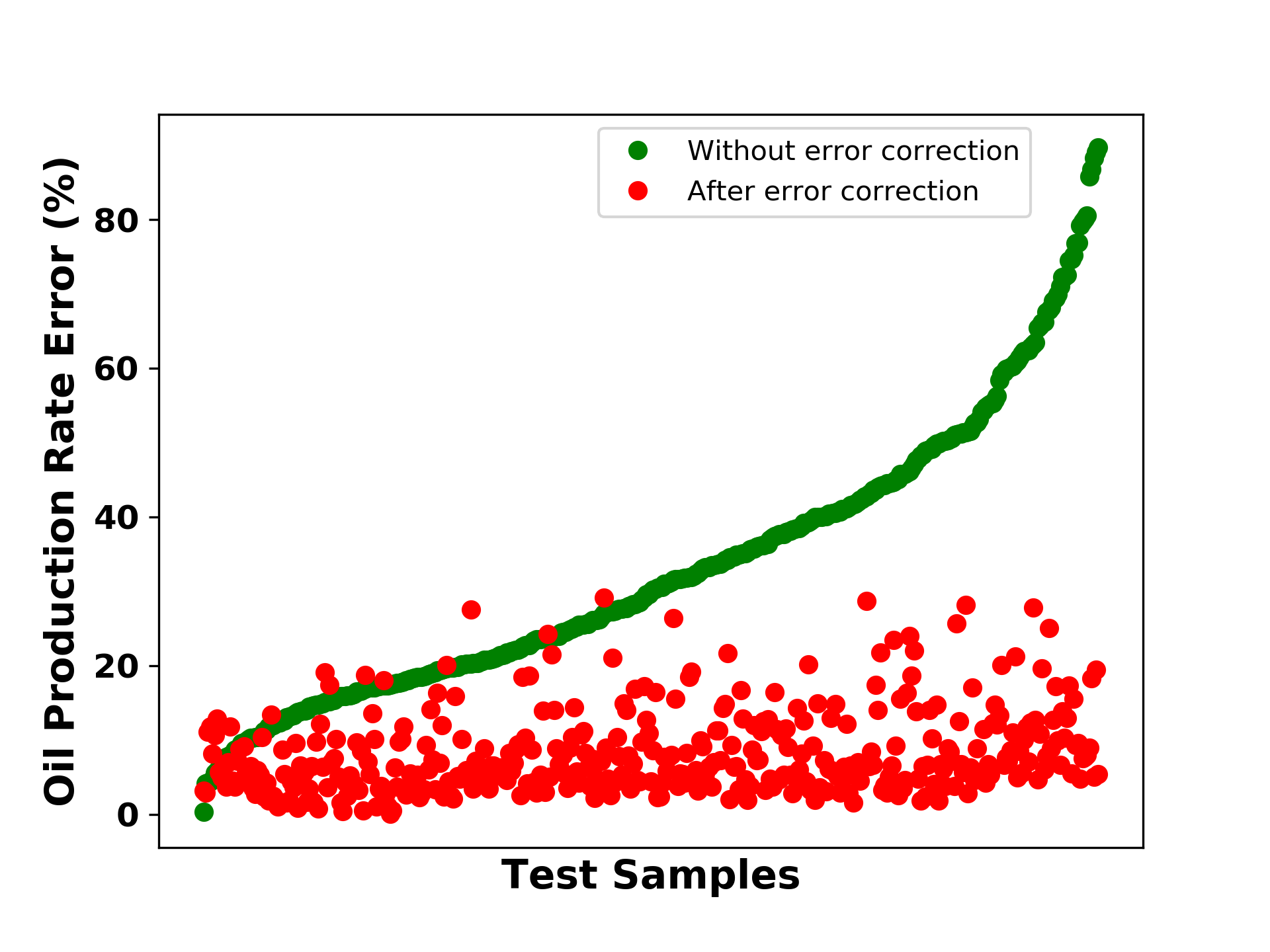}
		\caption{}
		\label{fig:5_36_1}
	\end{subfigure}%
	~
	\centering
	\begin{subfigure}{0.45\textwidth}
		\centering
		\includegraphics[width=\textwidth]{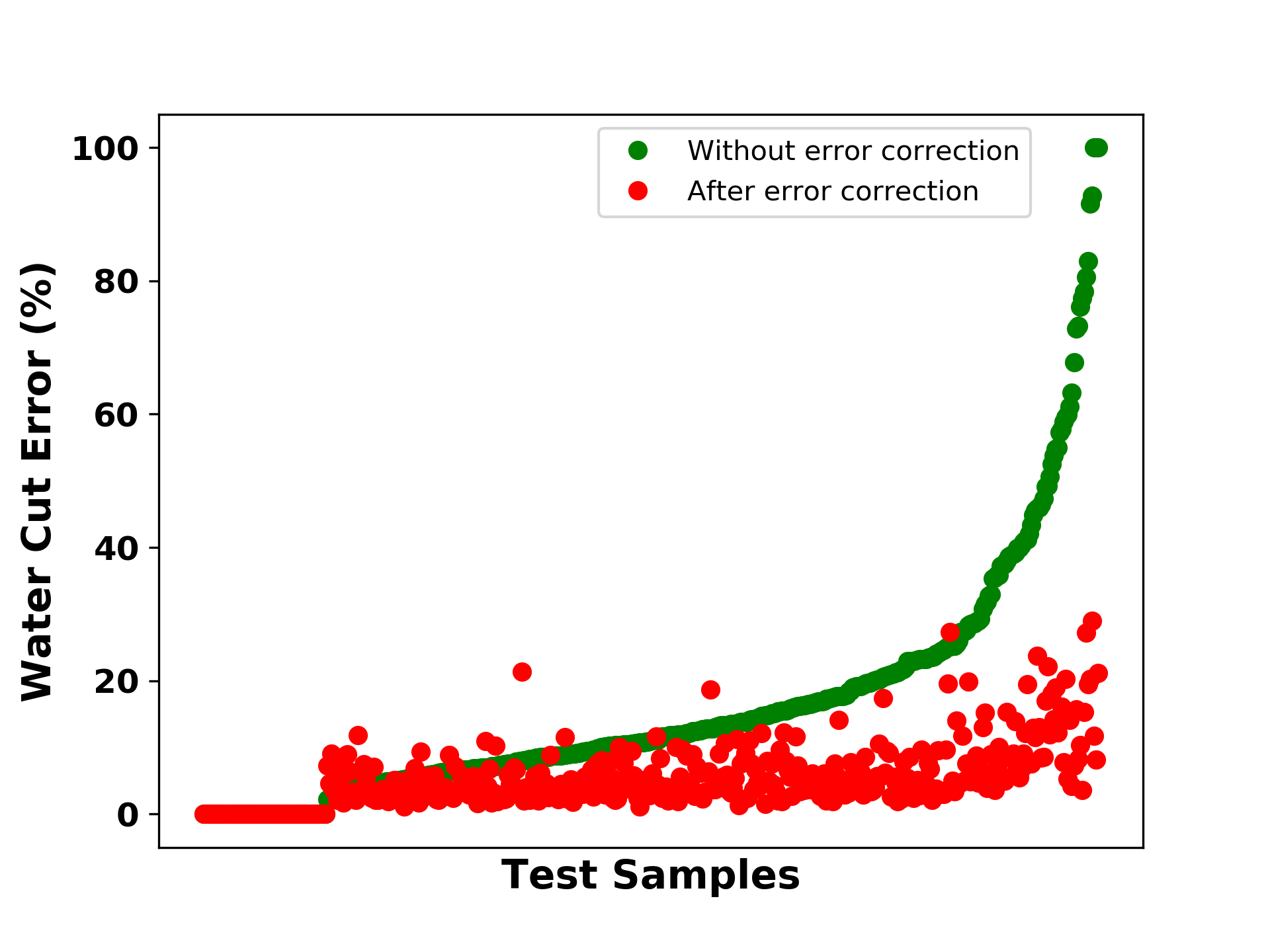}
		\caption{}
		\label{fig:5_36_2}
	\end{subfigure}%
	\caption{Error in prediction of (a) Oil Production Rate (b) Water Cut, for all the test cases before and after the error correction of the solutions }
	\label{fig:5_36}
\end{figure}

The case shown above correspond to pressure and saturation basis dimension selected using energy criteria of 99\% and 90\% respectively. Thus, we now evaluate, if the error correction model can also produce reasonable results if lower energy basis dimensions criteria is chosen. For this reason we consider another case with pressure and saturation basis dimension selected using energy criteria of 95\% and 80\% respectively (Note that these energy are chosen such that they capture most of the high energy singular values as shown in Figure \ref{fig:5_23_2}. If we choose even lower, the predicted basis coefficients are not expected to capture the physics of states behavior adequately and hence lead to inaccurate solutions). Figures \ref{fig:5_34} and \ref{fig:5_35} show two test cases comparing the oil production rate and water cut before and after error correction. In general, the solutions are a good match after error correction. The error in prediction, for all the test samples, before and after error correction is shown in Figures \ref{fig:5_36_1} and \ref{fig:5_36_2}. It can be seen that the error from predicted coefficients now increase at a higher rate which is expected due to much lower basis dimensions. However, after error correction, most of the test cases show much better accuracies. The average accuracy for oil production rate and water cut is shown in Table \ref{tab:hetero_acc2}. It is lower than that predicted in the case of higher basis dimension (Table \ref{tab:hetero_acc1}), but still not significantly less. Thus, it is up to one to choose appropriate basis dimensions such that significant computational speed ups are obtained at the cost of losing accuracy. 

\begin{table}[htb!]
	\begin{center}
		
		\begin{tabular}{lclcl}
			\hline\noalign{\smallskip}
			\textbf{} & \textbf{Average Accuracy (\%)} \\
			\noalign{\smallskip}\hline\noalign{\smallskip}
			\textbf{Oil production rate }& 91.74\\
			\textbf{Water cut} & 94.6\\
			\noalign{\smallskip}\hline
		\end{tabular}
	\end{center}
	\caption{Heterogeneous reservoir case 2: Average accuracy of oil production rate and water cut for all test samples}
	\label{tab:hetero_acc2}       
	\vspace*{-1em}
\end{table}

Table \ref{tab:time_1well_lowdim} shows the time comparison in seconds for these cases between fine scale simulations run and the proposed PMOR model with error correction, both run on a local 8 core machine. We see the speedups of about 100$\times$ for these test cases since the model outputs have lower dimensions and hence less complex. 

\begin{table}[htb!]
	\begin{center}
		
		\begin{tabular}{lclclc|}
			\hline\noalign{\smallskip}
			\textbf{} & \textbf{Fine scale simulation} & \textbf{PMOR + Error correction}  \\
			\noalign{\smallskip}\hline\noalign{\smallskip}
			\textbf{Test Case 1}& 56 seconds & \enspace \enspace\enspace \enspace\enspace \enspace0.7 seconds\\
			\textbf{Test Case 2} & 46 seconds & \enspace \enspace\enspace \enspace\enspace \enspace0.5 seconds\\
			\noalign{\smallskip}\hline
		\end{tabular}
	\end{center}
	\caption{Time (seconds) comparison for the two test cases (lower dimensional basis) between fine scale simulation and reduced order model with error correction}
	\label{tab:time_1well_lowdim}       
	\vspace*{-1em}
\end{table}

\clearpage
\subsection{Heterogeneous reservoir models - multiple wells case}

The results for a single producer and injector shows the validity of the proposed methodology, but for practical implementation, we need to see if this method shows such promising results. This is mainly the future work that needs to be addressed, but we show a simple case study as a beginning for future directions. We know from the MOR literature that, MOR techniques when used within optimization workflows, that require many simulation runs corresponding to new parameter sets during each iteration, the training set of parameters for MOR development is designed such that the parameters are not drastically different from those expected to be encountered in the test set. For well locations, as we see from the above results, it is very difficult to determine how different is one location to the other, especially for heterogeneous reservoirs like the section of SPE10 model considered. Also, the parameter domain for well locations increase significantly for large reservoir models, in that each grid block in the reservoir is a potential well location. This becomes even more challenging as we have more wells where we can have different combinations of well locations across all the gridblocks. In these scenarios, it just becomes impractical to devise PMOR strategies that can satisfy all possible combinations of well locations in the reservoir model. Thus, training a PMOR model must be tied to the optimization strategy to be used and training samples must be designed accordingly. For example, well location optimization being a very challenging problem given numerous possible well configurations, researchers resort to quick evaluations or diagnostics of the reservoir to rule out many areas in the reservoir as potential well locations \cite{Taware2012, Moyner2015}. And then, given this information, complex optimization routines using simulation runs can be sought for a systematic decision making process. 

\begin{figure}[!ht]
	\centering
	\includegraphics[scale=0.40]{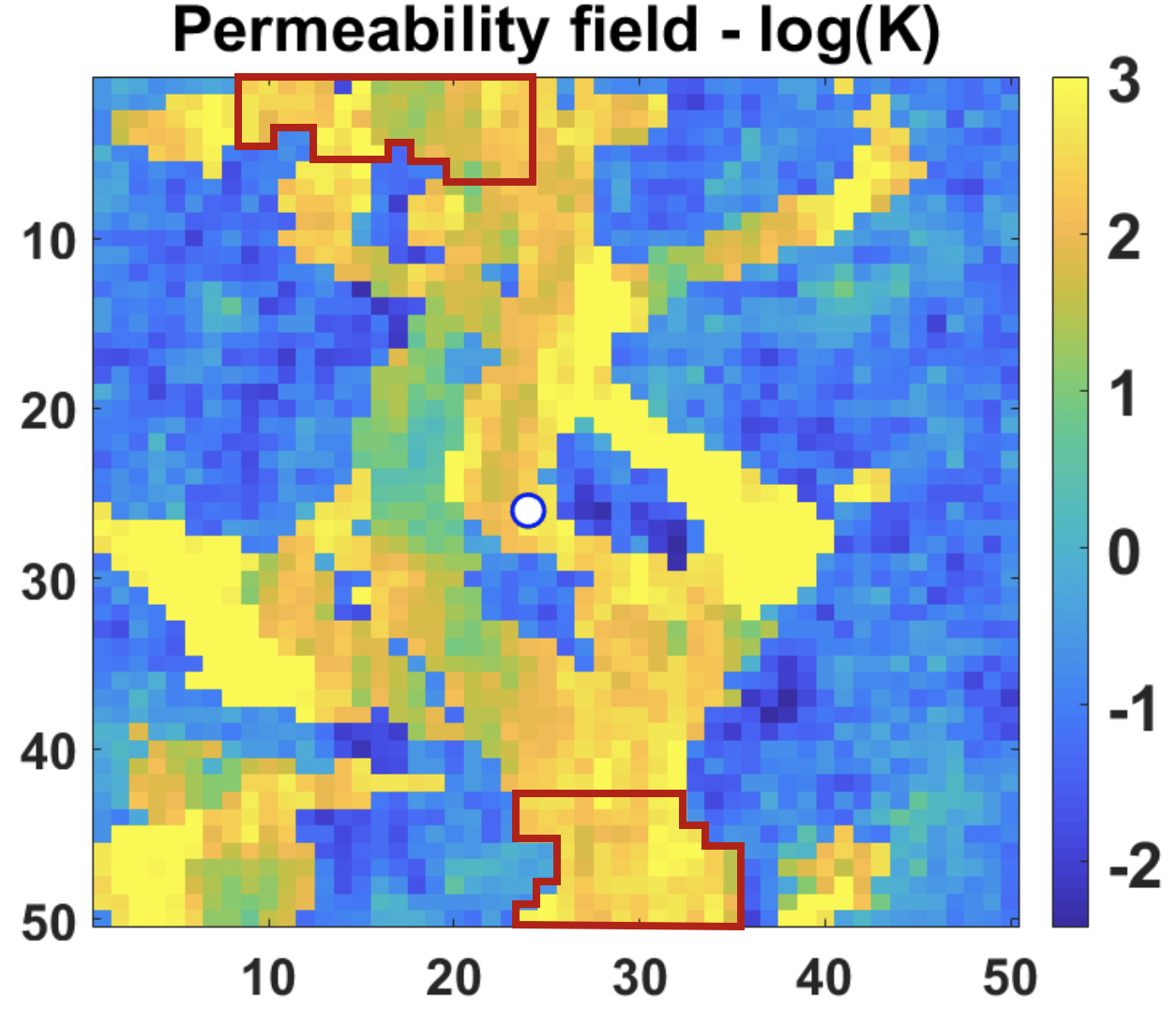}
	\caption{Reservoir model with on injector at location (24,26) and red regions showing the regions from where producer 1 (top of the reservoir) and producer 2 (bottom of the reservoir) locations are sampled}
	\label{fig:5_37}
\end{figure}

The method currently proposed, is suitable for any well location in the reservoir, and hence further adds to the training effort by collecting more samples. But in the future, when this workflow is applied with optimization routines, the training strategy should be designed keeping the optimization strategy in mind. As a preliminary idea, we show a simple case for the same model as considered for single well case, but now we consider 2 producers and 1 injector. We keep the injector location fixed and producers can change locations. With 2500 gridblocks and injector location fixed, we can have $2498 \times 2498 = 6,240,004$ number of producer combinations. If both the producers have the same well properties and produced with the same BHP schedule, i.e. both the well are equivalent, we have $ 6,240,004/2 = 3,120,002$ different well configurations. Thus, the optimization strategies are sought such that this possible combinations are reduced a priori by setting some constraints. It is thus impractical to develop the proposed PMOR strategy that is expected to work for all these combinations. We only consider a simplistic case here as a beginning to validate the method for multiple wells. Figure \ref{fig:5_37} shows the reservoir model with an injector and 2 regions (shown in red) where producer 1 and producer 2 locations are randomly sampled. We choose 20 samples in each region and hence we can have 400 combinations of producer 1 and producer 2 pairs. The BHP schedules for both producers are kept the same and have identical well properties. The simulation is ran for a duration of 500 days. ML model is trained to predict the POD basis coefficients for pressure and saturation using 50 sample points randomly chosen out of 400. 100 training samples are randomly chosen for error correction ANN model. Note that, the change here as compared to single well formulation lies in the input featured defined for ML models. Now the inputs include geometric and flow diagnostics based features for both the wells in ML model to predict the basis coefficients and the error correction model has the same features representing both producer wells along with the reduced order states and pore volumes injected. 

\begin{figure}[htb!]
	\centering
	\begin{subfigure}{0.40\textwidth}
		\centering
		\includegraphics[width=\textwidth]{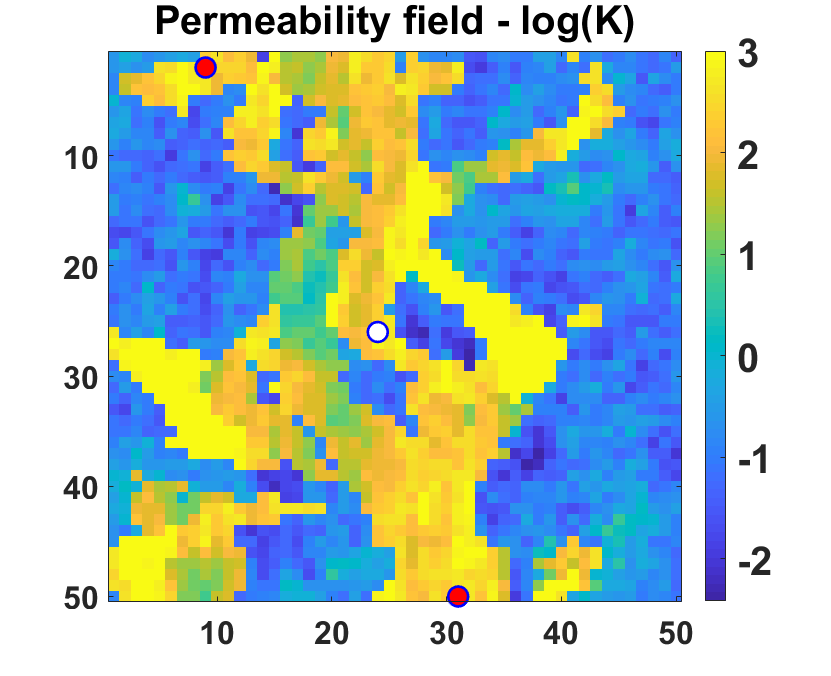}
		\caption{}
		\label{fig:5_38_1}
	\end{subfigure}%
	~\\
	\centering
	\begin{subfigure}{0.45\textwidth}
		\centering
		\includegraphics[width=\textwidth]{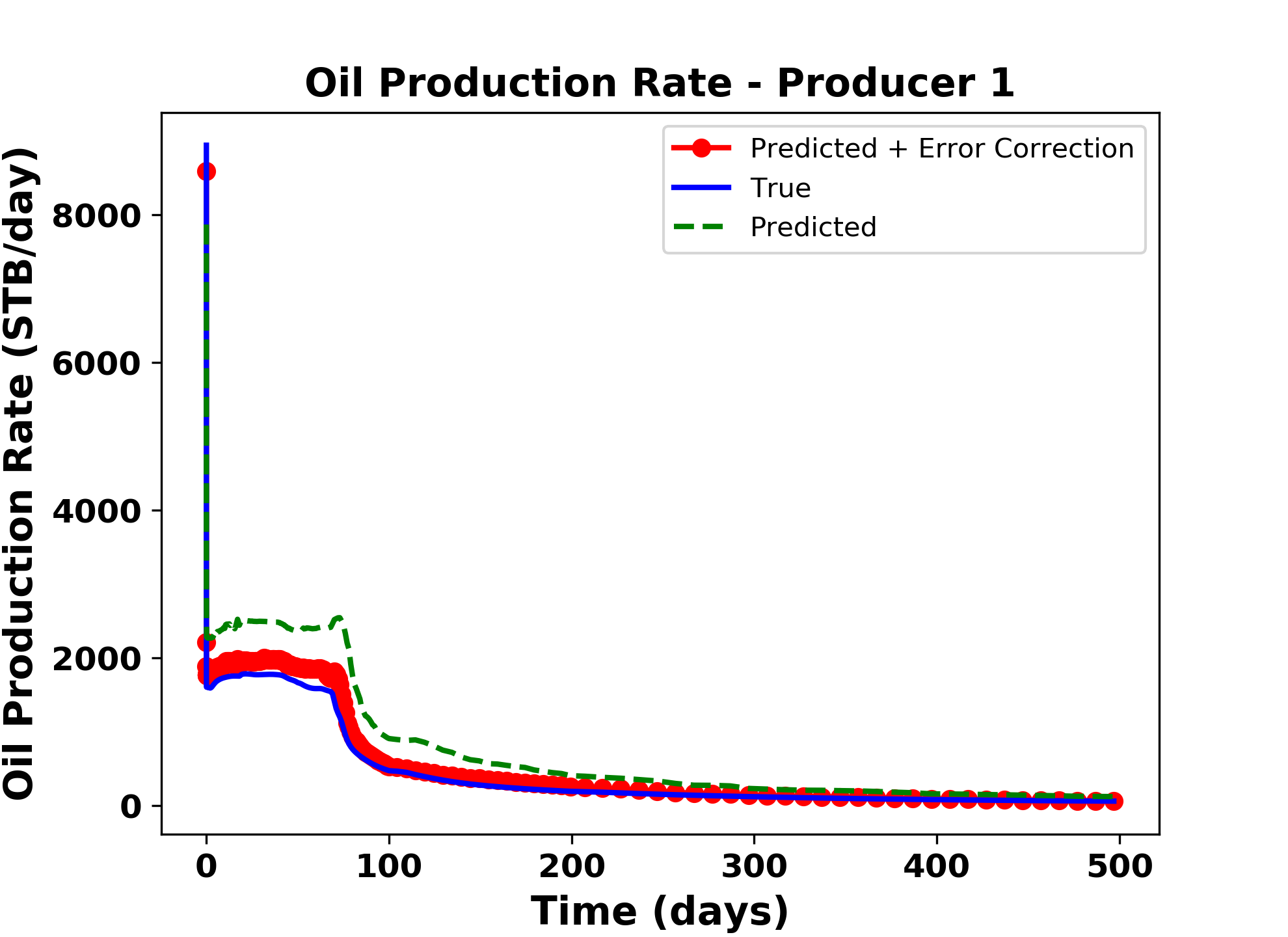}
		\caption{}
		\label{fig:5_38_2}
	\end{subfigure}
	~
	\centering
	\begin{subfigure}{0.45\textwidth}
		\centering
		\includegraphics[width=\textwidth]{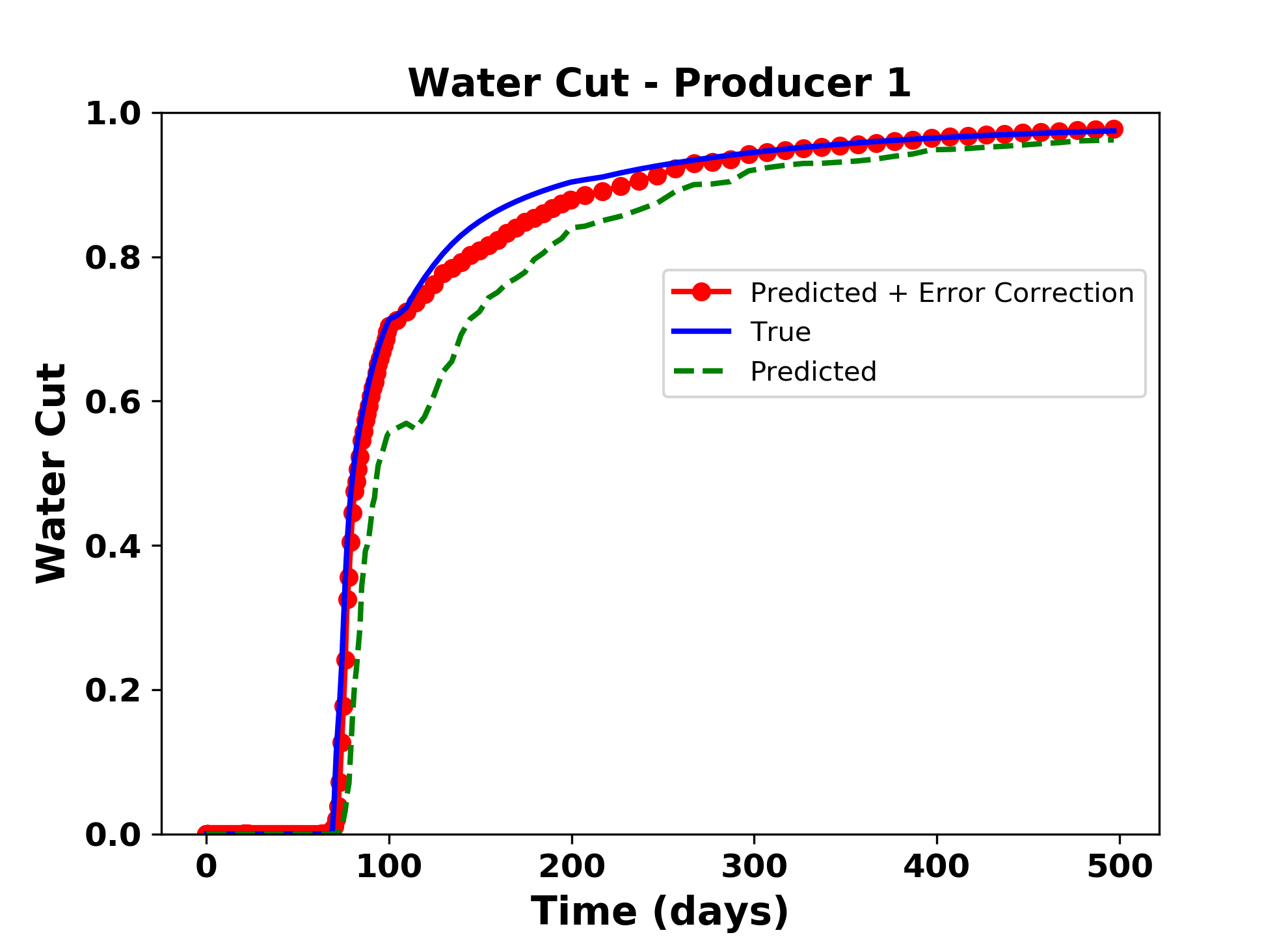}
		\caption{}
		\label{fig:5_38_4}
	\end{subfigure}%
	~\\
	\centering
	\begin{subfigure}{0.45\textwidth}
		\centering
		\includegraphics[width=\textwidth]{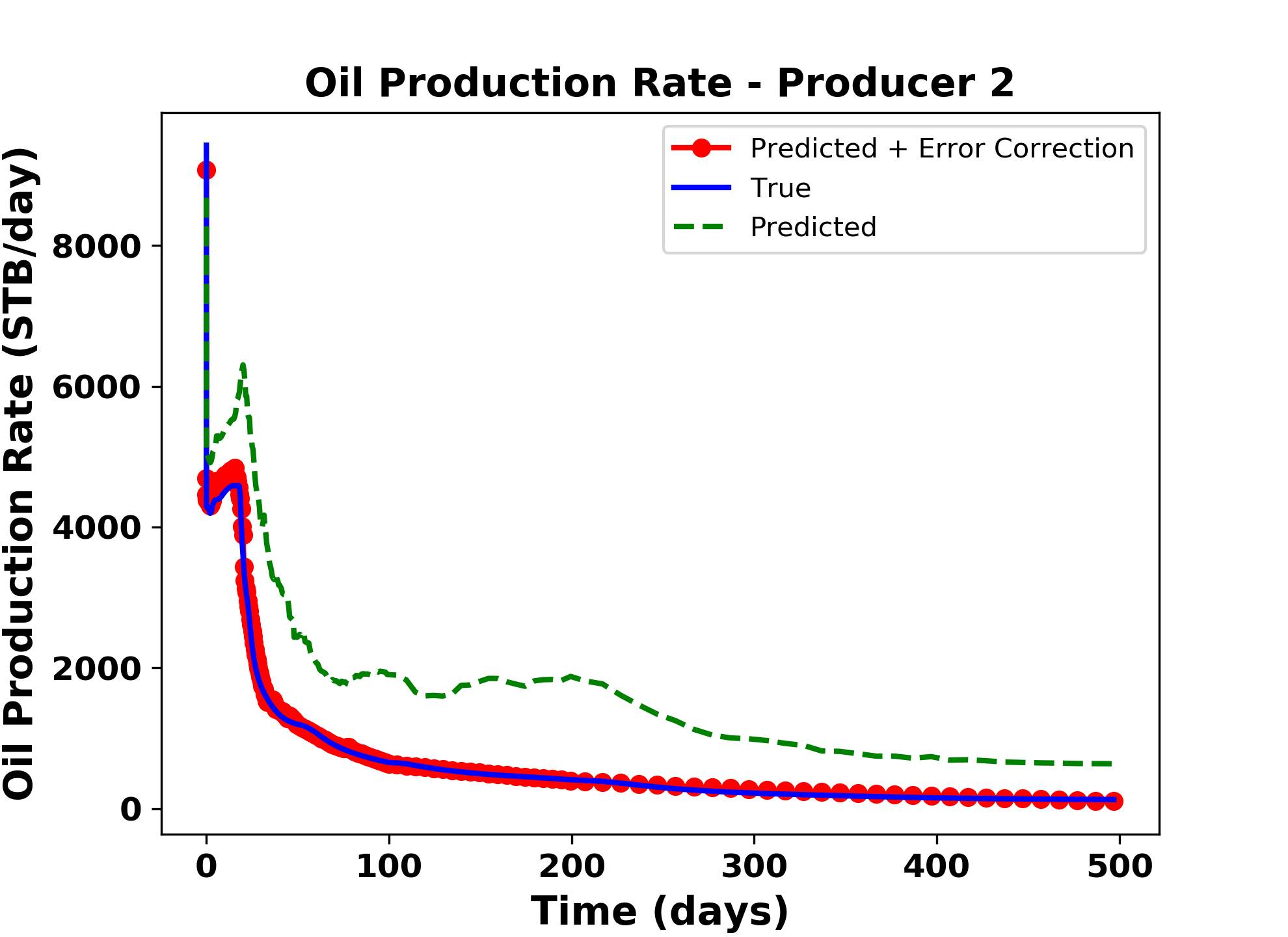}
		\caption{}
		\label{fig:5_38_3}
	\end{subfigure}
	~
	\centering
	\begin{subfigure}{0.45\textwidth}
		\centering
		\includegraphics[width=\textwidth]{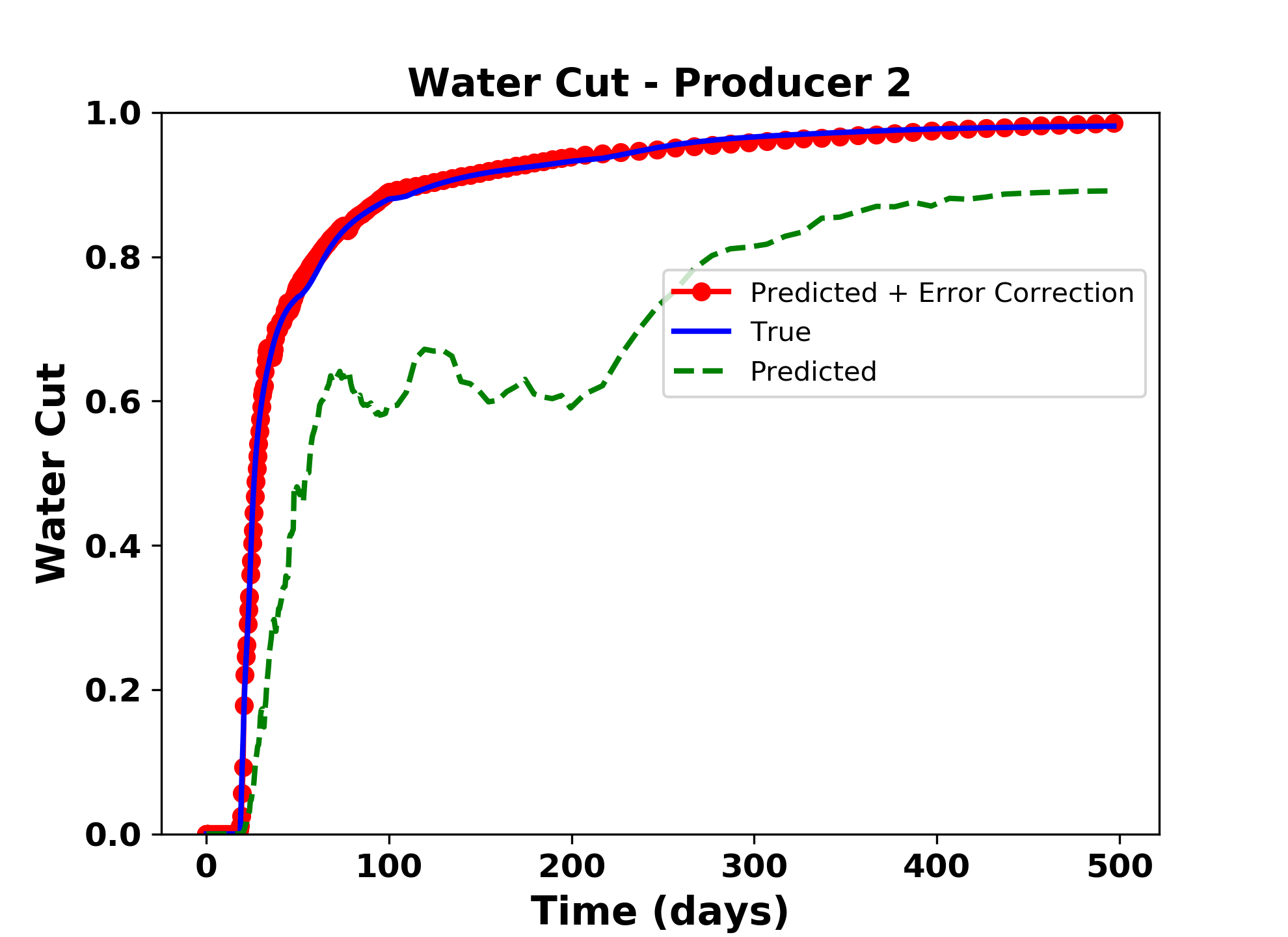}
		\caption{}
		\label{fig:5_38_5}
	\end{subfigure}%
	\caption{(a) Test case 1 with producer 1 at (9,2) and producer 2 at (31,50) (b) Comparison of oil production rate for producer 1, (c) Comparison of water cut for producer 1, (d) Comparison of oil production rate for producer 2, (e) Comparison of water cut for producer 2, predicted using global PMOR method alone (dotted green line) and after implementation of error correction model (red circled line) with the true solution (blue line)}
	\label{fig:5_38}
\end{figure}

\begin{figure}[htb!]
	\centering
	\begin{subfigure}{0.40\textwidth}
		\centering
		\includegraphics[width=\textwidth]{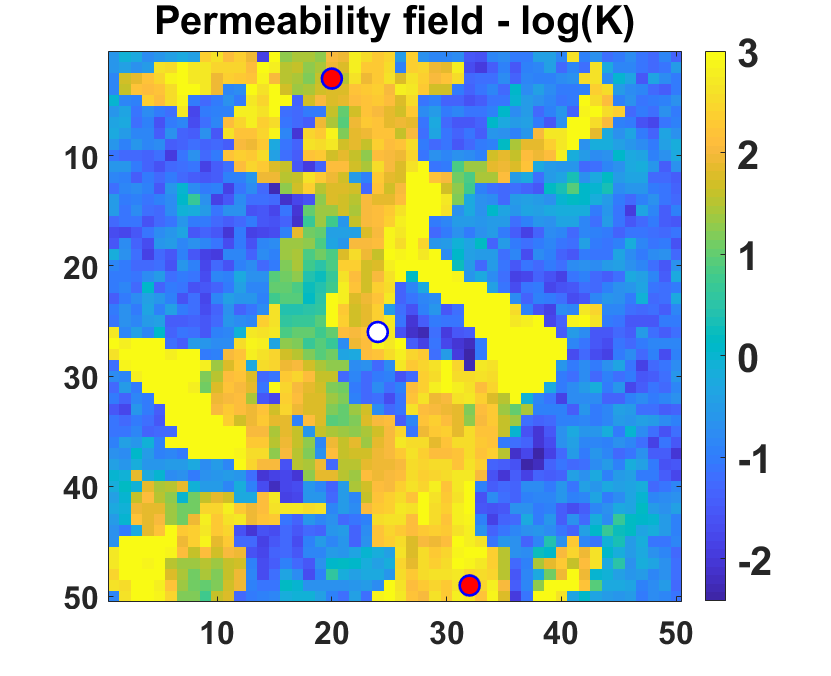}
		\caption{}
		\label{fig:5_39_1}
	\end{subfigure}%
	~\\
	\centering
	\begin{subfigure}{0.45\textwidth}
		\centering
		\includegraphics[width=\textwidth]{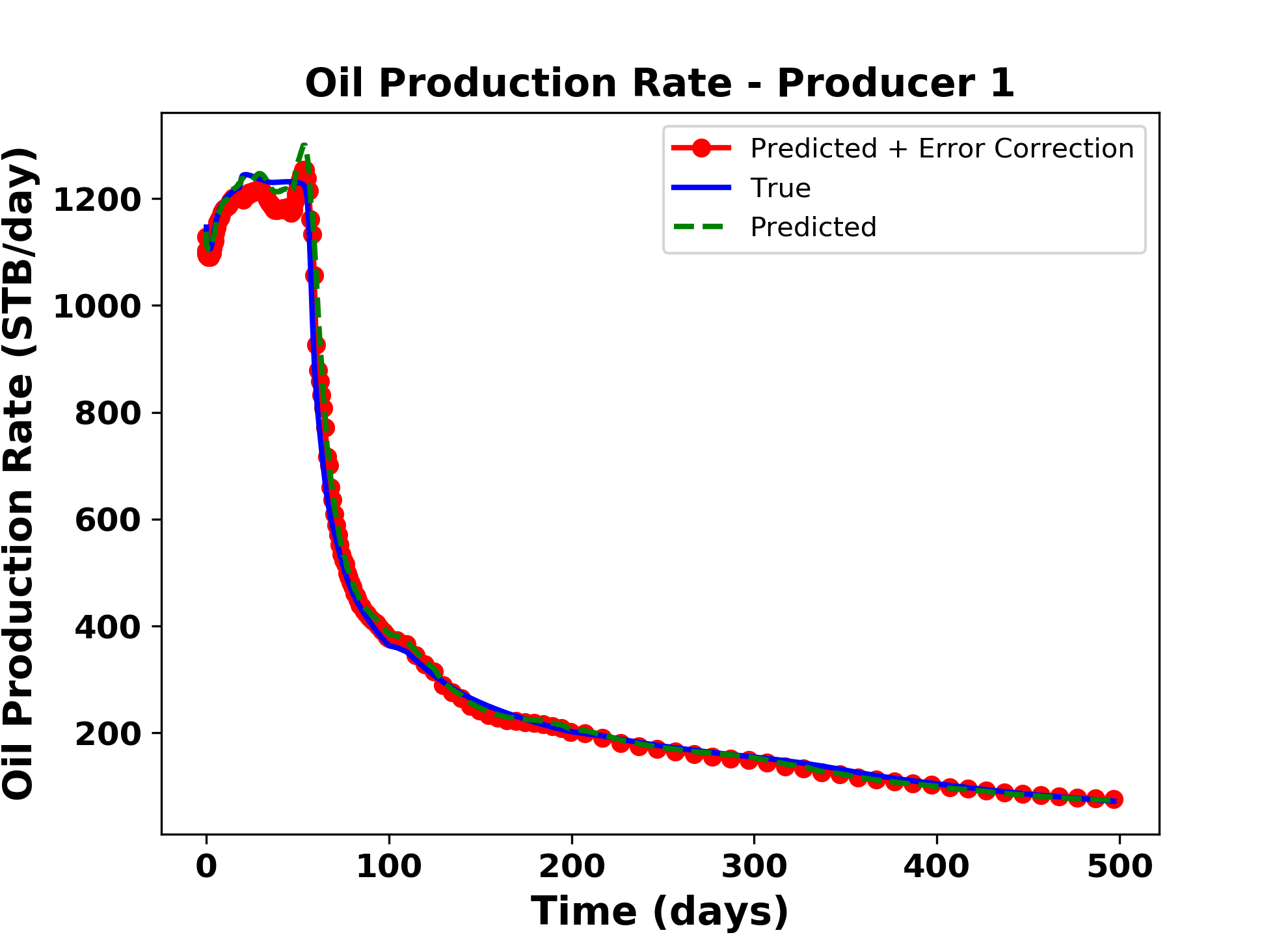}
		\caption{}
		\label{fig:5_39_2}
	\end{subfigure}
	~
	\centering
	\begin{subfigure}{0.45\textwidth}
		\centering
		\includegraphics[width=\textwidth]{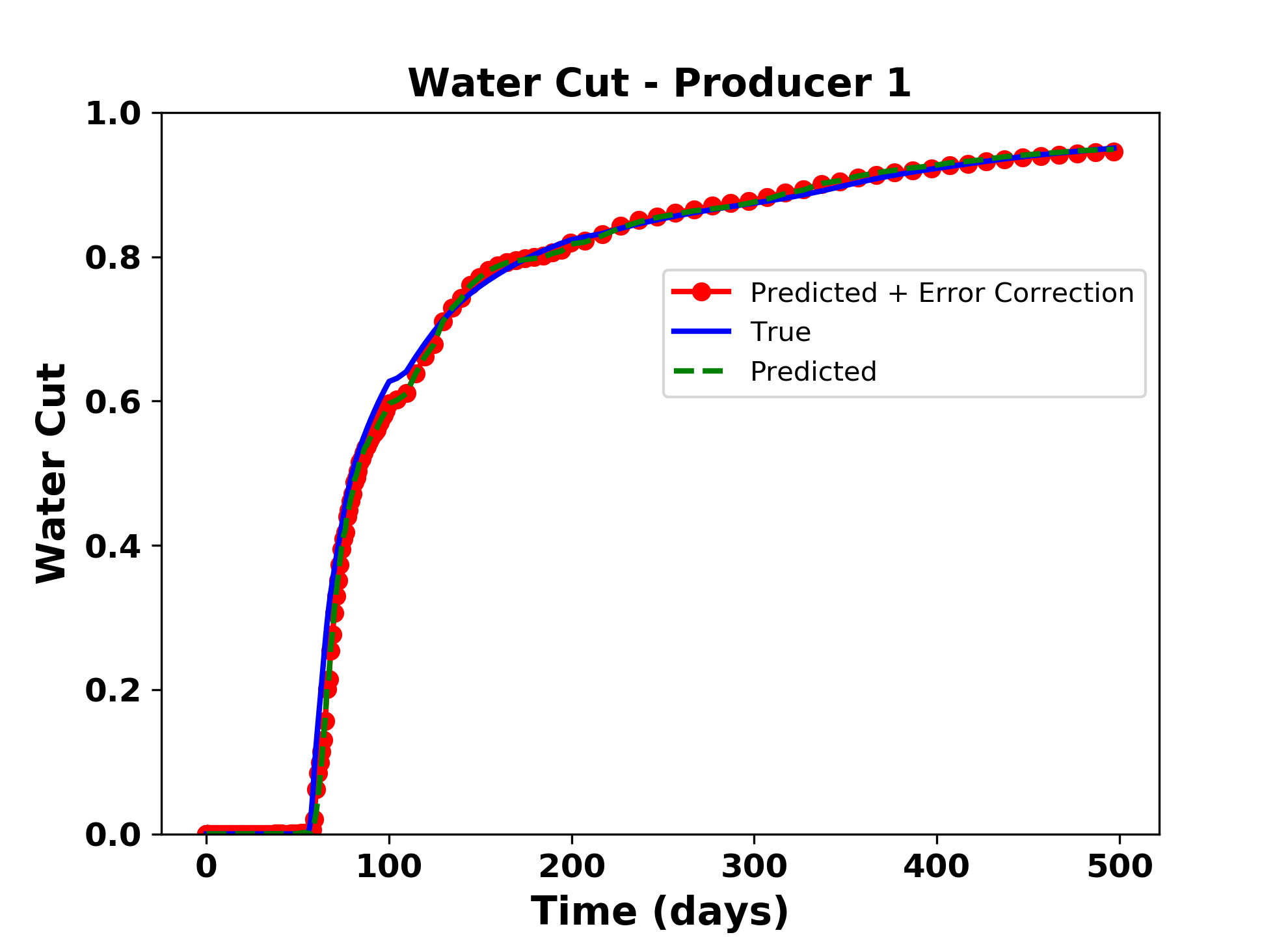}
		\caption{}
		\label{fig:5_39_4}
	\end{subfigure}%
	~\\
	\centering
	\begin{subfigure}{0.45\textwidth}
		\centering
		\includegraphics[width=\textwidth]{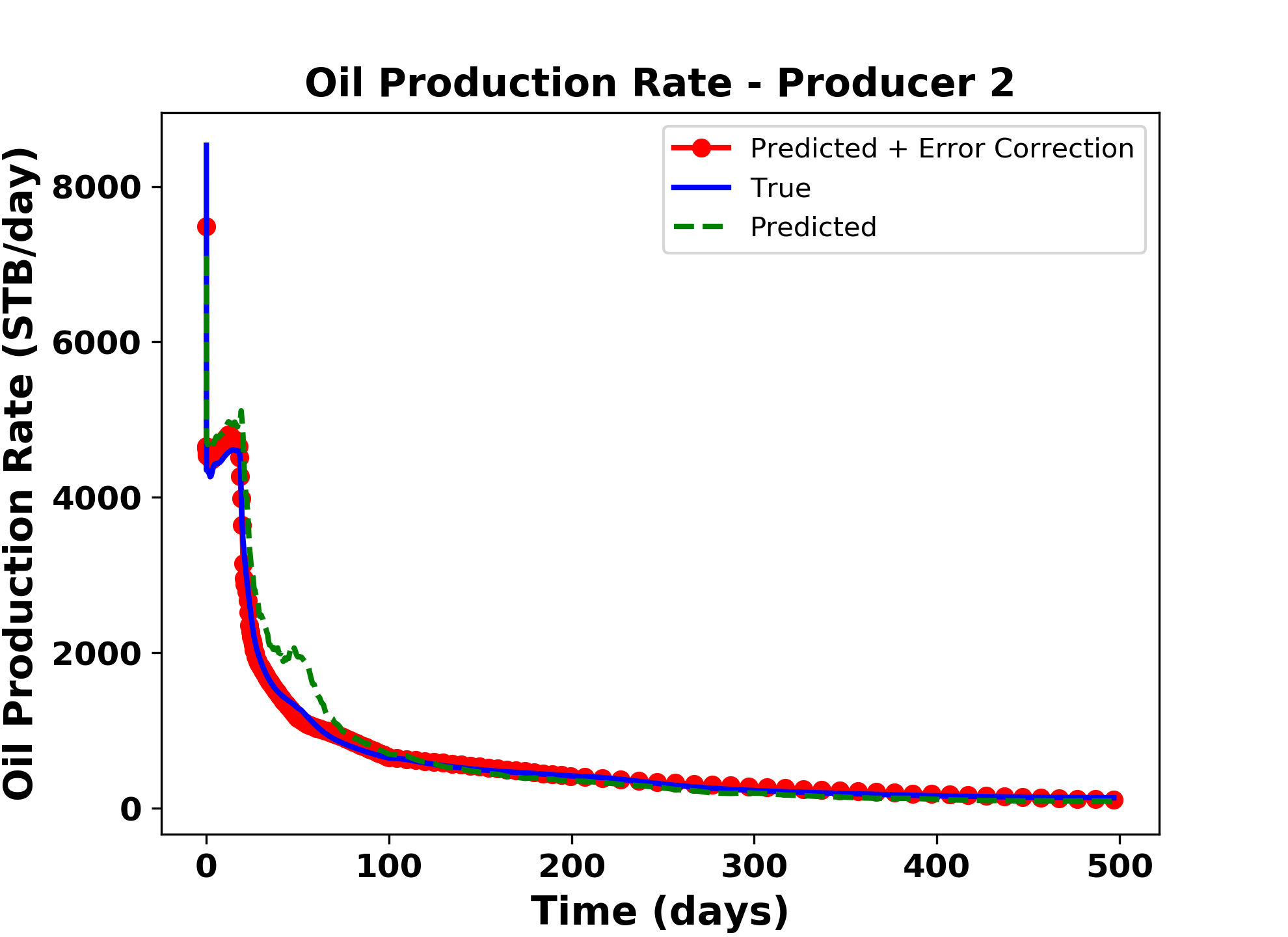}
		\caption{}
		\label{fig:5_39_3}
	\end{subfigure}
	~
	\centering
	\begin{subfigure}{0.45\textwidth}
		\centering
		\includegraphics[width=\textwidth]{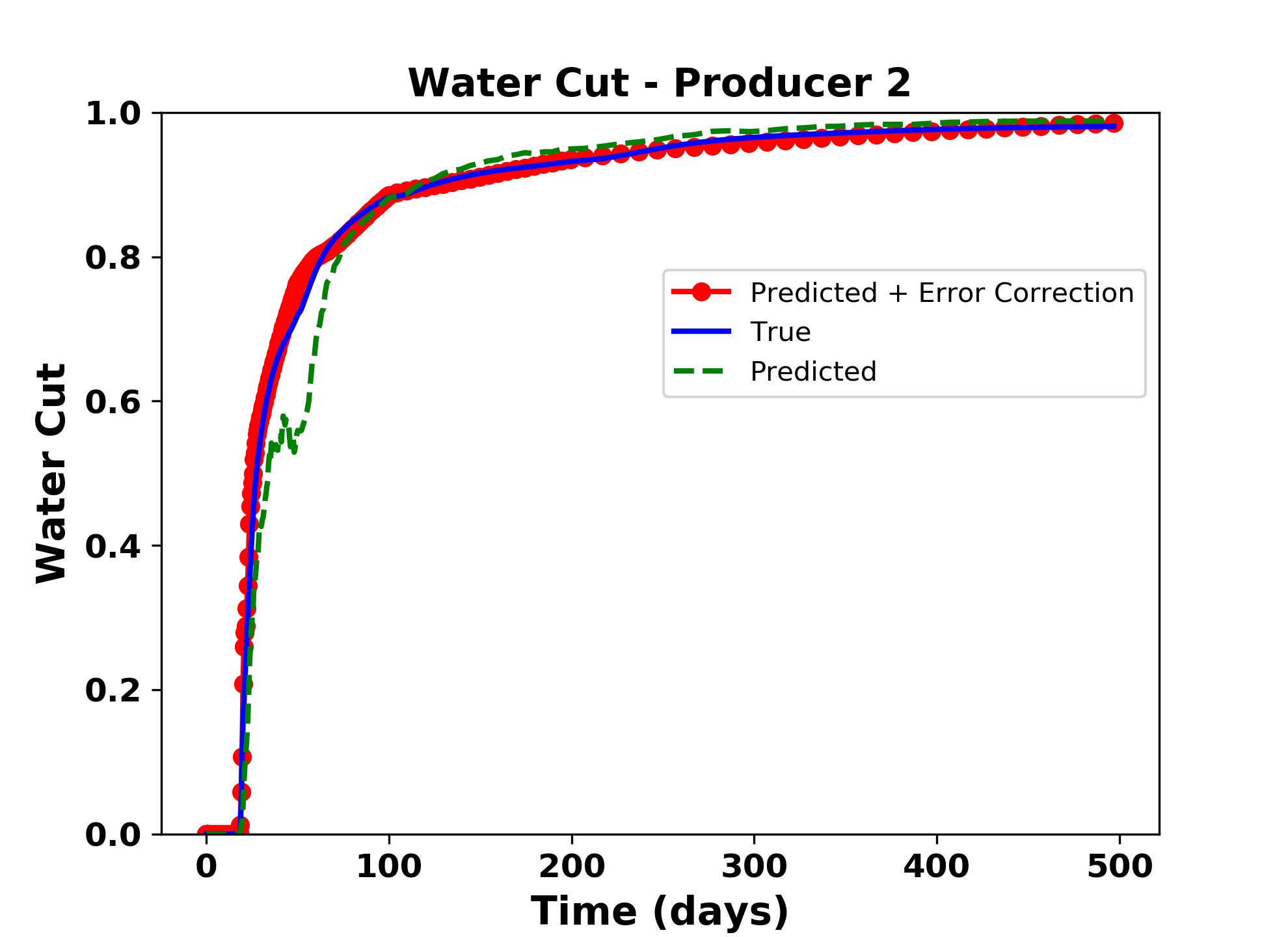}
		\caption{}
		\label{fig:5_39_5}
	\end{subfigure}%
	\caption{(a) Test case 2 with producer 1 at (20,3) and producer 2 at (32,49) (b) Comparison of oil production rate for producer 1, (c) Comparison of water cut for producer 1, (d) Comparison of oil production rate for producer 2, (e) Comparison of water cut for producer 2, predicted using global PMOR method alone (dotted green line) and after implementation of error correction model (red circled line) with the true solution (blue line)}
	\label{fig:5_39}
\end{figure}

\begin{figure}[htb!]
	\centering
	\begin{subfigure}{0.45\textwidth}
		\centering
		\includegraphics[width=\textwidth]{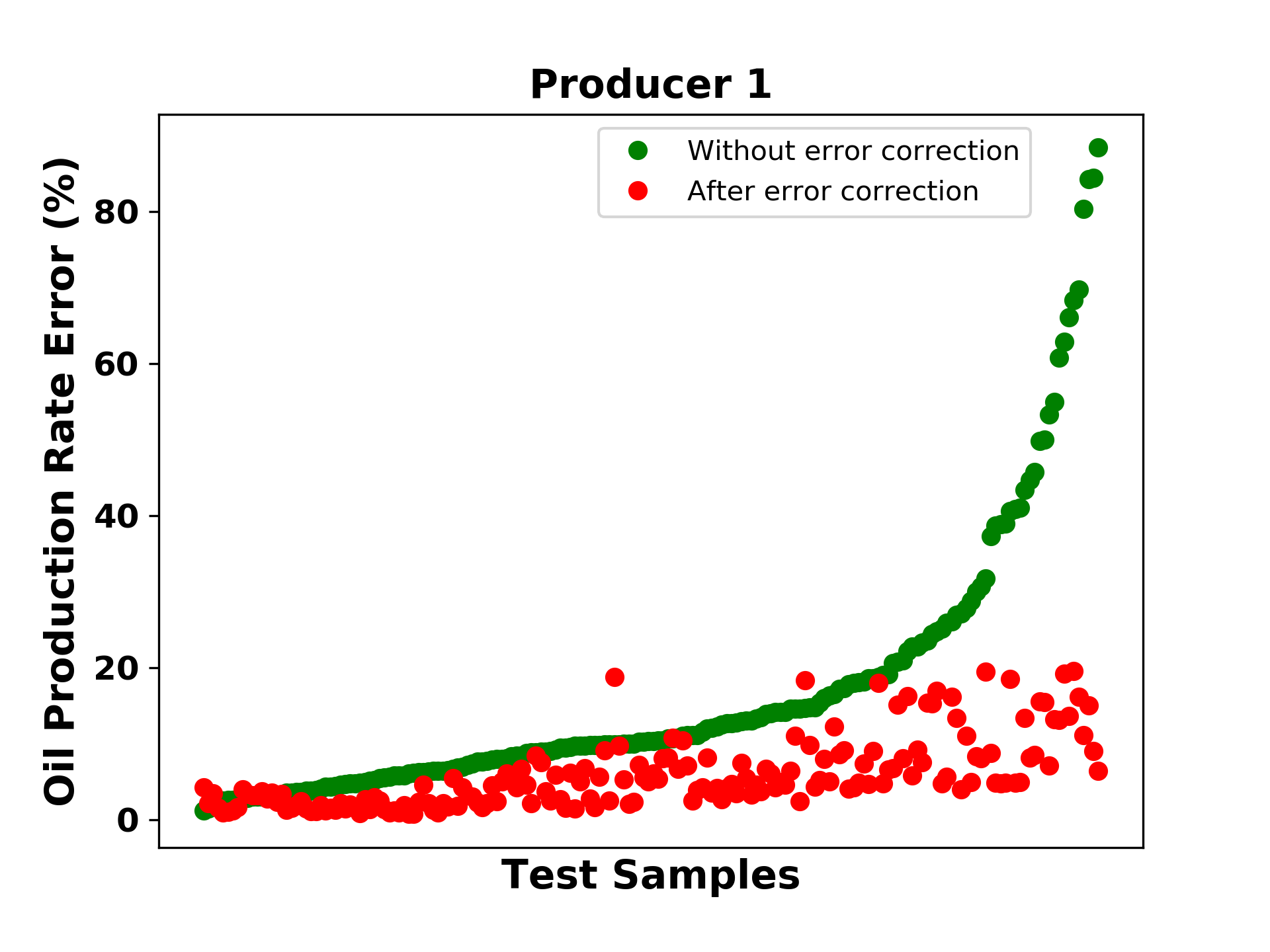}
		\caption{}
		\label{fig:5_40_1}
	\end{subfigure}%
	~
	\centering
	\begin{subfigure}{0.45\textwidth}
		\centering
		\includegraphics[width=\textwidth]{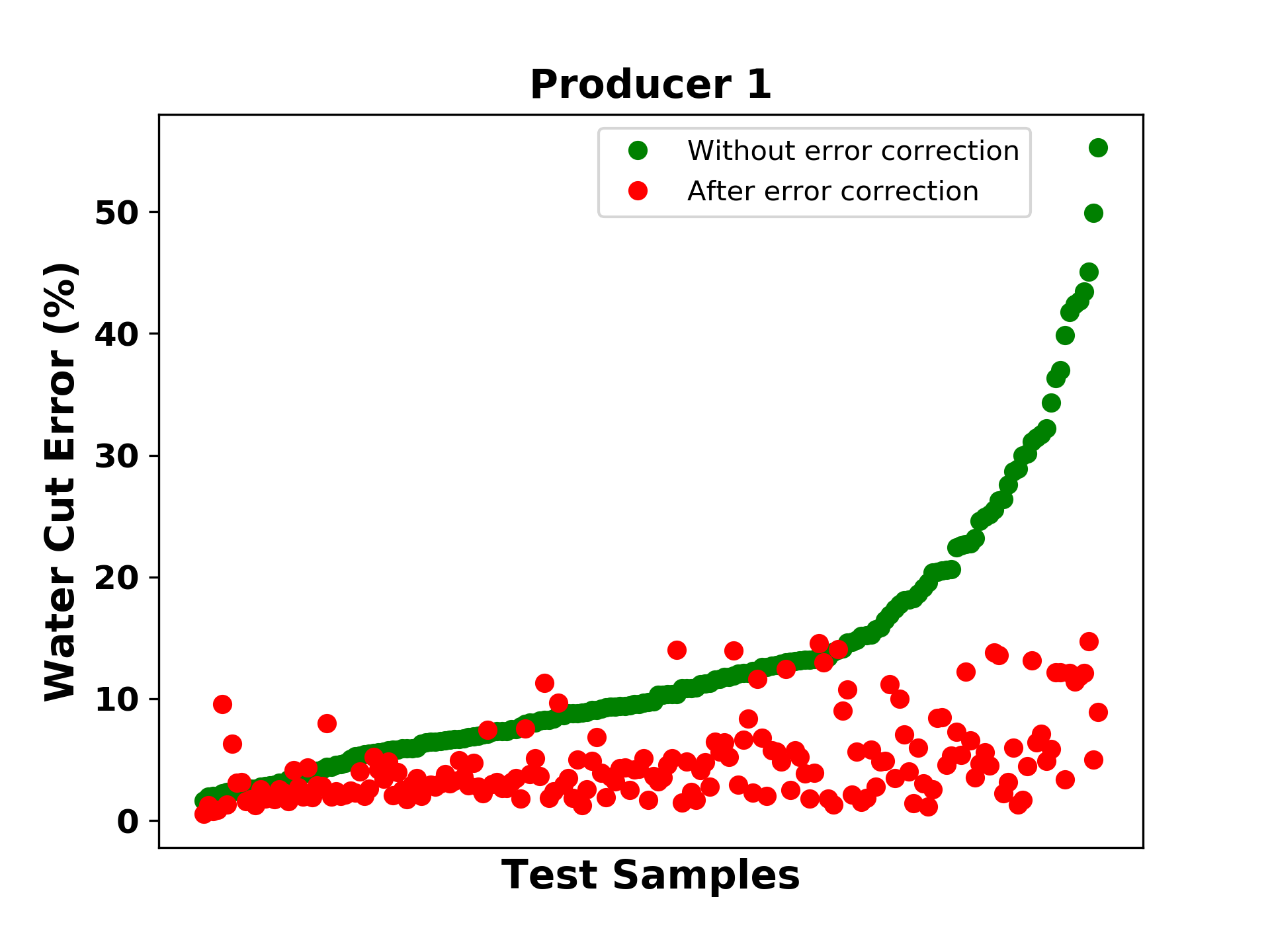}
		\caption{}
		\label{fig:5_40_3}
	\end{subfigure}%
	~\\
	\centering
	\begin{subfigure}{0.45\textwidth}
		\centering
		\includegraphics[width=\textwidth]{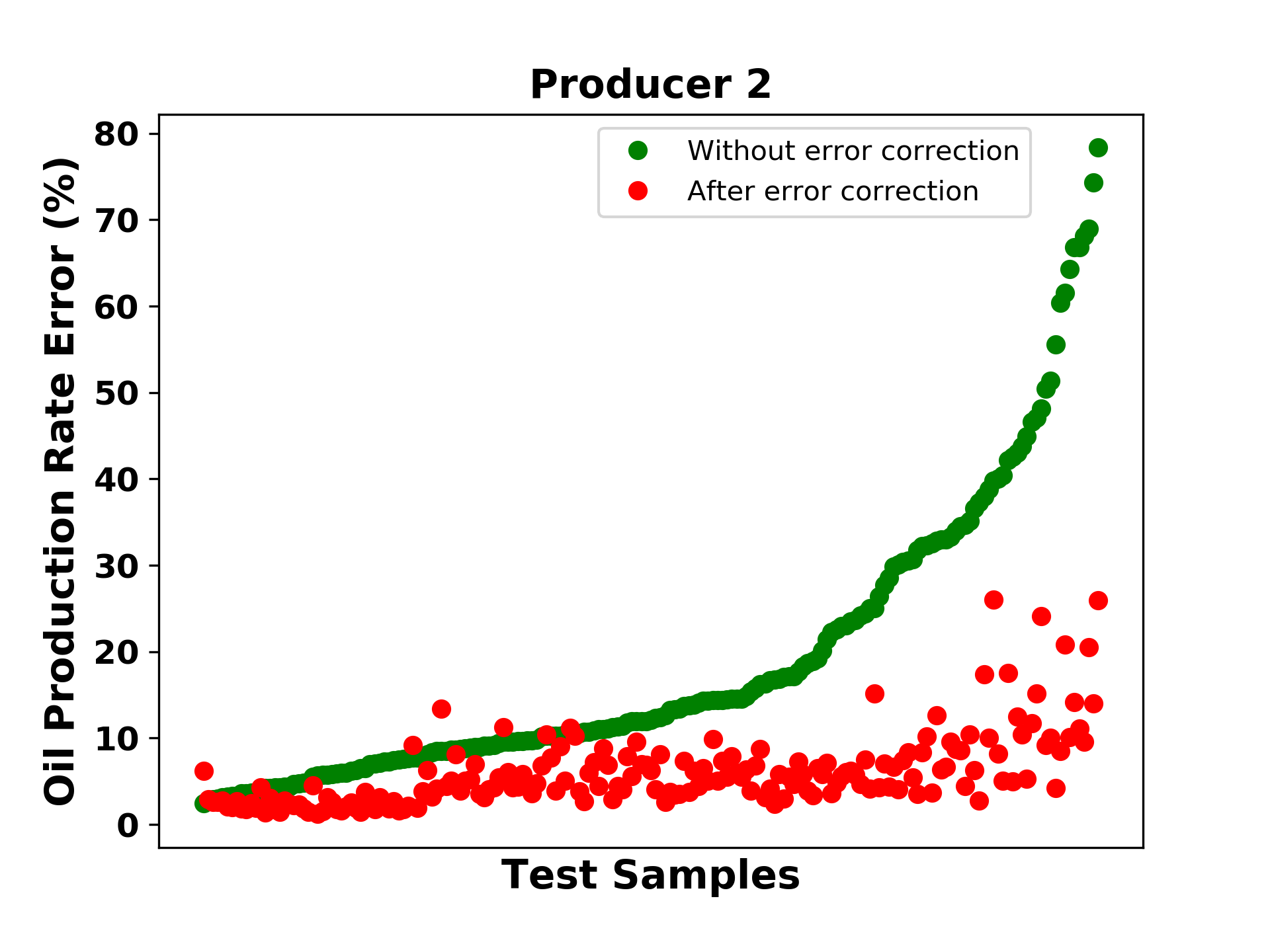}
		\caption{}
		\label{fig:5_40_2}
	\end{subfigure}%
	~
	\centering
	\begin{subfigure}{0.45\textwidth}
		\centering
		\includegraphics[width=\textwidth]{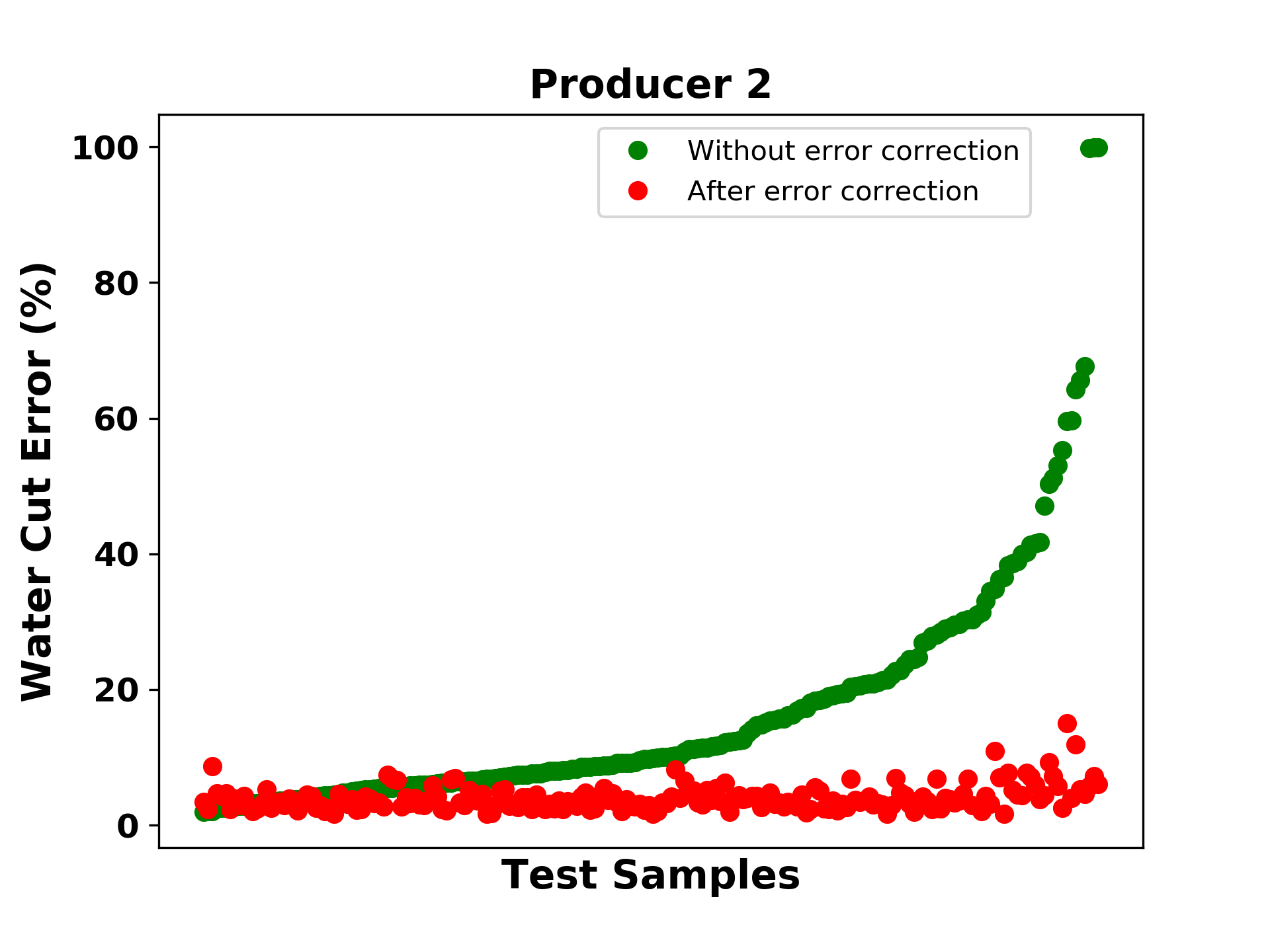}
		\caption{}
		\label{fig:5_40_4}
	\end{subfigure}%
	\caption{Error in prediction of (a) Oil Production Rate and (b) Water Cut for producer 1, (c) Oil Production Rate and (d) Water Cut for producer 2, for all the test cases before and after the error correction of the solutions}
	\label{fig:5_40}
\end{figure}

Figures \ref{fig:5_38} and \ref{fig:5_39} shows two of the test cases with different producer 1 and 2 pairs than used in training. For test case 1, the results show a good accuracy after error correction which shows erratic behavior for producer 2 using just the predicted basis coefficients. Similarly, the test case 2 is an example where even without error correction the results are a good match to the true solutions, which improve further after error correction. This shows that even though the global basis may not be a good quality basis for many parameters, we can employ error correction model to account for the solution discrepancies. Figure \ref{fig:5_40} shows the test accuracies for all the 250 test samples of oil production rates and water cut for both the producers. As we see, the accuracy after error correction improves significantly for many test cases. There are very few test cases again, with higher error after error correction that we deem as a result of random sampling procedure. 

Table \ref{tab:time_2wells} shows the comparison of time between fine scale simulation and the proposed PMOR model with error correction. The speedup obtained for these cases was about 100$\times$. Note that these times are just for the test runs. It does not include the training time, which must be the area of focus for reduction in the future. 

\begin{table}[htb!]
	\begin{center}
		
		\begin{tabular}{lclclc|}
			\hline\noalign{\smallskip}
			\textbf{} & \textbf{Fine scale simulation} & \textbf{PMOR + Error correction}  \\
			\noalign{\smallskip}\hline\noalign{\smallskip}
			\textbf{Test Case 1}& 102 seconds & \enspace \enspace\enspace \enspace\enspace \enspace1 second\\
			\textbf{Test Case 2} & 90 seconds & \enspace \enspace\enspace \enspace\enspace \enspace0.8 seconds\\
			\noalign{\smallskip}\hline
		\end{tabular}
	\end{center}
	\caption{Time (seconds) comparison for the two test cases between fine scale simulation and reduced order model with error correction}
	\label{tab:time_2wells}       
	\vspace*{-1em}
\end{table}

From these results, we validate the applicability of proposed workflow as a starting point for the case of multiple wells. We believe, the explanation on its use with the optimization procedure described above, should be taken into a consideration for practical implementation. As we move to the cases with many wells, the feature space becomes high dimensional, that might lead to slower training procedures. A possible way to avoid this, would be the observation that, time of flight, well block permeability, Lorenz coefficients and the indices of well gridblocks, which are the most important features representing well locations, be used for POD basis coefficients prediction and then during error correction step, we may use all the geometric features along with physics based features. This can reduced the dimensionality of feature space and hence faster training procedure. 

\subsection{Advantages and Limitations}

The method described above requires many fine scale simulations to train the ML models but can easily be used with high performance computing facilities. With the advent of parallel computing resources and cloud architecture, the training procedure can be extremely fast. It is true that one may argue running just fine scale simulations, but it should also be considered that the proposed technique has been designed for well location optimization problems, where, each optimization iteration may be dependent on the solution of previous iteration, depending on the optimization strategy used, and hence cannot make full use of parallel resources. Keeping the expensive training procedure in mind one may also design better sampling methods that reduce the number of sampling points significantly. We also note here that the methodology described is used for any parameter in the parametric domain. That is, the PMOR method is strictly global in the sense that the test cases may be significantly different from the training cases which is easily encountered in the new well locations as a new well location can change the dynamics of the problem significantly. This is not usually the case with PMOR methods developed for well control optimization problem, where, the training procedure is not designed such that it is drastically different from the test cases or they may fail or lead to unstable solutions. 

The advantages of the proposed method are manyfold. One, this PMOR strategy is completely non-intrusive in nature as it just requires data (pressure and saturation states in our cases) from any simulator and does not require evaluation of any non-linear functions like Jacobian and residuals. Second, the ML procedure described here is extremely fast in predicting the solutions at a given time in simulation run. For the case of heterogeneous reservoir above, computational speed up of around $50 \times$ (case 1 with large basis dimensions) to about $100\times$ (case 2 with lower basis dimensions) was observed for the test examples when run on a local 8-core machine. Each forward ML prediction (coefficients and error correction) is very fast and the proposed method also allows multiple timesteps computed in parallel. Since each timestep enters as a feature to the ML model, the solution at the current timestep is independent of the solution at previous timestep as in reservoir simulation. Thus, we do not rely on the time stepping for solution convergence and running the timesteps in series with the proposed method. This also helps to avoid convergence issues that may occur due to a few bad predictions at some timesteps. However, one limitation of this method is that it does not necessarily obey mass conservation and should be a research focus in the future to handle it. 

\section{Summary} \label{sec:1.10}
In this paper, we introduced a non-intrusive global parametric model order reduction method for changing well locations based on machine learning techniques. This method requires just the pressure and saturation states at different time steps and a global basis is constructed from the representative well locations in the reservoir. The machine learning model is trained to learn the coefficients of the POD basis for the well locations as defined by a set of geometric and flow diagnostics based features. We also introduce time as a feature to account for temporal behavior of the coefficients. The trained ML model is then used to predict the POD coefficients of the states for the new well locations. The ML model used here is Random Forest, which is much faster to train compared to other algorithms like ANN especially for large dimensional outputs, and has the ability to define complex input-output relations. ML model predicts the basis coefficients extremely fast with speed ups of about $50\times$(for high dimensional basis) - $100\times$ (for low dimensional basis). However, the ML model could capture the overall solution trend but the states predicted at the well block showed a significant bias as can be attributed to the controllability theory. But due to a systematic discrepancy in the solution, in that it captures the solution behavior, we considered fixing this discrepancy through error correction models. These error correction model is a ML based model that captures the state disparity given the reduced model solutions. ANN was used in this case which is faster to train due to this being a single output model and proved to reduced the error of prediction drastically for most test cases. The proposed workflow has a limitation in that it requires many training samples to define the global parameter space. This should be looked in the future research using efficient sampling techniques, that can reduce the number of samples required significantly and using local parameter regions to develop multiple local models in a similar way as proposed. We also note that, as mentioned in the case study for multiple wells, the proposed method must be tied with the optimization strategy used that would help design a better training strategy, as it is impractical to expect the model to be accurate for all possible well configurations in the reservoir.

\bibliographystyle{unsrt}  
\bibliography{ThesisRefs}  

\begin{thebibliography}{10}

\bibitem{Jansen2009}
Jan-Dirk Jansen, Roald Brouwer, and Sippe~G. Douma.
\newblock {Closed Loop Reservoir Management}.
\newblock In {\em SPE Reservoir Simulation Symposium}, 2009.

\bibitem{Shirangi2015}
Mehrdad~G. Shirangi and Louis~J. Durlofsky.
\newblock {Closed-Loop Field Development Under Uncertainty by Use of
  Optimization With Sample Validation}.
\newblock {\em SPE Journal}, 20(05):0908--0922, 2015.

\bibitem{Antoulas2005}
A.C. Antoulas.
\newblock {\em {Approximation of Large-scale Dynamical Systems}}.
\newblock SIAM Advances in Design and Control, 2005.

\bibitem{Hinze2005}
M~Hinze and S~Volkwein.
\newblock {\em {Dimension Reduction of Large-Scale Systems}}, volume~45.
\newblock 2005.

\bibitem{King1998}
M~J King, D~G Macdonald, S~P Todd, and H~Leung.
\newblock {SPE 50643-Application of Novel Upscaling Approaches to the Magnus
  and Andrew Reservoirs}.
\newblock In {\em European Petroleum Conference}, 1998.

\bibitem{March2012}
Andrew March and Karen Willcox.
\newblock {Provably Convergent Multi fi delity Optimization Algorithm Not
  Requiring High-Fidelity Derivatives}.
\newblock {\em AIAA Journal}, 50(5), 2012.

\bibitem{Kennedy2001}
Marc~C Kennedy and Anthony~O Hagan.
\newblock {Bayesian calibration of computer models}.
\newblock {\em Journal of the Royal Statistical Society}, 63(Part 3):425--464,
  2001.

\bibitem{Knill1999}
Duane~L Knill, Anthony~A Giunta, Chuck~A Baker, Bernard Grossman, William~H
  Mason, Raphael~T Haftka, and Layne~T Watsonft.
\newblock {Response Surface Models Combining Linear and Euler Aerodynamics for
  Supersonic Transport Design}.
\newblock {\em Journal of Aircraft}, 36(1), 1999.

\bibitem{Doren2006}
Jorn F M~Van Doren and Renato Markovinovic.
\newblock {Reduced-order optimal control of water flooding using proper
  orthogonal decomposition}.
\newblock {\em Computational Geosciences}, 10:137--158, 2006.

\bibitem{Cardoso2008}
M~A Cardoso, L~J Durlofsky, and P~Sarma.
\newblock {Development and application of reduced-order modeling procedures for
  subsurface flow simulation}.
\newblock {\em International Journal for Numerical Methods in Engineering},
  pages 1322--1350, 2008.

\bibitem{Cardoso2009}
Marco~Antonio Cardoso and Louis~J. Durlofsky.
\newblock {Use of Reduced-Order Modeling Procedures for Production
  Optimization}.
\newblock In {\em SPE Reservoir Simulation Symposium}, number February, pages
  2--4, 2009.

\bibitem{He2011}
J.~He, J.~S{\ae}trom, and L.~J. Durlofsky.
\newblock {Enhanced linearized reduced-order models for subsurface flow
  simulation}.
\newblock {\em Journal of Computational Physics}, 230(23):8313--8341, 2011.

\bibitem{Trehan2016a}
Sumeet Trehan and Louis~J. Durlofsky.
\newblock {Trajectory piecewise quadratic reduced-order model for subsurface
  flow, with application to PDE-constrained optimization}.
\newblock {\em Journal of Computational Physics}, 326:446--473, 2016.

\bibitem{Chaturantabut2010}
S~Chaturantabut and D~Sorensen.
\newblock {Nonlinear Model Reduction via Discrete Empirical Interpolation}.
\newblock {\em SIAM Journal on Scientific Computing}, 32(5):2737--2764, jan
  2010.

\bibitem{Gildin2013}
Eduardo Gildin, Mohammadreza Ghasemi, Anastasiya Protasov, and Yalchin
  Efendiev.
\newblock {SPE 163618 Nonlinear Complexity Reduction for Fast Simulation of
  Flow in Heterogenous Porous Media}.
\newblock In {\em Reservoir Simulation Symposium}, 2013.

\bibitem{Ghasemi2015}
Mohammadreza Ghasemi, Yanfang Yang, Eduardo Gildin, Yalchin Efendiev, and
  Victor Calo.
\newblock {Fast Multiscale Reservoir Simulations using POD-DEIM Model
  Reduction}.
\newblock In {\em SPE Reservoir Simulation Symposium}, 2015.

\bibitem{Sorek2017}
N~Sorek, H~Zalavadia, and E~Gildin.
\newblock {SPE-182652-MS Model Order Reduction and Control Polynomial
  Approximation for Well-Control Production Optimization}.
\newblock In {\em SPE Reservoir Simulation Conference}, 2017.

\bibitem{Tan2017}
Xiaosi Tan, Eduardo Gildin, Sumeet Trehan, Yahan Yang, and Nazish Hoda.
\newblock {Trajectory-Based DEIM TDEIM Model Reduction Applied to Reservoir
  Simulation}.
\newblock In {\em SPE Reservoir Simulation Conference}, 2017.

\bibitem{Ghasemi2014}
Mohammadreza Ghasemi, Ashraf Ibrahim, and Eduardo Gildin.
\newblock {Reduced Order Modeling In Reservoir Simulation Using the Bilinear
  Approximation Techniques}.
\newblock In {\em SPE Latin America and Caribbean Petroleum Engineering
  Conference}, 2014.

\bibitem{Ghasemi2016}
Mohammadreza Ghasemi and Eduardo Gildin.
\newblock {Model Order Reduction in Porous Media Flow Simulation using
  Quadratic Bilinear Formulation}.
\newblock {\em Computational Geosciences}, 20(3):723--735, 2016.

\bibitem{Jiang2019}
Rui Jiang and Louis~J. Durlofsky.
\newblock {Implementation and detailed assessment of a GNAT reduced-order model
  for subsurface flow simulation}.
\newblock {\em Journal of Computational Physics}, 379:192--213, 2019.

\bibitem{Zalavadia2019}
Hardikkumar Zalavadia, Sathish Sankaran, Mustafa Kara, Wenyue Sun, and Eduardo
  Gildin.
\newblock {A Hybrid Modeling Approach to Production Control Optimization Using
  Dynamic Mode Decomposition}, 2019.

\bibitem{Centilmen1999}
A~Centilmen, T~Ertekin, and A~S Grader.
\newblock {SPE 56433 Applications of Neural Networks in Multiwell Field
  Development}.
\newblock In {\em SPE Annual Technical Conference {\&} Exhibition}, 1999.

\bibitem{Taware2012}
Satyajit~Vijay Taware, Han-young Park, Akhil Datta-Gupta, Shyamal Bhattacharya,
  A.K. Tomar, Munil Kumar, and H.S. Rao.
\newblock {Well Placement Optimization in a Mature Carbonate Waterflood using
  Streamline-based Quality Maps}.
\newblock In {\em SPE Oil and Gas India Conference and Exhibition, 28-30 March,
  Mumbai, India}, 2012.

\bibitem{Casenave2015}
Fabien Casenave, Alexandre Ern, and Tony Leli{\`{e}}vre.
\newblock {A nonintrusive reduced basis method applied to aeroacoustic
  simulations}.
\newblock {\em Advances in Computational Mathematics}, 41(5):961--986, 2015.

\bibitem{Barthelmann2000}
Volker Barthelmann, Erich Novak, and Klaus Ritter.
\newblock {High dimensional polynomial interpolation on sparse grids}.
\newblock {\em Advances in Computational Mathematics}, 12(4):273--288, 2000.

\bibitem{Amsallem2011}
David Amsallem and Charbel Farhat.
\newblock {An Online Method for Interpolating Linear Parametric Reduced-Order
  Models}.
\newblock {\em SIAM Journal on Scientific Computing}, 33(5):2169--2198, 2011.

\bibitem{Hesthaven2018}
J.~S. Hesthaven and S.~Ubbiali.
\newblock {Non-intrusive reduced order modeling of nonlinear problems using
  neural networks}.
\newblock {\em Journal of Computational Physics}, 363:55--78, 2018.

\bibitem{Swischuk2018}
Renee Swischuk, Laura Mainini, Benjamin Peherstorfer, and Karen Willcox.
\newblock {Projection-based model reduction: Formulations for physics-based
  machine learning}.
\newblock {\em Computers and Fluids}, 0:1--14, 2018.

\bibitem{Zalavadia2018}
H~Zalavadia and E~Gildin.
\newblock {Th A2 08 Parametric Model Order Reduction For Adaptive Basis
  Selection Using Machine Learning Techniques During Well Location Opt}.
\newblock {\em 15th European Conference on the Mathematics of Oil Recovery
  (ECMOR XV)}, 2018.

\bibitem{Ertekin2001}
T~Ertekin, J~H Abou-Kassem, and G~R King.
\newblock {\em {Basic Applied Reservoir Simulation}}.
\newblock Society of Petroleum Engineers, Richardson, Tex., 2001.

\bibitem{Aziz1979}
AZIZ and K.
\newblock {Petroleum Reservoir Simulation}.
\newblock {\em Applied Science Publishers}, 476, 1979.

\bibitem{Chen2006}
Z~Chen, G~Huan, and Y~Ma.
\newblock {\em {Computational Methods for Multiphase Flows in Porous Media}}.
\newblock Society for Industrial and Applied Mathematics, jan 2006.

\bibitem{Lie2018}
Knut-andreas Lie.
\newblock {An Introduction to Reservoir Simulation Using MATLAB / GNU Octave
  User Guide for the Matlab Reservoir Simulation}.
\newblock 2018.

\bibitem{Carlberg2017}
Kevin Carlberg, Matthew Barone, and Harbir Antil.
\newblock {Galerkin v. least-squares Petrov–Galerkin projection in nonlinear
  model reduction}.
\newblock {\em Journal of Computational Physics}, 330:693--734, 2017.

\bibitem{He2015}
J~He and L~J Durlofsky.
\newblock {Constraint reduction procedures for reduced-order subsurface flow
  models based on POD–TPWL}.
\newblock {\em International Journal for Numerical Methods in Engineering},
  103(1):1--30, jul 2015.

\bibitem{Bao2017}
A.~Bao and E.~Gildin.
\newblock {Data-Driven Model Reduction Based on Sparsity-Promoting Methods for
  Multiphase Flow in Porous Media}.
\newblock In {\em SPE Latin America and Caribbean Petroleum Engineering
  Conference}, 2017.

\bibitem{Amsallem2009}
David Amsallem, Julien Cortial, and Charbel Farhat.
\newblock {On-Demand CFD-Based Aeroelastic Predictions Using a Database of
  Reduced-Order Bases and Models}.
\newblock {\em 47th AIAA Aerospace Sciences Meeting Including The New Horizons
  Forum and Aerospace Exposition}, (January):Orlando, Florida, 2009.

\bibitem{Choi2015}
Y~Choi, D~Amsallem, and C~Farhat.
\newblock {Gradient-based Constrained Optimization Using a Database of Linear
  Reduced-Order Models}.
\newblock {\em arXiv preprint arXiv:1506.07849}, pages 1--28, 2015.

\bibitem{Breiman2001}
L~E~O Breiman.
\newblock {Random Forests}.
\newblock {\em Machine Learning}, 45:5--32, 2001.

\bibitem{Datta-Gupta2007}
Akhil Datta-Gupta and Michael~J King.
\newblock {\em {Streamline simulation : theory and practice}}.
\newblock Society of Petroleum Engineers, Richardson, TX, 2007.

\bibitem{Datta-Gupta1995}
Akhil Datta-Gupta and Michael~J. King.
\newblock {A semianalytic approach to tracer flow modeling in heterogeneous
  permeable media}.
\newblock {\em Advances in Water Resources}, 18(1):9--24, 1995.

\bibitem{Shahvali2012}
M~Shahvali, B~Mallison, K~Wei, H~Gross, Chevron Energy, and Technology Company.
\newblock {An Alternative to Streamlines for Flow Diagnostics on Structured and
  Unstructured Grids}.
\newblock {\em SPE Journal}, (September):768--778, 2012.

\bibitem{Moyner2015}
Olav M{\o}yner, Stein Krogstad, and Knut-Andreas Lie.
\newblock {The Application of Flow Diagnostics for Reservoir Management}.
\newblock {\em SPE Journal}, 20(02):306--323, 2015.

\bibitem{Natvig2006}
Jostein~R Natvig and Knut-Andreas Lie.
\newblock {Fast Computation of Multiphase Flow in Porous Media by Implicit
  Discontinuous Galerkin Schemes With Optimal Ordering of Elements}.
\newblock {\em Journal of Computational Physics}, 227(24):10108--10124, 2008.

\bibitem{Shook2009}
G~Michael Shook and Kameron~M Mitchell.
\newblock {A Robust Measure of Heterogeneity for Ranking Earth Models: The F
  PHI Curve and Dynamic Lorenz Coefficient}, 2009.

\bibitem{Kohavi1997}
Ron Kohavi and George~H. John.
\newblock {Wrappers for Feature Subset Selection}.
\newblock {\em Artificial Intelligence}, 97(1-2):273--324, 1997.

\bibitem{Trehan2016_ML}
Sumeet Trehan, Louis Durlofsky, and Kevin Carlberg.
\newblock {Error Estimation for Surrogate Models of Dynamical Systems Using
  Machine Learning}.
\newblock {\em International Journal for Numerical Methods in Engineering},
  00:1--31, 2016.

\bibitem{Goodfellow2016}
Ian~J. Goodfellow, Yoshua Bengio, and Aaron Courville.
\newblock {\em {Deep Learning}}.
\newblock MIT Press, 2016.

\end{thebibliography}

\end{document}